\documentclass[aip,jcp,preprint,superscriptaddress,nofootinbib]{revtex4-1}
\usepackage{silence}
\usepackage[english]{babel}
\usepackage[T1]{fontenc}
\usepackage[utf8]{inputenc}
\usepackage{microtype}

\usepackage{amsmath}
\usepackage{amssymb}
\usepackage{textcomp}
\usepackage{textgreek}
\usepackage{physics}
\usepackage{upgreek}
\usepackage{mathtools}
\usepackage{bbm}
\usepackage{bm}
\usepackage{mathrsfs}
\usepackage{nicefrac}

\usepackage{graphicx}
\usepackage{xcolor}
\usepackage{float}
\usepackage{booktabs}
\usepackage{enumitem}
\usepackage[colorlinks]{hyperref}
\usepackage{braket}
\usepackage{array}
\usepackage{siunitx}
\usepackage{threeparttable}
\usepackage{lipsum}
\usepackage{caption}
\usepackage{subcaption}
\newcommand{\bottomstrut}[1]{\rule[#1]{0pt}{0pt}}

\newcommand\Tstrut{\rule{0pt}{2.4ex}}

\newcommand\TTstrut{\rule{0pt}{3.0ex}}

\newcommand{\mrm}[1]{\mathrm{#1}}
\newcommand{\mbf}[1]{\mathbf{#1}}
\newcommand{\nn}{\nonumber\\}
\newcommand{\trans}{\mspace{-3mu}\mathsf{T}}
\newcommand{\subA}{\mathrm{\scriptscriptstyle A}}
\newcommand{\subS}{\mathrm{\scriptscriptstyle S}}
\newcommand{\bigbraket}[2]{\Big\langle #1 \mspace{2mu} \Big\vert \mspace{1.5mu} #2 \Big\rangle}

\newcommand{\2}{\mspace{-2mu}}
\newcommand{\elm}[3]{\langle #1 \vert #2\vert #3 \rangle}

\DeclareRobustCommand{\raisedrho}{{\mathpalette\irrho\relax}}
\newcommand{\irrho}[2]{\raisebox{1pt}{$#1\rho$}}

\makeatletter

\def\env@dcases#1{%
  \let\@ifnextchar\new@ifnextchar
  \left\lbrace\def\arraystretch{1.2}%
  \array{@{}#1@{\quad}l@{}}}
\makeatother

\newcommand{\Eplain}[3]{E^{#1}_{{#2}^{#1} {#3}^{#1}}}
\newcommand{\Etilde}[3]{\tilde{E}^{#1}_{{#2}^{#1} {#3}^{#1}}}
\newcommand{\Eplainprime}[3]{E^{#1'}_{{#2}^{#1^{\smash{\prime}}} \! {#3}^{#1^{\smash{\prime}}}}}
\newcommand{\Etildeprime}[3]{\tilde{E}^{#1'}_{{#2}^{#1^{\smash{\prime}}} \! {#3}^{#1^{\smash{\prime}}}}}

\newcommand{\gplain}[3]{g^{#1}_{{#2}^{#1} {#3}^{#1}}}
\newcommand{\gplainprime}[3]{g^{#1'}_{{#2}^{#1^{\smash{\prime}}} \! {#3}^{#1^{\smash{\prime}}}}}

\newcommand{\FFcheckplain}[1]{\check{\mbf{F}}^{#1}}
\newcommand{\FFcheckprime}[1]{\check{\mbf{F}}^{\prime#1}}
\newcommand{\FFcheckplaindagger}[1]{\check{\mbf{F}}^{#1\dagger}}
\newcommand{\FFcheckprimedagger}[1]{\check{\mbf{F}}^{\prime#1\dagger}}
\newcommand{\FFtildeplain}[1]{\tilde{\mbf{F}}^{#1}}
\newcommand{\FFtildeprime}[1]{\tilde{\mbf{F}}^{\prime#1}}
\newcommand{\FFtildeplaindagger}[1]{\tilde{\mbf{F}}^{#1\dagger}}
\newcommand{\FFtildeprimedagger}[1]{\tilde{\mbf{F}}^{\prime#1\dagger}}

\newcommand{\Fcheckplain}[3]{\check{F}^{#1}_{{#2}^{#1} {#3}^{#1}}}
\newcommand{\Fcheckplainconj}[3]{\check{F}^{#1*}_{{#2}^{#1} {#3}^{#1}}}
\newcommand{\Fcheckprime}[3]{\check{F}^{\prime#1}_{{#2}^{#1} {#3}^{#1}}}
\newcommand{\Ftildeplain}[3]{\tilde{F}^{#1}_{{#2}^{#1} {#3}^{#1}}}
\newcommand{\Ftildeprime}[3]{\tilde{F}^{\prime#1}_{{#2}^{#1} {#3}^{#1}}}
\newcommand{\Fplain}[3]{F^{#1}_{{#2}^{#1} {#3}^{#1}}}

\newcommand{\fbarprime}[3]{\bar{f}^{\prime}_{(#1 \, {#2}^{#1} {#3}^{#1})}}
\newcommand{\fbarprimeconj}[3]{\bar{f}^{\prime*}_{(#1 \, {#2}^{#1} {#3}^{#1})}}

\newcommand{\rrho}[3]{\rho^{#1}_{\mspace{-0mu}#2^{#1} #3^{#1}}}
\newcommand{\rrhodot}[3]{\dot{\rho}^{#1}_{\mspace{-0mu}#2^{#1} #3^{#1}}}
\newcommand{\rrhoconj}[3]{\rho^{#1*}_{\mspace{-0mu}#2^{#1} #3^{#1}}}

\newcommand{\Cplain}[6]{C_{ ({#1} \, {#2}^{#1}  {#3}^{#1}) ({#4'} {#5}^{#4^{\smash{\prime}}} \! {#6}^{#4^{\smash{\prime}}}) } }
\newcommand{\Cbarplain}[6]{\bar{C}_{ ({#1} \, {#2}^{#1}  {#3}^{#1}) ({#4'} {#5}^{#4^{\smash{\prime}}} \! {#6}^{#4^{\smash{\prime}}}) } }
\newcommand{\Cbarprime}[6]{\bar{C}^{\prime}_{ ({#1} \, {#2}^{#1}  {#3}^{#1}) ({#4'} {#5}^{#4^{\smash{\prime}}} \! {#6}^{#4^{\smash{\prime}}}) } }
\newcommand{\Cbarconj}[6]{\bar{C}^{*}_{ ({#1} \, {#2}^{#1}  {#3}^{#1}) ({#4'} {#5}^{#4^{\smash{\prime}}} \! {#6}^{#4^{\smash{\prime}}}) } }
\newcommand{\Cbarprimeconj}[6]{\bar{C}^{\prime*}_{ ({#1} \, {#2}^{#1}  {#3}^{#1}) ({#4'} {#5}^{#4^{\smash{\prime}}} \! {#6}^{#4^{\smash{\prime}}}) } }

\newcommand{\Aprime}[4]{A_{ {#1} ({#2'} {#3}^{#2^{\smash{\prime}}} \! {#4}^{#2^{\smash{\prime}}}) } }
\newcommand{\Aplain}[4]{A_{ {#1} ({#2} \, {#3}^{#2} {#4}^{#2}) } }

\newcommand{\Splain}[6]{S_{ ({#1} \, {#2}^{#1}  {#3}^{#1}) ({#4'} {#5}^{#4^{\smash{\prime}}} \! {#6}^{#4^{\smash{\prime}}}) } }

\newcommand{\crea}[2]{a^{#1\mspace{-0.5mu}\raisebox{0.3ex}{$\scriptstyle\dagger$}}_{\mspace{-0mu}#2^{#1}}}
\newcommand{\creaprime}[2]{a^{#1^{\smash{\prime}}\mspace{-0.5mu}\raisebox{0.6ex}{$\scriptstyle\dagger$}}_{\mspace{-0mu}#2^{#1^{\smash{\prime}}}}}
\newcommand{\creatilde}[2]{\tilde{a}^{#1\mspace{-0.5mu}\raisebox{0.3ex}{$\scriptstyle\dagger$}}_{\mspace{-0mu}#2^{#1}}}
\newcommand{\anni}[2]{a^{#1}_{\mspace{-0mu}#2^{#1}}}
\newcommand{\anniprime}[2]{a^{#1^{\smash{\prime}}}_{\mspace{-1mu}#2^{#1^{\smash{\prime}}}}}
\newcommand{\annitilde}[2]{\tilde{a}^{#1}_{\mspace{-0mu}#2^{#1}}}

\newcommand{\vplain}[3]{V^{#1}_{\mspace{-2mu} #2^{#1} #3^{#1}}}
\newcommand{\vplainconj}[3]{V^{#1*}_{\mspace{-2mu} #2^{#1} #3^{#1}}}
\newcommand{\vplaindot}[3]{\dot{V}^{#1}_{\mspace{-2mu} #2^{#1} #3^{#1}}}
\newcommand{\vplaindotconj}[3]{\dot{V}^{#1*}_{\mspace{-2mu} #2^{#1} #3^{#1}}} \usepackage{acro}

\DeclareAcronym{mctdh}{
   short = MCTDH ,
   long = multiconfiguration time-dependent Hartree ,
}

\DeclareAcronym{nomctdh}{
   short = NOMCTDH ,
   long = non-orthogonal \ac{mctdh} ,
}

\DeclareAcronym{gmctdh}{
   short = G-MCTDH ,
   long = Gaussian-based \ac{mctdh} ,
}

\DeclareAcronym{mlgmctdh}{
   short = ML-GMCTDH ,
   long = multilayer Gaussian-based \ac{mctdh} ,
}

\DeclareAcronym{mlmctdh}{
   short = ML-MCTDH ,
   long = multilayer \ac{mctdh} ,
}

\DeclareAcronym{mpsmctdh}{
   short = MPS-MCTDH ,
   long = matrix product state \ac{mctdh} ,
}

\DeclareAcronym{vmcg}{
   short = vMCG ,
   long = variational multiconfiguration Gaussian ,
}

\DeclareAcronym{ms}{
   short = MS ,
   long = multiple spawning ,
}

\DeclareAcronym{ccs}{
   short = CCS ,
   long = coupled coherent states ,
}

\DeclareAcronym{mctdhn}{
   short = MCTDH[\textit{n}] ,
   long = systematically truncated multiconfiguration time-dependent Hartree ,
}

\DeclareAcronym{mrmctdhn}{
  short = MR-MCTDH[\textit{n}] ,
  long = multi-reference truncated multiconfiguration time-dependent Hartree ,
}

\DeclareAcronym{tdh}{
   short = TDH ,
   long = time-dependent Hartree ,
}

\DeclareAcronym{dmrg}{
   short = DMRG ,
   long = density matrix renormalization group,
}

\DeclareAcronym{tddmrg}{
   short = TD-DMRG ,
   long = time-dependent density matrix renormalization group,
}

\DeclareAcronym{scf}{
   short = SCF ,
   long = self-consistent field ,
}

\DeclareAcronym{casscf}{
   short = CASSCF ,
   long = complete active space self-consistent field ,
}

\DeclareAcronym{tdcasscf}{
   short = TD-CASSCF ,
   long = time-dependent \acl{casscf} ,
}

\DeclareAcronym{gasscf}{
   short = CASSCF ,
   long = generalized active space self-consistent field ,
}

\DeclareAcronym{tdgasscf}{
   short = TD-GASSCF ,
   long = time-dependent \acl{gasscf} ,
}

\DeclareAcronym{rasscf}{
   short = RASSCF ,
   long = restricted active space self-consistent field ,
}

\DeclareAcronym{tdrasscf}{
   short = TD-RASSCF ,
   long = time-dependent \acl{rasscf} ,
}

\DeclareAcronym{ormas}{
   short = ORMAS ,
   long = occupation-restricted multiple active space ,
}
\DeclareAcronym{tdormas}{
   short = TD-ORMAS ,
   long = time-dependent \acl{ormas} ,
}

\DeclareAcronym{mctdhf}{
   short = MCTDHF ,
   long = multiconfiguration time-dependent Hartree-Fock ,
}

\DeclareAcronym{occ}{
   short = OCC ,
   long = orbital-optimized coupled cluster ,
}

\DeclareAcronym{tdocc}{
   short = TD-OCC ,
   long = time-dependent \acl{occ} ,
}

\DeclareAcronym{nocc}{
   short = NOCC ,
   long = non-orthogonal orbital-optimized coupled cluster ,
}

\DeclareAcronym{oatdcc}{
   short = OATDCC ,
   long = orbital-adaptive time-dependent coupled cluster ,
}

\DeclareAcronym{fci}{
   short = FCI ,
   long = full configuration interaction ,
}

\DeclareAcronym{cud}{
   short = CUD ,
   long = closed under de-exciation ,
}

\DeclareAcronym{fsmr}{
   short = FSMR ,
   long = full-space matrix representation ,
}

\DeclareAcronym{hh}{
   short = HH ,
   long = H\'enon-Heiles ,
}

\DeclareAcronym{ho}{
   short = HO ,
   long = harmonic oscillator ,
}

\DeclareAcronym{dop853}{
   short = DOP853 ,
   long = Dormand-Prince 8{(5,3)} ,
}

\DeclareAcronym{sm}{
   short = SM ,
   long = supplementary material ,
}

\DeclareAcronym{vscf}{
   short = VSCF ,
   long = vibrational self-consistent field ,
}

\DeclareAcronym{eom}{
   short = EOM ,
   long = equation of motion ,
   short-plural-form = EOMs ,
   long-plural-form = equations of motion ,
}

\DeclareAcronym{tdvp}{
   short = TDVP ,
   long = time-dependent variational principle
}

\DeclareAcronym{tdse}{
   short = TDSE ,
   long = time-dependent Schr{\"o}dinger equation ,
}

\DeclareAcronym{cc}{
   short = CC ,
   long = coupled cluster ,
}

\DeclareAcronym{bcc}{
   short = BCC ,
   long = Brueckner coupled cluster ,
}

\DeclareAcronym{vcc}{
   short = VCC ,
   long = vibrational coupled cluster ,
}

\DeclareAcronym{tdvcc}{
   short = TDVCC ,
   long = time-dependent vibrational coupled cluster ,
}

\DeclareAcronym{tdvci}{
   short = TDVCI ,
   long = time-dependent vibrational configuration interaction ,
}

\DeclareAcronym{vci}{
   short = VCI ,
   long = vibrational configuration interaction ,
}

\DeclareAcronym{ci}{
   short = CI ,
   long = configuration interaction ,
}

\DeclareAcronym{tdci}{
   short = CI ,
   long = time-dependent \acl{ci} ,
}

\DeclareAcronym{sq}{
   short = SQ ,
   long = second quantization ,
}

\DeclareAcronym{fq}{
   short = FQ ,
   long = first quantization ,
}

\DeclareAcronym{mc}{
   short = MC ,
   long = mode combination ,
}

\DeclareAcronym{mcr}{
   short = MCR ,
   long = mode combination range ,
   long-plural = s ,
}

\DeclareAcronym{pes}{
   short = PES ,
   long = potential energy surface
}

\DeclareAcronym{svd}{
   short = SVD ,
   long = singular value decomposition ,
}
\DeclareAcronym{adga}{
   short = ADGA ,
   long = adaptive density-guided approach ,
}

\DeclareAcronym{rhs}{
   short = RHS ,
   long = right-hand side ,
}

\DeclareAcronym{lhs}{
   short = LHS ,
   long = left-hand side ,
}

\DeclareAcronym{ivr}{
   short = IVR ,
   long = intramolecular vibrational energy redistribution ,
}

\DeclareAcronym{fft}{
   short = FFT ,
   long = fast Fourier transform ,
}

\DeclareAcronym{spf}{
   short = SPF ,
   long = single-particle function ,
}

\DeclareAcronym{lls}{
   short = LLS ,
   long = linear least squares ,
}

\DeclareAcronym{itnamo}{
   short = ItNaMo ,
   long = iterative natural modal ,
}

\DeclareAcronym{hf}{
   short = HF ,
   long = Hartree-Fock ,
}

\DeclareAcronym{mcscf}{
   short = MCSCF ,
   long = multi-configurational self-consistent field ,
}

\DeclareAcronym{sop}{
   short = SOP ,
   long = sum-of-products ,
}
\DeclareAcronym{midascpp}{
   short = MidasCpp ,
   long = Molecular Interactions{,} Dynamics And Simulations Chemistry Program Package ,
   tag = abbrev ,
}

\DeclareAcronym{mpi}{
   short = MPI ,
   long = message passing interface ,
}

\DeclareAcronym{ode}{
   short = ODE ,
   long  = ordinary differential equation ,
   short-plural = s ,
   long-plural = s ,
   short-indefinite = an ,
   long-indefinite = an ,
   tag = abbrev ,
}

\DeclareAcronym{bch}{
   short = BCH ,
   long = Baker-Campbell-Hausdorff ,
}

\DeclareAcronym{sr}{
   short = SR ,
   long = single-reference ,
}
\DeclareAcronym{mr}{
   short = MR ,
   long = multi-reference ,
}

\DeclareAcronym{dof}{
   short = DOF ,
   long = degree of freedom ,
   short-plural-form = DOFs ,
   long-plural-form = degrees of freedom ,
}

\DeclareAcronym{hp}{
   short = HP ,
   long = Hartree product ,
}

\DeclareAcronym{tdbvp}{
   short = TDBVP ,
   long  = time-dependent bivariational principle ,
   short-plural = s ,
   long-plural = s ,
   short-indefinite = a ,
   long-indefinite = a ,
   tag = abbrev ,
}

\DeclareAcronym{dfvp}{
   short = DFVP ,
   long  = Dirac-Frenkel variational principle ,
}

\DeclareAcronym{ele}{
   short = ELE ,
   long  = Euler-Lagrange equation ,
   short-plural = s ,
   long-plural = s ,
   tag = abbrev ,
}

\DeclareAcronym{mrcc}{
   short = MRCC ,
   long = multi-reference coupled cluster ,
}

\DeclareAcronym{tdfvci}{
   short = TDFVCI ,
   long = time-dependent full vibrational configuration interaction ,
}

\DeclareAcronym{tdfci}{
   short = TDFCI ,
   long = time-dependent full configuration interaction ,
}

\DeclareAcronym{tdevcc}{
   short = TDEVCC ,
   long  = time-dependent extended vibrational coupled cluster ,
   short-plural = s ,
   long-plural = s ,
   short-indefinite = a ,
   long-indefinite = a ,
   tag = abbrev ,
}

\DeclareAcronym{holc}{
   short = HOLC ,
   long = hybrid optimized and localized vibrational coordinate ,
}

\DeclareAcronym{acf}{
   short = ACF ,
   long = autocorrelation function ,
}

\DeclareAcronym{fwhm}{
   short = FWHM ,
   long  = full width at half maximum ,
   short-plural = s ,
   long-plural = full widths at half maxima ,
   short-indefinite = an ,
   long-indefinite = a ,
   tag = abbrev ,
}

\DeclareAcronym{tdmvcc}{
   short = TDMVCC ,
   long = time-dependent modal vibrational coupled cluster ,
}

\DeclareAcronym{otdmvcc}{
   short = oTDMVCC ,
   long = orthogonal time-dependent modal vibrational coupled cluster ,
}

\DeclareAcronym{midas}{
   short = MidasCpp ,
   long = Molecular Interactions{,} Dynamics and Simulations Chemistry Program Package ,
}

\newcommand{\au}{Department of Chemistry, Aarhus University, Langelandsgade 140, 8000 Aarhus C, Denmark}
\newcommand{\upo}{Dipartimento di Scienze e Innovazione Tecnologica, Universit\`a  del Piemonte Orientale (UPO), Via T. Michel 11, 15100 Alessandria, Italy}

\begin{document}

\title{Time-dependent coupled cluster with orthogonal adaptive basis functions: General formalism and application to the vibrational problem}

\author{Mads Greisen Højlund}
\email{madsgh@chem.au.dk}
\affiliation{\au}

\author{Alberto Zoccante}
\email{alberto.zoccante@uniupo.it}
\affiliation{\upo}

\author{Ove Christiansen}
\email{ove@chem.au.dk}
\affiliation{\au}

\hypersetup{pdftitle={Time-dependent coupled cluster with orthogonal adaptive basis functions: General formalism and application to the vibrational problem}}
\hypersetup{pdfauthor={M.~G.~Højlund, et al.}}
\hypersetup{bookmarksopen=true}

\date{\today}


\begin{abstract}
    We derive equations of motion for bivariational wave functions with orthogonal adaptive
basis sets and specialize the formalism to the coupled cluster ansatz. 
The equations are related to the biorthogonal case in a
transparent way, and similarities and differences are analyzed.
We show that the amplitude equations are identical in the
orthogonal and biorthogonal formalisms, while the linear
equations that determined the basis set time evolution differ by symmetrization.
Applying the orthogonal framework to the nuclear dynamics problem, we
introduce and implement the orthogonal time-dependent modal vibrational coupled cluster 
(oTDMVCC) method and benchmark it
against exact reference results for four triatomic molecules as well as
a 5D \textit{trans}-bithiophene model. We confirm numerically that
the biorthogonal TDMVCC hierarchy converges to the exact solution,
while oTDMVCC does not.
The differences between TDMVCC and oTDMVCC are found to be small for
three of the five cases, but we also identify one case where the formal deficiency of the oTDMVCC approach
results in clear and visible errors relative to the exact result.
For the remaining example, oTDMVCC exhibits rather modest but visible errors. \end{abstract}

\maketitle

\acresetall

\section{Introduction} \label{sec:introduction}
The \ac{cc} method is a highly useful approach
for computing the electronic and vibrational structure of molecules.
Its benefits include polynomial-scaling cost, 
size extensivity and fast convergence
of the \ac{cc} hierarchy, leading in many cases to
a favorable balance between cost and accuracy. These advantages
ultimately stem from the exponential \ac{cc} parameterization.
However, it is also well known that the \ac{cc} ansatz only
works well when the amplitudes are sufficiently small and the 
reference describes a large part of the wave
function. Conversely, if the overlap between the wave function
and the reference decreases, the amplitudes grow
and the ansatz tends to break down.
This kind of situation is easily encountered
in explicitly time-dependent
or dynamical settings, where large-amplitude motion such as
ionization (in electronic structure) or 
dissociation (in vibrational structure) is commonplace.
It is quite obvious that a static reference is ill-suited
for describing such processes. However, much less
violent phenomena, e.g. the \ac{ivr} of water, 
can also lead to the breakdown of the \ac{cc} ansatz.\cite{madsenTimedependentVibrationalCoupled2020}
This weakness of the \ac{cc} approach can sometimes be alleviated
by choosing a dynamical single-particle basis, 
which in turn induces a dynamical reference that adapts
to the wave function at any given time. 

Historically, the idea of optimizing the basis set
in a \ac{cc} computation emerged in ground state theory
with the so-called 
\ac{bcc}\cite{chilesElectronPairOperator1981,handySizeconsistentBruecknerTheory1989,raghavachariSizeconsistentBruecknerTheory1990,hampelComparisonEfficiencyAccuracy1992} method. 
Here, the basis is optimized such that the singles projections vanish
(in other words, the singles vanish identically in the Brueckner basis).
The \ac{bcc} theory attracted considerable attention in the 1990s,
in part due to a perceived robustness towards symmetry breaking.\cite{stantonChoiceOrbitalsSymmetry1992,barnesSymmetryBreakingApplication1994,xieOxywaterRadicalCation1996} 
It was later discovered that this
robustness is not universal\cite{crawfordSurprisingFailuresBrueckner2000}
and that the \ac{bcc} response function contains spurious second-order poles\cite{aigaFrequencydependentHyperpolarizabilitiesBrueckner1994,kochBruecknerCoupledCluster1994}.
\ac{bcc} has since fallen somewhat out of fashion.

A related idea is to optimize the basis such
that the \ac{cc} energy is minimized.
\cite{purvisFullCoupledclusterSingles1982,scuseriaOptimizationMolecularOrbitals1987,sherrillEnergiesAnalyticGradients1998,krylovSizeconsistentWaveFunctions1998,pedersenGaugeInvariantCoupled1999}
When unitary (orthogonal) basis set transformations are used, we will refer to this
method as \ac{occ} (similar acronyms such as OO-CC are also encountered in the literature). 
In \ac{occ}, the single excitations are excluded
from the outset (i.e. $T = T_2 + T_3 + \cdots$), since $\exp(T_1)$ is redundant
with the basis set rotations.
Although \ac{bcc} and \ac{occ} are conceptually quite similar,
it turns out that the \ac{occ} hierarchy does not converge to the \ac{fci} limit\cite{kohnOrbitaloptimizedCoupledclusterTheory2005},
which is obviously a disadvantage. For the examples studied by Köhn and Olsen\cite{kohnOrbitaloptimizedCoupledclusterTheory2005} (ozone and $\mathrm{CH_2}$),
this deficiency of \ac{occ} starts to show at the quadruples (OCCDTQ) or quintuples (OCCDTQ5) level. At the doubles (OCCD) and triples
(OCCDT) levels, \ac{occ} and \ac{bcc} appear to be comparable in accuracy.
Pedersen et al.\cite{pedersenGaugeInvariantCoupled2001} later introduced
\ac{nocc}. The purpose of using a non-unitary (non-orthogonal) basis set transformation
was to simplify response equations, but it was later shown by Myhre\cite{myhreDemonstratingThatNonorthogonal2018} that
\ac{nocc} does in fact recover the \ac{fci} limit.

The concept of using optimized or adaptive basis functions for simulating real-time dynamics has
a long history in the nuclear dynamics community, where the \ac{mctdh}\cite{meyerMulticonfigurationalTimedependentHartree1990,beckMulticonfigurationTimedependentHartree2000}
method has been very successful. \ac{mctdh} employs a complete expansion inside
an adaptive active space and thus yields the exact solution for the given choice of space.
The analogous electron dynamics method is denoted 
\ac{mctdhf}.\cite{zanghelliniMCTDHFApproachMulti2003,katoTimedependentMulticonfigurationTheory2004,nestMulticonfigurationTimedependentHartree2005,caillatCorrelatedMultielectronSystems2005}
Both of these methods involve an exponentially scaling computational effort, so it is highly relevant to
investigate lower-scaling alternatives, e.g. based on the \ac{cc} ansatz.
Real-time time-dependent \ac{cc} with static basis functions has been considered in the literature
for vibrational,\cite{hansenTimedependentVibrationalCoupled2019,hansenExtendedVibrationalCoupled2020,madsenGeneralImplementationTimedependent2020}
electron,\cite{schonhammerTimedependentApproachCalculation1978,huberExplicitlyTimedependentCoupled2011,pedersenSymplecticIntegrationPhysical2019,skeidsvollTimedependentCoupledclusterTheory2020} 
and nucleon\cite{hoodbhoyTimedependentCoupledclusterApproximation1978,hoodbhoyTimedependentCoupledclusterApproximation1979,piggTimedependentCoupledclusterMethod2012}
dynamics (see also Ref.~\citenum{sverdrupofstadTimedependentCoupledclusterTheory2023a} for a recent review), but we will focus specifically
on combining adaptive basis functions with the time-dependent \ac{cc} ansatz.
This idea was first taken up in 2012 by Kvaal, who introduced the \ac{oatdcc}\cite{kvaalInitioQuantumDynamics2012} method
for electron dynamics.
\Ac{oatdcc} uses 
biorthogonal adaptive orbitals 
and also allows the basis to be split
into an active and a secondary part (only the active orbitals are correlated).
This yields a highly flexible ansatz that converges to the exact limit, i.e. \ac{mctdhf}.
In 2018, Sato et al. proposed the time-dependent OCC (\acs{tdocc})\cite{satoCommunicationTimedependentOptimized2018} method,
which uses orthogonal orbitals and presumably does not converge to the exact limit. 
However, TD-OCCDT calculations seem to agree very well with higher-level
calculations,
\cite{satoCommunicationTimedependentOptimized2018,pathakTimedependentOptimizedCoupledcluster2020a,pathakTimedependentOptimizedCoupledcluster2021}
which indicates that the use of an orthogonal basis does not introduce large errors in practice.

In the context of nuclear or vibrational dynamics, our group has introduced the
\ac{tdmvcc}\cite{madsenTimedependentVibrationalCoupled2020} method, which uses an adaptive active space inspired by \ac{mctdh}
(formally speaking, \ac{tdmvcc} can be considered a vibrational analogue of \ac{oatdcc}).
Again, the use of a biorthogonal basis guarantees the convergence to the exact solution, i.e. \ac{mctdh}.

Splitting a biorthogonal basis into active and secondary parts leads to the
peculiar situation that the active ket and bra bases are allowed
to span different spaces. Although it is consistent with the formalism,
we have found that this feature sometimes leads to numerical instability.\cite{hojlundBivariationalTimedependentWave2022}
In Ref.~\citenum{hojlundBivariationalTimedependentWave2022} we proposed
a scheme that effectively locks the active bra and ket spaces together, while still allowing
non-unitary transformations within the active space. Although this scheme was shown to
solve the stability problem without sacrificing accuracy, we certainly feel there is
more to be learned about time-dependent \ac{cc} with adaptive basis functions.
In this paper we therefore consider the use of orthogonal, adaptive basis functions
in vibrational \ac{cc}. The resulting method (which is
analogous to \ac{tdocc}) is denoted orthogonal \ac{tdmvcc}, or \acs{otdmvcc} for short.

Time-dependent \ac{cc} equations are often derived using Arponen's
\ac{tdbvp}\cite{arponenVariationalPrinciplesLinkedcluster1983}.
The original version of this principle uses a complex-valued
action, which must be a holomorphic or complex
analytic function of the wave function parameters,
as explained by Kvaal\cite{kvaalInitioQuantumDynamics2012}.
We will see that the use of an orthogonal basis leads to
a non-holomorphic action, which has some important
mathematical consequences that are best explained using the terminology of complex analysis\cite{steinComplexAnalysis2003}.
For the convenience of the reader and for the clarity of our exposition, we provide
a brief overview of some aspects of complex analysis (see Appendix~\ref{appendix:complex_analysis})
that we will use throughout the paper.

The paper is organized as follows:
Section~\ref{sec:theory} covers the theory,
including the \ac{tdbvp} for holomorphic and non-holomorphic
parameterizations and derivations of the \acp{eom}.
This is followed by a brief description of our 
computer implementation in Sec.~\ref{sec:implementation}
and a few numerical examples in Sec.~\ref{sec:results}.
Section~\ref{sec:summary} summarizes our findings
and concludes with an outlook on future work. 

\section{Theory} \label{sec:theory}

\subsection{The time-dependent bivariational principle} \label{sec:tdbvp}
In Ref.~\citenum{hojlundBivariationalTimedependentWave2022}, we considered a
complex bivariational Lagrangian,
\begin{align}
    \mathcal{L} = \elm{\Psi'}{(i \partial_t - H)}{\Psi} = \mathcal{L}(\mbf{y}, \dot{\mbf{y}}, t),
\end{align}
where the bra and ket states are formally independent. We showed the
well-known fact that stationary points ($\delta \mathcal{S} = 0$) of
the action-like functional
\begin{align}
    \mathcal{S} = \int_{t_0}^{t_1} \mathcal{L} \dd{t}
\end{align}
correspond to the solutions of a set of \acp{ele},
\begin{align} \label{eq:eles}
    0 = \pdv{\mathcal{L}}{y_i} - \dv{t} \pdv{\mathcal{L}}{\dot{y}_i},
\end{align}
for the wave function parameters $y_i$. This formalism requires the Lagrangian to
be a complex analytic or holomorphic\cite{steinComplexAnalysis2003} function of the complex parameters $y_i$, implying
that complex conjugation cannot appear in the wave function parameterization.\cite{kvaalInitioQuantumDynamics2012} 
This restriction rules out the use of
orthogonal basis sets, since in that case the bra basis functions are simply the complex conjugate of the ket basis functions.
Instead, one should a biorthogonal basis. 

One might be tempted to ignore these formal considerations and simply
write down $\mathcal{L}$ with an orthogonal basis set, which would then make $\mathcal{L}$ a complex-valued, non-holomorphic function.
As explained in Appendix~\ref{appendix:complex_analysis}, we can consider such a function $\mathcal{L}$ as depending on 
$y_i$ and $y_i^*$, which are treated as independent variables.
Making the corresponding action-like functional stationary again leads to a set of \acp{ele},
\begin{subequations} \label{eq:eles_direct_and_conj_params_inconsistent}
    \begin{align} 
        0 &= \pdv{\mathcal{L}}{y_i} - \dv{t} \pdv{\mathcal{L}}{\dot{y}_i}
        ,    \label{eq:eles_direct_and_conj_params_inconsistent_a} \\
        0 &= \pdv{\mathcal{L}}{y_i^*} - \dv{t} \pdv{\mathcal{L}}{\dot{y}_i^*} \label{eq:eles_direct_and_conj_params_inconsistent_b}.
    \end{align}
\end{subequations}
The trouble is that these equations are not each other's complex conjugate since $\mathcal{L}$ is complex.
As an example, note that
\begin{align}
    \left( \pdv{\mathcal{L}}{y_i} \right)^{\!*} = \pdv{\mathcal{L}^*}{y_i^*} \neq \pdv{\mathcal{L}}{y_i^*}.
\end{align}
The consequence is that Eqs.~\eqref{eq:eles_direct_and_conj_params_inconsistent} do not have a
consistent solution (this kind of situation is also explained in Appendix~\ref{appendix:complex_analysis}). 
We can attempt to solve Eqs.~\eqref{eq:eles_direct_and_conj_params_inconsistent} while treating $y_i$ and $y_i^*$ as truly independent,
but the solution will not
respect the obvious requirement that $(\mrm{d} y_i / \mrm{d} t)^* = (\mrm{d} y_i^* / \mrm{d} t)$, i.e. complex conjugate pairs do not
remain each other's complex conjugate. Such a formalism
is thus inconsistent, as also noted by Kvaal\cite{kvaalInitioQuantumDynamics2012}.

In this paper we consider instead
a manifestly real Lagrangian,
\begin{align} \label{eq:real_lagrangian}
    \bar{\mathcal{L}} = \tfrac{1}{2} (\mathcal{L} + \mathcal{L}^*) = \bar{\mathcal{L}}(\mbf{y}, \dot{\mbf{y}}, \mbf{y}^*, \dot{\mbf{y}}^*, t),
\end{align}
that depends on the wave function parameters, the complex conjugate parameters and the time derivative of both.
As usual, we define an action-like functional,
\begin{align}
    \bar{\mathcal{S}} = \int_{t_0}^{t_1} \bar{\mathcal{L}} \dd{t},
\end{align}
which is made stationary ($\delta \bar{\mathcal{S}} = 0$) with respect
to variations in the parameters $y_i$ and the complex conjugate parameters
$y_i^*$. The resulting \acp{ele} are:
\begin{subequations} \label{eq:eles_direct_and_conj_params}
    \begin{align} 
        0 &= \pdv{\bar{\mathcal{L}}}{y_i}   - \dv{t} \pdv{\bar{\mathcal{L}}}{\dot{y}_i}  , \label{eq:eles_direct_and_conj_params_a} \\
        0 &= \pdv{\bar{\mathcal{L}}}{y_i^*} - \dv{t} \pdv{\bar{\mathcal{L}}}{\dot{y}_i^*}. \label{eq:eles_direct_and_conj_params_b}
    \end{align}
\end{subequations}
Since $\bar{\mathcal{L}}$ is real, the two sets of \acp{ele} are exactly each other's complex conjugate ensuring
that the \acp{eom} for $y_i$ and $y_i^*$ are also each other's complex conjugate in the sense that
$(\mrm{d} y_i / \mrm{d} t)^* = (\mrm{d} y_i^* / \mrm{d} t)$.
Having settled this point, we only need to solve either Eq.~\eqref{eq:eles_direct_and_conj_params_a} or \eqref{eq:eles_direct_and_conj_params_b}.
Real Lagrangians like Eq.~\eqref{eq:real_lagrangian} have previously been considered in the literature.
\cite{pedersenCoupledClusterResponse1997,pedersenTimedependentLagrangianApproach1998,satoCommunicationTimedependentOptimized2018,kristiansenLinearNonlinearOptical2022}

It is quite possible some parameters $\alpha_i \in \mbf{y}$ appear only as $\alpha_i$ and never as
$\alpha_i^*$ in the bra and ket states, meaning $\mathcal{L}$ is a holomorphic function of $\alpha_i$.
In that case the \acp{ele} of Eq.~\eqref{eq:eles_direct_and_conj_params_a} simplify in the following way:
\begin{align}
    0 &= \pdv{\bar{\mathcal{L}}}{\alpha_i} - \dv{t} \pdv{\bar{\mathcal{L}}}{\dot{\alpha}_i} \nn
    &= \frac{1}{2} \left( \pdv{\mathcal{L}}{\alpha_i} - \dv{t} \pdv{\mathcal{L}}{\dot{\alpha}_i} \right) +
    \frac{1}{2} \left( \pdv{\mathcal{L}}{\alpha_i^*} - \dv{t} \pdv{\mathcal{L}}{\dot{\alpha}_i^*} \right)^* \nn
    &= \frac{1}{2} \left( \pdv{\mathcal{L}}{\alpha_i} - \dv{t} \pdv{\mathcal{L}}{\dot{\alpha}_i} \right) \label{eq:simplified_eles}
\end{align}
We recover a set of \acp{ele} based on the complex Lagrangian
$\mathcal{L}$, which we will refer to as being \textit{complex bivariational}.
Similarly, we will use the term \textit{real bivariational} when referring to
equations based on the real Lagrangian.
Equation~\eqref{eq:simplified_eles} then shows that the real and complex bivariational principles are
equivalent for the special subset of parameters $\alpha_i$. For parameterizations that
are holomorphic in all parameters, e.g. time-dependent \ac{cc} with static basis functions,
the two principles yield fully equivalent \acp{eom}. 
\subsection{Parameterization} \label{sec:parameterization}
\subsubsection{Time-dependent basis sets}

The wave function is parameterized using time-dependent orthonormal 
basis functions (denoted modals or orbitals), that are in turn expanded in a primitive
orthonormal basis.
Rather than working directly with the concrete basis functions, we employ
a \ac{sq} formalism for vibrational structure\cite{christiansenSecondQuantizationFormulation2004}
based on creation and annihilation operators (not ladder operators) that obey the following commutator relations:
\begin{subequations}
    \begin{align} \label{eq:basic_commutator_vib_ortho}
        [\anniprime{m}{\alpha}, \crea{m}{\beta}] &= \delta_{mm'} \delta_{\alpha^m \beta^m} \\
        [\anniprime{m}{\alpha}, \anni{m}{\beta}] &= 0 \\
        [\creaprime{m}{\alpha}, \crea{m}{\beta}] &= 0.
    \end{align}
\end{subequations}
We use indices $m$ and $m'$ to enumerate the vibrational modes (which are distinguishable degrees of freedom), while
the indices $\alpha^m$ and $\beta^m$ indicate the modal basis functions for each individual mode.
The well-known electronic structure \ac{sq} formalism\cite{helgakerMolecularElectronicstructureTheory2000}
is defined in terms of anti-commutators rather than commutators, which reflects the very different physical nature
of electronic and nuclear motion. However, the one-mode shift operators,
\begin{align}
    \Eplain{m}{\alpha}{\beta} = \crea{m}{\alpha} \anni{m}{\beta},
\end{align}
satisfy essentially the same commutator as their electronic counterparts:
\begin{align}
    [\Eplain{m}{\alpha}{\beta}, \Eplainprime{m}{\gamma}{\delta}] = \delta_{mm'} (\delta_{\beta^m \gamma^m} \Eplain{m}{\alpha}{\delta} - \delta_{\alpha^m \delta^m} \Eplain{m}{\gamma}{\beta}).
\end{align}
This means that our derivations carry over to electronic structure after removing mode indices as appropriate.
\begin{align}
    \creatilde{m}{p} = \sum_{\alpha} \crea{m}{\alpha} \vplain{m}{\alpha}{p}.
\end{align}
The tilde on the left-hand side indicates the time-dependence of the creation operator.
We use $\alpha^m$, $\beta^m$ to denote primitive indices, while $p^m$, $q^m$, $r^m$, $s^m$
denote generic time-dependent indices. The time-dependent basis is
split into an active basis indexed by $t^m$, $u^m$, $v^m$, $w^m$ and a secondary basis indexed by $x^m$, $y^m$.
We use the symbols $N$ and $N_{\2\subA}$ for the total number of basis functions per mode and the number
of active basis functions per mode, respectively.
Collecting the basis set coefficients in square matrices $\mathbf{V}^m$,
we obtain the following block structure:
\begin{align}
    \mbf{V}^m &=
    \left[
    \begin{array}{c | c} 
    {\mbf{V}^m_{\!\!\subA} \,} & \, \mbf{V}^m_{\!\!\subS}  \bottomstrut{-1.0ex}
    \end{array}\right].
\end{align}  
Using the same matrix notation, we write the orthonormality
constraint as
\begin{align}
    \mathbf{V}^{m \dagger} \, \mathbf{V}^m = \mathbf{1}^m.
\end{align}
We also need to consider the corresponding consistency conditions\cite{diracGeneralizedHamiltonianDynamics1950,ohtaTimedependentVariationalPrinciple2000,ohtaTimedependentVariationalPrinciple2004},
i.e. the time derivative of the constraints:
\begin{align}
    \mathbf{0} = \dot{\mathbf{V}}^{m \dagger} \, \mathbf{V}^m
    + \mathbf{V}^{m \dagger} \, \dot{\mathbf{V}}^m.
\end{align}
The consistency conditions are trivially satisfied if
\begin{subequations} \label{eq:consistency}
\begin{align}
    \dot{\mathbf{V}}^{m \dagger} \, \mathbf{V}^m &= +i \mathbf{G}^m, \label{eq:consistency_a} \\
    \mathbf{V}^{m \dagger} \, \dot{\mathbf{V}}^m &= -i \mathbf{G}^m, \label{eq:consistency_b}
\end{align}
\end{subequations}
where $\mathbf{G}^m$ is a Hermitian but otherwise arbitrary time-dependent matrix, which
is typically denoted the constraint matrix. The corresponding constraint operator
is defined as
\begin{align} \label{eq:constraint_operator}
    g = \sum_m g^m = \sum_m \sum_{p^m q^m} \gplain{m}{p}{q} \Etilde{m}{p}{q}.
\end{align}
Using Eq.~\eqref{eq:consistency_b} and the unitarity of $\mathbf{V}^m$ now yields the \acp{eom} for the basis set coefficients,
\begin{align} \label{eq:V_dot_general}
    i \dot{\mathbf{V}}^m = \mathbf{V}^m \mathbf{G}^m.
\end{align}
This shows that the time evolution of the basis set is generated by the constraint matrices,
which must be determined from the basis set \acp{ele}.
 
\subsubsection{Wave function expansion}
We consider wave function expansions of the form
\begin{subequations}
\begin{align}
    \ket{\Psi}  &= \ket{\Psi (\bm{\alpha}, \mbf{V})}, \label{eq:ket_state_param} \\
    \bra{\Psi'} &= \bra{\Psi'(\bm{\alpha}, \mbf{V}^{*})} \label{eq:bra_state_param}
\end{align}
\end{subequations}
where $\bm{\alpha}$ denotes a vector of complex configurational parameters
as opposed to the basis set parameters $\mbf{V} = \{ \mbf{V}^m \}$.
We make the restriction from the outset that the bra and ket states depend only on $\bm{\alpha}$
and not on $\bm{\alpha}^*$. 
The ket state depends on the basis set coefficients $\mbf{V}$ themselves,
while the bra state depends only on the complex conjugate basis set coefficients,
as indicated in Eq.~\eqref{eq:bra_state_param}.
Finally, the wave function is required to
be contained in the active space, i.e. the space spanned by the active basis functions.
Although this means that the secondary basis functions are not explicitly present in the wave function,
we find it instructive to derive \acp{eom} for the full matrices $\mbf{V}^m$, including the active \textit{and} secondary blocks.
 
\subsection{Equations of motion} \label{sec:eom}
\subsubsection{Configurational parameters}

The Lagrangian is (by ansatz) a holomorphic function of the configurational parameters $\alpha_i$,
so the real bivariational \acp{ele} (based on $\bar{\mathcal{L}}$) simplify to a set of complex bivariational \acp{ele} (based on $\mathcal{L}$)
as shown in Eq.~\eqref{eq:simplified_eles}. The complex bivariational case was considered in Ref.~\citenum{hojlundBivariationalTimedependentWave2022},
so rather than deriving the \acp{eom} from scratch, we simply cite the result:
\begin{align}
    i \dot{\alpha}_i 
    &= \sum_{j} (\mbf{M}^{-1})_{ij} \bigg( h_j
    - \sum_{m} \sum_{p^m q^m} \Aplain{j}{m}{p}{q}  \gplain{m}{p}{q} \bigg) \label{eq:alpha_EOM_expanded}
\end{align}
Here, we have defined an anti-symmetric matrix $\mbf{M}$ with elements
\begin{align}
    M_{ij} =
    \bigbraket{\pdv{\Psi'}{\alpha_i}}{\pdv{\Psi}{\alpha_j}} -
    \bigbraket{\pdv{\Psi'}{\alpha_j}}{\pdv{\Psi}{\alpha_i}}
    \label{eq:M_matrix_definition}
\end{align}
and a vector $\mbf{h}$ that contains energy derivatives:
\begin{align}
    h_i  &= \pdv{\mathcal{H}}{\alpha_i}, \quad \mathcal{H} = \elm{\Psi'}{H}{\Psi}.
    \label{eq:H_plain_def}
\end{align}
The matrix $\mbf{A}$ holds derivatives of the one-mode density matrices,
\begin{align} \label{eq:A_matrix}
    \Aplain{j}{m}{p}{q} &= \pdv{\rrho{m}{q}{p}}{\alpha_j}, \quad \rrho{m}{q}{p} = \elm{\Psi'}{\Etilde{m}{p}{q}}{\Psi}.
\end{align}
The notation $(m \, p^m q^m)$ is understood to denote a single compound index.

\subsubsection{Basis set parameters}
Before considering the basis set equations derived from the real Lagrangian $\bar{\mathcal{L}}$,
we recall the analogous equations from Ref.~\citenum{hojlundBivariationalTimedependentWave2022}:
\begin{align} 
    \mbf{0} = \mbf{F}^m  - \big( \bm{\rho}^m \mbf{G}^m  - \mbf{G}^m \bm{\rho}^m  \big)
    + i \dot{\bm{\rho}}^m,  \label{eq:g_equations_nonsym}
\end{align}
The matrix $\mbf{F}^m$ has elements
\begin{align} \label{eq:F_matrix_element}
    \Fplain{m}{q}{p} &= \elm{\Psi'}{[H, \Etilde{m}{p}{q}]}{\Psi} \nn
    &= \elm{\Psi'}{[H, \creatilde{m}{p}] \annitilde{m}{q}}{\Psi} - \elm{\Psi'}{\creatilde{m}{p} [\annitilde{m}{q}, H]}{\Psi}
\end{align}
and is thus the difference between the generalized mean-field or Fock matrices
$\FFtildeprime{m}$ and $\FFtildeplain{m}$:
\begin{subequations} \label{eq:F_tilde_elements}
    \begin{align}
        \Ftildeprime{m}{q}{p} 
        &= \elm{\Psi'}{[H, \creatilde{m}{p}] \annitilde{m}{q}}{\Psi}, \\
       \Ftildeplain{m}{q}{p} 
        &= \elm{\Psi'}{\creatilde{m}{p} [\annitilde{m}{q}, H]}{\Psi}.
    \end{align}
\end{subequations}

We showed in Ref.~\citenum{hojlundBivariationalTimedependentWave2022} that the element indexed by $(q^m p^m)$
of the complex bivariational constraint equations, Eq.~\eqref{eq:g_equations_nonsym},
can be written as
\begin{multline} \label{eq:g_equations_elementwise}
    \sum_{m'} \sum_{r^{m'} s^{m'}} \bigg[ 
        \Cplain{m}{p}{q}{m}{r}{s} +
        \sum_{ij} \Aplain{i}{m}{p}{q} (\mbf{M}^{-1})_{ij} \Aprime{j}{m}{r}{s}
    \bigg]
    \gplainprime{m}{r}{s} \\
    = 
    \elm{\Psi'}{ [H, \Etilde{m}{p}{q}] }{\Psi} + 
    \sum_{ij} \Aplain{i}{m}{p}{q} (\mbf{M}^{-1})_{ij} h_j
\end{multline}
with the definition
\begin{align}
    \Cplain{m}{p}{q}{m}{r}{s} &= \elm{\Psi'}{[ \Etildeprime{m}{r}{s}, \Etilde{m}{p}{q}]}{\Psi} \nn
    &= \delta_{mm'} ( 
        \delta_{s^m p^m} \rrho{m}{q}{r}
        - \delta_{r^m q^m} \rrho{m}{s}{p} 
    ). \label{eq:C_element}
\end{align}
In matrix notation, Eq.~\eqref{eq:g_equations_elementwise} reads
\begin{align}
    (\mbf{C} + \mbf{A}^{\trans} \mbf{M}^{-1} \mbf{A}) \mbf{g} = \mbf{f} + \mbf{A}^{\trans} \mbf{M}^{-1} \mbf{h}
\end{align}
or, even more compactly, 
\begin{align} \label{eq:compact_constraint_bivariational}
    \mbf{C}' \mbf{g} = \mbf{f}' \mbf{h}.
\end{align}
The vector $\mbf{g}$ simply contains all constraint elements for all modes.
Equation~\eqref{eq:compact_constraint_bivariational} leads to generic or non-Hermitian (i.e. neither Hermitian nor anti-Hermitian) constraint matrices
and biorthonormal basis functions.

The real Lagrangian considered in this paper leads to Hermitian constraint matrices and
orthonormal basis functions. The derivation of the constraint equations is not
complicated and can be found in Appendix~\ref{appendix:basis_set_eoms}.
The result simply reads
\begin{align} 
    \mbf{0} &= \mathbb{A} \Big[ \mbf{F}^m  - \big( \bm{\rho}^m \mbf{G}^m  - \mbf{G}^m \bm{\rho}^m  \big)
    + i \dot{\bm{\rho}}^m \Big], \label{eq:g_equations_antiherm}
\end{align}
where $\mathbb{A}[\,\cdot\,]$ denotes the anti-Hermitian part of a square matrix
(later, we will use $\mathbb{H}[\,\cdot\,]$ for the Hermitian part).
It is interesting to note that Eq.~\eqref{eq:g_equations_antiherm} (derived from a real Lagrangian) is exactly the anti-Hermitian
part of Eq.~\eqref{eq:g_equations_nonsym} (derived from a complex Lagrangian).
This shows very clearly how the symmetrization of the Lagrangian induces a symmetrization of the constraint equations.

The elementwise representation of Eq.~\eqref{eq:g_equations_antiherm} is obtained
by taking element $(q^m p^m)$ of Eq.~\eqref{eq:g_equations_elementwise}, subtracting from it the
complex conjugate of element $(p^m q^m)$ and multiplying the result by one half. Using the fact that the constraint matrices
are Hermitian, the result reads
\begin{multline} \label{eq:g_equations_antiherm_elementwise}
    \sum_{m'} \sum_{r^{m'} s^{m'}} 
    \begin{alignedat}[t]{2}
        \frac{1}{2}
        \bigg[ \; &\Cplain{m}{p}{q}{m}{r}{s} {} + {}
        &&\sum_{ij} \Aplain{i}{m}{p}{q} (\mbf{M}^{-1})_{ij} \Aprime{j}{m}{r}{s} \\
        -&\Cplain{m}{q}{p}{m}{s}{r}^* {} - {}
        &&\sum_{ij} \Aplain{i}{m}{q}{p}^* (\mbf{M}^{-1})^*_{ij} \Aprime{j}{m}{s}{r}^* \bigg] \gplainprime{m}{r}{s}
    \end{alignedat}
     \\
    \begin{alignedat}[t]{3}
        = \; \frac{1}{2} \bigg[ &\elm{\Psi'}{ [H, \Etilde{m}{p}{q}] }{\Psi} {}&&+{} 
        &&\sum_{ij} \Aplain{i}{m}{p}{q} (\mbf{M}^{-1})_{ij} h_j \\
        - &\elm{\Psi'}{ [H, \Etilde{m}{q}{p}] }{\Psi}^* {}&&-{} 
        &&\sum_{ij} \Aplain{i}{m}{q}{p}^* (\mbf{M}^{-1})_{ij}^* h_j^* \bigg].
    \end{alignedat}
\end{multline}
Note how the modal indices are exchanged in the complex conjugate terms:
\begin{align}
    (p^m q^m) &\rightarrow (q^m p^m), \nn
    (r^{m'} s^{m'}) &\rightarrow (s^{m'} r^{m'}).
\end{align}
This exchange of indices can be performed by a block diagonal permutation matrix $\mbf{S}$ with elements
\begin{subequations}
    \begin{align}
        \Splain{m}{p}{q}{m}{r}{s} &= \delta_{mm'} S^{m}_{(p^m q^m)(r^m s^m)}, \\
        S^{m}_{(p^m q^m)(r^m s^m)} &= \delta_{p^m s^m} \delta_{q^m r^m}.
    \end{align}
\end{subequations}
This allows us to write Eq.~\eqref{eq:g_equations_antiherm_elementwise} as
\begin{align} \label{eq:g_equations_antiherm_matrix}
    \tfrac{1}{2} (\mbf{C}' - \mbf{S} \mbf{C}^{\prime*} \mbf{S}) \mbf{g} = \tfrac{1}{2}( \mbf{f}' - \mbf{S} \mbf{f}^{\prime *} )
\end{align}
or, with obvious definitions,
\begin{align} \label{eq:g_equations_antiherm_matrix_compact}
    \bar{\mbf{C}}' \mbf{g} = \bar{\mbf{f}}'.
\end{align}
It is easy to verify that $\mbf{S} = \mbf{S}^{-1} = \mbf{S}^{\trans}$, which
enables a concise statement of the symmetries of Eq.~\eqref{eq:g_equations_antiherm_matrix_compact}:
\begin{subequations} \label{eq:symmetry_of_C_and_f}
    \begin{align}
        \mbf{S} \bar{\mbf{C}}' \mbf{S} &= - \mbf{C}^{\prime*}, \\
        \mbf{S} \bar{\mbf{f}}'         &= - \mbf{f}^{\prime*}.
    \end{align}
\end{subequations}
These symmetries embody the Hermiticity of the constraint matrices.
Table~\ref{tab:symmetrization} provides an overview of the effect of symmetrizing the Lagrangian.
\setlength{\tabcolsep}{8pt}
\begin{table}[H]
    \centering
    \caption{Effect of symmetrizing the Lagrangian. 
    See text for definitions.}
    \footnotesize
    \begin{tabular}{
    l
    c
    c
    }
\toprule
    & Complex Lagrangian & Real Lagrangian  \\
\midrule
Lagrangian & $\mathcal{L}$ & $\tfrac{1}{2}(\mathcal{L} + \mathcal{L}^*)$ \\
Basis set  & Biorthogonal & Orthogonal \\
Constraint matrix & Generic & Hermitian \\
Ket basis evolution & $\dot{\mbf{U}}^m = -i\mbf{U}^m \mbf{G}^m $ & $\dot{\mbf{V}}^m = -i\mbf{V}^m \mbf{G}^m$  \\
Bra basis evolution & $\dot{\mbf{W}}^m = +i\mbf{G}^m \mbf{U}^m $ & $\dot{\mbf{V}}^{m\dagger} = +i\mbf{G}^m \mbf{V}^{m\dagger} $ \\
Constraint eqs. (a) & 
$\mbf{0} = \mbf{F}^m  - \big( \bm{\rho}^m \mbf{G}^m  - \mbf{G}^m \bm{\rho}^m  \big) + i \dot{\bm{\rho}}^m$
& 
$\mbf{0} = \mathbb{A} \big[ \mbf{F}^m  - \big( \bm{\rho}^m \mbf{G}^m  - \mbf{G}^m \bm{\rho}^m  \big) + i \dot{\bm{\rho}}^m \big]$
\\
Constraint eqs. (b) & 
$\mbf{C}' \mbf{g} = \mbf{f}' \mbf{h}$ 
& 
$(\mbf{C}' - \mbf{S} \mbf{C}^{\prime*} \mbf{S}) \mbf{g} = \mbf{f}' - \mbf{S} \mbf{f}^{\prime *}$
\\
\bottomrule           
\end{tabular}
\label{tab:symmetrization}%
\end{table}%

\subsubsection{Analysis of the basis set equations}

\begin{figure}[h]
    \centering
    \includegraphics[width = 0.8\textwidth]{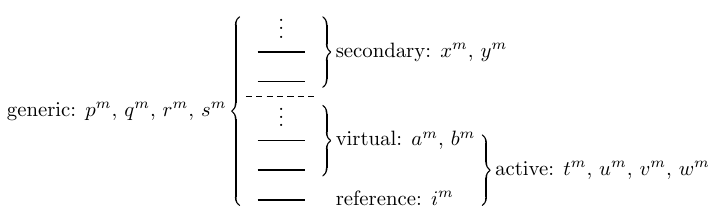}
    \caption{Summary of index conventions used throughout the text.}
    \label{fig:index_convention}
\end{figure}

\begin{figure}[h]
    \centering
    \includegraphics[width = 0.5\textwidth]{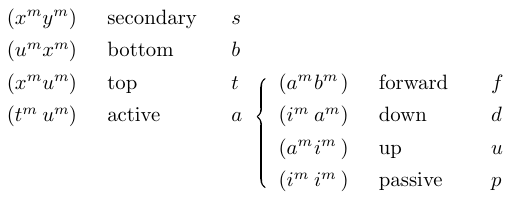}
    \caption{Summary of index pair conventions used throughout the text.}
    \label{fig:index_pair_convention}
\end{figure}

Before performing the analysis, it is convenient to introduce 
some notation to denote various kinds of indices and pairs of indices; see
Figs.~\ref{fig:index_convention} and \ref{fig:index_pair_convention}.
The vibrational case differs from the electronic case by
having a separate basis set for each mode (and thus a separate set of indices)
and by having only one reference (or occupied) index per mode. These differences have some
consequences in terms of the concrete appearance of the final working equations.
The general analysis, however, remains valid in both cases.

With the notation in place, we need to consider the different blocks
of Eq.~\eqref{eq:g_equations_antiherm_elementwise}. The first important
point is that
\begin{align}
    \Aplain{i}{m}{p}{q} = \pdv{\rrho{m}{q}{p}}{\alpha_i} = 0 \quad \text{if $p^m$ or $q^m$ is secondary}
\end{align}
since the wave function is completely contained within the active space.
The second important point is that the $\Cplain{m}{p}{q}{m}{r}{s}$
elements vanish in many cases (this is easily shown using second
quantization commutator relations and killer conditions).
Using these simplications, Eq.~\eqref{eq:g_equations_antiherm_matrix_compact}
reduces to
\begin{align}
	\left[
	\begin{array}{c | c | c | c} 
	{^{aa}\bar{\mbf{C}}'}  & \mbf{0} & \mbf{0}          & \mbf{0} \TTstrut
	\\ \hline
	\mbf{0}     	         & \mbf{0} & \; {^{tb}\bar{\mbf{C}}} \;{} & \mbf{0} \TTstrut
	\\ \hline
	\mbf{0}     	         & \; {^{bt}\bar{\mbf{C}}} \;{} & \mbf{0} & \mbf{0} \TTstrut
	\\ \hline
	\mbf{0}     	         & \mbf{0}                  & \mbf{0} & \hspace{2ex}\mbf{0}{\hspace{1.8ex}} \TTstrut
	\end{array}\right] 
    \left[
    \begin{array}{c} 
    {^{a}\mbf{g}} \TTstrut
    \\ \hline
    {^{t}\mbf{g}} \TTstrut
    \\ \hline
    {^{b}\mbf{g}} \TTstrut
    \\ \hline
    {^{s}\mbf{g}} \TTstrut
    \end{array}\right] 
	=
	\left[
	\begin{array}{c} 
	{^a\bar{\mbf{f}}'} \TTstrut
	\\ \hline
	{^t\bar{\mbf{f}}} \TTstrut
	\\ \hline
	{^b\bar{\mbf{f}}} \TTstrut
	\\ \hline
	\mbf{0} \TTstrut
	\end{array}\right] 
\end{align}
or, equivalently,
\begin{subequations}
\begin{align}
    {^{aa}\bar{\mbf{C}}'} \; {^{a}\mbf{g}} &= {^a\bar{\mbf{f}}'}, \label{eq:g_active_eq} \\
    {^{tb}\bar{\mbf{C}}}  \; {^{b}\mbf{g}} &= {^t\bar{\mbf{f}}},  \label{eq:g_bottom_eq} \\
    {^{bt}\bar{\mbf{C}}}  \; {^{t}\mbf{g}} &= {^b\bar{\mbf{f}}}.  \label{eq:g_top_eq}
\end{align}
\end{subequations}
Several comments are in order: 
(i) the ${^{s}\mbf{g}}$ elements are redundant and may be chosen 
freely (in a Hermitian fashion); 
(ii) the concrete structure of Eq.~\eqref{eq:g_active_eq} (which is important from an implementation point of view) depends
on the wave function type, e.g. coupled cluster;
(iii) the absence of primes in Eqs.~\eqref{eq:g_bottom_eq} and \eqref{eq:g_top_eq} indicates 
the absence of terms involving the $\Aplain{i}{m}{p}{q}$ elements; and
(iv) only one of Eqs.~\eqref{eq:g_bottom_eq} and \eqref{eq:g_top_eq} needs to be solved
since the constraint matrices are Hermitian. 

It is a simple matter (see Appendix~\ref{appendix:gtop_elementwise}) to show that Eq.~\eqref{eq:g_top_eq} reduces to
\begin{subequations}
    \begin{align}
        {^t\mbf{G}^m} \; \mathbb{H}[{^{a\!}}\bm{\raisedrho}^m]
        &= \tfrac{1}{2} ( {^t\FFtildeplain{m}} + {^b\FFtildeprimedagger{m}}) \\
        &= \tfrac{1}{2} \mbf{V}^{m\dagger}_{\!\!\subS} ( \FFcheckplain{m}_{\!\!\subA} + \FFcheckprimedagger{m}_{\!\!\subA}), \label{eq:Gt_equation_halftrans}
    \end{align}
\end{subequations}
where we have introduced the half-transformed mean-field matrices $\FFcheckplain{m}$ and $\FFcheckprime{m}$:
\begin{subequations}
    \begin{align}
        \Fcheckprime{m}{q}{\alpha} 
        &= \elm{\Psi'}{[H, \crea{m}{\alpha}] \annitilde{m}{q}}{\Psi}, \\
        \Fcheckplain{m}{\alpha}{q}
        &= \elm{\Psi'}{\creatilde{m}{q} [\anni{m}{\alpha}, H]}{\Psi}.
    \end{align}
\end{subequations}

\subsubsection{Secondary-space projection}
Writing the active and secondary parts explicitly (and dropping the mode index for simplicity),
Eq.~\eqref{eq:V_dot_general} reads
\begin{align}
    i
    \left[
	\begin{array}{c | c}
        \dot{\mbf{V}}_{\!\!\subA} {\,} & {\,} \dot{\mbf{V}}_{\!\!\subS} \Tstrut
	\end{array} \right]
    =
    \left[
	\begin{array}{c | c}
        \mbf{V}_{\!\!\subA} {\,} & {\,} \mbf{V}_{\!\!\subS} \Tstrut
	\end{array} \right]
    \left[
	\begin{array}{c | c}
	{^a\mbf{G}} {\,} & {\,} {^b\mbf{G}} \Tstrut
	\\ \hline
	{^t\mbf{G}} {\,} & {\,} {^s\mbf{G}}   \Tstrut
	\end{array} \right].
\end{align}
Assuming that we are able to solve Eq.~\eqref{eq:g_active_eq} for ${^a\mbf{G}}$,
the time derivative of the active basis functions becomes
\begin{align} \label{eq:VA_dot}
    i \dot{\mbf{V}}_{\!\!\subA} 
    &= \mbf{V}_{\!\!\subA} \, {^a\mbf{G}} 
      + \mbf{V}_{\!\!\subS} \, {^t\mbf{G}} \nn
    &= \mbf{V}_{\!\!\subA}  \, {^a\mbf{G}} 
     + \mbf{V}_{\!\!\subS}^{\phantom{\dagger}} \2 \mbf{V}_{\!\!\subS}^{\dagger}  \tfrac{1}{2} ( \check{\mbf{F}}_{\!\!\subA} + \check{\mbf{F}}^{\prime \dagger}_{\!\!\subA}) \, (\mathbb{H}[^{a\!}\bm{\raisedrho}\,])^{-1} \nn
     &= \mbf{V}_{\!\!\subA}  \, {^a\mbf{G}} 
     + \mbf{Q} \tfrac{1}{2} ( \check{\mbf{F}}_{\!\!\subA} + \check{\mbf{F}}^{\prime \dagger}_{\!\!\subA}) \, (\mathbb{H}[^{a\!}\bm{\raisedrho}\,])^{-1}
\end{align}
with the secondary-space projector $\mbf{Q}$ given by
\begin{align} \label{eq:Q_def}
    \mbf{Q} = \mbf{V}_{\!\!\subS}^{\phantom{\dagger}} \2 \mbf{V}_{\!\!\subS}^{\dagger} = \mbf{1} - \mbf{V}_{\!\!\subA}^{\phantom{\dagger}} \2 \mbf{V}_{\!\!\subA}^{\dagger}.
\end{align}
Equations~\eqref{eq:VA_dot} and \eqref{eq:Q_def} allow us to propagate the active basis functions without reference to secondary-space quantities.
For an \ac{mctdh} state, the density matrices are Hermitian and $\check{\mbf{F}} = \check{\mbf{F}}^{\prime \dagger}$, in which case
Eq.~\eqref{eq:VA_dot} reduces to the standard \ac{mctdh} expression\cite{beckMulticonfigurationTimedependentHartree2000}. 
\subsection{Application to coupled cluster} \label{sec:application_to_tdmvcc}
In this section we consider \acl{cc} expansions
of the form
\begin{subequations}
    \begin{align}
        \ket{\Psi}  &= \exp(T) \ket{\Phi} \\
        \bra{\Psi'} &= \bra{\Phi} L \exp(-T)
    \end{align}
\end{subequations}
where
\begin{subequations}
\begin{alignat}{2}
    T &= \sum_{\mu} t_{\mu} \tilde{\tau}_{\mu}            &&= t_0 + T_2 + T_3 + \ldots \\
    L &= \sum_{\mu} l_{\mu} \tilde{\tau}_{\mu}^{\dagger} &&= l_0 + L_2 + L_3 + \ldots
\end{alignat}
\end{subequations}
Note that single excitations are not included in the wave function since they are redundant
with the basis set transformations.\cite{kvaalInitioQuantumDynamics2012,satoCommunicationTimedependentOptimized2018,madsenTimedependentVibrationalCoupled2020,pedersenGaugeInvariantCoupled1999,pedersenGaugeInvariantCoupled2001}
Ordering the amplitudes like $\bm{\alpha} = (\mbf{s}, \mbf{l})$, one easily applies Eq.~\eqref{eq:M_matrix_definition} to show that
\begin{align} \label{eq:M_and_M_inv}
    \mbf{M} = 
    \left[
	\begin{array}{c | c} 
	\mbf{0}  & - \mbf{1} 
	\\ \hline
	+\mbf{1}  & \mbf{0}  
	\end{array} \right]
    , \quad
    \mbf{M}^{-1} = 
    \left[
	\begin{array}{c | c} 
	\mbf{0}  & +\mbf{1} 
	\\ \hline
	-\mbf{1}  & \mbf{0}  
	\end{array} \right],
\end{align}
while Eqs.~\eqref{eq:H_plain_def} and \eqref{eq:A_matrix} yield
\begin{subequations} \label{eq:CC_H_derivatives}
    \begin{alignat}{2}
        h'_{t_{\mu} } &= \pdv{\mathcal{H}'}{t_{\mu} } &&= \elm{\Psi'}{[H - g, \tilde{\tau}_{\mu}]}{\Psi},  \\
        h'_{l_{\mu} } &= \pdv{\mathcal{H}'}{l_{\mu} } &&= \elm{\mu'}{e^{-T}(H - g)}{\Psi},
    \end{alignat}
\end{subequations}
\begin{subequations} \label{eq:CC_rho_derivatives}
    \begin{alignat}{2}
        \Aplain{t_\mu }{m}{p}{q} &= \pdv{\rrho{m}{q}{p}}{t_{\mu} } &&= \elm{\Psi'}{[\Etilde{m}{p}{q}, \tilde{\tau}_{\mu}]}{\Psi},  \\
        \Aplain{l_\mu }{m}{p}{q} &= \pdv{\rrho{m}{q}{p}}{l_{\mu} } &&= \elm{\mu'}{e^{-T}\Etilde{m}{p}{q}}{\Psi}.
    \end{alignat}
\end{subequations}
The amplitude \acp{eom} follow from Eq.~\eqref{eq:alpha_EOM_expanded}:
\begin{subequations}
\begin{align}
    i \dot{t}_{\mu} &= + \elm{\mu'}{e^{-T}(H - g)}{\Psi}, \\
    i \dot{l}_{\mu} &= - \elm{\Psi'}{[H - g, \tilde{\tau}_{\mu}]}{\Psi}.
\end{align}
\end{subequations}
Our main concern is the active-space constraint equations. Referring to Eqs.~\eqref{eq:g_equations_antiherm_elementwise}
and \eqref{eq:g_equations_antiherm_matrix_compact} and using the equations above, the active-space elements read
\begin{multline} \label{eq:tdmvcc_active_space_c}
    2 \Cbarprime{m}{t}{u}{m}{v}{w} =
    \elm{\Psi'}{[\Etildeprime{m}{v}{w}, \Etilde{m}{t}{u}]}{\Psi} -
    \elm{\Psi'}{[\Etildeprime{m}{w}{v}, \Etilde{m}{u}{t}]}{\Psi}^* \\
    \begin{alignedat}[t]{2}
        &+ &&\sum_\mu \Big( 
        \elm{\Psi}{[\Etilde{m}{t}{u}, \tau_\mu]}{\Psi}   \elm{\mu}{e^{-T} \Etildeprime{m}{v}{w}}{\Psi} 
        - \elm{\Psi}{[\Etildeprime{m}{v}{w}, \tau_\mu]}{\Psi}   \elm{\mu}{e^{-T} \Etilde{m}{t}{u}}{\Psi}
        \Big) \\
        &- &&\sum_\mu \Big( 
        \elm{\Psi}{[\Etilde{m}{u}{t}, \tau_\mu]}{\Psi}   \elm{\mu}{e^{-T} \Etildeprime{m}{w}{v}}{\Psi} 
        - \elm{\Psi}{[\Etildeprime{m}{w}{v}, \tau_\mu]}{\Psi}   \elm{\mu}{e^{-T} \Etilde{m}{u}{t}}{\Psi}
        \Big)^*
    \end{alignedat}
\end{multline}
and
\begin{multline} \label{eq:tdmvcc_active_space_f}
    2 \fbarprime{m}{t}{u} = 
    \elm{\Psi'}{[H, \Etilde{m}{t}{u}]}{\Psi} -
    \elm{\Psi'}{[H, \Etilde{m}{u}{t}]}{\Psi}^* \\
    \begin{alignedat}[t]{2}
        &+ &&\sum_\mu \Big( 
        \elm{\Psi}{[\Etilde{m}{t}{u}, \tau_\mu]}{\Psi}   \elm{\mu}{e^{-T} H}{\Psi} 
        - \elm{\Psi}{[H, \tau_\mu]}{\Psi}   \elm{\mu}{e^{-T} \Etilde{m}{t}{u}}{\Psi}
        \Big) \\
        &- &&\sum_\mu \Big( 
        \elm{\Psi}{[\Etilde{m}{u}{t}, \tau_\mu]}{\Psi}   \elm{\mu}{e^{-T} H}{\Psi} 
        - \elm{\Psi}{[H, \tau_\mu]}{\Psi}   \elm{\mu}{e^{-T} \Etilde{m}{u}{t}}{\Psi}
        \Big)^*.
    \end{alignedat}
\end{multline}
The full expressions appear quite complicated, but it turns out that
many elements simplify considerably or vanish completely. The full
analysis is somewhat tedious, but we can luckily reuse the main points
from Ref.~\citenum{madsenTimedependentVibrationalCoupled2020}.
Here, it was shown that blocks having at least one index pair of type passive ($i^m i^m$) or
forward $(a^m b^m)$ vanish; see Fig.~\eqref{fig:index_pair_convention} for details on the index pair nomenclature.
This property is conserved after symmetrization, so the overall structure of the active-space equations is
\begin{align} \label{eq:tdmvcc_constraint_equations_structure}
	\left[
	\begin{array}{c | c | c | c} 
    {^{uu}\bar{\mbf{C}}}'  & {^{ud}\bar{\mbf{C}}}  & \mbf{0} & \mbf{0} \TTstrut
	\\ \hline
	{^{du}\bar{\mbf{C}}}   & {^{dd}\bar{\mbf{C}}}' & \mbf{0} & \mbf{0} \TTstrut
	\\ \hline
	\mbf{0}     	                & \mbf{0} & \mbf{0} & \mbf{0} \TTstrut
	\\ \hline
	\mbf{0}     	                & \mbf{0} & {\hspace{1.7ex}}\mbf{0}{\hspace{1.7ex}} & \hspace{1.7ex}\mbf{0}{\hspace{1.5ex}} \TTstrut
	\end{array}\right] 
    \left[
    \begin{array}{c} 
    {^{u}\mbf{g}} \TTstrut
    \\ \hline
    {^{d}\mbf{g}} \TTstrut
    \\ \hline
    {^{f\!}\mbf{g}} \TTstrut
    \\ \hline
    {^{p}\mbf{g}} \TTstrut
    \end{array}\right] 
	=
	\left[
	\begin{array}{c} 
	{^u\bar{\mbf{f}}'} \TTstrut
	\\ \hline
	{^d\bar{\mbf{f}}'} \TTstrut
	\\ \hline
	\mbf{0} \TTstrut
	\\ \hline
	\mbf{0} \TTstrut
	\end{array}\right].
\end{align}
We see that the forward and passive constraint elements are redundant while the up and down elements (which mix the occupied and virtual spaces)
are non-redundant. The non-zero elements in Eq.~\eqref{eq:tdmvcc_constraint_equations_structure} can be computed
by noting that
\begin{subequations} \label{eq:zero_shift_oper_expressions}
\begin{align}
    \elm{\Psi}{[\Etilde{m}{a}{i}, \tau_\mu]}{\Psi} &= 0, \\ 
    \elm{\mu}{e^{-T} \Etilde{m}{a}{i}}{\Psi}  &= 0.
\end{align}
\end{subequations}
The former holds trivially since the two excitation operators $\Etilde{m}{a}{i}$ and $\tau_\mu$ commute,
while the latter holds since single excitations are excluded from the wave function (see the appendix of Ref.~\citenum{madsenTimedependentVibrationalCoupled2020} for a proof).
Combining these properties with Eqs.~\eqref{eq:tdmvcc_active_space_c} and \eqref{eq:tdmvcc_active_space_f}
now yields
\begin{align} \label{eq:C_ud_TDMVCC}
    {^{ud}\Cbarplain{m}{a}{i}{m}{i}{b}} 
    &=
    \frac{1}{2} \Big(
    \elm{\Psi'}{[\Etildeprime{m}{i}{b}, \Etilde{m}{a}{i}]}{\Psi} -
    \elm{\Psi'}{[\Etildeprime{m}{b}{i}, \Etilde{m}{i}{a}]}{\Psi}^*  \Big) \nn
    &= \delta_{mm'} \big( \delta_{a^m b^m} \mathbb{H}[\bm{\raisedrho}^m]_{i^m i^m} - \mathbb{H}[\bm{\raisedrho}^m]_{b^m a^m} \big)
\end{align}
\begin{align} \label{eq:C_dd_TDMVCC}
    {^{dd}\Cbarprime{m}{i}{a}{m}{i}{b}} &=
    \frac{1}{2}
    \begin{aligned}[t]
        \sum_\mu \Big( 
            &\elm{\Psi}{[\Etilde{m}{i}{a}, \tau_\mu]}{\Psi}   \elm{\mu}{e^{-T} \Etildeprime{m}{i}{b}}{\Psi} \\
            - &\elm{\Psi}{[\Etildeprime{m}{i}{b}, \tau_\mu]}{\Psi}   \elm{\mu}{e^{-T} \Etilde{m}{i}{a}}{\Psi}
            \Big),
    \end{aligned}
\end{align}
\begin{multline} \label{eq:f_d_TDMVCC}
    {^{d\!}\fbarprime{m}{i}{a}} =
    \frac{1}{2} \Big(
    \elm{\Psi'}{[H, \Etilde{m}{i}{a}]}{\Psi} -
    \elm{\Psi'}{[H, \Etilde{m}{a}{i}]}{\Psi}^* \Big) \\
    \begin{alignedat}[t]{2}
        {}+ \frac{1}{2} \sum_\mu \Big( 
        \elm{\Psi}{[\Etilde{m}{i}{a}, \tau_\mu]}{\Psi}   \elm{\mu}{e^{-T} H}{\Psi} 
        - \elm{\Psi}{[H, \tau_\mu]}{\Psi}   \elm{\mu}{e^{-T} \Etilde{m}{i}{a}}{\Psi}
        \Big).
    \end{alignedat}
\end{multline}
The remaining elements are given by symmetry; see Eqs.~\eqref{eq:g_equations_antiherm_elementwise} and \eqref{eq:symmetry_of_C_and_f}:
\begin{subequations} 
    \begin{align}
        {^{du}\Cbarplain{m}{i}{a}{m}{b}{i}} &= -{^{ud}\Cbarconj{m}{a}{i}{m}{i}{b}}, \\
        {^{uu}\Cbarprime{m}{a}{i}{m}{b}{i}} &= -{^{dd}\Cbarprimeconj{m}{i}{a}{m}{i}{b}}, \\
        {^{u\!}\fbarprime{m}{a}{i}} &= -{^{d\!}\fbarprimeconj{m}{i}{a}}.
    \end{align}
\end{subequations}
We remark that the summations over $\mu$ in Eqs.~\eqref{eq:C_dd_TDMVCC} and \eqref{eq:f_d_TDMVCC} vanish if $T$ and $L$ are truncated after the doubles.\cite{madsenTimedependentVibrationalCoupled2020}
This has the effect that $\bar{\mbf{C}}'$ becomes block diagonal in the mode index,
i.e. the oTDMVCC$[2]$ constraint equations can be solved one mode at a time, which is a significant simplification
that is also observed for TDMVCC$[2]$.
In other words, the oTDMVCC$[2]$ and TDMVCC$[2]$ methods involve the same computational effort.
For excitation levels higher than $n=2$, the oTDMVCC$[n]$ equations involve mode-mode coupling,
while the TDMVCC$[n]$ equations can still be solved mode by mode.\cite{madsenTimedependentVibrationalCoupled2020}
Effectively, the symmetrization
of the constraint equations replaces many small sets of linear equations (one set for each mode)
with one large set. The oTDMVCC$[n]$ hierarchy is thus more involved in terms of implementation and computational effort.

In the electronic structure case, the detailed expressions look slightly different.
They are stated in Appendix~\ref{appendix:constraint_eqs_electronic_structure} for the interested reader and can be compared to the
work on \ac{tdocc} by Sato et al.\cite{satoCommunicationTimedependentOptimized2018} 

\section{Implementation} \label{sec:implementation}
The \ac{otdmvcc} method has been implemented in the \ac{midas}\cite{MidasCpp2023040}.
At the two-mode excitation level and for Hamiltonians with one- and two-mode couplings
(oTDMVCC[2]/H2), the code uses the efficient TDMVCC[2]/H2 implementation of Ref.~\citenum{jensenEfficientTimedependentVibrational2023},
which allows computations on large systems.
The only modification necessary was the symmetrization of mean-field and density matrices,
which carries negligible cost.

For higher excitation and coupling levels, the implementation is based on the \ac{fsmr}
framework introduced in Ref.~\citenum{hansenExtendedVibrationalCoupled2020}. This
is essentially a \ac{fci} code, so the computational effort scales exponentially with respect to the number
of modes. 
To solve the active-space constraint equations, Eq.~\eqref{eq:g_active_eq},
we compute the matrix ${^{aa}\bar{\mbf{C}}'}$ and perform a \ac{svd}:
\begin{align}
    {^{aa}\bar{\mbf{C}}'} = \bm{\mathcal{U}} \mbf{\Sigma} \bm{\mathcal{V}}^{\dagger}.
\end{align}
The \ac{svd} is regularized to avoid singularities before the inverse is computed:
\begin{align}
    {^{aa}\bar{\mbf{C}}'_\mrm{reg}} = \bm{\mathcal{U}} \big( \mbf{\Sigma} + \exp(-\mbf{\Sigma} / \epsilon_\mrm{reg}) \big) \bm{\mathcal{V}}^{\dagger}.
\end{align}
Here, $\epsilon_\mrm{reg}$ is a small regularization parameter (typically, $\epsilon_\mrm{reg} = 10^{-8}$).
The constraint elements are finally obtained as
\begin{align}
    {^{a}\mbf{g}} 
    = 
    \left[
    \begin{array}{c} 
    {^{u}\mbf{g}} 
    \\ \hline
    {^{d}\mbf{g}} 
    \end{array}\right] 
    = 
    \big({^{aa}\bar{\mbf{C}}'_\mrm{reg}} \big)^{-1} \; {^a\bar{\mbf{f}}'}.
\end{align}
The symmetries of ${^{aa}\bar{\mbf{C}}'}$ ensure that ${^{u}\mbf{g}} = {^{d}\mbf{g}^*}$ to numerical precision
so that the constraint matrices are properly Hermitian.
However, this also means that we are, in a sense, solving twice for the same constraint elements.
It is clear that an efficient, scalable implementation should 
(i) utilize the symmetries at hand and
(ii) use iterative solvers. Such refinements are, however, beyond the scope of this paper.

To improve numerical stability, the secondary-space projector in Eq.~\eqref{eq:Q_def} is replaced by a modified
projector
\begin{align} 
    \mbf{Q}_\mrm{mod} = \mbf{1} - \mbf{V}_{\!\!\subA}^{\phantom{\dagger}} (\mbf{V}_{\!\!\subA}^{\dagger} \mbf{V}_{\!\!\subA}^{\phantom{\dagger}})^{-1} \, \mbf{V}_{\!\!\subA}^{\dagger}.
\end{align}
This procedure, which is commonly used in the \ac{mctdh}\cite{beckMulticonfigurationTimedependentHartree2000} community, ensures that $\mbf{Q}_\mrm{mod}$
remains a proper projector, even if the basis is not strictly orthonormal due to numerical noise and integration error.

Proving the correctness of the implementation is somewhat challenging since the \ac{otdmvcc} method
does not formally converge to the \ac{tdfvci} or \ac{mctdh} limit. Perfect agreement with an exact reference calculation
can therefore not be expected. The \ac{tdmvcc} hierarchy, on the other hand, does converge to the exact solution,
and we have shown that the \ac{tdmvcc} and \ac{otdmvcc}
methods differ only by symmetrization of the constraint equations.
We have thus written a new implementation for the \ac{tdmvcc} constraint equations that
explicitly constructs and inverts the full matrix $^{aa}\mbf{C}'$ (rather than solving
the equations mode by mode). This code was validated against
a \ac{tdfvci} calculation. The \ac{otdmvcc} code computes $^{aa}\mbf{C}'$
and explicitly applies the symmetrization to obtain $^{aa}\bar{\mbf{C}}'$. 
\section{Numerical examples} \label{sec:results}
We consider a few numerical examples in order to study the convergence of the \ac{otdmvcc} 
hierarchy relative to the \ac{tdfvci} ($N = N_{\2\subA}$, i.e. no basis splitting) and \ac{mctdh} ($N > N_{\2\subA}$) limits.
Two examples from Refs.~\citenum{madsenTimedependentVibrationalCoupled2020} and \citenum{hojlundGeneralExponentialBasis2023a}
are studied in detail, and three additional examples are assessed.
We also compare with the \ac{tdmvcc}\cite{madsenTimedependentVibrationalCoupled2020,hojlundBivariationalTimedependentWave2022} hierarchy, 
which is known to converge correctly to \ac{tdfvci} and \ac{mctdh}.
The original \ac{tdmvcc} method is not always numerically stable when $N > N_{\2\subA}$
for reasons that were analyzed in Ref.~\citenum{hojlundBivariationalTimedependentWave2022}.
For the present calculations we observed no problems, but we note that the so-called restricted polar \ac{tdmvcc}
approach of Ref.~\citenum{hojlundBivariationalTimedependentWave2022}
restores stability in difficult cases.

The first example is the \ac{ivr} of water after excitation of the symmetric stretch to $n=2$.
The initial state is the harmonic oscillator state $[0,2,0]$ corresponding to the harmonic part
of the \ac{pes}. The wave function is then propagated on the full anharmonic and coupled \ac{pes}
at the oTDMVCC[2--3], TDMVCC[2--3] and \ac{tdfvci} levels. The calculations use $N = 8$ primitive
and $N_{\2\subA} = 8$ active basis function per mode,
and the propagation time is \SI{10000}{au} ($\sim$\SI{242}{fs}).

We have furthermore considered the \ac{ivr} of ozone, sulfur dioxide and hydrogen sulfide using the \acp{pes}
of Ref.~\citenum{majlandOptimizingNumberMeasurements2023}. The computational setup is identical to
that of the water calculation, apart from the simulation time, which has been adjusted
to reflect the different characteristic time scales of the molecules. Ozone and sulfur dioxide
have thus been propagated for \SI{30000}{au} ($\sim$\SI{363}{fs}), while hydrogen sulfide
has been propagated for \SI{15000}{au} ($\sim$\SI{726}{fs}).

The final example is the $S_1 \rightarrow S_0$ emission of the
5D \textit{trans}-bithiophene model from Ref.~\citenum{madsenVibrationallyResolvedEmission2019}.
The molecule has 42 vibrational modes in total; the model includes
the normal coordinates $Q_{10}$, $Q_{12}$,
$Q_{19}$, $Q_{34}$ and $Q_{51}$.
The initial state is taken as the \ac{vscf} ground state of the $S_1$ electronic surface.
The wave packed is then placed on the $S_0$ surface and allowed to propagate
for a total time of \SI{10000}{au} ($\sim$\SI{242}{fs}).
The calculation is repeated at the oTDMVCC[2--5], TDMVCC[2--5] and \ac{mctdh} levels
using $N = 30$ and $N_{\2\subA} = 4$.

In all cases we used the \ac{dop853}\cite{hairerSolvingOrdinaryDifferential2009b} integrator 
with tight absolute and relative tolerances $\tau_\mrm{abs} = \tau_\mrm{rel} = 10^{-14}$. A regularization parameter
of $\epsilon_\mrm{reg} = 10^{-8}$ was used in all computations. 
We report \aclp{acf},
\begin{align}
    S(t) = \braket{\Psi'(0) | \Psi(t)},
\end{align}
and expectation values of the
displacement coordinates $Q$. For a bivariational ansatz such as \ac{cc},
we take the physical expectation
value of a Hermitian operator $\Omega$ to be
\begin{align}
    \langle \Omega \rangle = \mathrm{Re}\elm{\Psi'}{\Omega}{\Psi}.
\end{align}
The imaginary part is generally non-zero (except when the wave function
is exact) and is typically discarded since it has no physical meaning.
It can be useful, however,
as a diagnostic for the sensibleness of the wave function.
In the supplementary material we show $\mathrm{Im}\elm{\Psi'}{Q}{\Psi}$
and find that it is generally small and converges to zero for the \ac{tdmvcc} hierarchy.
For the \ac{otdmvcc} hierarchy, the imaginary part is typically larger, and does
not go to zero. The supplementary material also contains
additional details on, e.g., energy conservation.
Generally, we find that the physical energy $E = \mathrm{Re} (\mathcal{H}) = \mathrm{Re} \elm{\Psi'}{H}{\Psi}$
is conserved as one should expect when the Hamiltonian is time-independent.
\Ac{tdmvcc} (complex action) also conserves $\mathrm{Im} (\mathcal{H})$,
while \ac{otdmvcc} (real action) does not, except for certain special
cases that are discussed in Sec.~\ref{ssec:otdmvcc_tdmvcc_equivalence}.

For calculations with $N = N_{\2\subA}$, we also consider the
Hilbert space angle between the ket state $\ket{\Psi}$
and the \ac{tdfvci} state, i.e.
\begin{align}
    \theta = \arccos( \frac{\abs{ \braket{\Psi_\textsc{TDFVCI} | \Psi} }} { \sqrt{ \braket{\Psi_\textsc{TDFVCI} | \Psi_\textsc{TDFVCI}}  \braket{\Psi | \Psi   } } }  ).
\end{align}
One can similarly define a Hilbert space bra angle, which is generally different from the ket angle. Bra angles are shown in
the supplementary material.

\subsection{Water}

\subsubsection{Hilbert space angles}
\begin{figure}[H]
    \centering
    \includegraphics[width=0.80\textwidth]{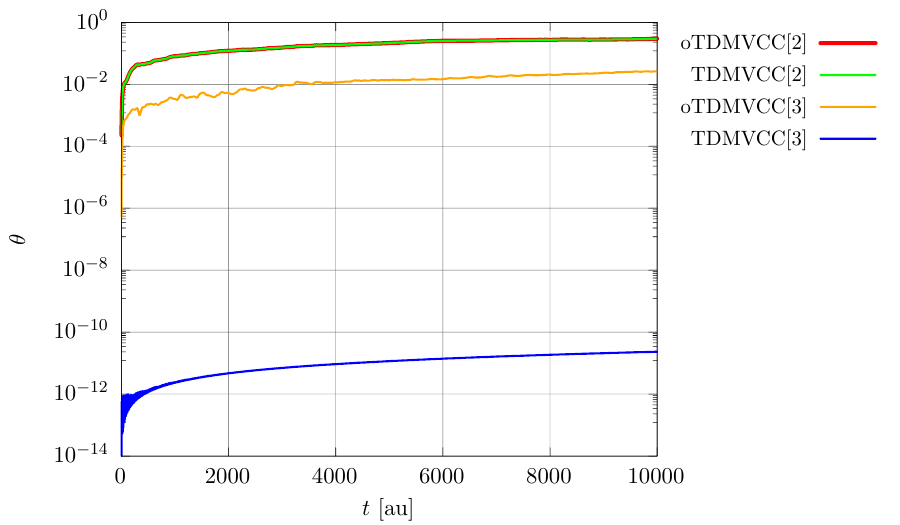}
    \caption{Hilbert space ket angles for water at the oTDMVCC[2--3] and TDMVCC[2--3] levels
    with $N = 8$ and $N_{\2\subA} = 8$ for all modes. The angles are computed relative to \ac{tdfvci}.}
    \label{fig:water_ketangle}
\end{figure}

Looking at the Hilbert space angles in Fig.~\ref{fig:water_ketangle}, it is evident
that oTDMVCC[3] is not equivalent to \ac{tdfvci}, i.e. the \ac{otdmvcc} hierarchy does
not converge to the exact limit in contrast to the \ac{tdmvcc} hierarchy.
We also note that oTDMVCC[2] and TDMVCC[2] perform identically (up to noise from the numerical integration).

%
%

\subsubsection{\Aclp{acf}}

\begin{figure}[H]
    \centering

    \begin{subfigure}[c]{0.80\textwidth}
        \centering
        \caption{Autocorrelation function}
        \includegraphics[width=\textwidth]{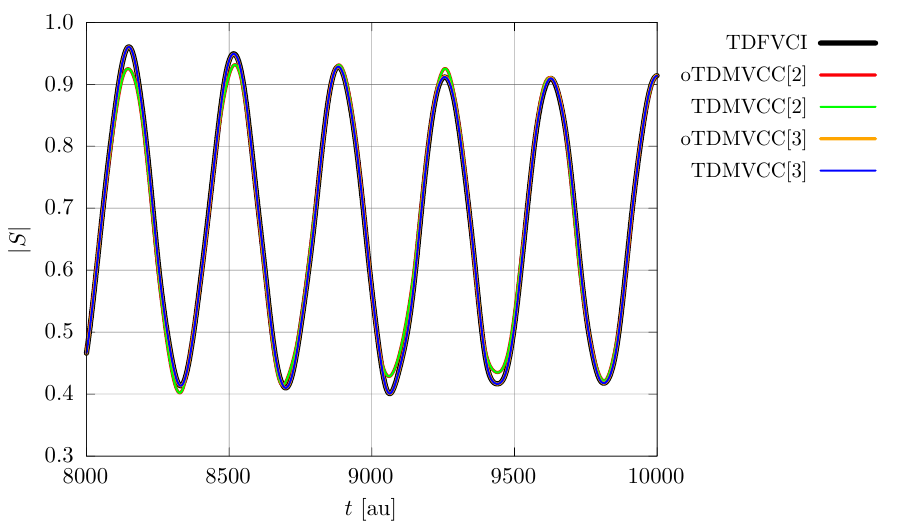}
        \label{fig:water_autocorr}
    \end{subfigure}


    \begin{subfigure}[c]{0.80\textwidth}
        \centering
        \caption{Absolute error in autocorrelation function}
        \includegraphics[width=\textwidth]{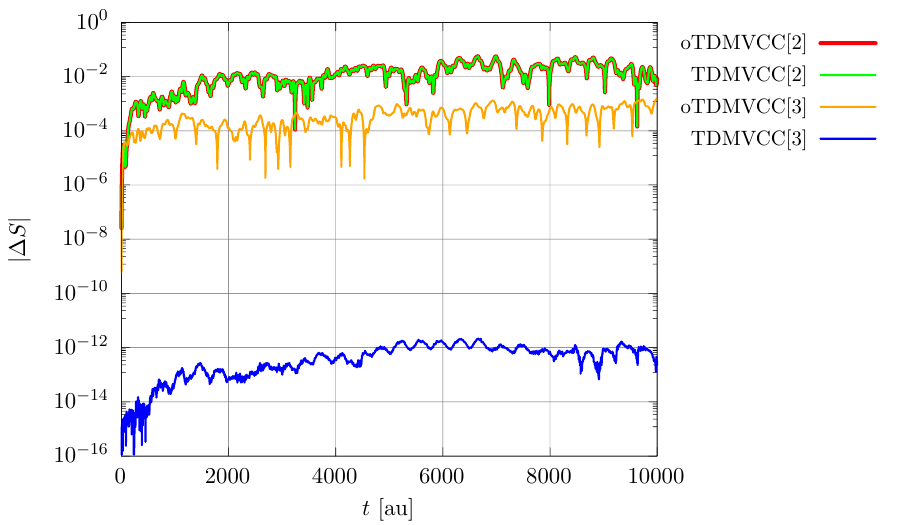}
        \label{fig:water_autocorr_diff}
    \end{subfigure}

    \caption{\Ac{ivr} of water at the oTDMVCC[2--3] and TDMVCC[2--3] levels
    with $N = 8$ and $N_{\2\subA} = 8$ for all modes. (a) \Acl{acf}. (b) Absolute error in the \acl{acf} (relative to \ac{tdfvci}).}
    \label{fig:water_autocorr_and_diff}
\end{figure}

Figure~\ref{fig:water_autocorr} shows that oTDMVCC[3] and TDMVCC[3] are
both visually converged with respect to the \ac{tdfvci} reference,
while oTDMVCC[2] and TDMVCC[2] exhibit a visible but rather modest error.
Although oTDMVCC[3] produces visually converged \aclp{acf},
the absolute error (Fig.~\ref{fig:water_autocorr_diff}) clearly shows that the \ac{tdfvci} result
is not exactly reproduced. 

%
%

\subsubsection{Expectation values}

\begin{figure}[H]
    \centering

    \begin{subfigure}[c]{0.55\textwidth}
        \centering
        \caption{}
        \includegraphics[width=\textwidth]{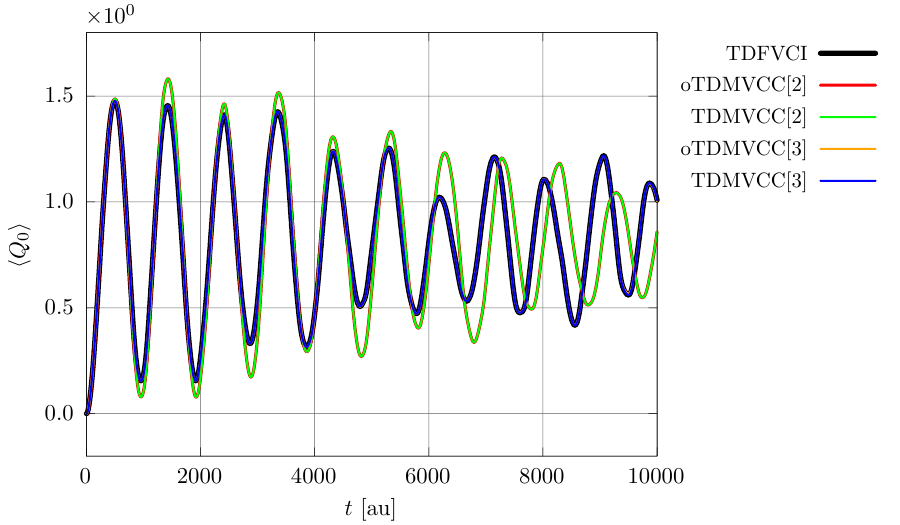}
        \label{fig:water_Q0}
    \end{subfigure}


    \begin{subfigure}[c]{0.55\textwidth}
        \centering
        \caption{}
        \includegraphics[width=\textwidth]{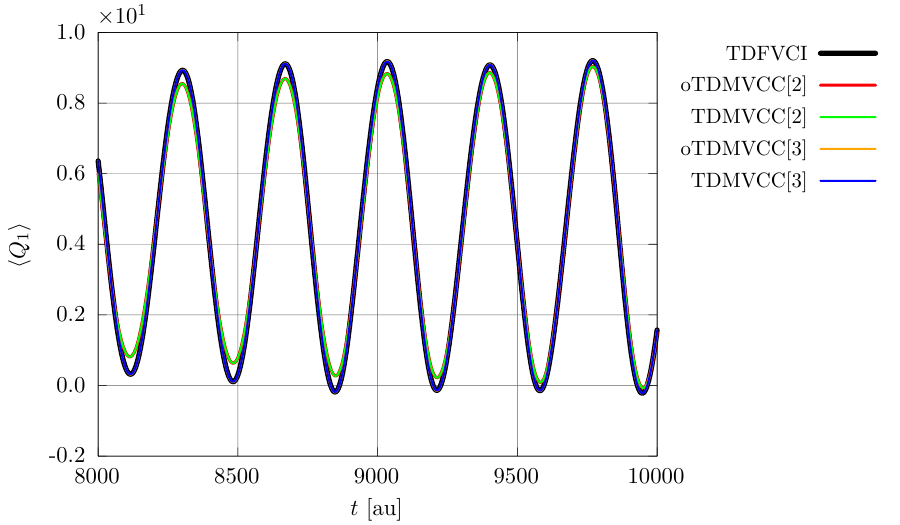}
        \label{fig:water_Q1}
    \end{subfigure}

    \caption{\Ac{ivr} of water at the oTDMVCC[2--3] and TDMVCC[2--3] levels
    with $N = 8$ and $N_{\2\subA} = 8$ for all modes. (a) Expectation value of $Q_0$ (bend). (b) Expectation value of $Q_1$ (symmetric stretch).}
    \label{fig:water_Q0_Q1}
\end{figure}

\begin{figure}[H]
    \centering

    \begin{subfigure}[c]{0.75\textwidth}
        \centering
        \caption{}
        \includegraphics[width=\textwidth]{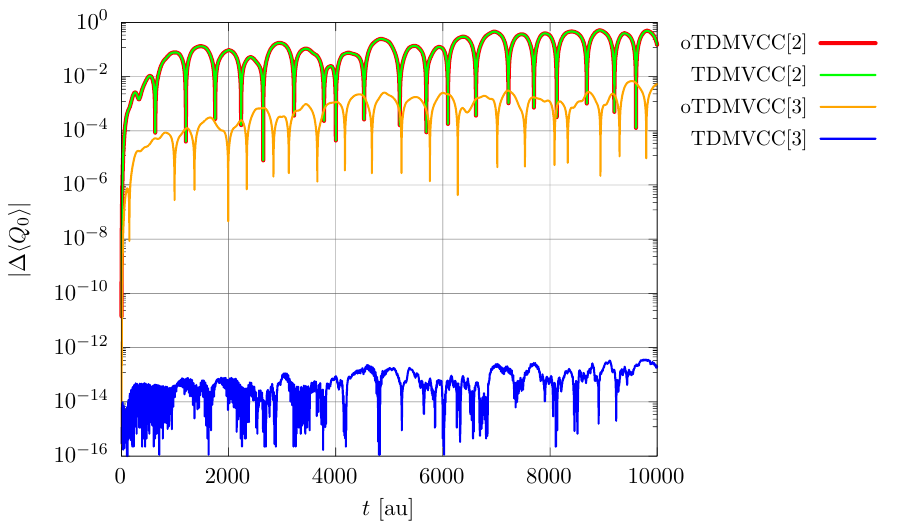}
        \label{fig:water_Q0_diff}
    \end{subfigure}


    \begin{subfigure}[c]{0.75\textwidth}
        \centering
        \caption{}
        \includegraphics[width=\textwidth]{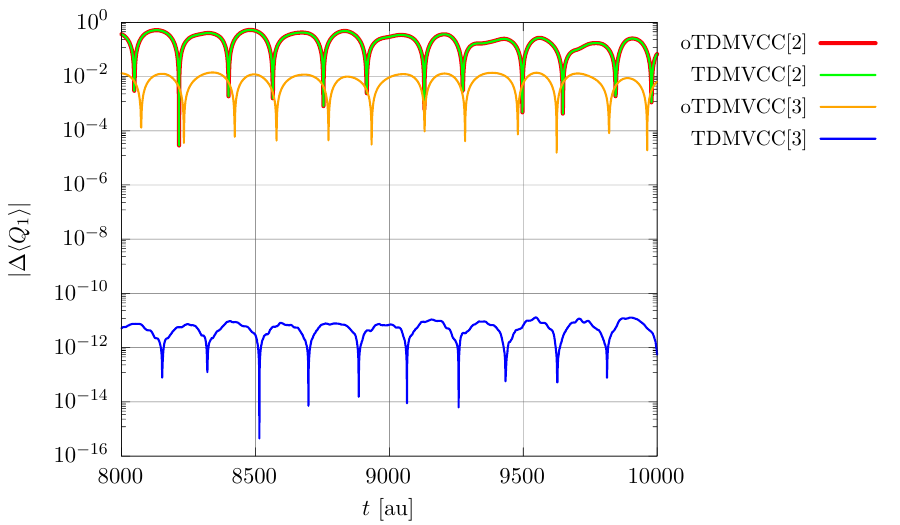}
        \label{fig:water_Q1_diff}
    \end{subfigure}

    \caption{\Ac{ivr} of water at the oTDMVCC[2--3] and TDMVCC[2--3] levels
    with $N = 8$ and $N_{\2\subA} = 8$ for all modes. (a) Absolute error in the expectation value of $Q_0$ (bend). (b) Absolute error in the expectation value of $Q_1$ (symmetric stretch).}
    \label{fig:water_Q0_Q1_diff}
\end{figure}

Figure~\ref{fig:water_Q0_Q1} shows the expectation value of the
displacement coordinates $Q_0$ (bend) and $Q_1$ (symmetric stretch).
The remaining mode $Q_2$ (asymmetric stretch) is not shown
since it couples only very weakly to $Q_0$ and $Q_1$ and is barely displaced during the simulation.
Again, we observe that oTDMVCC[3] and TDMVCC[3]
are visually identical to \ac{tdfvci}. oTDMVCC[2] and TDMVCC[2]
show small errors in $\langle Q_1 \rangle$ (Fig.~\ref{fig:water_Q1})
and somewhat larger errors in $\langle Q_0 \rangle$, especially at later times.
The absolute errors (Fig.~\ref{fig:water_Q0_Q1_diff})
again demonstrate the non-convergence of the \ac{otdmvcc} hierarchy.

%
%

%
%

\subsection{Ozone and sulfur dioxide}
These molecules behave similarly to water in terms of convergence
to \ac{tdfvci} (the results can be found in the supplementary material).
oTDMVCC[3] is always visually converged for sulfur dioxide, while
small errors can sometimes be seen for ozone.

%
\subsection{Hydrogen sulfide}

\subsubsection{\Aclp{acf}}

\begin{figure}[H]
    \centering
    \includegraphics[width=0.80\textwidth]{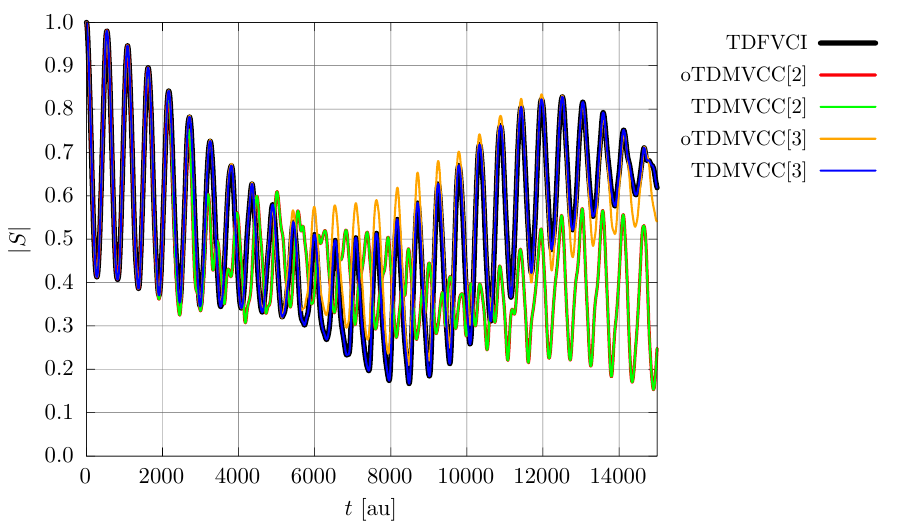}
    \caption{\Acl{acf} for the \ac{ivr} of hydrogen sulfide at the oTDMVCC[2--3] and TDMVCC[2--3] levels
    with $N = 8$ and $N_{\2\subA} = 8$ for all modes.}
    \label{fig:H2S_autocorr}
\end{figure}

The hydrogen sulfide case is interesting because it stands out from the remaining
triatomic molecules (water, ozone and sulfur dioxide). 
We note, for example, that the oTDMVCC[2] and TDMVCC[2] \aclp{acf}
in Fig.~\ref{fig:H2S_autocorr} are quite far from the exact result. The
prediction at the doubles level is in fact qualitatively wrong, which
suggests that the validity of the \ac{cc} ansatz is challenged.
We remark that oTDMVCC[2] and TDMVCC[2] appear to be exactly
equivalent, even in this seemingly difficult case. This surprising fact
is discussed in more detail in Sec.~\ref{ssec:otdmvcc_tdmvcc_equivalence}.

oTDMVCC[3] restores qualitative agreement with \ac{tdfvci},
but errors are still clearly visible. Only TDMVCC[3] succeeds in
reproducing the correct result, which we see as an indication
that the formal deficiency of the orthogonal formalism can have
practical consequences.

%
%

\subsubsection{Expectation values}

\begin{figure}[H]
    \centering

    \begin{subfigure}[c]{0.75\textwidth}
        \centering
        \caption{}
        \includegraphics[width=\textwidth]{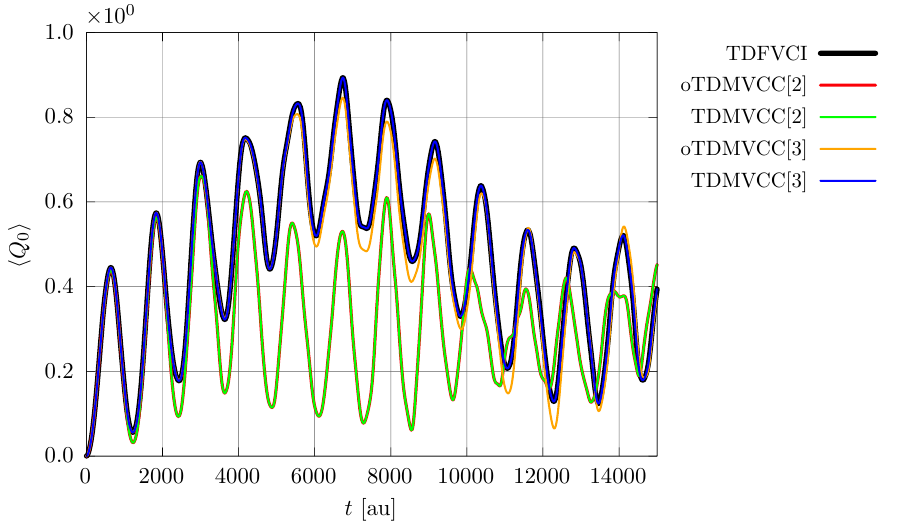}
        \label{fig:H2S_Q0}
    \end{subfigure}


    \begin{subfigure}[c]{0.75\textwidth}
        \centering
        \caption{}
        \includegraphics[width=\textwidth]{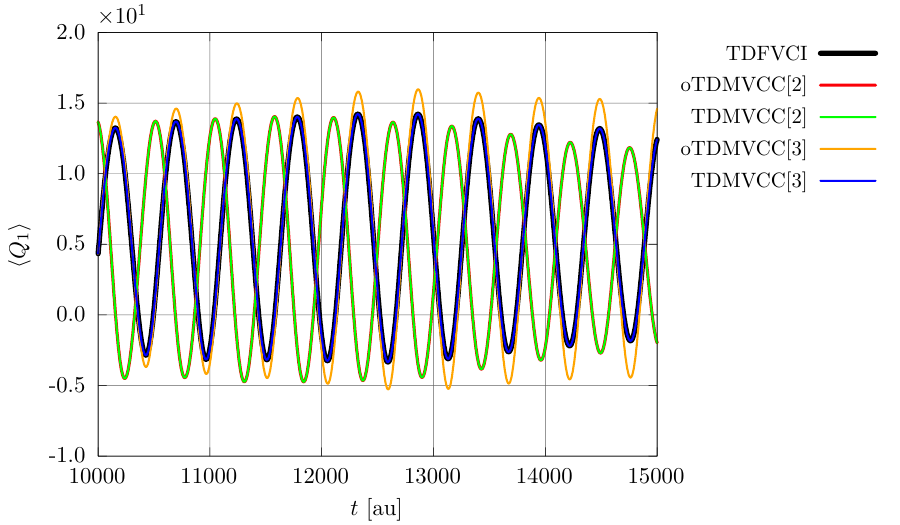}
        \label{fig:H2S_Q1}
    \end{subfigure}

    \caption{\Ac{ivr} of hydrogen sulfide at the oTDMVCC[2--3] and TDMVCC[2--3] levels
    with $N = 8$ and $N_{\2\subA} = 8$ for all modes. (a) Expectation value of $Q_0$ (bend). (b) Expectation value of $Q_1$ (symmetric stretch).}
    \label{fig:H2S_Q0_Q1}
\end{figure}

\begin{figure}[H]
    \centering

    \begin{subfigure}[c]{0.75\textwidth}
        \centering
        \caption{}
        \includegraphics[width=\textwidth]{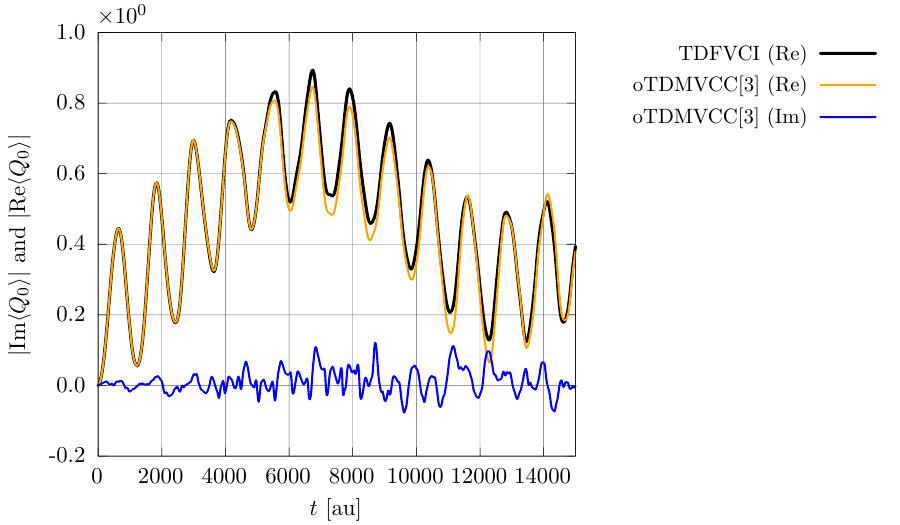}
        \label{fig:H2S_Q0_re_im_simple}
    \end{subfigure}


    \begin{subfigure}[c]{0.75\textwidth}
        \centering
        \caption{}
        \includegraphics[width=\textwidth]{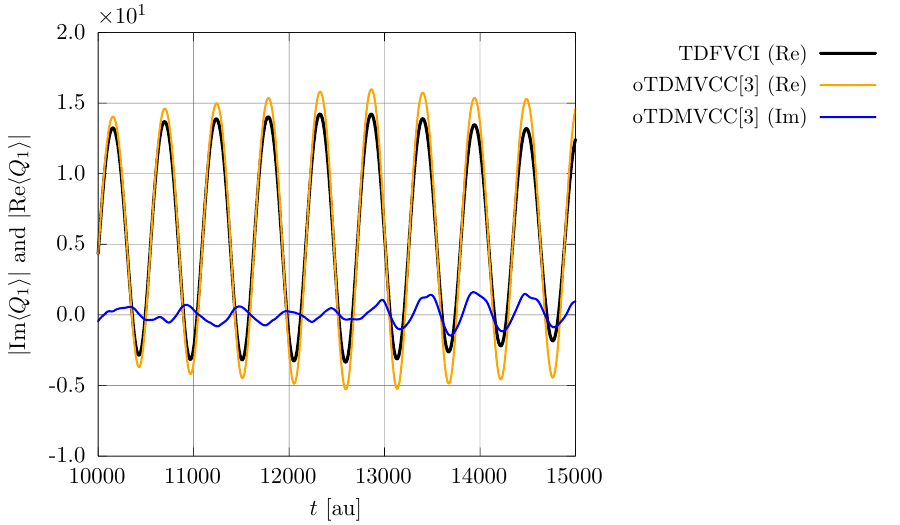}
        \label{fig:H2S_Q1_re_im_simple}
    \end{subfigure}

    \caption{\Ac{ivr} of hydrogen sulfide at the oTDMVCC[3] level
    with $N = 8$ and $N_{\2\subA} = 8$ for all modes. 
    $\mathrm{Re}\elm{\Psi'}{Q}{\Psi}$ and $\mathrm{Im}\elm{\Psi'}{Q}{\Psi}$ for (a) $Q_0$ (bend) and (b) $Q_1$ (symmetric stretch).}
    \label{fig:H2S_Q0_Q1_re_im_simple}
\end{figure}

For the $Q_0$ and $Q_1$ expectation values (Fig.~\ref{fig:H2S_Q0_Q1}),
the picture is much the same as for the \aclp{acf}. The oTDMVCC[2]/TDMVCC[2]
errors are rather large for $\langle Q_0 \rangle$ (Fig.~\ref{fig:H2S_Q0}),
while the error in $\langle Q_1 \rangle$ (Fig.~\ref{fig:H2S_Q1}) is less striking. In the latter case,
the overall shape of the curve is quite reasonable, but the period of the motion
is not correct. oTDMVCC[3] agrees reasonably well with the \ac{tdfvci} result,
which could be takes as an indication that the wave function is in some sense well-behaved,
in spite of the apparent error. However, Fig.~\ref{fig:H2S_Q0_Q1_re_im_simple} 
shows that $\mathrm{Im}\elm{\Psi'}{Q}{\Psi}$ is in fact
quite large compared to $\langle Q \rangle = \mathrm{Re}\elm{\Psi'}{Q}{\Psi}$ for 
$Q_0$ and $Q_1$. Although a large imaginary part has no experimental meaning, we take it as
yet another clear sign that the oTDMVCC wave function is generally not able to reproduce the exact wave function.
We will only comment explicitly on the unphysical imaginary part for the hydrogen sulfide case, where it
is largest. However, as shown in the supplementary material, we also find significant imaginary parts
in \ac{otdmvcc} calculations at full excitation level for some of the other molecules.
\newpage

%
%
\newpage

%
%

\newpage

\subsubsection{Amplitude norms}

\begin{figure}[H]
    \centering

    \begin{subfigure}[c]{0.85\textwidth}
        \centering
        \caption{$T$ amplitude norm}
        \includegraphics[width=\textwidth]{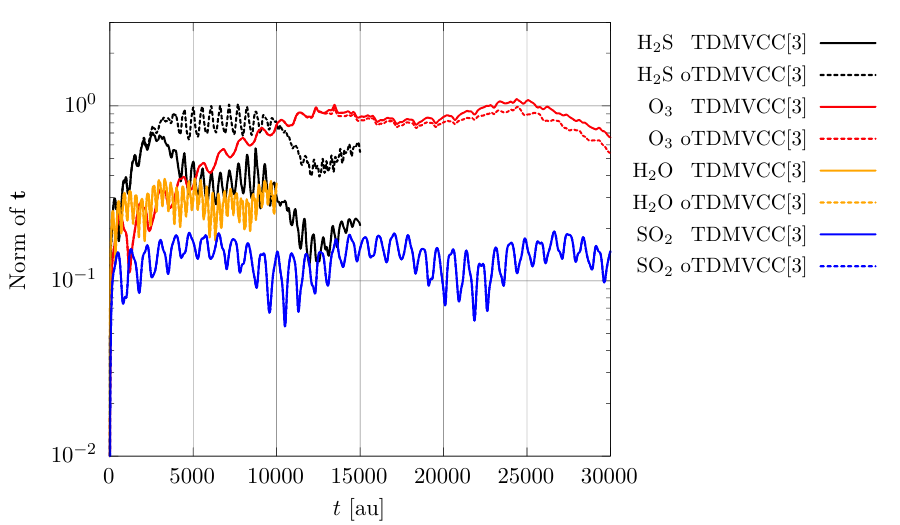}
        \label{fig:H2S_amplitude_normt}
    \end{subfigure}


    \begin{subfigure}[c]{0.85\textwidth}
        \centering
        \caption{$L$ amplitude norm}
        \includegraphics[width=\textwidth]{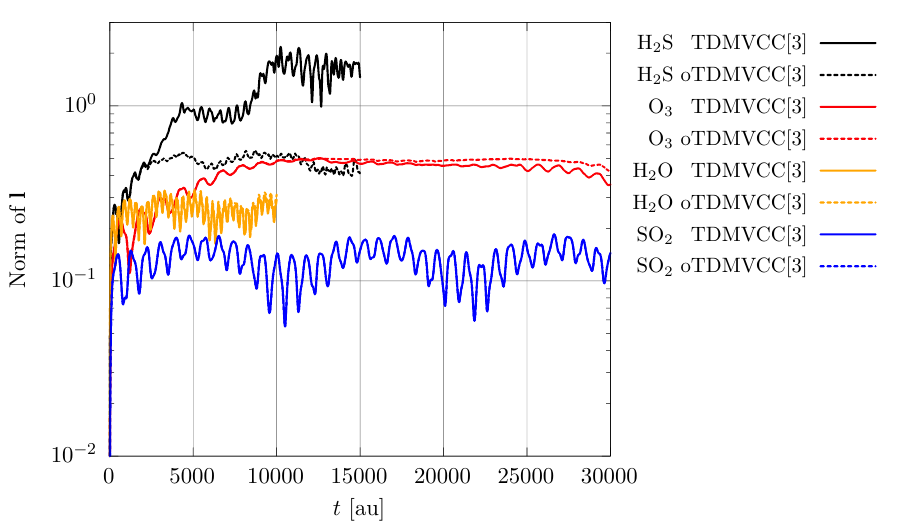}
        \label{fig:H2S_amplitude_norml}
    \end{subfigure}

    \caption{oTDMVCC[3] and TDMVCC[3] amplitude vector norms for the \ac{ivr} of water, ozone, sulfur dioxide and hydrogen sulfide. Note that full and dotted lines
    may coincide.}
    \label{fig:H2S_amplitude_norm}
\end{figure}

The fact that oTDMVCC[3] deviates visibly from TDMVCC[3] is reflected by the amplitude
norms in Fig.~\ref{fig:H2S_amplitude_norm} (the remaining triatomic molecules are also shown for comparison).
We note the following:
(i) For water and sulfur dioxide, the amplitude norms
are small and there is good agreement between $T$ and $L$, 
and between oTDMVCC[3] and TDMVCC[3];
(ii) For ozone, the amplitudes are comparatively large
and differ visibly; and 
(iii) For hydrogen sulfide, the amplitudes are again large
and differ by a significant amount (the difference is particularly large between the TDMVCC[3] $T$ and $L$ amplitudes).
At the same time, the TDMVCC[3] basis set shows considerable non-orthogonality in the hydrogen sulfide case (see Figs.~S48--S50 in the supplementary material).
We interpret this as a symptom that the \ac{cc} ansatz is straining to describe the hydrogen sulfide dynamics
correctly. The TDMVCC[3] is able to reproduce \ac{tdfvci} (as it should), but only by 
using rather large amplitudes and the full
flexibility of having a biorthogonal basis set. The oTDMVCC[3] ansatz is simply not sufficiently flexible in this particular case.

\newpage

%
%

\newpage

\subsection{5D \textit{trans}-bithiophene}

\subsubsection{\Aclp{acf}}

\begin{figure}[H]
    \centering

    \begin{subfigure}[c]{0.85\textwidth}
        \centering
        \caption{Full time interval}
        \includegraphics[width=\textwidth]{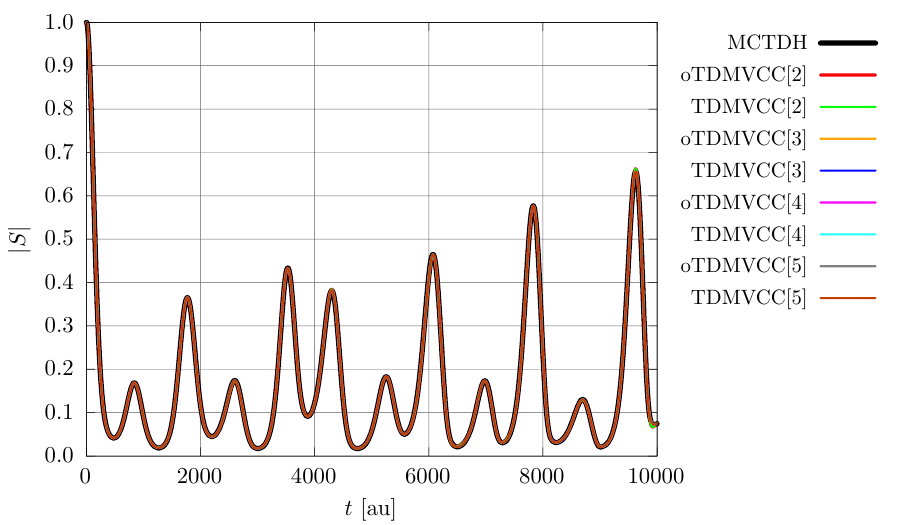}
        \label{fig:tbithio2345_autocorr_zoom_A}
    \end{subfigure}


    \begin{subfigure}[c]{0.85\textwidth}
        \centering
        \caption{Excerpt}
        \includegraphics[width=\textwidth]{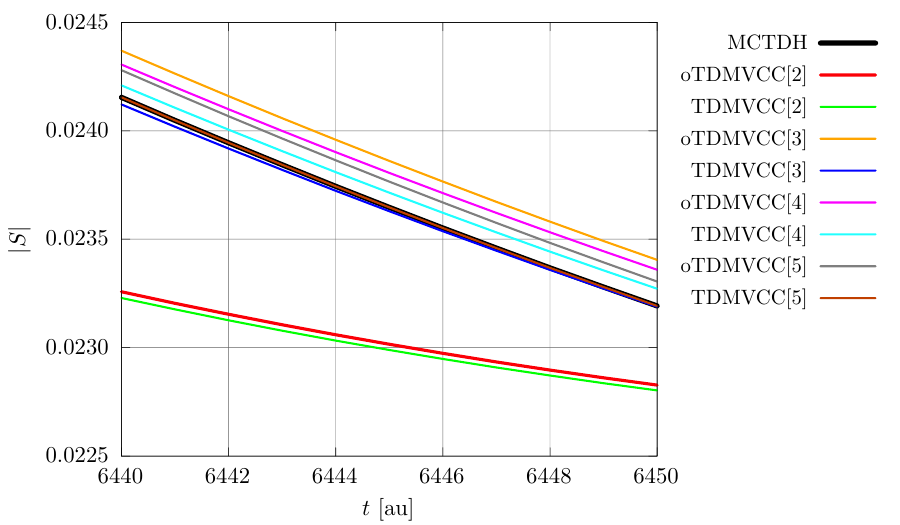}
        \label{fig:tbithio2345_autocorr_zoom_B}
    \end{subfigure}

    \caption{\Aclp{acf} for a 5D \textit{trans}-bithiophene model at the oTDMVCC[2--5] and TDMVCC[2--5] levels
    with $N = 30$ and $N_{\2\subA} = 4$ for all modes.}
    \label{fig:tbithio2345_autocorr_zoom_AB}
\end{figure}

\begin{figure}[H]
    \centering
    \includegraphics[width=0.8\textwidth]{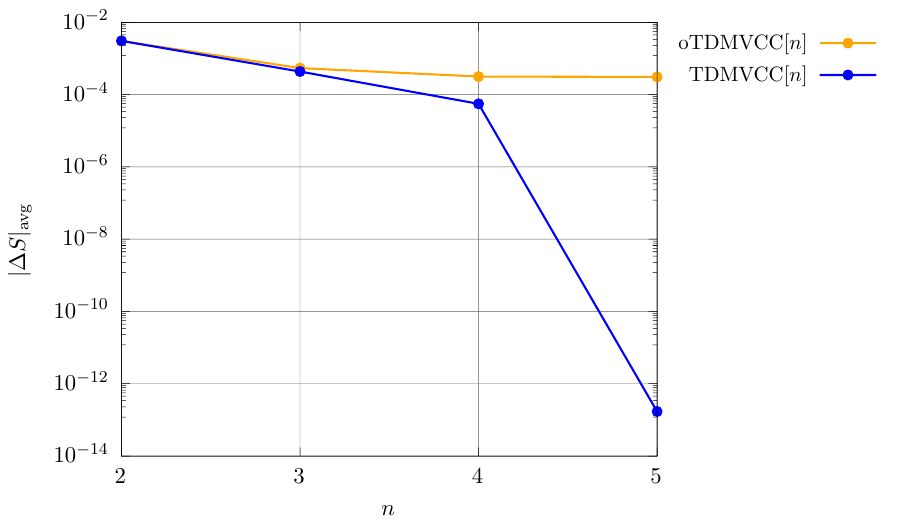}
    \caption{Average error in the \acl{acf} for a 5D \textit{trans}-bithiophene model at the oTDMVCC[2--5] and TDMVCC[2--5] levels
    with $N = 30$ and $N_{\2\subA} = 4$ for all modes. The errors are computed relative to \ac{mctdh} 
    (the first \SI{100}{au} have been excluded from the average).}
    \label{fig:tbithio2345_autocorr_zoom_A_diff_avg}
\end{figure}

For \textit{trans}-bithiophene, the \acl{acf} is almost
visually converged already at the oTDMVCC[2] and TDMVCC[2]
levels (Fig.~\ref{fig:tbithio2345_autocorr_zoom_A}).
The differences between the various levels only become
visible when looking rather closely
(Fig.~\ref{fig:tbithio2345_autocorr_zoom_B}), and
it is revealed that oTDMVCC[2] and TDMVCC[2]
are very similar (though not quite identical),
while showing a small error relative to the \ac{mctdh} reference.
TDMVCC[5] lies exactly on top
of the \ac{mctdh} trace, while the remaining calculations
are very close to it. The precise ranking is given
in terms of the average absolute 
error in Fig.~\ref{fig:tbithio2345_autocorr_zoom_A_diff_avg}.
We note that the TDMVCC$[n]$
error is less than or equal to the oTDMVCC$[n]$ error
for all $n$, and that the errors are always small.

\newpage

%
%

\newpage

%
%

\subsubsection{Expectation values}

\begin{figure}[H]
    \centering

    \begin{subfigure}[c]{0.85\textwidth}
        \centering
        \caption{Expectation value}
        \includegraphics[width=\textwidth]{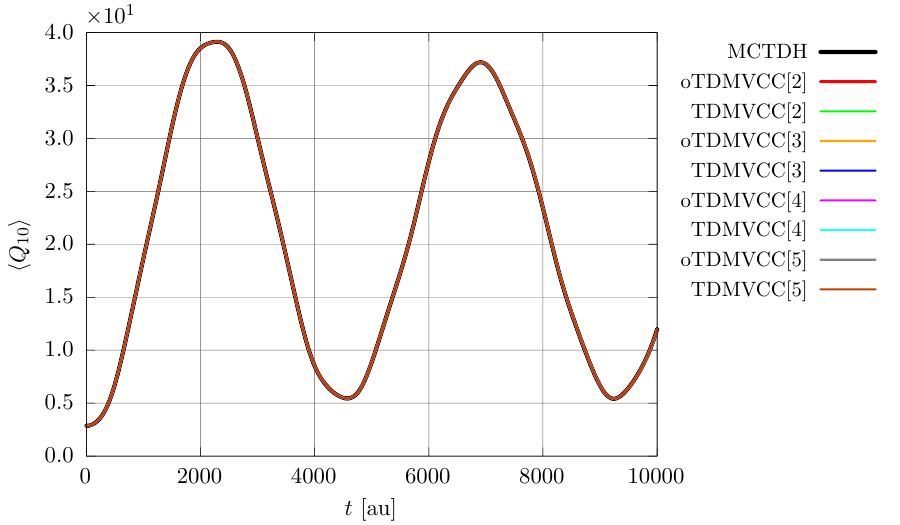}
        \label{fig:tbithio2345_Q10}
    \end{subfigure}


    \begin{subfigure}[c]{0.85\textwidth}
        \centering
        \caption{Absolute error in expectation value}
        \includegraphics[width=\textwidth]{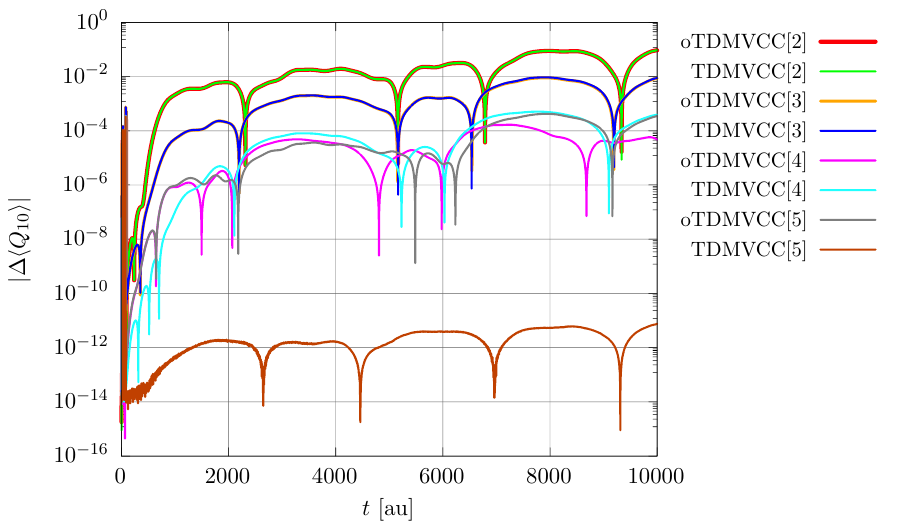}
        \label{fig:tbithio2345_Q10_diff}
    \end{subfigure}

    \caption{5D \textit{trans}-bithiophene model at the oTDMVCC[2--5] and TDMVCC[2--5] levels
    with $N = 30$ and $N_{\2\subA} = 4$ for all modes. (a) Expectation value of $Q_{10}$. (b) Absolute error in the expectation value (relative to \ac{mctdh}).}
    \label{fig:tbithio2345_Q10_and_diff}
\end{figure}

\begin{figure}[H]
    \centering
    \includegraphics[width=0.8\textwidth]{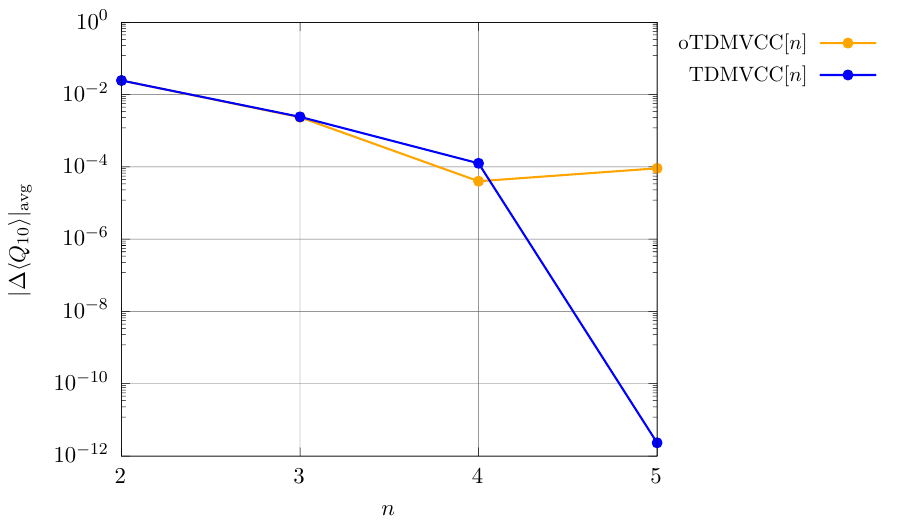}
    \caption{Average absolute error in the expectation value of $Q_{10}$ for a 5D \textit{trans}-bithiophene model at the oTDMVCC[2--5] and TDMVCC[2--5] levels
    with $N = 4$ and $N_{\2\subA} = 4$ for all modes. The errors are computed relative to \ac{mctdh}
    (the first \SI{100}{au} have been excluded from the average).}
    \label{fig:tbithio2345_Q10_diff_avg}
\end{figure}

Figure~\ref{fig:tbithio2345_Q10_and_diff} displays the expectation
value $\langle Q_{10} \rangle$ and the absolute error relative
to \ac{mctdh}. The results all appear converged to the unaided eye
(Fig.~\ref{fig:tbithio2345_Q10}), 
but the absolute error (Fig.~\ref{fig:tbithio2345_Q10_diff}) shows
that this is not exactly the case. It is clear that
oTDMVCC[2]/TDMVCC[2] and oTDMVCC[3]/TDMVCC[3] yield pairwise
near-identical values of $\langle Q_{10} \rangle$
and that TDMVCC[5] is fully converged.
The relative quality of oTDMVCC[4], TDMVCC[4] and oTDMVCC[5]
is not obvious from Fig.~\ref{fig:tbithio2345_Q10_diff},
but it is resolved quite clearly by Fig.~\ref{fig:tbithio2345_Q10_diff_avg}:
oTDMVCC[4] performs slightly better than both TDMVCC[4] and,
surprisingly, oTDMVCC[5].
We note the strong oscillations in the absolute error (Fig.~\ref{fig:tbithio2345_Q10_diff}) near $t = 0$.
These are caused by the singular initial density matrices (due to the \ac{vscf} initial state),
which make the \acp{eom} somewhat difficult to integrate. We have thus chosen to exclude
the first \SI{100}{au} from the computation of the average shown in Fig.~\ref{fig:tbithio2345_Q10_diff_avg}.

The expectation values for the remaining modes in the
5D model system are provided in the supplementary material
(see Figs.~S58--S65). These modes behave in much the same
way as $Q_{10}$, also with respect to the ranking
of the various levels of theory.

%
%

\newpage

%
%

\subsection{Are oTDMVCC[2] and TDMVCC[2] equivalent?} \label{ssec:otdmvcc_tdmvcc_equivalence}

For the triatomic molecules, our results show that oTDMVCC[2] are TDMVCC[2]
are identical up to numerical precision (this is confirmed by inspecting the Hilbert space angles between the oTDMVCC[2] and TDMVCC[2] wave functions; see Fig.~S51). 
It is not surprising that the two methods should yield
similar predictions in many cases, but we find the agreement rather striking, in particular
for the demanding case of hydrogen sulfide. Here, one might expect to see a clear
manifestation of the difference between the orthogonal and biorthogonal formalisms,
especially since such a difference is visible at the triples level.

We explain this surprising result with the following observations:
First, we note that the oTDMVCC[2] and TDMVCC[2] expansions assume
a very particular form for two- and three-mode systems:
\begin{subequations}
    \begin{align}
        \ket{\Psi}  
        &= e^T \ket{\Phi}       \nn
        &= e^{t_0} (1 + T_2) \ket{\Phi}, \\
        \bra{\Psi'} 
        &= \bra{\Phi'} L e^{-T} \nn
        &= \bra{\Phi'} (l_0 + L_2) (1 + T_2) e^{-t_0} \nn
        &= \bra{\Phi'} \big((l_0 + L_2 T_2) + L_2 \big) e^{-t_0} \nn
        &\equiv \bra{\Phi'} (l_0' + L_2) e^{-t_0}.
    \end{align}
\end{subequations}
This is essentially a CI-type expansion, although with a somewhat peculiar normalization.
The second observation is that the \ac{vscf} initial state has an orthogonal
basis and satisfies $\bra{\Psi'} = \ket{\Psi}^{\dagger}$.
We thus have an initial state with symmetry between bra and ket, and a parameterization
that allows this symmetry to be maintained. Numerically, we find
that the one-mode density matrices stay Hermitian 
and that the mean fields satisfy $\check{\mbf{F}} = \check{\mbf{F}}^{\prime \dagger}$ for each mode (see Figs.~S11, S22, S33 and S44).
In essence, we get variational rather than bivariational time evolution for this special case.
This also has the effect that expectation values like $\elm{\Psi'}{Q}{\Psi}$ and $\elm{\Psi'}{H}{\Psi}$
are strictly real, as can be seen in the supplementary material.


For \textit{trans}-bithiophene, oTDMVCC[2] are TDMVCC[2] are very similar, and one must
look closely (e.g. Fig~\ref{fig:tbithio2345_autocorr_zoom_B}) to see a difference.
It is of course difficult to define precisely what it means for two calculations to be strictly identical
due to the presence of integration error and numerical
noise. However, if we take the agreement between TDMVCC[5] and \ac{mctdh} (Fig~\ref{fig:tbithio2345_autocorr_zoom_B}) as a benchmark
for perfect agreement, then it is clear that oTDMVCC[2] are TDMVCC[2] are not exactly identical.
We are also not able to see any mathematical reason that the oTDMVCC[2] and TDMVCC[2] equations
should generally be equivalent. 
\section{Summary and outlook} \label{sec:summary}
The \acp{eom} for bivariational wave functions
with orthogonal, adaptive basis functions have been derived
from a \acl{tdbvp}. The use of an orthogonal basis
makes the parameterization non-holomorphic (non-analytic in the complex sense),
which necessitates the use of a manifestly real action functional in the derivations.
We relate the orthogonal formalism (real action) to the
corresponding biorthogonal formalism (complex action) in a
transparent way and analyze similarities and differences.
The general \acp{eom} are then specialized to the \ac{cc} ansatz
and implemented for the nuclear dynamics problem.
We denote the resulting method as \ac{otdmvcc} (orthogonal \ac{tdmvcc}) in order
to distinguish it from \ac{tdmvcc}, which uses a biorthogonal basis set.
The \ac{otdmvcc} amplitude equations are unchanged relative to the biorthogonal case, while the
linear equations that determine the basis set time
evolution (the so-called constraint equations)
are symmetrized in a particular manner.
Although the orthogonal and biorthogonal formalisms thus rely on
the same matrix elements, the symmetrization has some
consequences for the practical implementation of the constraint
equations. In particular, the orthogonal formalism leads
to constraint equations that couple all modes, in contrast
to the biorthogonal constraint equations that can be solved one mode
at a time. At the doubles level, certain simplification occur
so that the oTDMVCC[2] constraint equations can also be solved 
mode by mode. The computational cost of oTDMVCC[2] is thus identical
to that of TDMVCC[2]. At higher excitation levels, the orthogonal
formalism generally involves a larger computational and implementation effort.

It is known\cite{kohnOrbitaloptimizedCoupledclusterTheory2005} that \ac{cc} with orthogonal optimized or adaptive basis functions
does not converge to the exact solution, even when the cluster expansion
is complete. The precise convergence behavior is, however, not well understood, although electron dynamics studies by 
Sato and coworkers\cite{satoCommunicationTimedependentOptimized2018,pathakTimedependentOptimizedCoupledcluster2020a,pathakTimedependentOptimizedCoupledcluster2021}
indicate that no substantial error is introduced by using an orthogonal basis.
We benchmarked all members of the oTDMVCC and TDMVCC hierarchies against \ac{tdfvci} or \ac{mctdh}
for a number of triatomics (water, ozone, sulfur dioxide and hydrogen sulfide)
and for a 5D model of the \textit{trans}-bithiophene molecule.
It is confirmed very clearly that TDMVCC converges to the exact limit, while
oTDMVCC does not. 
For 5D \textit{trans}-bithiophene, the oTDMVCC[5] observables are
visually converged relative to the exact result, and
differences are only revealed on close inspection. Water and sulfur dioxide behave similarly in
the sense that the oTDMVCC[3] results appear fully converged to the unaided eye.
For ozone, small errors are visible at the oTDMVCC[3] level, while
hydrogen sulfide shows very clear differences between oTDMVCC[3] and TDMVCC[3]/TDFVCI (which are equivalent for a three-mode system).
This difference correlates with rather large amplitudes, which indicates
that the coupled cluster expansion is struggling in order to
describe the wave function. In this difficult case, only TDMVCC[3] is able to reproduce TDFVCI
due to the added flexibility of having a biorthogonal basis. 
Based on the examples at hand, our conclusion is thus that the orthogonal and biorthogonal formalisms
agree quite closely in many cases, although noticeable differences are certainly possible.

We have emphasized two drawbacks of the oTDMVCC hierarchy, namely the
lack of convergence to the exact limit and the fact that it involves
larger sets of linear equations compared to the TDMVCC method.
However, the TDMVCC method also has a drawback in the sense that it
can sometimes be prone to numerical
instability when the basis is split into and active
and a secondary basis.\cite{hojlundBivariationalTimedependentWave2022}
The so-called restricted polar scheme presented in Ref.~\citenum{hojlundBivariationalTimedependentWave2022} solves
the stability problem without deteriorating accuracy, but the scheme is not fully bivariational. It is thus highly
pertinent to investigate new formalisms that combine the following desirable properties
in a fully bivariational way:
(i) Numerical stability; 
(ii) convergence to the exact solution; and
(iii) simple linear equations.
Such a formalism is the subject of current research in our group. 
\section*{Supplementary material} \label{sec:supplementary_material}
The supplementary material contains results for ozone and sulfur dioxide, as
well as additional details for water, hydrogen sulfide and 5D \textit{trans}-bithiophene.

\section*{Acknowledgements}
O.C. acknowledges support from the Independent Research Fund Denmark through grant number 1026-00122B.
This work was funded by the Danish National Research Foundation (DNRF172) through the Center of Excellence for Chemistry of Clouds.

\section*{Author declarations}
\subsection*{Conflict of Interest}
The authors have no conflicts to disclose.

\subsection*{Author Contributions}
\textbf{Mads Greisen Højlund}: 
Conceptualization (equal);
Data curation (lead);
Formal analysis (equal);
Investigation (lead);
Software (lead);
Visualization (lead);
Writing -- original draft (lead);
Writing -- review \& editing (equal).
\textbf{Alberto Zoccante}:
Conceptualization (equal);
Formal analysis (equal);
Writing -- review \& editing (equal).
\textbf{Ove Christiansen}:
Conceptualization (equal);
Formal analysis (equal);
Funding acquisition (lead);
Project administration (lead);
Supervision (lead);
Writing -- review \& editing (equal).

\section*{Data availability}
The data that supports the findings of this study are available within the article and its supplementary material.

\appendix

\section{Complex analysis} \label{appendix:complex_analysis}
\setcounter{equation}{0}
\renewcommand{\theequation}{\ref{appendix:complex_analysis}\arabic{equation}}
This appendix covers a few basic aspects of complex analysis.
Our description is strongly inspired by Chapter 2.2 of Ref.~\citenum{steinComplexAnalysis2003} to which we
refer the reader for further details.
For simplicity of notation, we consider only functions of a single complex variable,
e.g. $f: \Omega \rightarrow \mathbb{C}$, where the domain $\Omega$ is an open set in $\mathbb{C}$.
We are free to write the function $f$ and its argument $z$ in terms of real and imaginary parts, i.e.
\begin{gather}
    z = x + iy, \\
    f(z) = f(x,y) = u(x,y) + i v(x,y),
\end{gather}
where $x,y,u,v$ are real.
As an entry point to our discussion, let us consider a common kind
of problem, namely the problem of making $f$ stationary:
\begin{align} \label{eq:f_stationary_xy}
    \pdv{f}{x} = 0, \quad \pdv{f}{y} = 0.
\end{align}
Here, $\partial/\partial x$ and $\partial/\partial y$ denote partial derivatives in the ordinary, real sense.
Separating $f$ into real and imaginary parts, this is obviously equivalent to
\begin{align} \label{eq:four_equations_two_unknowns}
    \pdv{u}{x} = 0, \quad \pdv{v}{x} = 0, \quad \pdv{u}{y} = 0, \quad \pdv{v}{y} = 0,
\end{align} 
which are four real equations with two real unknowns. Such a system 
cannot have a solution unless $f$ has some additional structure
that eliminates two of the equations. We will discuss two kinds of
structure that are relevant to our work, 
one which is mathematically trivial and one which
has far-reaching consequences. The first case
can be stated as
\begin{align} \label{eq:dressed_real_function}
    f(x,y) = \alpha + \beta g(x,y),
\end{align}
where $g$ is a real-valued function and $\alpha,\beta$ are complex numbers. 
This includes $f$ being real, which is of course a common situation, or purely imaginary. 
Making $f$ stationary with respect to $x$ and $y$
is now a matter of solving two real equations with two real unknowns:
\begin{align}
    \pdv{g}{x} = 0, \quad \pdv{g}{y} = 0.
\end{align}
A solution need not exist, of course, but we cannot rule out the possibility without more information.

In the second case, $f$ is holomorphic.
The function $f$
is holomorphic (or complex differentiable) at the point $z_0 \in \Omega$ if
the quotient
\begin{align}
    \frac{f(z_0 + h) - f(z_0)}{h}
\end{align}
converges to a limit when $h \rightarrow 0$. In that case,
the limit is denoted by $f'(z_0)$, and is called the derivative of $f$ at $z_0$:
\begin{align}
    f'(z_0) = \lim_{h \rightarrow 0} \frac{f(z_0 + h) - f(z_0)}{h}.
\end{align}
Although this definition looks exactly like the definition of a real derivative,
it should be emphasized that $h$ is complex number that may approach $0$
from any direction. This turns out to have profound implications.
A holomorphic function is, for example, infinitely differentiable, i.e.
the existence of the first derivative guarantees the existence of all higher derivatives.\cite{steinComplexAnalysis2003}
Holomorphic functions are also analytic in the sense that they are given (locally)
by a convergent power series expansion.\cite{steinComplexAnalysis2003}

One can easily show (by taking $h$ to be real and then purely imaginary) 
that the existence of $f'(z_0)$ implies
\begin{align} \label{eq:cauchy_riemann_precursor}
    f'(z_0) = \pdv{f}{x} \/ (z_0) = \frac{1}{i} \pdv{f}{y} \/ (z_0)
\end{align}
or, after separating real and imaginary parts,
\begin{align} \label{eq:cauchy_riemann}
    \pdv{u}{x} = \pdv{v}{y}, \quad \pdv{u}{y} = -\pdv{v}{x}.
\end{align}
These are the Cauchy-Riemann conditions, which connect real
and complex analysis. We note that the Cauchy-Riemann conditions eliminate
two equations in Eq.~\eqref{eq:four_equations_two_unknowns}, so that
we are left with two real equations with two real unknowns. 
Equation~\eqref{eq:cauchy_riemann_precursor} also implies that
\begin{alignat}{2}
    \pdv{f}{z}   \/ (z_0) &\equiv  \frac{1}{2} \left( \pdv{f}{x} \/ (z_0) + \frac{1}{i} \pdv{f}{y} \/ (z_0) \right) &&= f'(z_0), \label{eq:wirtinger_z_holomorphic} \\
    \pdv{f}{z^*} \/ (z_0) &\equiv  \frac{1}{2} \left( \pdv{f}{x} \/ (z_0) - \frac{1}{i} \pdv{f}{y} \/ (z_0) \right) &&= 0,       \label{eq:wirtinger_z_conj_holomorphic}
\end{alignat}
if $f$ is holomorphic.
Here, we have \textit{defined} the differential operators $\partial / \partial z$ and 
$\partial / \partial z^*$, which are sometimes called Wirtinger derivatives.
These derivatives are, in principle, nothing more that a shorthand 
for a certain combination of real derivatives, but they are nonetheless very convenient.
Equation~\eqref{eq:wirtinger_z_holomorphic} means that $\partial / \partial z$ agrees
with the complex partial derivative for holomorphic functions, while Eq.~\eqref{eq:wirtinger_z_conj_holomorphic}
states that holomorphic functions have no formal dependence on $z^*$.
For non-holomorphic functions
(where the complex derivative does not exist), the Wirtinger derivatives still
have meaning provided the real derivatives $\partial/\partial x$ and $\partial/\partial y$ exist.
With the definition of Wirtinger derivatives, the optimization problem in Eq.~\eqref{eq:f_stationary_xy} is equivalent to
\begin{align}
    \pdv{f}{z} = 0, \quad \pdv{f}{z^*} = 0,
\end{align}
where $z$ and $z^*$ are considered as independent variables.
For a holomorphic function $f$,
the latter equation is identically zero, and we are simply left with
\begin{align}
    \pdv{f}{z} = 0.
\end{align}
For a non-holomorphic function $f$, both equations must generally be considered.

Wirtinger derivatives have a number of pleasant properties that greatly simplify practical calculations:
\begin{align}
    \left( \pdv{f}{z}   \right)^{\!*} &= \pdv{f^*}{z^*} \\
    \left( \pdv{f}{z^*} \right)^{\!*} &= \pdv{f^*}{z}
\end{align}
\begin{alignat}{2}
    \pdv{}{z}   \/ (fg) &= \pdv{f}{z} g + f \pdv{g}{z}     \quad &&\text{(product rule)} \\
    \pdv{}{z^*} \/ (fg) &= \pdv{f}{z^*} g + f \pdv{g}{z^*} \quad &&\text{(product rule)}
\end{alignat}
\begin{alignat}{2}
    \pdv{}{z}   \/ (f \circ g) &= \left( \pdv{f}{z} \circ g \right) \pdv{g}{z}   + \left( \pdv{f}{z^*} \circ g \right) \pdv{g^*}{z}   \quad &&\text{(chain rule)}\\
    \pdv{}{z^*} \/ (f \circ g) &= \left( \pdv{f}{z} \circ g \right) \pdv{g}{z^*} + \left( \pdv{f}{z^*} \circ g \right) \pdv{g^*}{z^*} \quad &&\text{(chain rule)}
\end{alignat}
\begin{alignat}{2}
    \pdv{z}{z^*} &= \pdv{z^*}{z}   &&= 0, \\
    \pdv{z}{z}   &= \pdv{z^*}{z^*} &&= 1.
\end{alignat}
The mechanics of computing Wirtinger derivatives is thus essentially identical
to that of computing real derivatives, provided we think about $z$ and $z^*$ as independent variables.
In addition, we never need to think about real and imaginary parts explicitly.

An important mapping that is not holomorphic is complex conjugation, $f(z) = z^*$.
Indeed,
\begin{align}
    \frac{f(z_0 + h) - f(z_0)}{h} = \frac{h^*}{h},
\end{align}
which has no limit as $h \rightarrow 0$. This is easily checked by taking
$h$ to be real (in which case the quotient equals $1$) 
and then purely imaginary (in which case the quotient equals $-1$).
Another common function is the square modulus, $f(z) = \abs{z}^2 = zz^*$,
which is non-holomorphic due to the presence of $z^*$. In spite
of this, we can make $f$ stationary without resorting to real and imaginary parts
by making use of the Wirtinger derivatives:
\begin{align}
    \pdv{f}{z} = z^* = 0, \quad \pdv{f}{z^*} = z = 0.
\end{align}
The solution is obviously $z = 0$. It is noted that since $f$ is real, the two
equations are each other's complex conjugate, so that one equation
is effectively eliminated.

For non-holomorphic functions like Eq.~\eqref{eq:dressed_real_function}
we find that the optimization problem becomes 
\begin{align}
    \pdv{f}{z}   = \beta \pdv{g}{z}   = 0, \quad 
    \pdv{f}{z^*} = \beta \pdv{g}{z^*} = 0.
\end{align}
or, equivalently,
\begin{align}
    \pdv{g}{z}   = 0, \quad \pdv{g}{z^*} = 0.
\end{align}
Since $g$ is real, the two equations are 
simply each other's complex conjugate, so we only need to solve one of them.

For generic non-holomorphic functions we cannot
hope for a solution. As an example, consider $f(z) = z^2 + z^*$.
Attempting to make this function stationary yields
\begin{align}
    \pdv{f}{z} = 2z = 0, \quad \pdv{f}{z^*} = 1 = 0,
\end{align}
which has no solution. This mirrors the discussion after Eq.~\eqref{eq:four_equations_two_unknowns}. 
\section{Basis set \acp{eom}} \label{appendix:basis_set_eoms}
\setcounter{equation}{0}
\renewcommand{\theequation}{\ref{appendix:basis_set_eoms}\arabic{equation}}
We can reuse the derivation of the complex Lagrangian $\mathcal{L}$
from Ref.~\citenum{hojlundBivariationalTimedependentWave2022} since the wave function
does not depend on $\bm{\alpha}^*$. The modifications necessary to account for
an orthonormal (rather than biorthonormal) basis are straight forward and the result reads
\begin{align}
    \mathcal{L} &= \elm{\Psi'}{(i \partial_t - H)}{\Psi} \nn
    &= i \sum_j \dot{\alpha}_j m_j - \mathcal{H}'
\end{align}
where $m_j$ depends only on the configurational parameters:
\begin{align}
    m_j (\bm{\alpha}) = \bigbraket{\Psi'}{\pdv{\Psi}{\alpha_j}}.
\end{align}
The quantity $\mathcal{H}'$ is a modified energy function defined as
\begin{align} \label{eq:Hp_function}
    \mathcal{H}' (\bm{\alpha}, \mathbf{V}, \dot{\mathbf{V}}, \mbf{V}^*) =
    \elm{\Psi'}{(H - g)}{\Psi} = \mathcal{H} - \mathcal{G}.
\end{align}
This function contains a proper energy function,
\begin{align} \label{eq:H_function}
    \mathcal{H} (\bm{\alpha}, \mbf{V}, \mbf{V}^*) = \elm{\Psi'}{H}{\Psi},
\end{align}
and a constraint function,
\begin{subequations}
\begin{align}
    \mathcal{G} (\bm{\alpha}, \dot{\mbf{V}}, \mbf{V}^*) &= \elm{\Psi'}{g}{\Psi} \\
    &= \sum_m \sum_{p^m q^m} \rrho{m}{q}{p} \gplain{m}{p}{q} \\
    &= i \sum_m \sum_{p^m q^m} \rrho{m}{q}{p} \Big( 
    \sum_{\alpha^m} \vplainconj{m}{\alpha}{p} \vplaindot{m}{\alpha}{q}    
    \Big). \label{eq:G_function_c}
\end{align}
\end{subequations}
We have used Eqs.~\eqref{eq:constraint_operator} and \eqref{eq:consistency_b} and
defined a one-mode density matrix $\bm{\rho}^m$ with elements
\begin{align}
    \rrho{m}{q}{p} (\bm{\alpha}) = \elm{\Psi'}{\Etilde{m}{p}{q}}{\Psi}.
\end{align}
Note the reversed indices. Having determined $\mathcal{L}$,
we are ready to compute $\bar{\mathcal{L}}$ as
\begin{align} \label{eq:real_lagrangian_concrete_expression}
    \bar{\mathcal{L}} &= \tfrac{1}{2} (\mathcal{L} + \mathcal{L}^*) \nn
    &= \tfrac{i}{2}  \sum_j (\dot{\alpha}_j m_j - \dot{\alpha}_j^* m_j^*)
    - \tfrac{1}{2} (\mathcal{H}' + \mathcal{H}^{\prime *})
\end{align}
Using the real Lagrangian from Eq.~\eqref{eq:real_lagrangian_concrete_expression},
the basis set \acp{ele} read
\begin{align}
    0 &= \pdv{\bar{\mathcal{L}}}{\vplain{m}{\alpha}{q}}
    - \dv{t} \pdv{\bar{\mathcal{L}}}{\vplaindot{m}{\alpha}{q}} \nn
    &= \frac{1}{2} \dv{t} \pdv{(\mathcal{H}' + \mathcal{H}^{\prime *})}{\vplaindot{m}{\alpha}{q}} 
    - \frac{1}{2} \pdv{(\mathcal{H}' + \mathcal{H}^{\prime *})}{\vplain{m}{\alpha}{q}}.
\end{align}
The four terms are easily computed using Eqs.~\eqref{eq:Hp_function}, \eqref{eq:H_function} 
and \eqref{eq:G_function_c}. One finds that
\begin{align}
    \dv{t} \pdv{\mathcal{H}'}{\vplaindot{m}{\alpha}{q}} 
    &= -i \dv{t} \sum_{p^m} \rrho{m}{q}{p} \vplainconj{m}{\alpha}{p} \nn
    &= -i \sum_{p^m} \big( \rrhodot{m}{q}{p} \vplainconj{m}{\alpha}{p} 
    + \rrho{m}{q}{p} \vplaindotconj{m}{\alpha}{p} 
    \big),
\end{align}
\begin{align}
    \dv{t} \pdv{\mathcal{H}^{\prime *}}{\vplaindot{m}{\alpha}{q}} 
    &= 0,
\end{align}
\begin{align} \label{eq:SPF_ELE_c}
    \pdv{\mathcal{H}'}{\vplain{m}{\alpha}{q}}
    &= \pdv{\mathcal{H}}{\vplain{m}{\alpha}{q}} \nn
    &= \Fcheckprime{m}{q}{\alpha},
\end{align}
\begin{align} \label{eq:SPF_ELE_d}
    \pdv{\mathcal{H}^{\prime *}}{\vplain{m}{\alpha}{q}}
    &= \bigg( \pdv{\mathcal{H}}{\vplainconj{m}{\alpha}{q}} \bigg)^{\! *}
    +i \sum_{p^m} \rrhoconj{m}{p}{q} \vplaindotconj{m}{\alpha}{p} \nn
    &= \Fcheckplainconj{m}{\alpha}{q}
    +i \sum_{p^m} \rrhoconj{m}{p}{q} \vplaindotconj{m}{\alpha}{p} 
\end{align}
Equations~\eqref{eq:SPF_ELE_c} and \eqref{eq:SPF_ELE_d} introduce the
half-transformed mean-field matrices $\FFcheckprime{m}$ and $\FFcheckplain{m}$
with elements
\begin{subequations}
    \begin{alignat}{2}
        \Fcheckprime{m}{q}{\alpha} 
         &= \pdv{\mathcal{H}}{\vplain{m}{\alpha}{q}} 
        &&= \elm{\Psi'}{[H, \crea{m}{\alpha}] \annitilde{m}{q}}{\Psi}, \\
        \Fcheckplain{m}{\alpha}{q}, 
         &= \pdv{\mathcal{H}}{\vplainconj{m}{\alpha}{q}} 
        &&= \elm{\Psi'}{\creatilde{m}{q} [\anni{m}{\alpha}, H]}{\Psi}.
    \end{alignat}
\end{subequations}
The concrete expressions in terms of commutators hold in the vibrational 
case\cite{madsenTimedependentVibrationalCoupled2020} and in the electronic case\cite{hojlundBivariationalTimedependentWave2022}
after removal of mode indices. The corresponding fully transformed mean-field matrices
are given by
\begin{subequations} \label{eq:F_tilde_def}
    \begin{align}
        \FFtildeprime{m} &= \FFcheckprime{m} \mbf{V}^m, \\
        \FFtildeplain{m} &= \mbf{V}^{m\dagger} \FFcheckplain{m}
    \end{align}
\end{subequations}
with elements
\begin{subequations} \label{eq:F_tilde_elements_appendix}
\begin{align}
    \Ftildeprime{m}{q}{p} 
    &= \elm{\Psi'}{[H, \creatilde{m}{p}] \annitilde{m}{q}}{\Psi}, \\
   \Ftildeplain{m}{q}{p} 
    &= \elm{\Psi'}{\creatilde{m}{p} [\annitilde{m}{q}, H]}{\Psi}.
\end{align}
\end{subequations}
The \acp{ele} can now be written in matrix notation as
\begin{align} \label{eq:modal_eles}
    \mbf{0} &= 
    - \frac{1}{2} \big( \FFcheckprime{m} + \FFcheckplaindagger{m} \big) 
    - i \mathbb{H}[ \bm{\rho}^m ] \dot{\mbf{V}}^{m\dagger}
    - \frac{i}{2} \dot{\bm{\rho}}^m \mbf{V}^{m\dagger}
\end{align}
where $\mathbb{H}[\,\cdot\,]$ denotes the Hermitian part of a square matrix. Multiplication of Eq.~\eqref{eq:modal_eles}
by $\mbf{V}^m$ from the right then yields
\begin{align} \label{eq:modal_eles_times_v_dagger}
    \mbf{0} &= 
    - \frac{1}{2} \big( \FFtildeprime{m} + \FFtildeplaindagger{m} \big) 
    + \mathbb{H}[ \bm{\rho}^m ] \mbf{G}^m
    - \frac{i}{2} \dot{\bm{\rho}}^m
\end{align}
where we have used the unitarity of $\mbf{V}^m$ as well as Eqs.~\eqref{eq:consistency_a} and \eqref{eq:F_tilde_def}.
In order to proceed, we subtract Eq.~\eqref{eq:modal_eles_times_v_dagger} from
the Hermitian conjugate of Eq.~\eqref{eq:modal_eles_times_v_dagger}:
\begin{align} 
    \mbf{0} &= \mathbb{A}(\mbf{F}^m)  - \big( \mathbb{H}[\bm{\rho}^m] \mbf{G}^m  - \mbf{G}^m  \mathbb{H}[\bm{\rho}^m]  \big)
    + i \mathbb{H}[\dot{\bm{\rho}}^m], \nn
            &= \mathbb{A} \Big[ \mbf{F}^m  - \big( \bm{\rho}^m \mbf{G}^m  - \mbf{G}^m \bm{\rho}^m  \big)
    + i \dot{\bm{\rho}}^m \Big]. \label{eq:g_equations_antiherm_appendix}
\end{align}
Here, $\mathbb{A}$ denotes the anti-Hermitian part of a square matrix. The matrix $\mbf{F}^m$ is defined as
\begin{align}
    \mbf{F}^m = \FFtildeprime{m} - \FFtildeplain{m}
\end{align}
and has the elements
\begin{align} \label{eq:F_matrix_element_appendix}
    \Fplain{m}{q}{p} = \elm{\Psi'}{[H, \Etilde{m}{p}{q}]}{\Psi}.
\end{align} 
\section{Simplification of Eq.~\eqref{eq:g_top_eq}} \label{appendix:gtop_elementwise}
\setcounter{equation}{0}
\renewcommand{\theequation}{\ref{appendix:gtop_elementwise}\arabic{equation}}
The element-wise form of Eq.~\eqref{eq:g_top_eq}
reads
\begin{multline}
    \frac{1}{2}
    \sum_{m'} \sum_{y^{m'} u^{m'}} 
        \Big[  \Cplain{m}{t}{x}{m}{y}{u} - \Cplain{m}{x}{t}{m}{u}{y}^*  \Big] \gplainprime{m}{y}{u} \\
    = \frac{1}{2} \Big[ \elm{\Psi'}{ [H, \Etilde{m}{t}{x}] }{\Psi}
    - \elm{\Psi'}{ [H, \Etilde{m}{x}{t}] }{\Psi}^* \Big].
\end{multline}
This left-hand side is simplified using Eqs.~\eqref{eq:C_element}, while the right-hand side 
is re-written by expanding the commutators and using the killer conditions:
\begin{multline}
    -\frac{1}{2}
    \sum_{m'} \sum_{y^{m'} u^{m'}} \delta_{mm'} \delta_{x^m y^m}
        \big(  \rrho{m}{u}{t} + \rrhoconj{m}{t}{u}  \big) \gplainprime{m}{y}{u} \\
    \begin{aligned}
        &=
        \begin{multlined}[t]
            \frac{1}{2} \Big[ 
        \elm{\Psi'}{[H, \creatilde{m}{t}] \annitilde{m}{x} }{\Psi}
    -   \elm{\Psi'}{\creatilde{m}{t} [\annitilde{m}{x}, H] }{\Psi} \\
    -   \elm{\Psi'}{[H, \creatilde{m}{x}] \annitilde{m}{t} }{\Psi}^*
    +   \elm{\Psi'}{\creatilde{m}{x} [\annitilde{m}{t}, H] }{\Psi}^*
        \Big]
        \end{multlined}
         \\
        &= -\frac{1}{2} \Big[  \elm{\Psi'}{\creatilde{m}{t} [\annitilde{m}{x}, H] }{\Psi} + \elm{\Psi'}{[H, \creatilde{m}{x}] \annitilde{m}{t} }{\Psi}^* \Big]
    \end{aligned}
\end{multline}
Reducing the sums and introducing the mean-field matrices from Eqs.~\eqref{eq:F_tilde_elements}
now yields
\begin{subequations}
    \begin{align}
        {^t\mbf{G}^m} \; \mathbb{H}[{^{a\!}}\bm{\raisedrho}^m]
        &= \tfrac{1}{2} ( {^t\FFtildeplain{m}} + {^b\FFtildeprimedagger{m}}) \\
        &= \tfrac{1}{2} \mbf{V}^{m\dagger}_{\!\!\subS} ( \FFcheckplain{m}_{\!\!\subA} + \FFcheckprimedagger{m}_{\!\!\subA}). \label{eq:Gt_equation_halftrans_appendix}
    \end{align}
\end{subequations}
The latter expression follows directly from Eq.~\eqref{eq:F_tilde_def}. 
\section{Constraint equations in electronic coupled cluster theory} \label{appendix:constraint_eqs_electronic_structure}
\setcounter{equation}{0}
\renewcommand{\theequation}{\ref{appendix:constraint_eqs_electronic_structure}\arabic{equation}}
The electronic structure equivalents of Eqs.~\eqref{eq:C_ud_TDMVCC}, \eqref{eq:C_dd_TDMVCC} and \eqref{eq:f_d_TDMVCC}
are obtained by substituting $i^m \rightarrow i$ and $i^{m'} \rightarrow j$ and then deleting all remaining mode indices:
\begin{align}
    {^{ud}\bar{C}_{(ai)(jb)}}
    &=
    \frac{1}{2} \Big(
    \elm{\Psi'}{[\tilde{E}_{jb}, \tilde{E}_{ai}]}{\Psi} -
    \elm{\Psi'}{[\tilde{E}_{bj}, \tilde{E}_{ia}]}{\Psi}^* \Big) \nn
    &= 
    \delta_{ab} \mathbb{H}[\bm{\raisedrho}]_{ij} - \delta_{ij} \mathbb{H}[\bm{\raisedrho}]_{ba}
\end{align}
\begin{align}
    {^{dd}\bar{C}'_{(ia)(jb)}} &=
        \frac{1}{2}
        \sum_\mu \Big( 
            \elm{\Psi}{[\tilde{E}_{ia}, \tau_\mu]}{\Psi}   \elm{\mu}{e^{-T} \tilde{E}_{jb}}{\Psi}
            - \elm{\Psi}{[\tilde{E}_{jb}, \tau_\mu]}{\Psi}   \elm{\mu}{e^{-T} \tilde{E}_{ia}}{\Psi}
            \Big),
\end{align}
\begin{multline}
    {^{d}\bar{f}'_{(ia)}} =
    \frac{1}{2} \Big(
    \elm{\Psi'}{[H, \tilde{E}_{ia}]}{\Psi} -
    \elm{\Psi'}{[H, \tilde{E}_{ai}]}{\Psi}^* \Big) \\
    \begin{alignedat}[t]{2}
        {}+ \frac{1}{2} \sum_\mu \Big( 
        \elm{\Psi}{[\tilde{E}_{ia}, \tau_\mu]}{\Psi}   \elm{\mu}{e^{-T} H}{\Psi} 
        - \elm{\Psi}{[H, \tau_\mu]}{\Psi}   \elm{\mu}{e^{-T} \tilde{E}_{ia}}{\Psi}
        \Big).
    \end{alignedat}
\end{multline}

\newpage

\end{document}


\title{Time-dependent coupled cluster with orthogonal adaptive basis functions: General formalism and application to the vibrational problem: Supplementary material}

\author{Mads Greisen Højlund}
\email{madsgh@chem.au.dk}
\affiliation{\au}

\author{Alberto Zoccante}
\email{alberto.zoccante@uniupo.it}
\affiliation{\upo}

\author{Ove Christiansen}
\email{ove@chem.au.dk}
\affiliation{\au}

\hypersetup{pdftitle={Time-dependent coupled cluster with orthogonal adaptive basis functions: General formalism and application to the vibrational problem: Supplementary material}}
\hypersetup{pdfauthor={M.~G.~Højlund, et al.}}
\hypersetup{bookmarksopen=true}

\date{\today}
\maketitle
\tableofcontents

\newpage
\section{Additional numerical data}
\renewcommand{\thefigure}{S\arabic{figure}}

For ease of comparison, we present the same data for all the triatomics, including
some that is already included in the main manuscript. Computational details can
be found in the main manuscript.
\newpage

\subsection{Water}

\subsubsection{Hilbert space angles}

%
%
\begin{figure}[H]
    \centering

    \begin{subfigure}[c]{0.75\textwidth}
        \centering
        \caption{Ket angle}
        \includegraphics[width=\textwidth]{angleket_water_tdmvcc23_vs_tdfvci_t10000_points10001.pdf}
        \label{fig:water_ketangle}
    \end{subfigure}


    \begin{subfigure}[c]{0.75\textwidth}
        \centering
        \caption{Bra angle}
        \includegraphics[width=\textwidth]{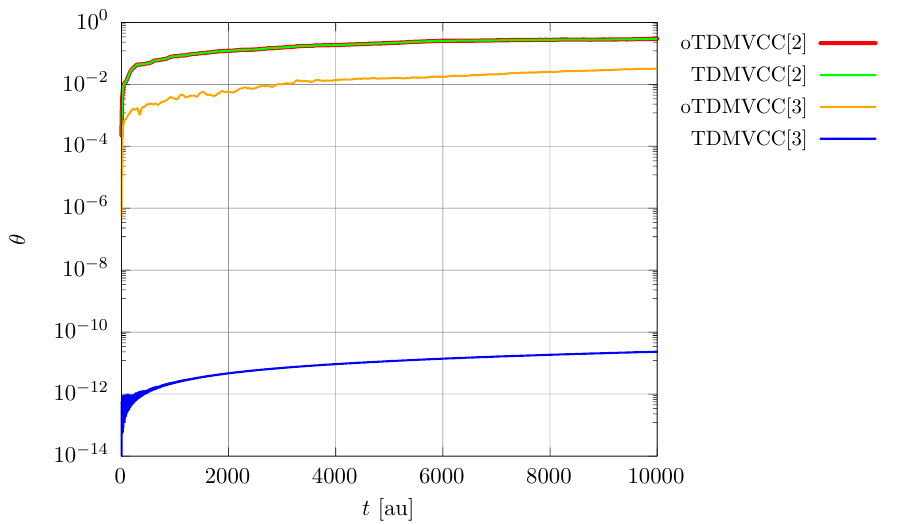}
        \label{fig:water_brangle}
    \end{subfigure}

    \caption{IVR of water at the oTDMVCC[2--3] and TDMVCC[2--3] levels
    with $N = 8$ and $N_{\2\subA} = 8$ for all modes. (a) Hilbert space ket angle. (b) Hilbert space bra angle.}
    \label{fig:water_ketbraangle}
\end{figure}

\newpage

\subsubsection{\Aclp{acf}}

%
%
\begin{figure}[H]
    \centering

    \begin{subfigure}[c]{0.75\textwidth}
        \centering
        \caption{Autocorrelation function}
        \includegraphics[width=\textwidth]{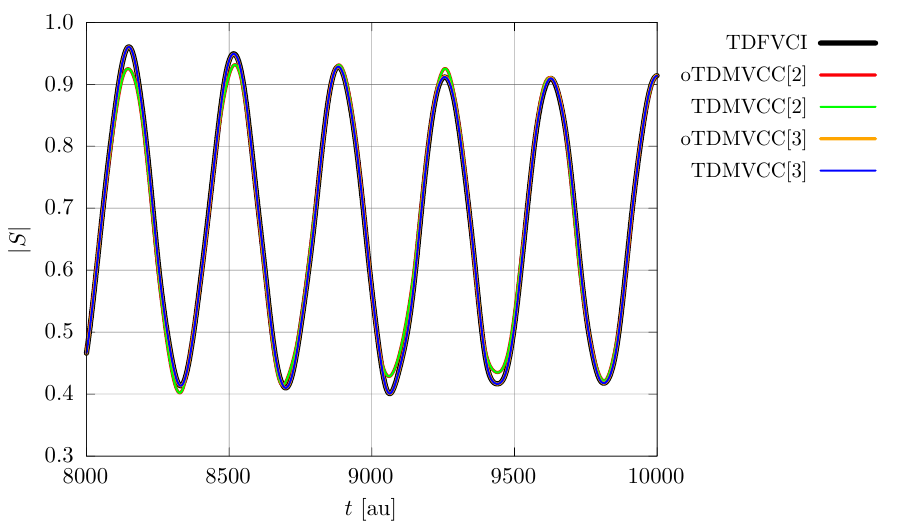}
        \label{fig:water_autocorr}
    \end{subfigure}


    \begin{subfigure}[c]{0.75\textwidth}
        \centering
        \caption{Absolute error in autocorrelation function}
        \includegraphics[width=\textwidth]{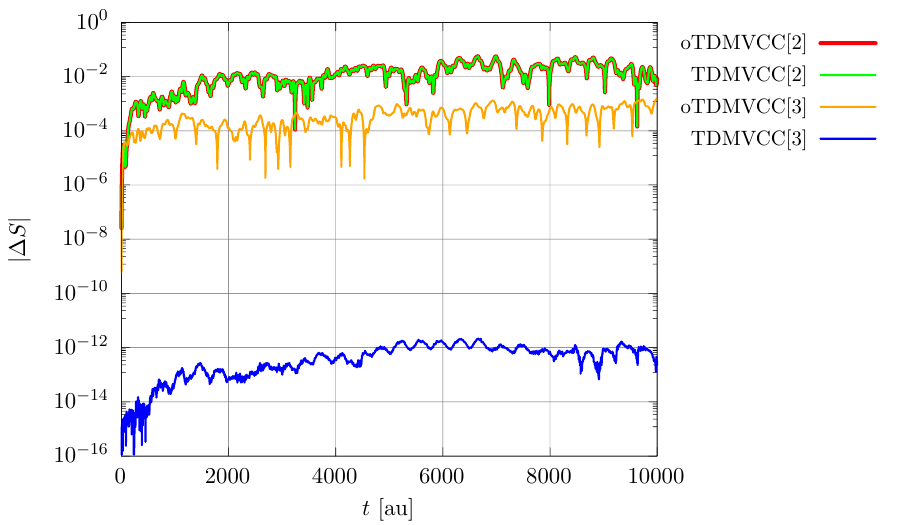}
        \label{fig:water_autocorr_diff}
    \end{subfigure}

    \caption{IVR of water at the oTDMVCC[2--3] and TDMVCC[2--3] levels
    with $N = 8$ and $N_{\2\subA} = 8$ for all modes. (a) \Acl{acf}. (b) Absolute error in the \acl{acf} (relative to TDFVCI).}
    \label{fig:water_autocorr_and_diff}
\end{figure}

\newpage

%
%
\begin{figure}[H]
    \centering
    \includegraphics[width=\textwidth]{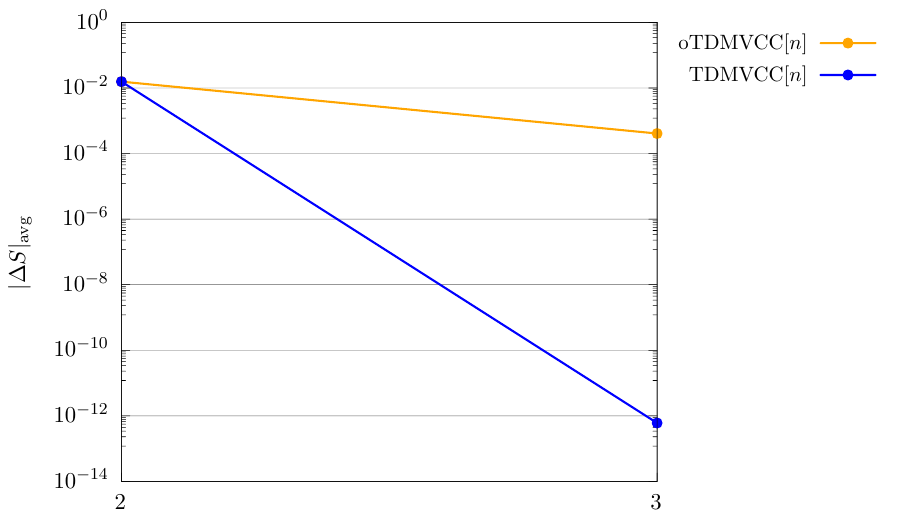}
    \caption{Average absolute error in the \acl{acf} for water at the oTDMVCC[2--3] and TDMVCC[2--3] levels
    with $N = 8$ and $N_{\2\subA} = 8$ for all modes. The errors are computed relative to TDFVCI.}
    \label{fig:water_autocorr_diff_avg}
\end{figure}

\newpage

\subsubsection{Expectation values}

%
%
\begin{figure}[H]
    \centering

    \begin{subfigure}[c]{0.75\textwidth}
        \centering
        \caption{}
        \includegraphics[width=\textwidth]{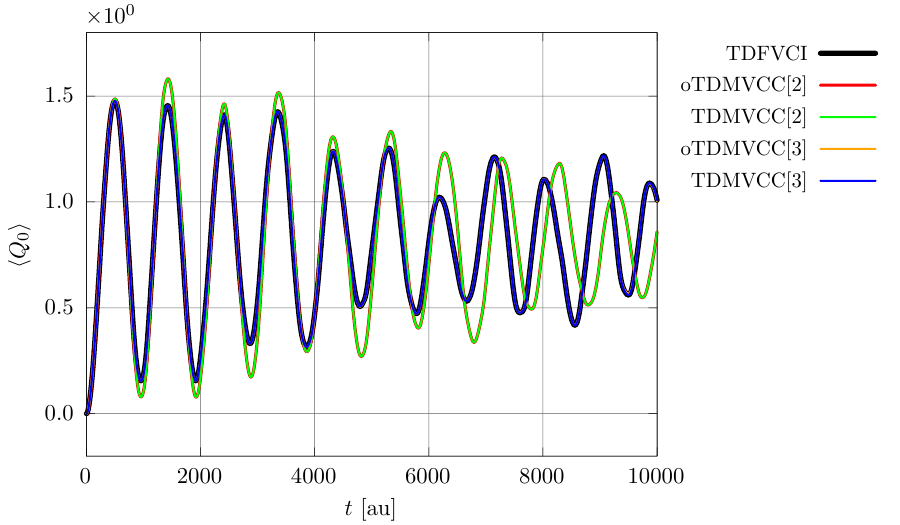}
        \label{fig:water_Q0}
    \end{subfigure}


    \begin{subfigure}[c]{0.75\textwidth}
        \centering
        \caption{}
        \includegraphics[width=\textwidth]{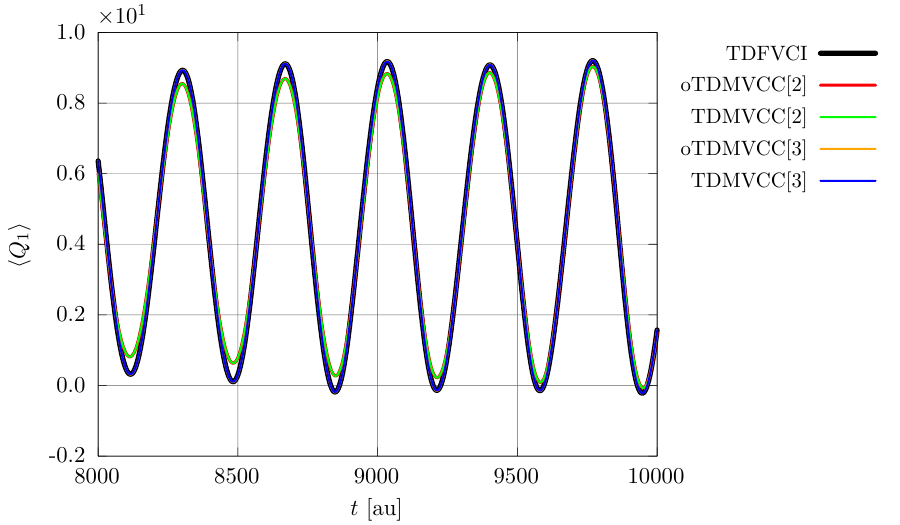}
        \label{fig:water_Q1}
    \end{subfigure}

    \caption{IVR of water at the oTDMVCC[2--3] and TDMVCC[2--3] levels
    with $N = 8$ and $N_{\2\subA} = 8$ for all modes. 
    Expectation value of (a) $Q_0$ (bend) and (b) $Q_1$ (symmetric stretch).
    Errors are calculated relative to TDFVCI.}

    \label{fig:water_Q0_Q1}
\end{figure}

\newpage

%
%
\begin{figure}[H]
    \centering

    \begin{subfigure}[c]{0.75\textwidth}
        \centering
        \caption{}
        \includegraphics[width=\textwidth]{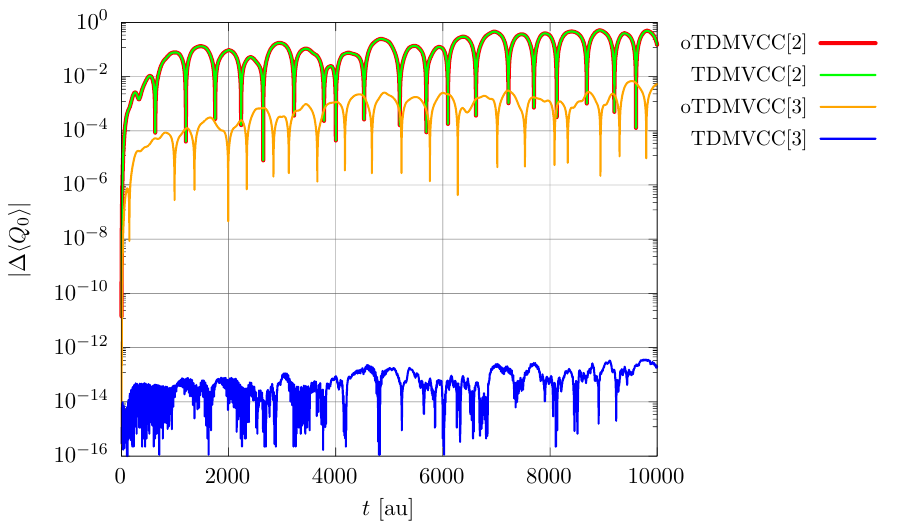}
        \label{fig:water_Q0_diff}
    \end{subfigure}


    \begin{subfigure}[c]{0.75\textwidth}
        \centering
        \caption{}
        \includegraphics[width=\textwidth]{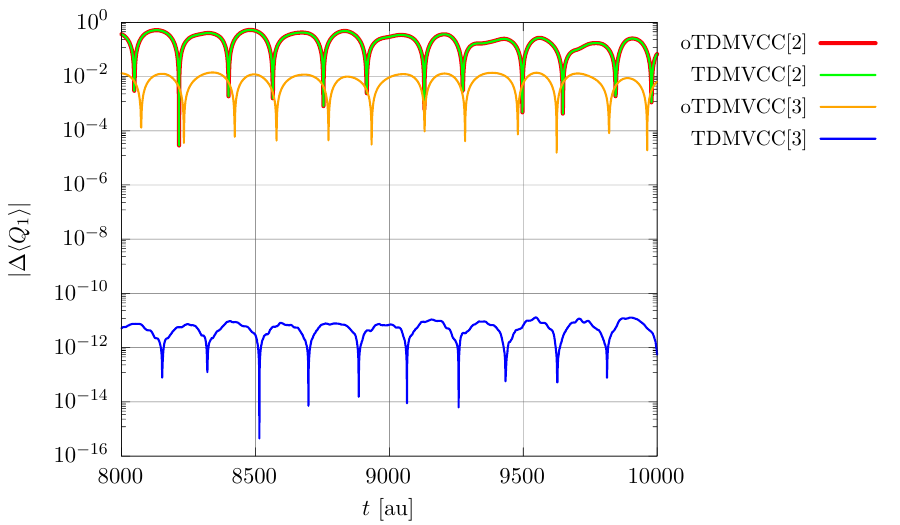}
        \label{fig:water_Q1_diff}
    \end{subfigure}

    \caption{IVR of water at the oTDMVCC[2--3] and TDMVCC[2--3] levels
    with $N = 8$ and $N_{\2\subA} = 8$ for all modes. 
    Absolute error in the expectation value of (a) $Q_0$ (bend) and (b) $Q_1$ (symmetric stretch).
    Errors are calculated relative to TDFVCI.}

    \label{fig:water_Q0_Q1_diff}
\end{figure}

\newpage

%
%
\begin{figure}[H]
    \centering

    \begin{subfigure}[c]{0.75\textwidth}
        \centering
        \caption{}
        \includegraphics[width=\textwidth]{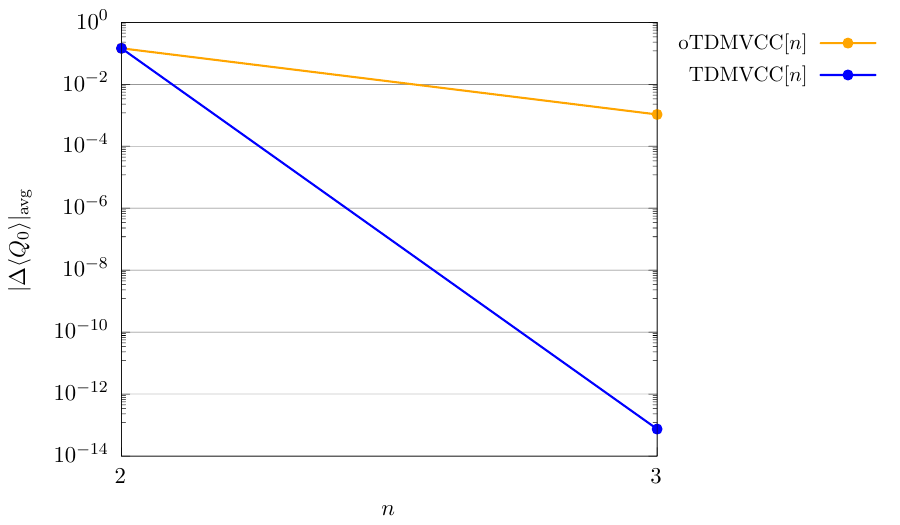}
        \label{fig:water_Q0_diff_avg}
    \end{subfigure}


    \begin{subfigure}[c]{0.75\textwidth}
        \centering
        \caption{}
        \includegraphics[width=\textwidth]{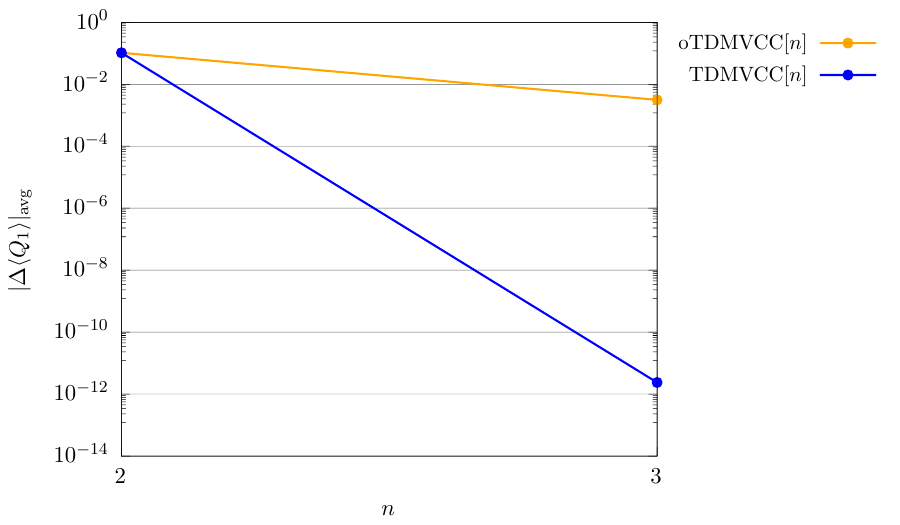}
        \label{fig:water_Q1_diff_avg}
    \end{subfigure}

    \caption{IVR of water at the oTDMVCC[2--3] and TDMVCC[2--3] levels
    with $N = 8$ and $N_{\2\subA} = 8$ for all modes. 
    Average absolute error in the expectation value of (a) $Q_0$ (bend) and (b) $Q_1$ (symmetric stretch).
    Errors are calculated relative to TDFVCI.}
    \label{fig:water_Q0_Q1_diff_avg}
\end{figure}

\newpage

\subsubsection{Imaginary part of expectation values}

%
%
\begin{figure}[H]
    \centering

    \begin{subfigure}[c]{0.75\textwidth}
        \centering
        \caption{}
        \includegraphics[width=\textwidth]{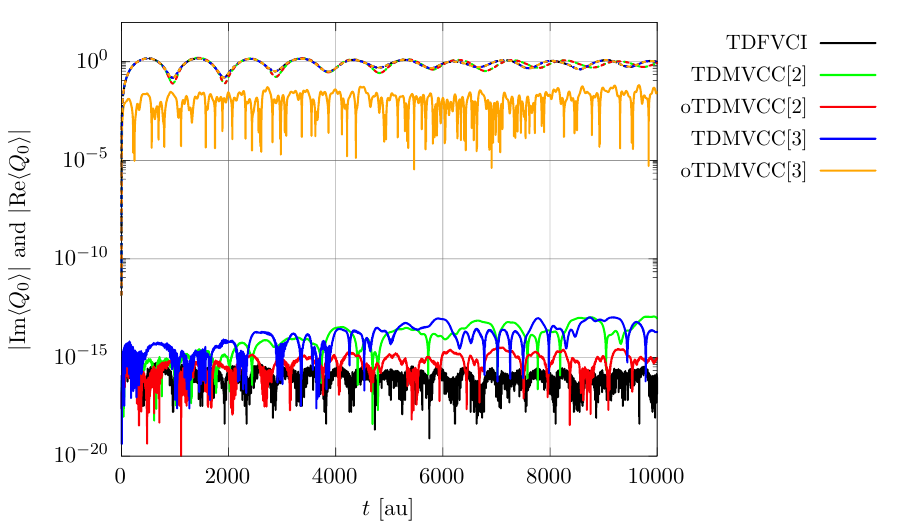}
        \label{fig:water_re_im_Q0}
    \end{subfigure}


    \begin{subfigure}[c]{0.75\textwidth}
        \centering
        \caption{}
        \includegraphics[width=\textwidth]{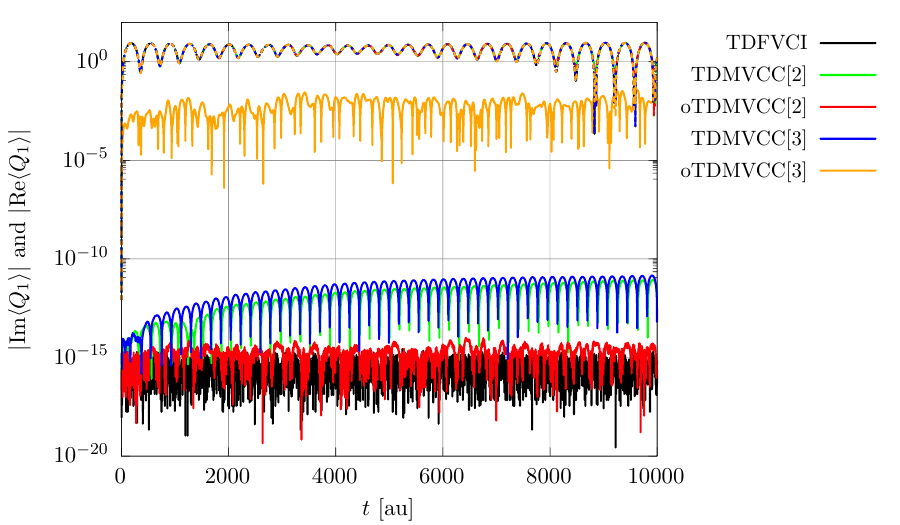}
        \label{fig:water_re_im_Q1}
    \end{subfigure}

    \caption{IVR of water at the oTDMVCC[2--3] and TDMVCC[2--3] levels
    with $N = 8$ and $N_{\2\subA} = 8$ for all modes. 
    Absolute value of $\mathrm{Re}\elm{\Psi'}{Q}{\Psi}$ (dashed) and $\mathrm{Im}\elm{\Psi'}{Q}{\Psi}$ (full) for (a) $Q_0$ (bend) and (b) $Q_1$ (symmetric stretch).
    Note that the dashed lines (real part) are almost on top of each other.}

    \label{fig:water_re_im_Q0_Q1}
\end{figure}

\newpage

%
%
\begin{figure}[H]
    \centering

    \begin{subfigure}[c]{0.75\textwidth}
        \centering
        \caption{}
        \includegraphics[width=\textwidth]{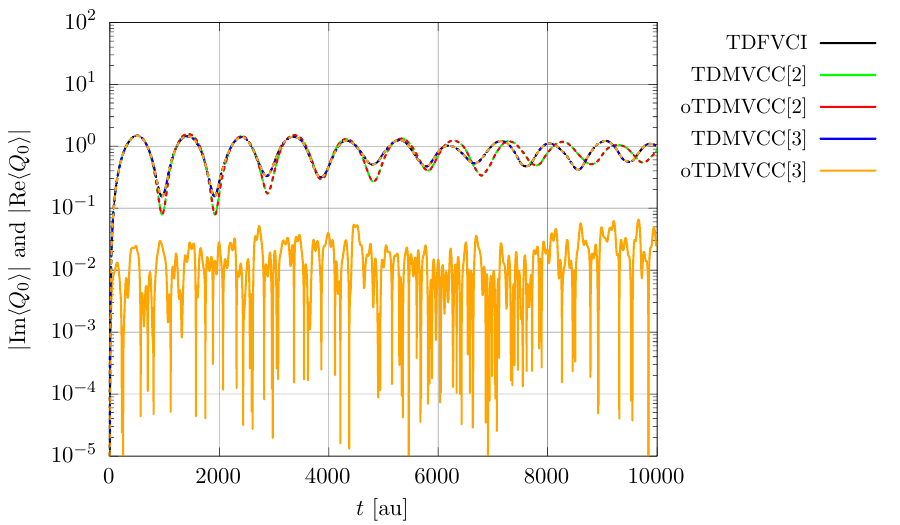}
        \label{fig:water_re_im_zoom_Q0}
    \end{subfigure}


    \begin{subfigure}[c]{0.75\textwidth}
        \centering
        \caption{}
        \includegraphics[width=\textwidth]{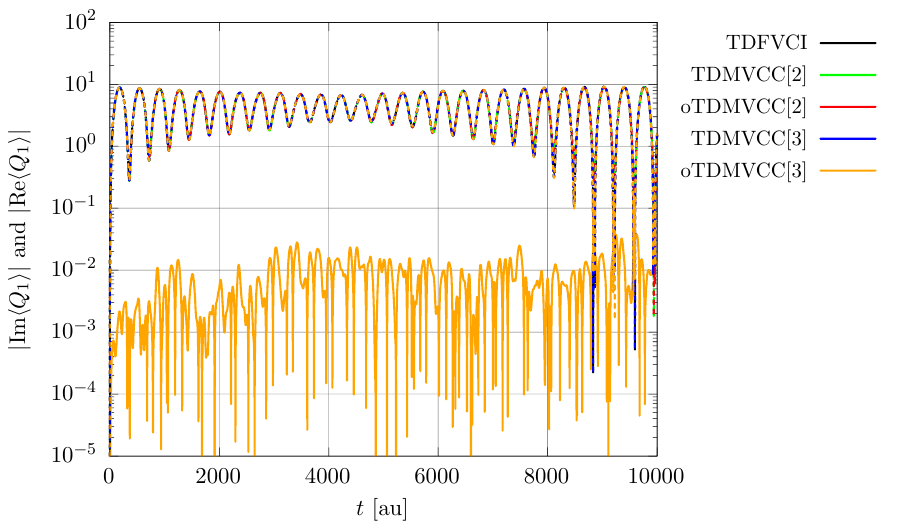}
        \label{fig:water_re_im_zoom_Q1}
    \end{subfigure}

    \caption{IVR of water at the oTDMVCC[2--3] and TDMVCC[2--3] levels
    with $N = 8$ and $N_{\2\subA} = 8$ for all modes. 
    Absolute value of $\mathrm{Re}\elm{\Psi'}{Q}{\Psi}$ (dashed) and $\mathrm{Im}\elm{\Psi'}{Q}{\Psi}$ (full) for (a) $Q_0$ (bend) and (b) $Q_1$ (symmetric stretch).
    Note that the dashed lines (real part) are almost on top of each other.}

    \label{fig:water_re_im_zoom_Q0_Q1}
\end{figure}

\newpage

%
%
\begin{figure}[H]
    \centering

    \begin{subfigure}[c]{0.75\textwidth}
        \centering
        \caption{}
        \includegraphics[width=\textwidth]{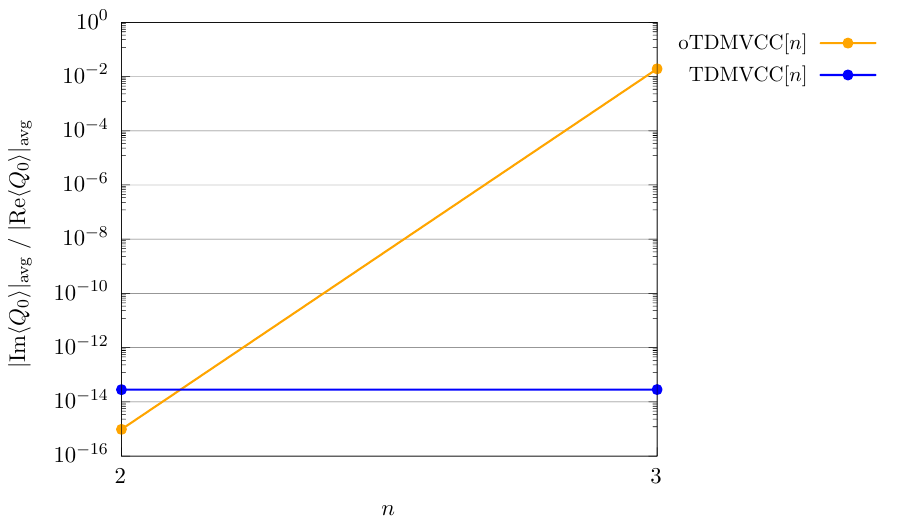}
        \label{fig:water_re_im_ratio_Q0}
    \end{subfigure}


    \begin{subfigure}[c]{0.75\textwidth}
        \centering
        \caption{}
        \includegraphics[width=\textwidth]{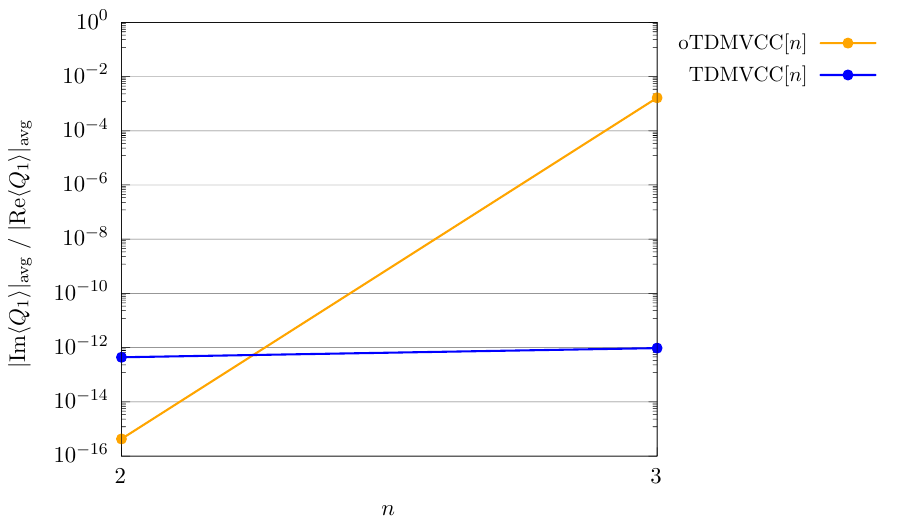}
        \label{fig:water_re_im_ratio_Q1}
    \end{subfigure}

    \caption{IVR of water at the oTDMVCC[2--3] and TDMVCC[2--3] levels
    with $N = 8$ and $N_{\2\subA} = 8$ for all modes. 
    Ratio of mean absolute values of $\mathrm{Re}\elm{\Psi'}{Q}{\Psi}$ (dashed) and $\mathrm{Im}\elm{\Psi'}{Q}{\Psi}$ (full) for (a) $Q_0$ (bend) and (b) $Q_1$ (symmetric stretch).}

    \label{fig:water_re_im_ratio_Q0_Q1}
\end{figure}

\newpage

\subsubsection{Energy conservation}

\begin{figure}[H]
    \centering

    \begin{subfigure}[c]{0.8\textwidth}
        \centering
        \caption{TDMVCC[2] and oTDMVCC[2]}
        \includegraphics[width=\textwidth]{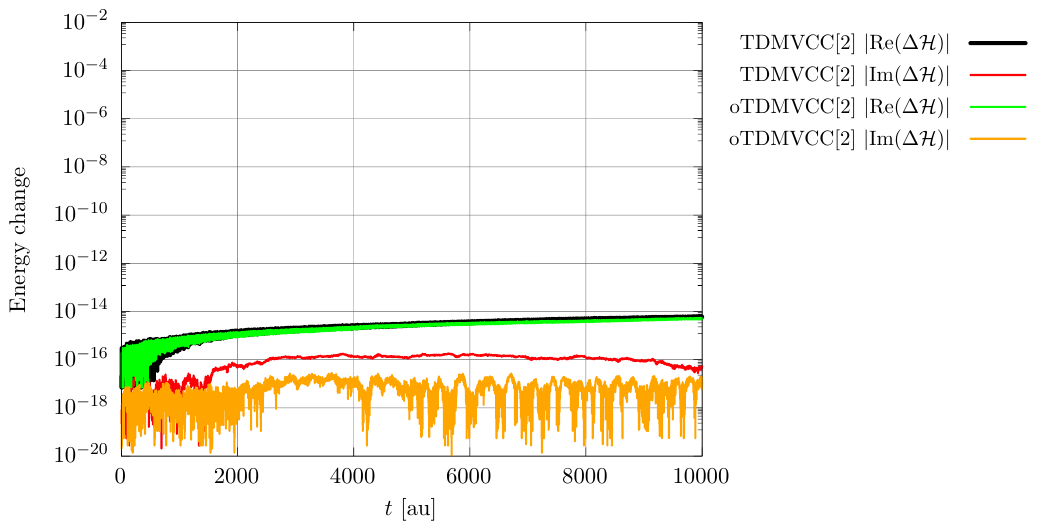}
        \label{fig:water_energy_tdmvcc2}
    \end{subfigure}


    \begin{subfigure}[c]{0.8\textwidth}
        \centering
        \caption{TDMVCC[3] and oTDMVCC[3]}
        \includegraphics[width=\textwidth]{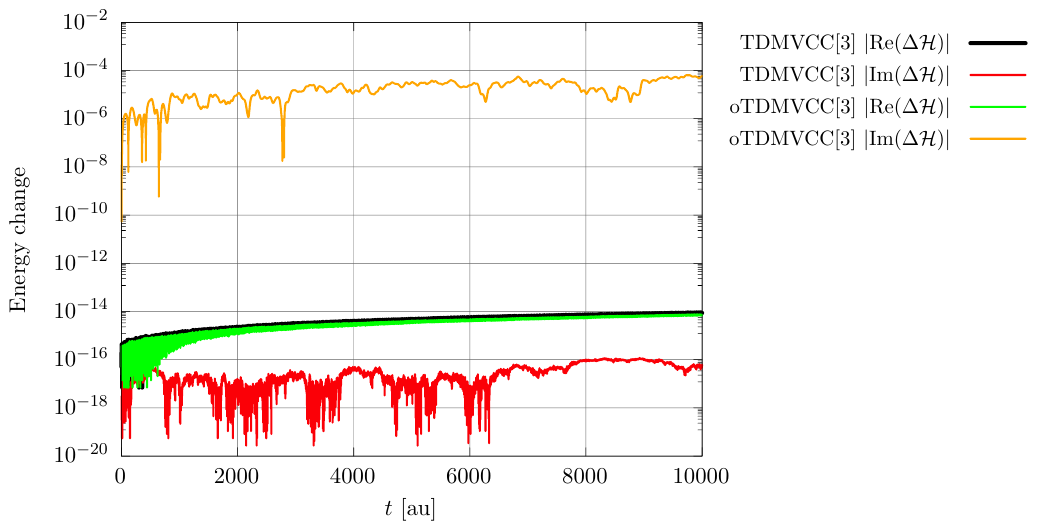}
        \label{fig:water_energy_tdmvcc3}
    \end{subfigure}

    \caption{IVR of water at the oTDMVCC[2--3] and TDMVCC[2--3] levels
    with $N = 8$ and $N_{\2\subA} = 8$ for all modes. 
    Absolute change in the energy function $\mathcal{H} = \elm{\Psi'}{H}{\Psi}$ for (a) TDMVCC[2]/oTDMVCC[2] and (b) TDMVCC[3]/oTDMVCC[3].}
    \label{fig:water_energy}
\end{figure}

\newpage

\subsubsection{Non-hermiticity in density matrices and mean fields}

\begin{figure}[H]
    \centering

    \begin{subfigure}[c]{0.8\textwidth}
        \centering
        \caption{Density}
        \includegraphics[width=\textwidth]{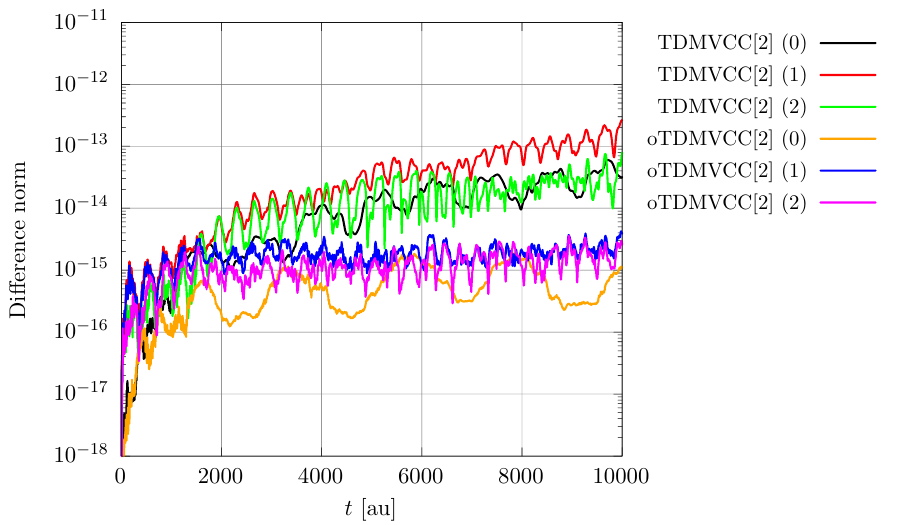}
        \label{fig:water_nonherm_tdmvcc2_density}
    \end{subfigure}


    \begin{subfigure}[c]{0.8\textwidth}
        \centering
        \caption{Mean field}
        \includegraphics[width=\textwidth]{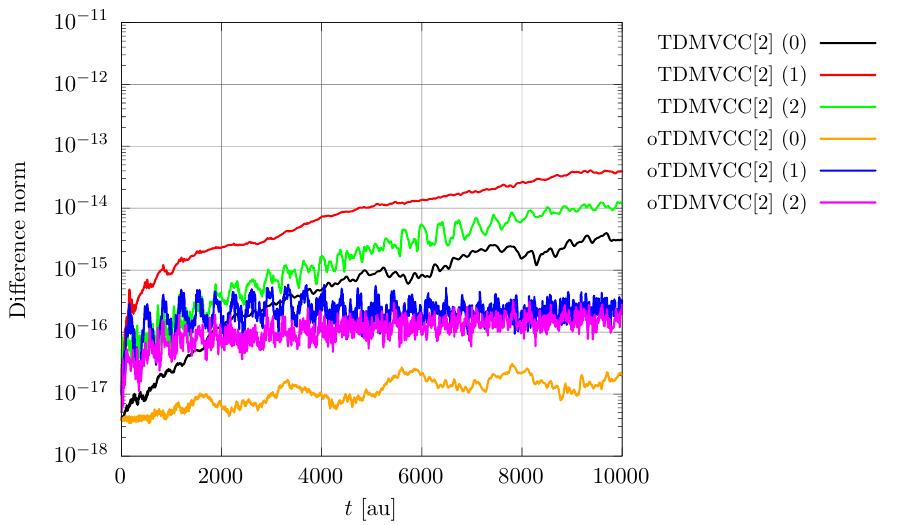}
        \label{fig:water_nonherm_tdmvcc2_meanfield}
    \end{subfigure}

    \caption{IVR of water at the oTDMVCC[2] and TDMVCC[2] levels
    with $N = 8$ and $N_{\2\subA} = 8$ for all modes. 
    Deviation from Hermiticity in 
    (a) one-mode density matrices ($|| \bm{\rho} -  \bm{\rho}^{\dagger} ||$) and 
    (b) half-transformed mean-field matrices ($|| \check{\mbf{F}} - \check{\mbf{F}}^{\prime \dagger} ||$).
    Mode indices are given in parentheses.}
    \label{fig:water_nonherm_tdmvcc2}
\end{figure}

\newpage 
\subsection{Ozone}

\subsubsection{Hilbert space angles}

%
%
\begin{figure}[H]
    \centering

    \begin{subfigure}[c]{0.75\textwidth}
        \centering
        \caption{Ket angle}
        \includegraphics[width=\textwidth]{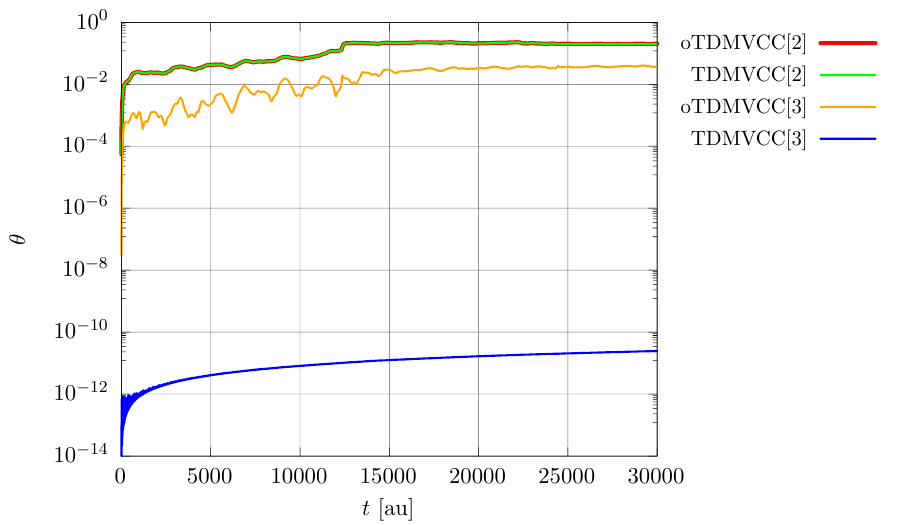}
        \label{fig:O3_ketangle}
    \end{subfigure}


    \begin{subfigure}[c]{0.75\textwidth}
        \centering
        \caption{Bra angle}
        \includegraphics[width=\textwidth]{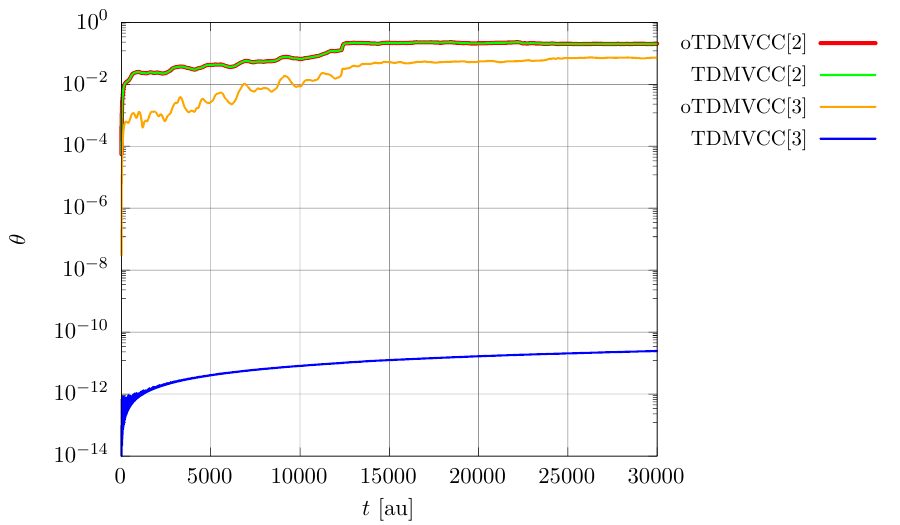}
        \label{fig:O3_brangle}
    \end{subfigure}

    \caption{IVR of ozone at the oTDMVCC[2--3] and TDMVCC[2--3] levels
    with $N = 8$ and $N_{\2\subA} = 8$ for all modes. (a) Hilbert space ket angle. (b) Hilbert space bra angle.}
    \label{fig:O3_ketbraangle}
\end{figure}

\newpage

\subsubsection{\Aclp{acf}}

%
%
\begin{figure}[H]
    \centering

    \begin{subfigure}[c]{0.75\textwidth}
        \centering
        \caption{Autocorrelation function}
        \includegraphics[width=\textwidth]{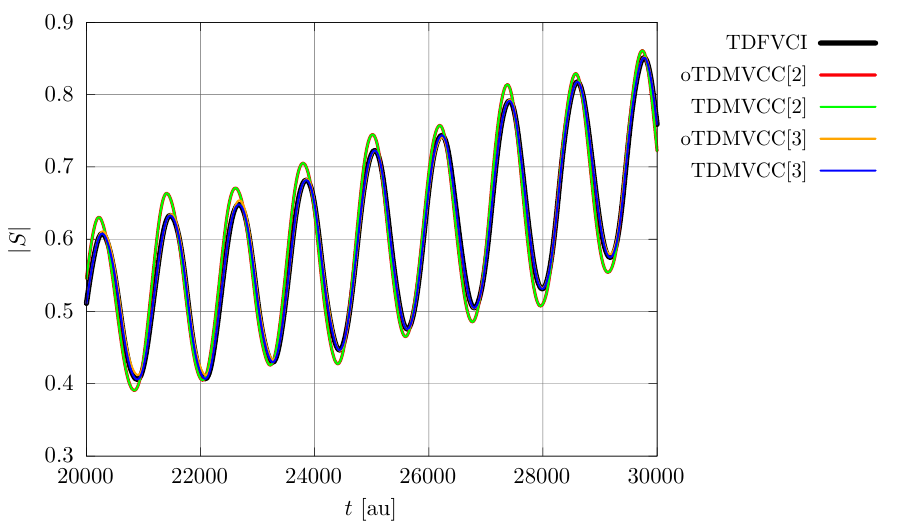}
        \label{fig:O3_autocorr}
    \end{subfigure}


    \begin{subfigure}[c]{0.75\textwidth}
        \centering
        \caption{Absolute error in autocorrelation function}
        \includegraphics[width=\textwidth]{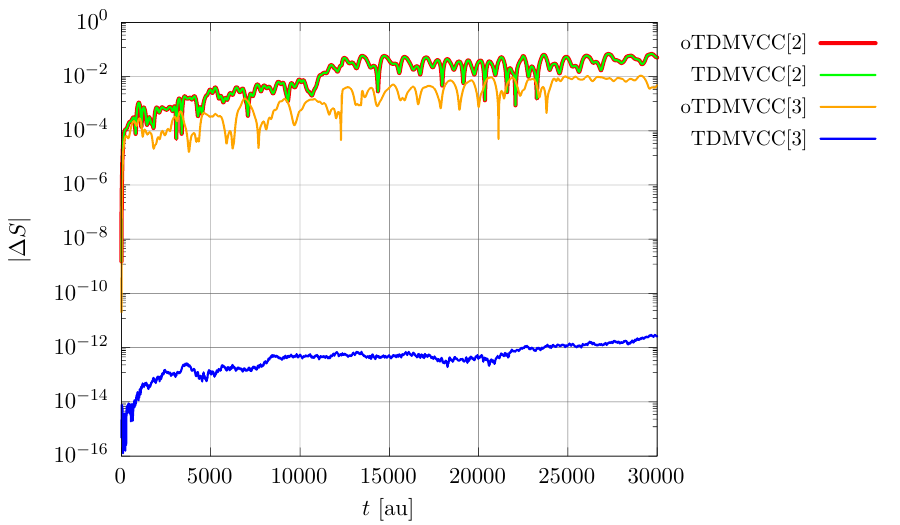}
        \label{fig:O3_autocorr_diff}
    \end{subfigure}

    \caption{IVR of ozone at the oTDMVCC[2--3] and TDMVCC[2--3] levels
    with $N = 8$ and $N_{\2\subA} = 8$ for all modes. (a) \Acl{acf}. (b) Absolute error in the \acl{acf} (relative to TDFVCI).}
    \label{fig:O3_autocorr_and_diff}
\end{figure}

\newpage

%
%
\begin{figure}[H]
    \centering
    \includegraphics[width=\textwidth]{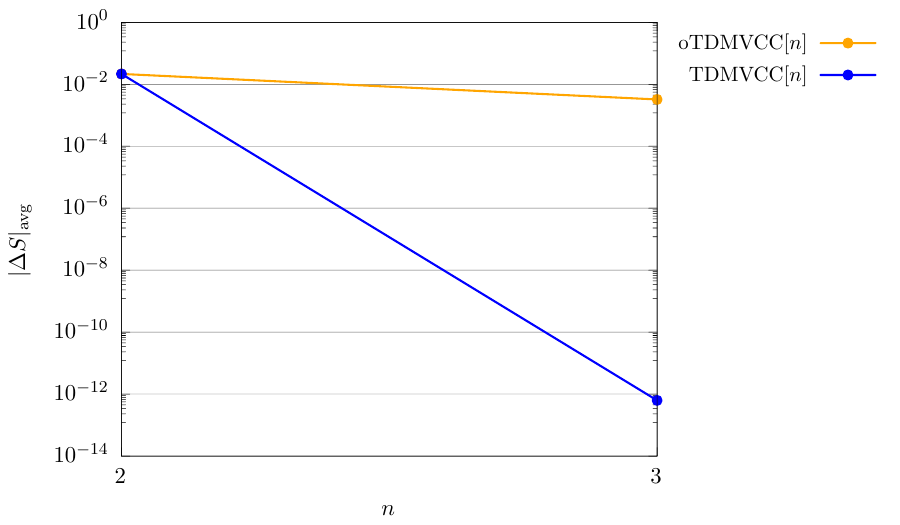}
    \caption{Average absolute error in the \acl{acf} for ozone at the oTDMVCC[2--3] and TDMVCC[2--3] levels
    with $N = 8$ and $N_{\2\subA} = 8$ for all modes. The errors are computed relative to TDFVCI.}
    \label{fig:O3_autocorr_diff_avg}
\end{figure}

\newpage

\subsubsection{Expectation values}

%
%
\begin{figure}[H]
    \centering

    \begin{subfigure}[c]{0.75\textwidth}
        \centering
        \caption{}
        \includegraphics[width=\textwidth]{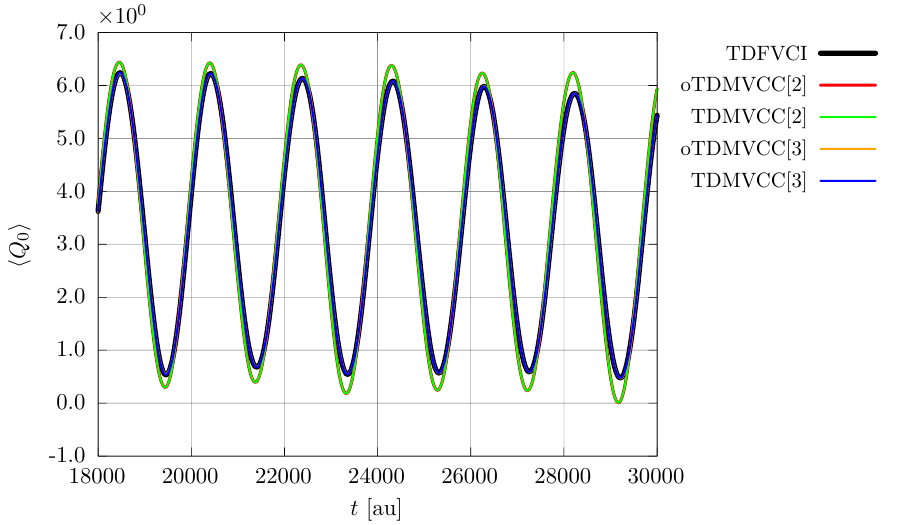}
        \label{fig:O3_Q0}
    \end{subfigure}


    \begin{subfigure}[c]{0.75\textwidth}
        \centering
        \caption{}
        \includegraphics[width=\textwidth]{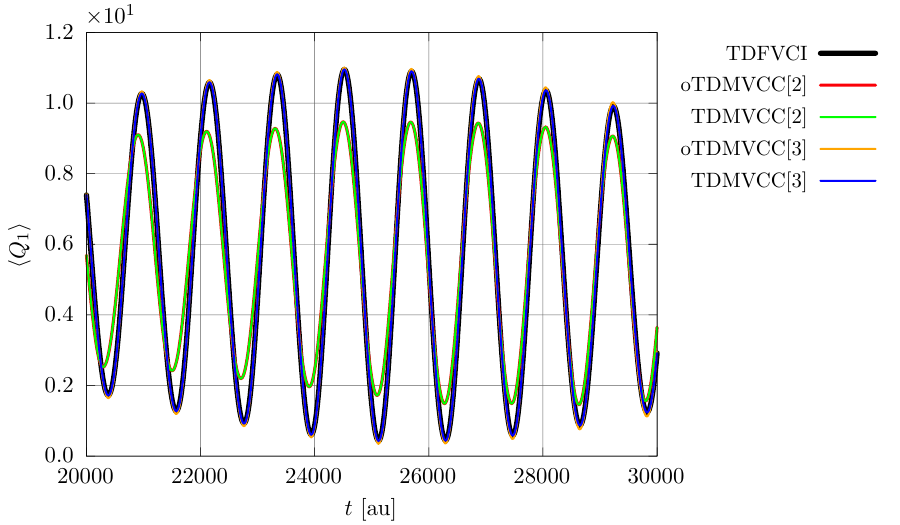}
        \label{fig:O3_Q1}
    \end{subfigure}

    \caption{IVR of ozone at the oTDMVCC[2--3] and TDMVCC[2--3] levels
    with $N = 8$ and $N_{\2\subA} = 8$ for all modes. 
    Expectation value of (a) $Q_0$ (bend) and (b) $Q_1$ (symmetric stretch).
    Errors are calculated relative to TDFVCI.}

    \label{fig:O3_Q0_Q1}
\end{figure}

\newpage

%
%
\begin{figure}[H]
    \centering

    \begin{subfigure}[c]{0.75\textwidth}
        \centering
        \caption{}
        \includegraphics[width=\textwidth]{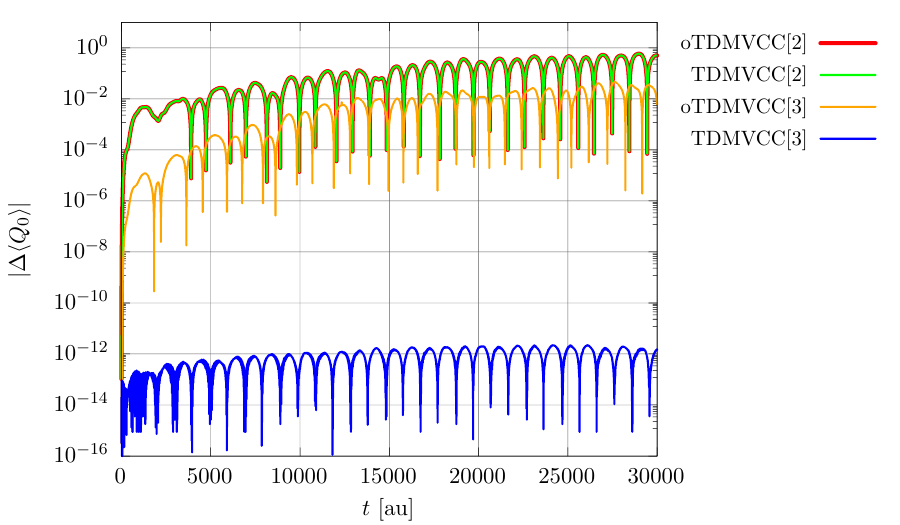}
        \label{fig:O3_Q0_diff}
    \end{subfigure}


    \begin{subfigure}[c]{0.75\textwidth}
        \centering
        \caption{}
        \includegraphics[width=\textwidth]{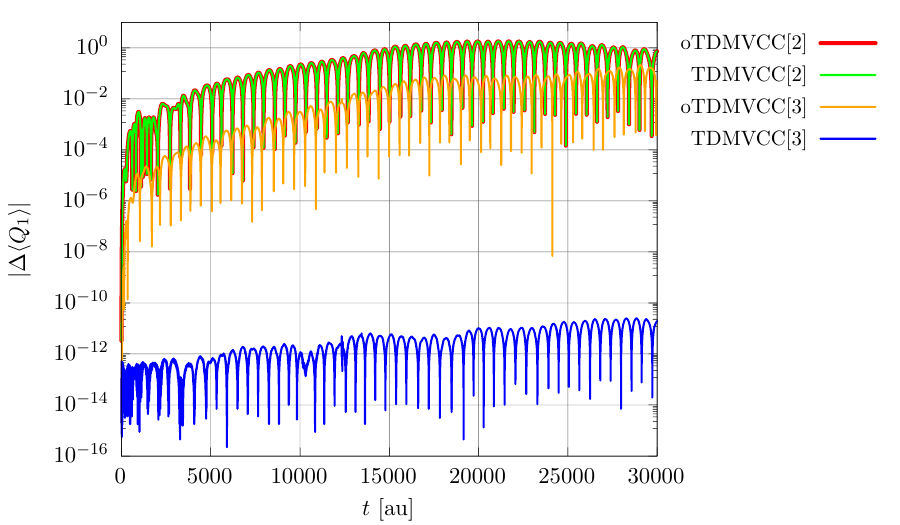}
        \label{fig:O3_Q1_diff}
    \end{subfigure}

    \caption{IVR of ozone at the oTDMVCC[2--3] and TDMVCC[2--3] levels
    with $N = 8$ and $N_{\2\subA} = 8$ for all modes. 
    Absolute error in the expectation value of (a) $Q_0$ (bend) and (b) $Q_1$ (symmetric stretch).
    Errors are calculated relative to TDFVCI.}

    \label{fig:O3_Q0_Q1_diff}
\end{figure}

\newpage

%
%
\begin{figure}[H]
    \centering

    \begin{subfigure}[c]{0.75\textwidth}
        \centering
        \caption{}
        \includegraphics[width=\textwidth]{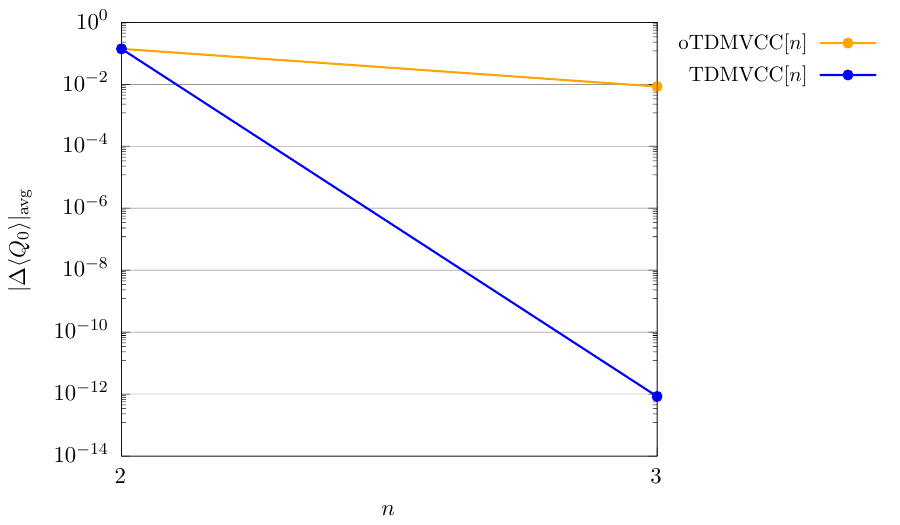}
        \label{fig:O3_Q0_diff_avg}
    \end{subfigure}


    \begin{subfigure}[c]{0.75\textwidth}
        \centering
        \caption{}
        \includegraphics[width=\textwidth]{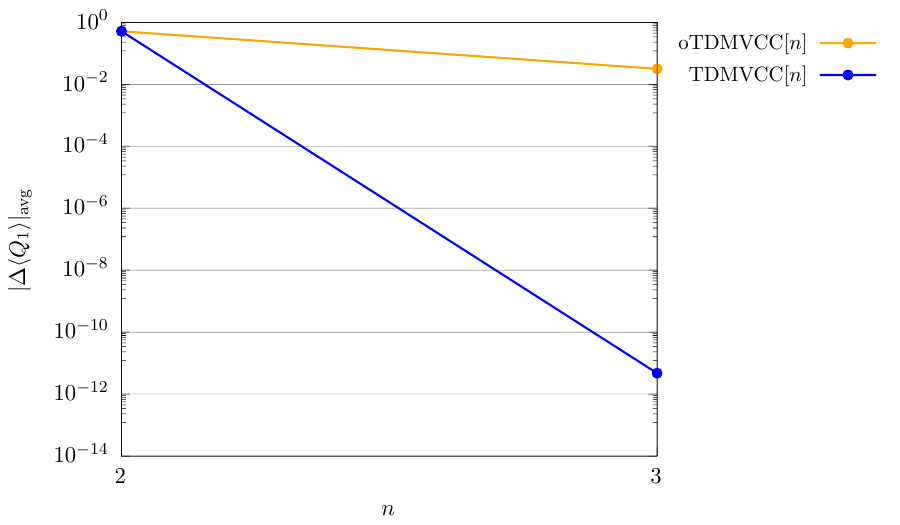}
        \label{fig:O3_Q1_diff_avg}
    \end{subfigure}

    \caption{IVR of ozone at the oTDMVCC[2--3] and TDMVCC[2--3] levels
    with $N = 8$ and $N_{\2\subA} = 8$ for all modes. 
    Average absolute error in the expectation value of (a) $Q_0$ (bend) and (b) $Q_1$ (symmetric stretch).
    Errors are calculated relative to TDFVCI.}
    \label{fig:O3_Q0_Q1_diff_avg}
\end{figure}

\newpage

\subsubsection{Imaginary part of expectation values}

%
%
\begin{figure}[H]
    \centering

    \begin{subfigure}[c]{0.75\textwidth}
        \centering
        \caption{}
        \includegraphics[width=\textwidth]{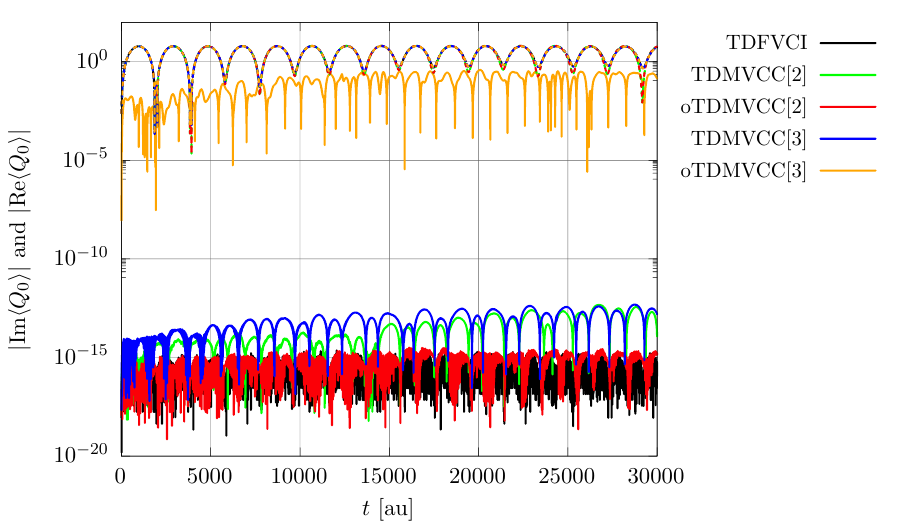}
        \label{fig:O3_re_im_Q0}
    \end{subfigure}


    \begin{subfigure}[c]{0.75\textwidth}
        \centering
        \caption{}
        \includegraphics[width=\textwidth]{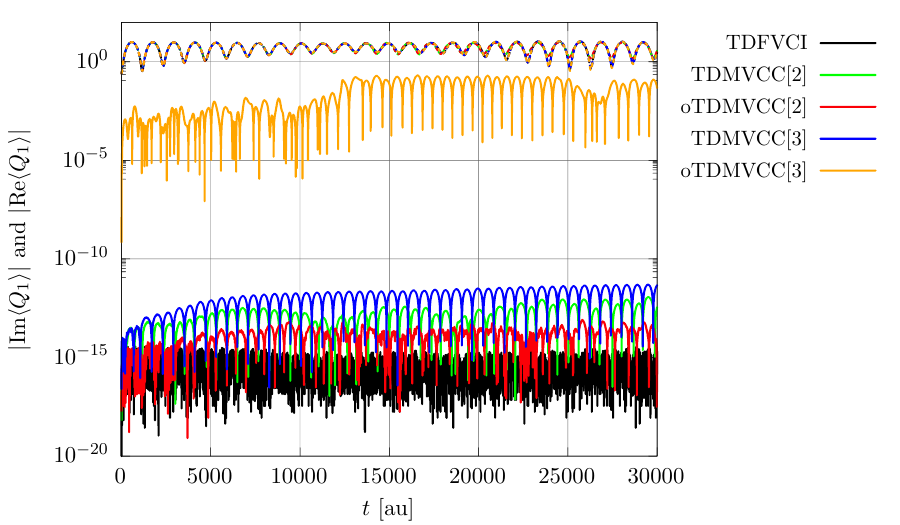}
        \label{fig:O3_re_im_Q1}
    \end{subfigure}

    \caption{IVR of ozone at the oTDMVCC[2--3] and TDMVCC[2--3] levels
    with $N = 8$ and $N_{\2\subA} = 8$ for all modes. 
    Absolute value of $\mathrm{Re}\elm{\Psi'}{Q}{\Psi}$ (dashed) and $\mathrm{Im}\elm{\Psi'}{Q}{\Psi}$ (full) for (a) $Q_0$ (bend) and (b) $Q_1$ (symmetric stretch).
    Note that the dashed lines (real part) are almost on top of each other.}

    \label{fig:O3_re_im_Q0_Q1}
\end{figure}

\newpage

%
%
\begin{figure}[H]
    \centering

    \begin{subfigure}[c]{0.75\textwidth}
        \centering
        \caption{}
        \includegraphics[width=\textwidth]{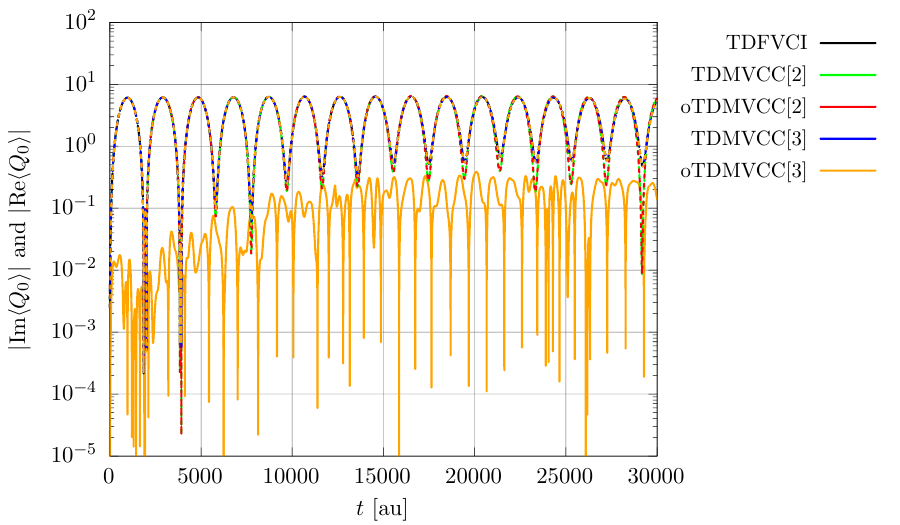}
        \label{fig:O3_re_im_zoom_Q0}
    \end{subfigure}


    \begin{subfigure}[c]{0.75\textwidth}
        \centering
        \caption{}
        \includegraphics[width=\textwidth]{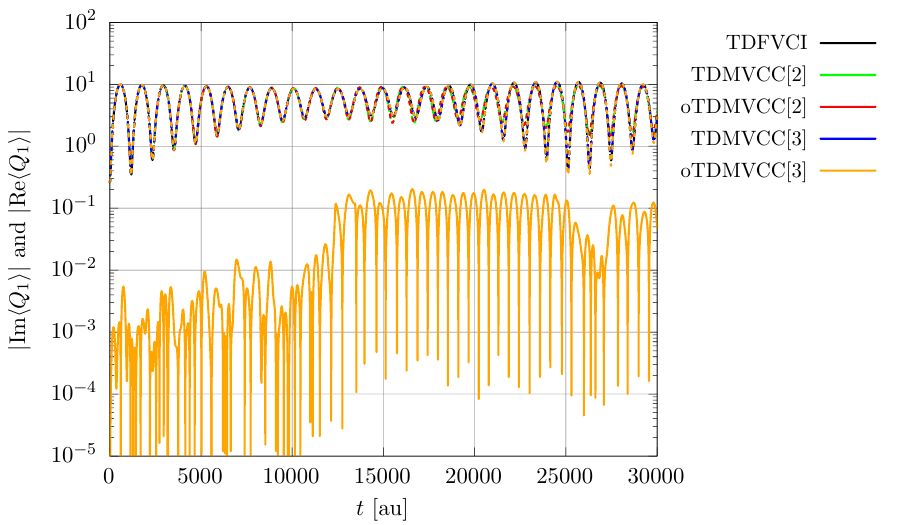}
        \label{fig:O3_re_im_zoom_Q1}
    \end{subfigure}

    \caption{IVR of ozone at the oTDMVCC[2--3] and TDMVCC[2--3] levels
    with $N = 8$ and $N_{\2\subA} = 8$ for all modes. 
    Absolute value of $\mathrm{Re}\elm{\Psi'}{Q}{\Psi}$ (dashed) and $\mathrm{Im}\elm{\Psi'}{Q}{\Psi}$ (full) for (a) $Q_0$ (bend) and (b) $Q_1$ (symmetric stretch).
    Note that the dashed lines (real part) are almost on top of each other.}

    \label{fig:O3_re_im_zoom_Q0_Q1}
\end{figure}

\newpage

%
%
\begin{figure}[H]
    \centering

    \begin{subfigure}[c]{0.75\textwidth}
        \centering
        \caption{}
        \includegraphics[width=\textwidth]{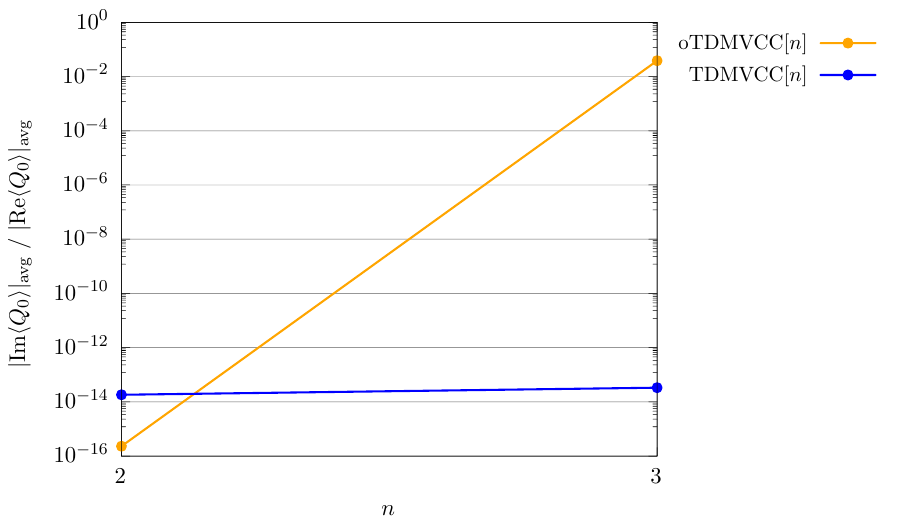}
        \label{fig:O3_re_im_ratio_Q0}
    \end{subfigure}


    \begin{subfigure}[c]{0.75\textwidth}
        \centering
        \caption{}
        \includegraphics[width=\textwidth]{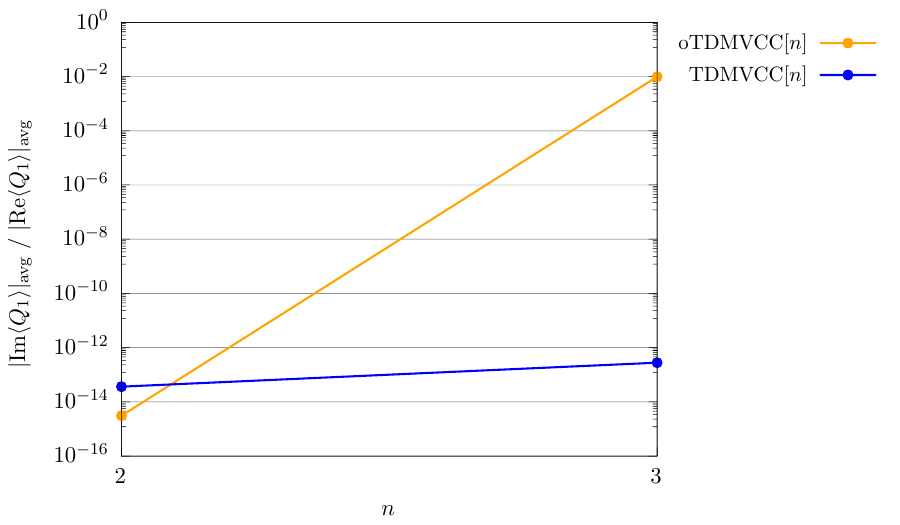}
        \label{fig:O3_re_im_ratio_Q1}
    \end{subfigure}

    \caption{IVR of ozone at the oTDMVCC[2--3] and TDMVCC[2--3] levels
    with $N = 8$ and $N_{\2\subA} = 8$ for all modes. 
    Ratio of mean absolute values of $\mathrm{Re}\elm{\Psi'}{Q}{\Psi}$ (dashed) and $\mathrm{Im}\elm{\Psi'}{Q}{\Psi}$ (full) for (a) $Q_0$ (bend) and (b) $Q_1$ (symmetric stretch).}

    \label{fig:O3_re_im_ratio_Q0_Q1}
\end{figure}

\newpage

\subsubsection{Energy conservation}

\begin{figure}[H]
    \centering

    \begin{subfigure}[c]{0.8\textwidth}
        \centering
        \caption{TDMVCC[2] and oTDMVCC[2]}
        \includegraphics[width=\textwidth]{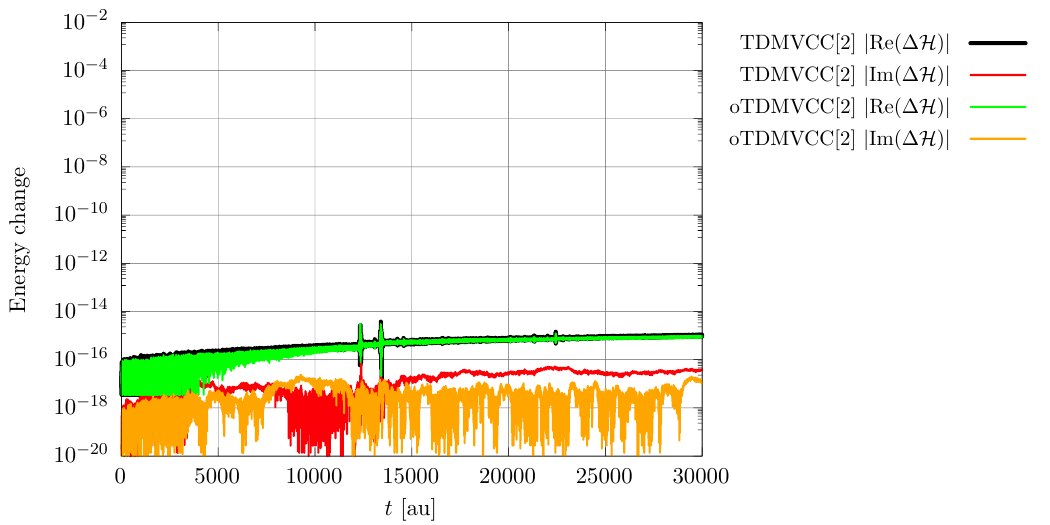}
        \label{fig:O3_energy_tdmvcc2}
    \end{subfigure}


    \begin{subfigure}[c]{0.8\textwidth}
        \centering
        \caption{TDMVCC[3] and oTDMVCC[3]}
        \includegraphics[width=\textwidth]{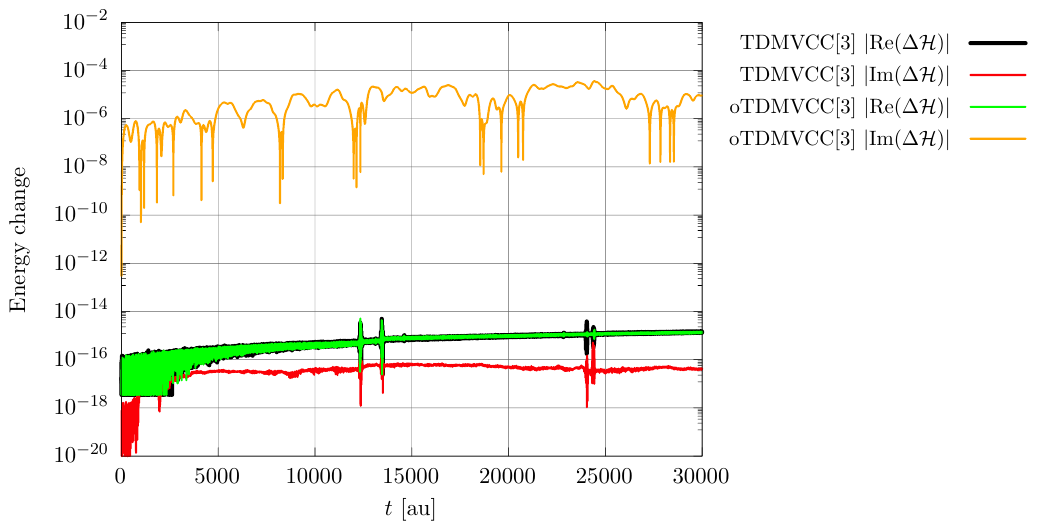}
        \label{fig:O3_energy_tdmvcc3}
    \end{subfigure}

    \caption{IVR of ozone at the oTDMVCC[2--3] and TDMVCC[2--3] levels
    with $N = 8$ and $N_{\2\subA} = 8$ for all modes. 
    Absolute change in the energy function $\mathcal{H} = \elm{\Psi'}{H}{\Psi}$ for (a) TDMVCC[2]/oTDMVCC[2] and (b) TDMVCC[3]/oTDMVCC[3].}
    \label{fig:O3_energy}
\end{figure}

\newpage

\subsubsection{Non-hermiticity in density matrices and mean fields}

\begin{figure}[H]
    \centering

    \begin{subfigure}[c]{0.8\textwidth}
        \centering
        \caption{Density}
        \includegraphics[width=\textwidth]{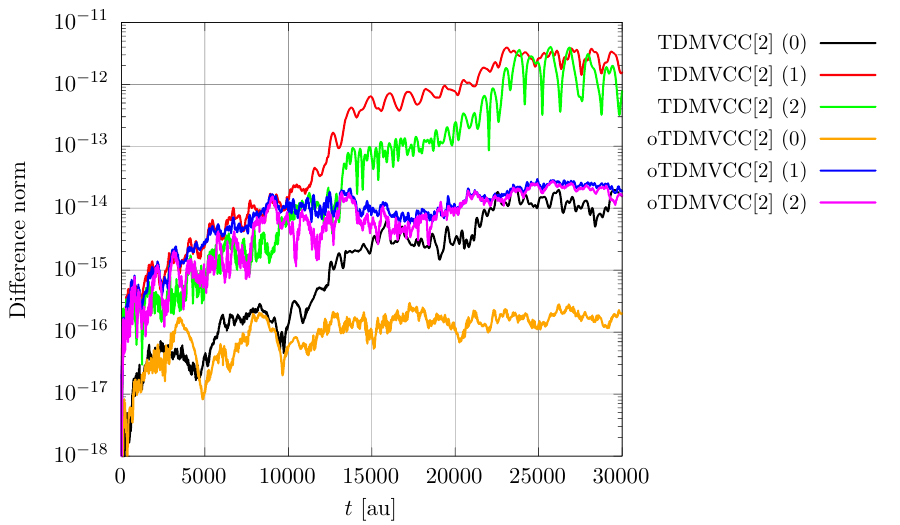}
        \label{fig:O3_nonherm_tdmvcc2_density}
    \end{subfigure}


    \begin{subfigure}[c]{0.8\textwidth}
        \centering
        \caption{Mean field}
        \includegraphics[width=\textwidth]{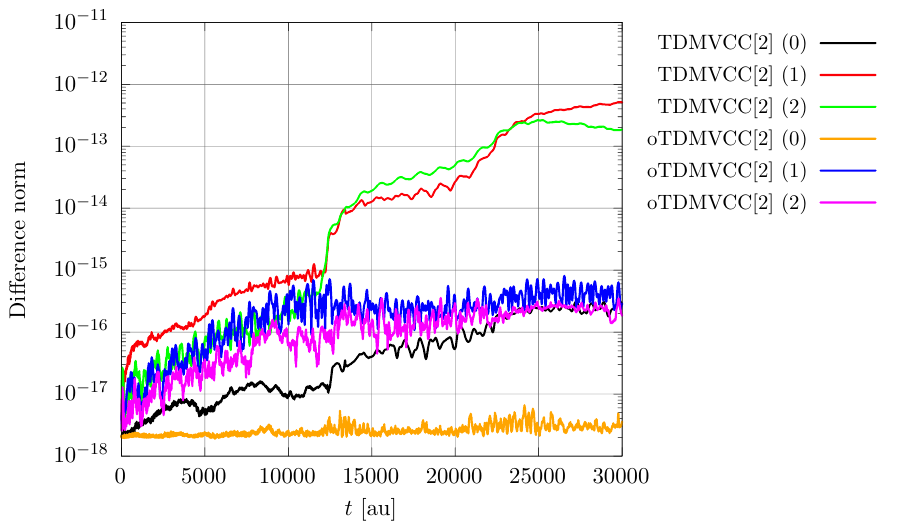}
        \label{fig:O3_nonherm_tdmvcc2_meanfield}
    \end{subfigure}

    \caption{IVR of ozone at the oTDMVCC[2] and TDMVCC[2] levels
    with $N = 8$ and $N_{\2\subA} = 8$ for all modes. 
    Deviation from Hermiticity in 
    (a) one-mode density matrices ($|| \bm{\rho} -  \bm{\rho}^{\dagger} ||$) and 
    (b) half-transformed mean-field matrices ($|| \check{\mbf{F}} - \check{\mbf{F}}^{\prime \dagger} ||$).
    Mode indices are given in parentheses.}
    \label{fig:O3_nonherm_tdmvcc2}
\end{figure}

\newpage 
\subsection{Sulfur dioxide}

\subsubsection{Hilbert space angles}

%
%
\begin{figure}[H]
    \centering

    \begin{subfigure}[c]{0.75\textwidth}
        \centering
        \caption{Ket angle}
        \includegraphics[width=\textwidth]{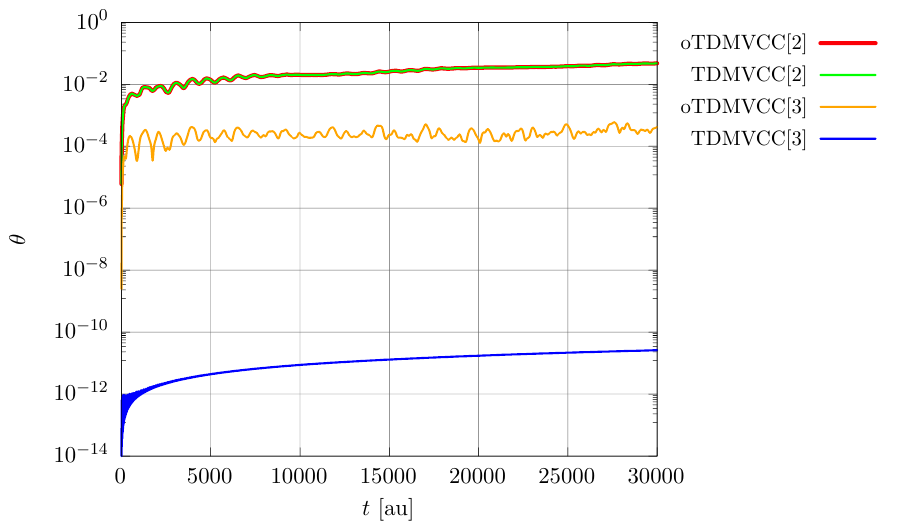}
        \label{fig:SO2_ketangle}
    \end{subfigure}


    \begin{subfigure}[c]{0.75\textwidth}
        \centering
        \caption{Bra angle}
        \includegraphics[width=\textwidth]{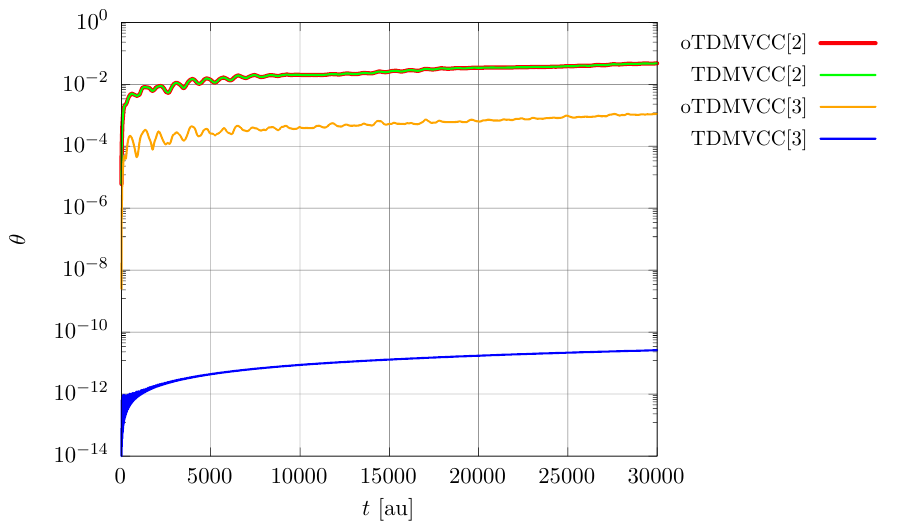}
        \label{fig:SO2_brangle}
    \end{subfigure}

    \caption{IVR of sulfur dioxide at the oTDMVCC[2--3] and TDMVCC[2--3] levels
    with $N = 8$ and $N_{\2\subA} = 8$ for all modes. (a) Hilbert space ket angle. (b) Hilbert space bra angle.}
    \label{fig:SO2_ketbraangle}
\end{figure}

\newpage

\subsubsection{\Aclp{acf}}

%
%
\begin{figure}[H]
    \centering

    \begin{subfigure}[c]{0.75\textwidth}
        \centering
        \caption{Autocorrelation function}
        \includegraphics[width=\textwidth]{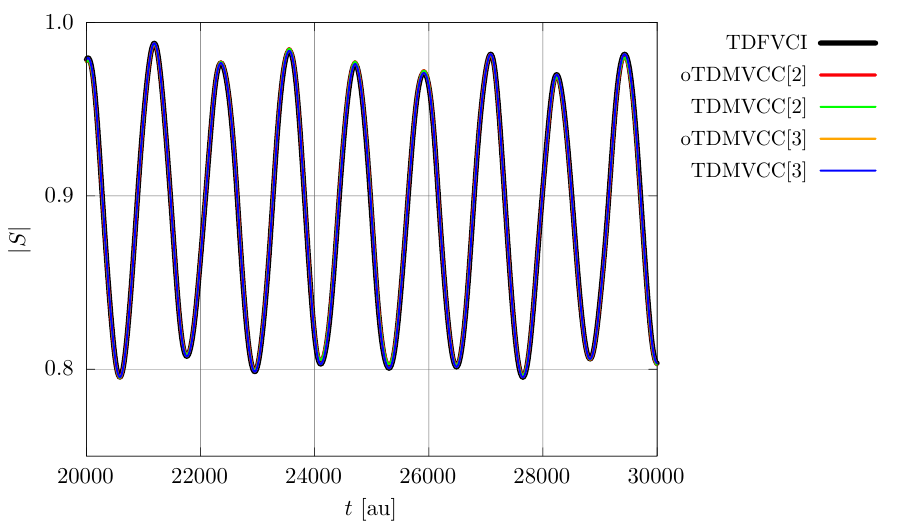}
        \label{fig:SO2_autocorr}
    \end{subfigure}


    \begin{subfigure}[c]{0.75\textwidth}
        \centering
        \caption{Absolute error in autocorrelation function}
        \includegraphics[width=\textwidth]{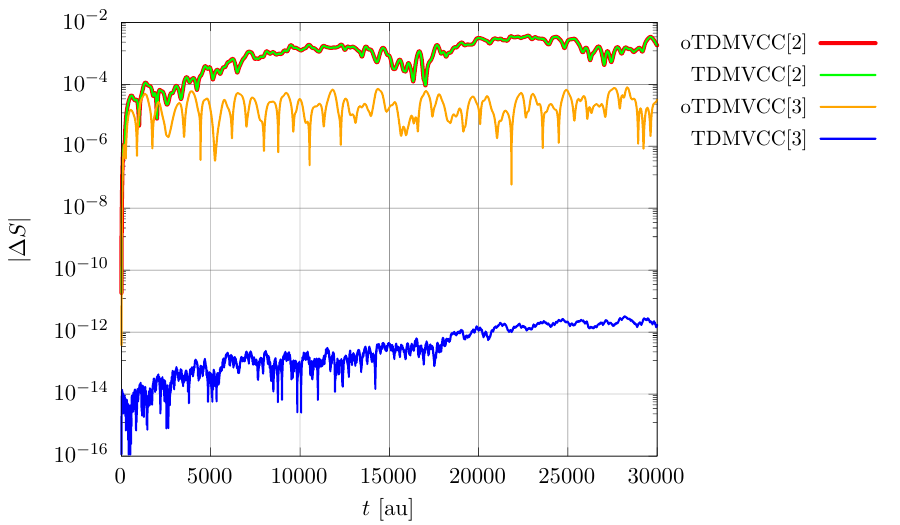}
        \label{fig:SO2_autocorr_diff}
    \end{subfigure}

    \caption{IVR of sulfur dioxide at the oTDMVCC[2--3] and TDMVCC[2--3] levels
    with $N = 8$ and $N_{\2\subA} = 8$ for all modes. (a) \Acl{acf}. (b) Absolute error in the \acl{acf} (relative to TDFVCI).}
    \label{fig:SO2_autocorr_and_diff}
\end{figure}

\newpage

%
%
\begin{figure}[H]
    \centering
    \includegraphics[width=\textwidth]{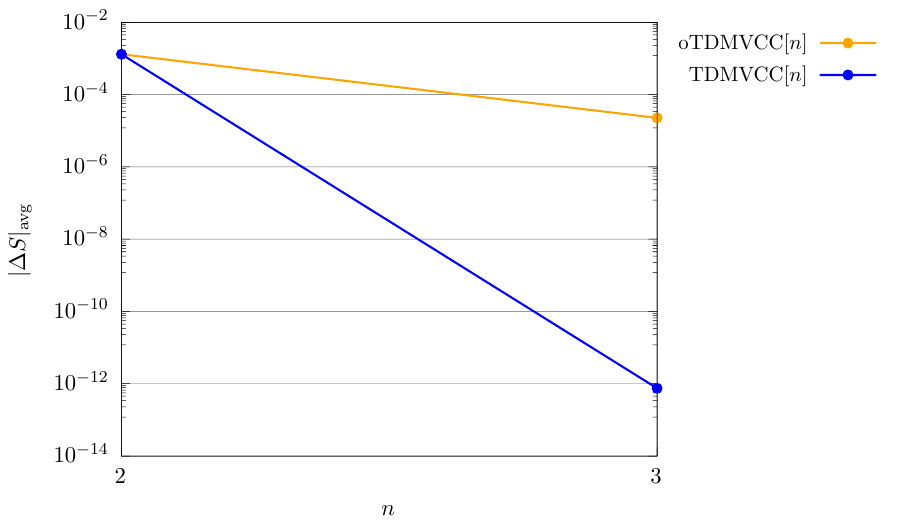}
    \caption{Average absolute error in the \acl{acf} for sulfur dioxide at the oTDMVCC[2--3] and TDMVCC[2--3] levels
    with $N = 8$ and $N_{\2\subA} = 8$ for all modes. The errors are computed relative to TDFVCI.}
    \label{fig:SO2_autocorr_diff_avg}
\end{figure}

\newpage

\subsubsection{Expectation values}

%
%
\begin{figure}[H]
    \centering

    \begin{subfigure}[c]{0.75\textwidth}
        \centering
        \caption{}
        \includegraphics[width=\textwidth]{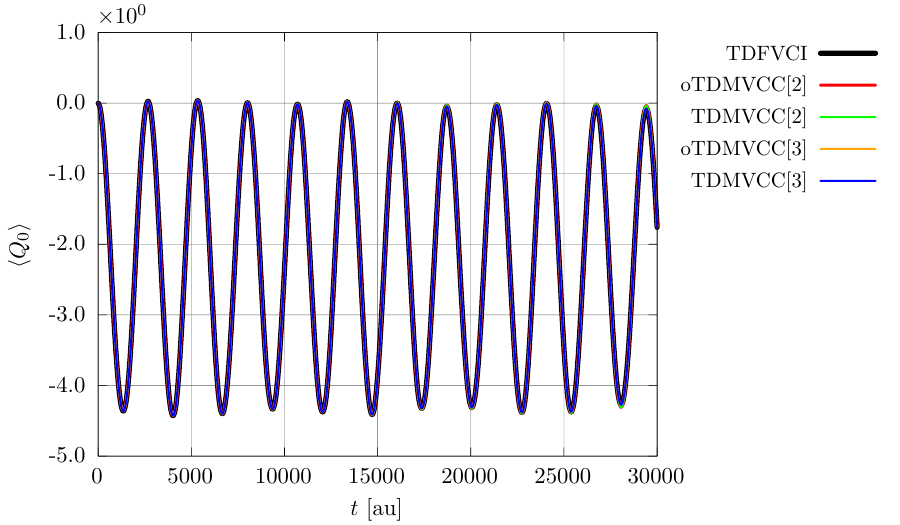}
        \label{fig:SO2_Q0}
    \end{subfigure}


    \begin{subfigure}[c]{0.75\textwidth}
        \centering
        \caption{}
        \includegraphics[width=\textwidth]{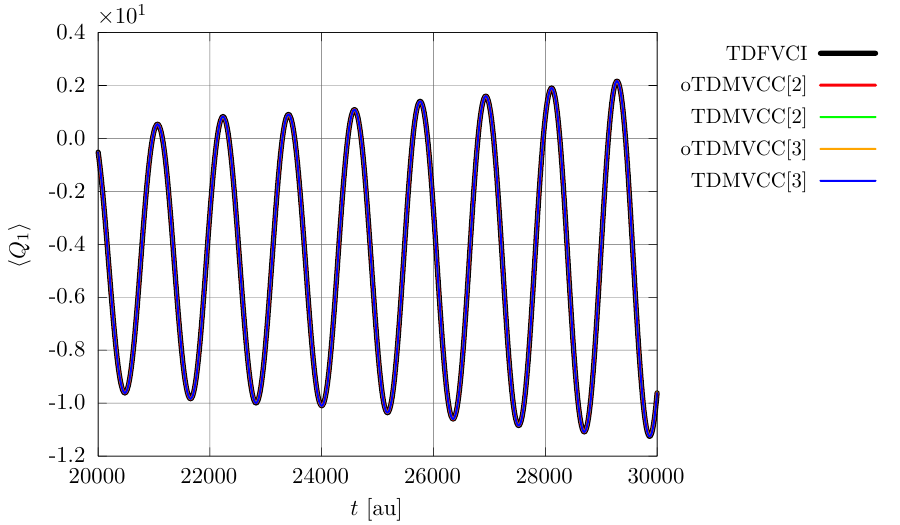}
        \label{fig:SO2_Q1}
    \end{subfigure}

    \caption{IVR of sulfur dioxide at the oTDMVCC[2--3] and TDMVCC[2--3] levels
    with $N = 8$ and $N_{\2\subA} = 8$ for all modes. 
    Expectation value of (a) $Q_0$ (bend) and (b) $Q_1$ (symmetric stretch).
    Errors are calculated relative to TDFVCI.}

    \label{fig:SO2_Q0_Q1}
\end{figure}

\newpage

%
%
\begin{figure}[H]
    \centering

    \begin{subfigure}[c]{0.75\textwidth}
        \centering
        \caption{}
        \includegraphics[width=\textwidth]{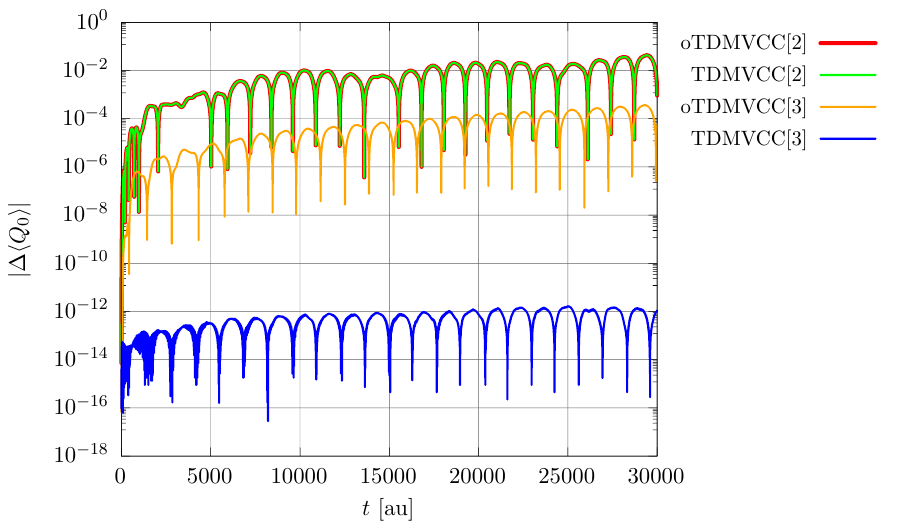}
        \label{fig:SO2_Q0_diff}
    \end{subfigure}


    \begin{subfigure}[c]{0.75\textwidth}
        \centering
        \caption{}
        \includegraphics[width=\textwidth]{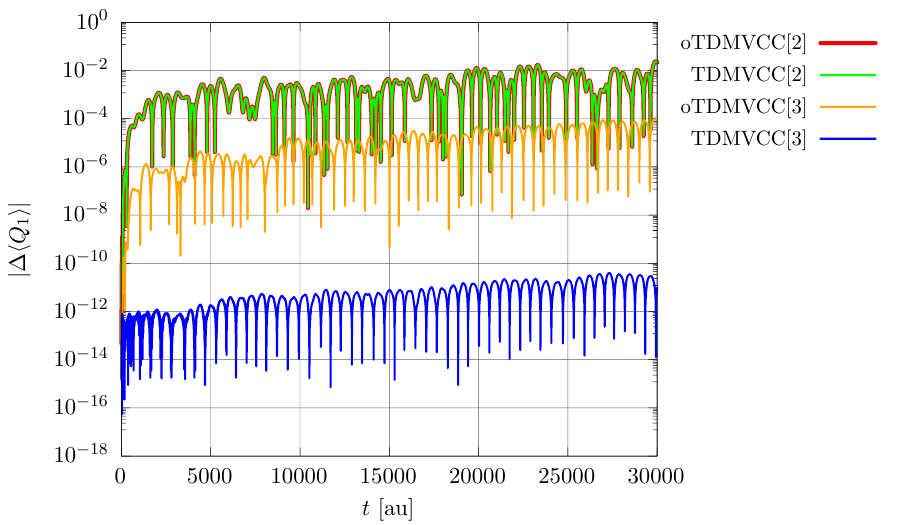}
        \label{fig:SO2_Q1_diff}
    \end{subfigure}

    \caption{IVR of sulfur dioxide at the oTDMVCC[2--3] and TDMVCC[2--3] levels
    with $N = 8$ and $N_{\2\subA} = 8$ for all modes. 
    Absolute error in the expectation value of (a) $Q_0$ (bend) and (b) $Q_1$ (symmetric stretch).
    Errors are calculated relative to TDFVCI.}

    \label{fig:SO2_Q0_Q1_diff}
\end{figure}

\newpage

%
%
\begin{figure}[H]
    \centering

    \begin{subfigure}[c]{0.75\textwidth}
        \centering
        \caption{}
        \includegraphics[width=\textwidth]{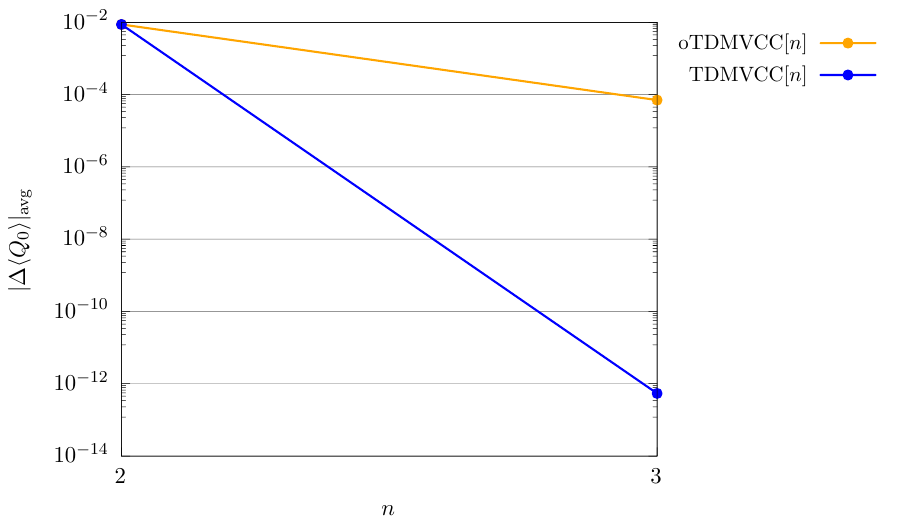}
        \label{fig:SO2_Q0_diff_avg}
    \end{subfigure}


    \begin{subfigure}[c]{0.75\textwidth}
        \centering
        \caption{}
        \includegraphics[width=\textwidth]{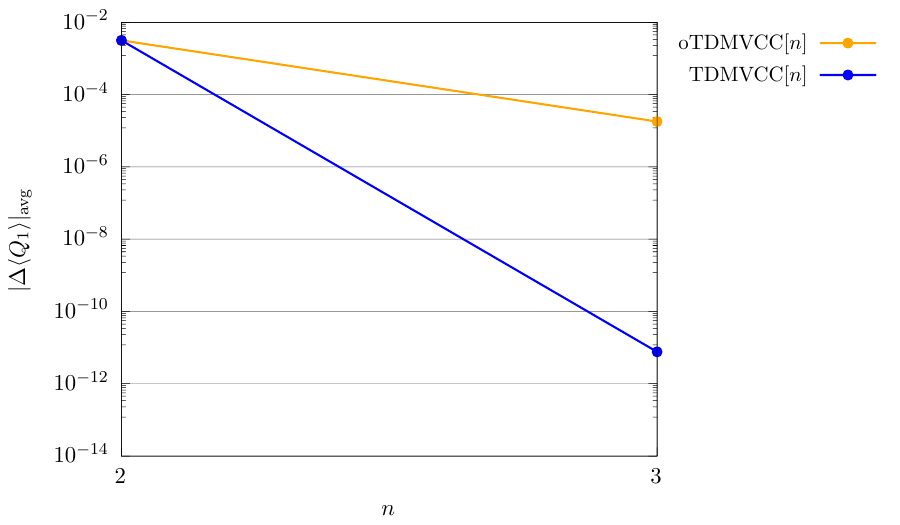}
        \label{fig:SO2_Q1_diff_avg}
    \end{subfigure}

    \caption{IVR of sulfur dioxide at the oTDMVCC[2--3] and TDMVCC[2--3] levels
    with $N = 8$ and $N_{\2\subA} = 8$ for all modes. 
    Average absolute error in the expectation value of (a) $Q_0$ (bend) and (b) $Q_1$ (symmetric stretch).
    Errors are calculated relative to TDFVCI.}
    \label{fig:SO2_Q0_Q1_diff_avg}
\end{figure}

\newpage

\subsubsection{Imaginary part of expectation values}

%
%
\begin{figure}[H]
    \centering

    \begin{subfigure}[c]{0.75\textwidth}
        \centering
        \caption{}
        \includegraphics[width=\textwidth]{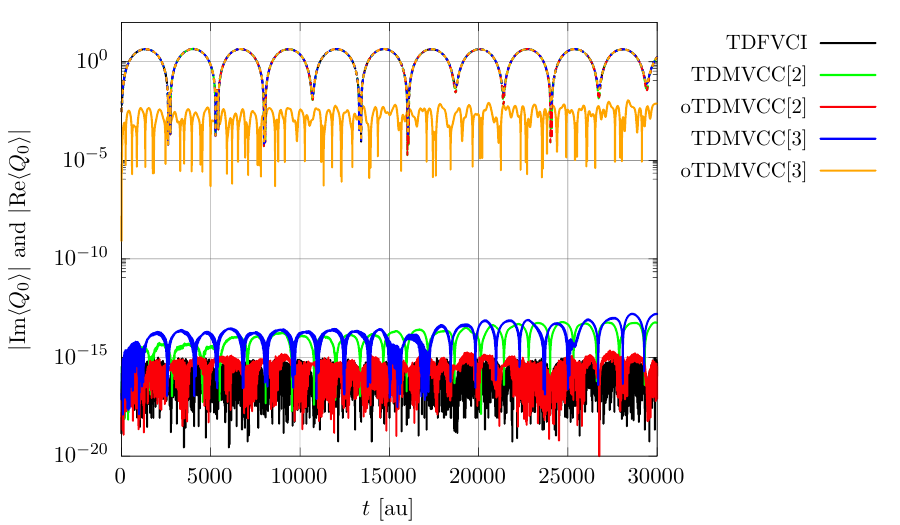}
        \label{fig:SO2_re_im_Q0}
    \end{subfigure}


    \begin{subfigure}[c]{0.75\textwidth}
        \centering
        \caption{}
        \includegraphics[width=\textwidth]{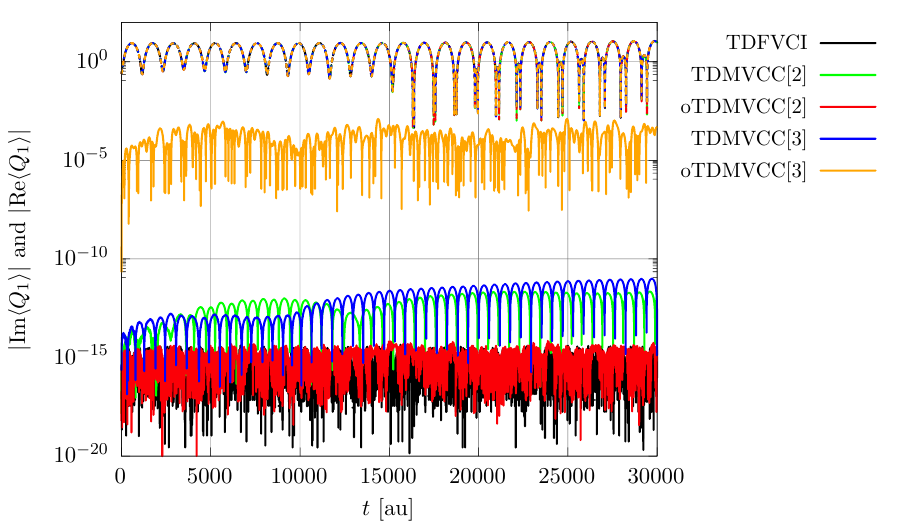}
        \label{fig:SO2_re_im_Q1}
    \end{subfigure}

    \caption{IVR of sulfur dioxide at the oTDMVCC[2--3] and TDMVCC[2--3] levels
    with $N = 8$ and $N_{\2\subA} = 8$ for all modes. 
    Absolute value of $\mathrm{Re}\elm{\Psi'}{Q}{\Psi}$ (dashed) and $\mathrm{Im}\elm{\Psi'}{Q}{\Psi}$ (full) for (a) $Q_0$ (bend) and (b) $Q_1$ (symmetric stretch).
    Note that the dashed lines (real part) are almost on top of each other.}

    \label{fig:SO2_re_im_Q0_Q1}
\end{figure}

\newpage

%
%
\begin{figure}[H]
    \centering

    \begin{subfigure}[c]{0.75\textwidth}
        \centering
        \caption{}
        \includegraphics[width=\textwidth]{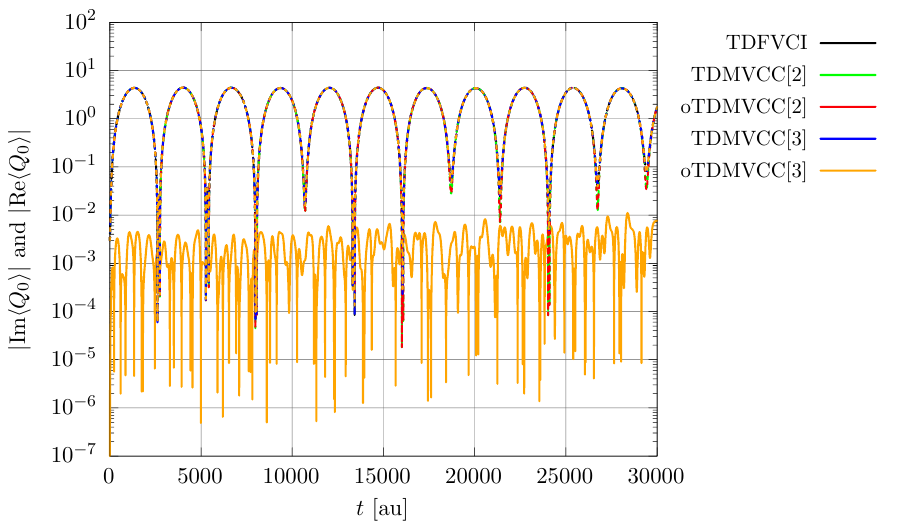}
        \label{fig:SO2_re_im_zoom_Q0}
    \end{subfigure}


    \begin{subfigure}[c]{0.75\textwidth}
        \centering
        \caption{}
        \includegraphics[width=\textwidth]{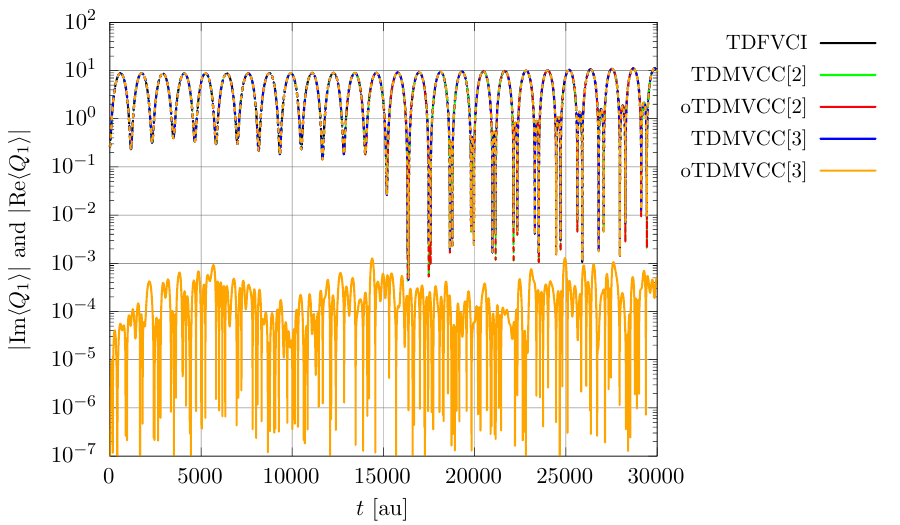}
        \label{fig:SO2_re_im_zoom_Q1}
    \end{subfigure}

    \caption{IVR of sulfur dioxide at the oTDMVCC[2--3] and TDMVCC[2--3] levels
    with $N = 8$ and $N_{\2\subA} = 8$ for all modes. 
    Absolute value of $\mathrm{Re}\elm{\Psi'}{Q}{\Psi}$ (dashed) and $\mathrm{Im}\elm{\Psi'}{Q}{\Psi}$ (full) for (a) $Q_0$ (bend) and (b) $Q_1$ (symmetric stretch).
    Note that the dashed lines (real part) are almost on top of each other.}

    \label{fig:SO2_re_im_zoom_Q0_Q1}
\end{figure}

\newpage

%
%
\begin{figure}[H]
    \centering

    \begin{subfigure}[c]{0.75\textwidth}
        \centering
        \caption{}
        \includegraphics[width=\textwidth]{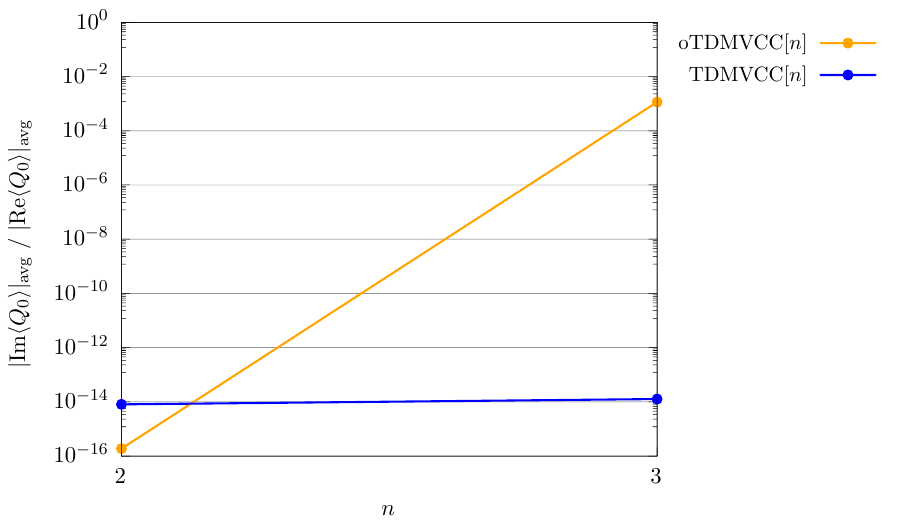}
        \label{fig:SO2_re_im_ratio_Q0}
    \end{subfigure}


    \begin{subfigure}[c]{0.75\textwidth}
        \centering
        \caption{}
        \includegraphics[width=\textwidth]{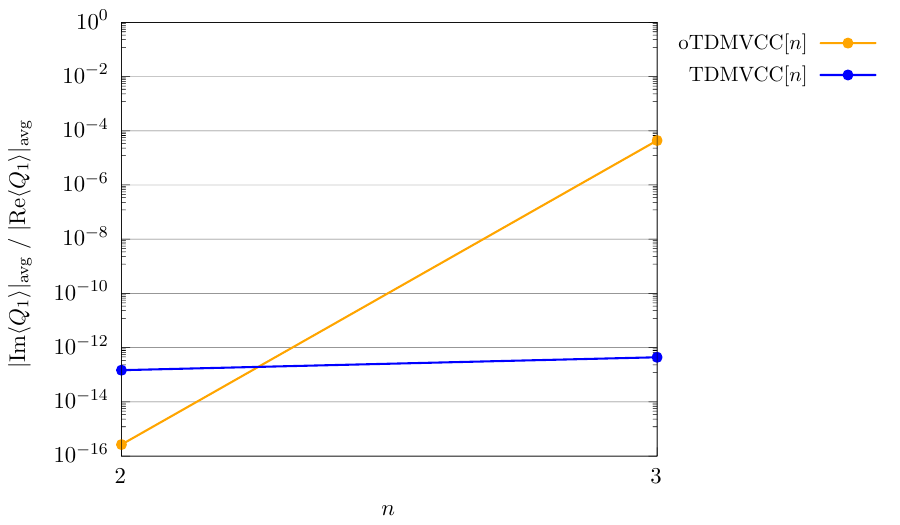}
        \label{fig:SO2_re_im_ratio_Q1}
    \end{subfigure}

    \caption{IVR of sulfur dioxide at the oTDMVCC[2--3] and TDMVCC[2--3] levels
    with $N = 8$ and $N_{\2\subA} = 8$ for all modes. 
    Ratio of mean absolute values of $\mathrm{Re}\elm{\Psi'}{Q}{\Psi}$ (dashed) and $\mathrm{Im}\elm{\Psi'}{Q}{\Psi}$ (full) for (a) $Q_0$ (bend) and (b) $Q_1$ (symmetric stretch).}

    \label{fig:SO2_re_im_ratio_Q0_Q1}
\end{figure}

\newpage

\subsubsection{Energy conservation}

\begin{figure}[H]
    \centering

    \begin{subfigure}[c]{0.8\textwidth}
        \centering
        \caption{TDMVCC[2] and oTDMVCC[2]}
        \includegraphics[width=\textwidth]{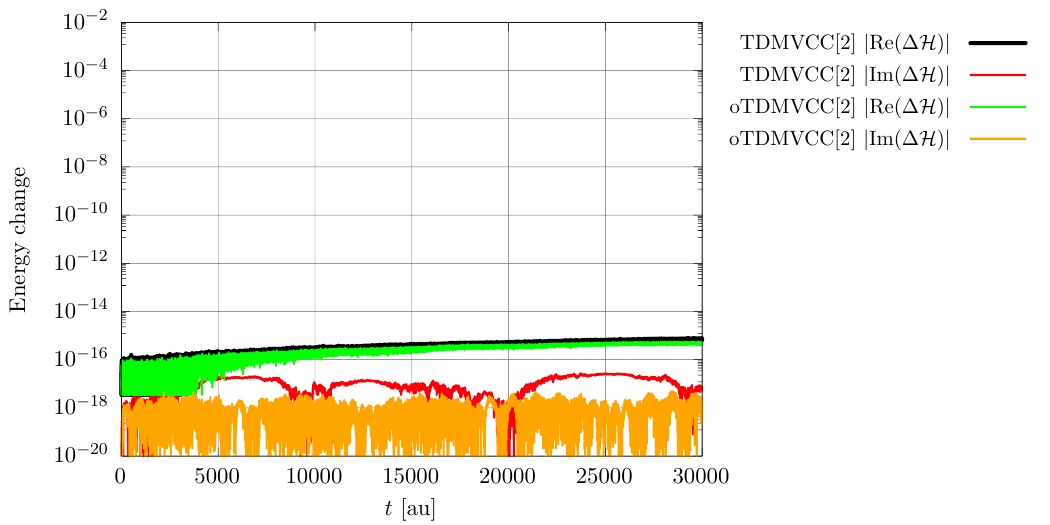}
        \label{fig:SO2_energy_tdmvcc2}
    \end{subfigure}


    \begin{subfigure}[c]{0.8\textwidth}
        \centering
        \caption{TDMVCC[3] and oTDMVCC[3]}
        \includegraphics[width=\textwidth]{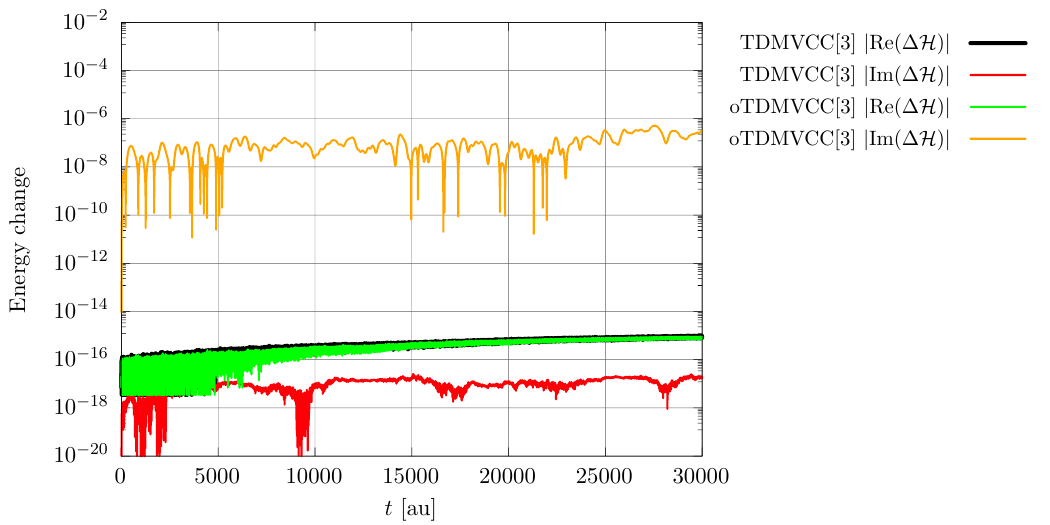}
        \label{fig:SO2_energy_tdmvcc3}
    \end{subfigure}

    \caption{IVR of sulfur dioxide at the oTDMVCC[2--3] and TDMVCC[2--3] levels
    with $N = 8$ and $N_{\2\subA} = 8$ for all modes. 
    Absolute change in the energy function $\mathcal{H} = \elm{\Psi'}{H}{\Psi}$ for (a) TDMVCC[2]/oTDMVCC[2] and (b) TDMVCC[3]/oTDMVCC[3].}
    \label{fig:SO2_energy}
\end{figure}

\newpage

\subsubsection{Non-hermiticity in density matrices and mean fields}

\begin{figure}[H]
    \centering

    \begin{subfigure}[c]{0.8\textwidth}
        \centering
        \caption{Density}
        \includegraphics[width=\textwidth]{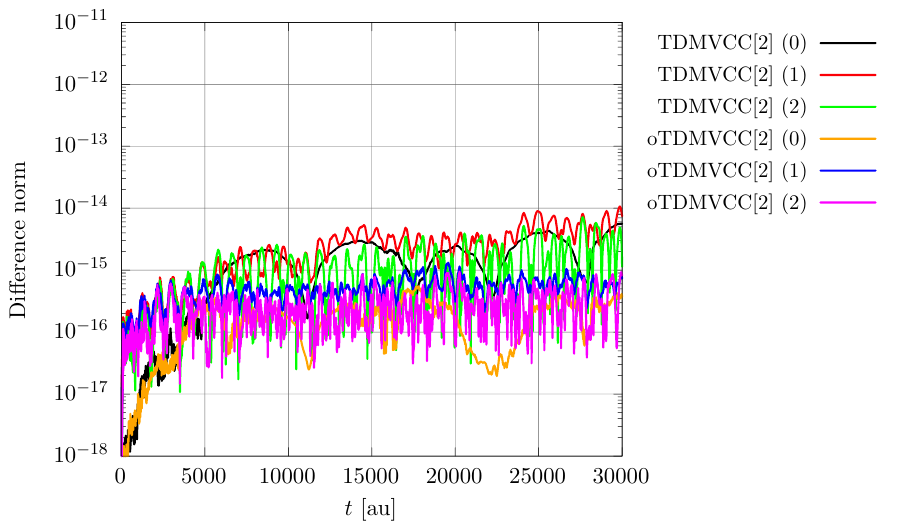}
        \label{fig:SO2_nonherm_tdmvcc2_density}
    \end{subfigure}


    \begin{subfigure}[c]{0.8\textwidth}
        \centering
        \caption{Mean field}
        \includegraphics[width=\textwidth]{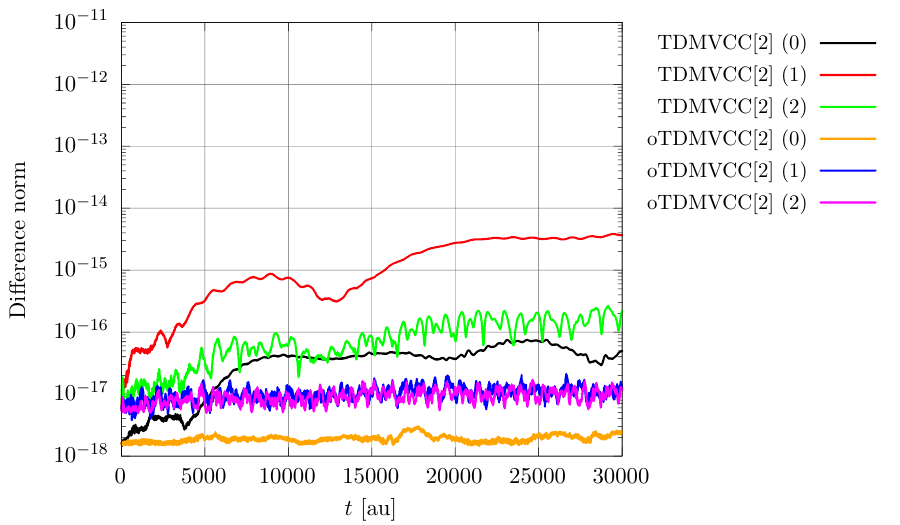}
        \label{fig:SO2_nonherm_tdmvcc2_meanfield}
    \end{subfigure}

    \caption{IVR of sulfur dioxide at the oTDMVCC[2] and TDMVCC[2] levels
    with $N = 8$ and $N_{\2\subA} = 8$ for all modes. 
    Deviation from Hermiticity in 
    (a) one-mode density matrices ($|| \bm{\rho} -  \bm{\rho}^{\dagger} ||$) and 
    (b) half-transformed mean-field matrices ($|| \check{\mbf{F}} - \check{\mbf{F}}^{\prime \dagger} ||$).
    Mode indices are given in parentheses.}
    \label{fig:SO2_nonherm_tdmvcc2}
\end{figure}

\newpage 
\subsection{Hydrogen sulfide}

\subsubsection{Hilbert space angles}

%
%
\begin{figure}[H]
    \centering

    \begin{subfigure}[c]{0.75\textwidth}
        \centering
        \caption{Ket angle}
        \includegraphics[width=\textwidth]{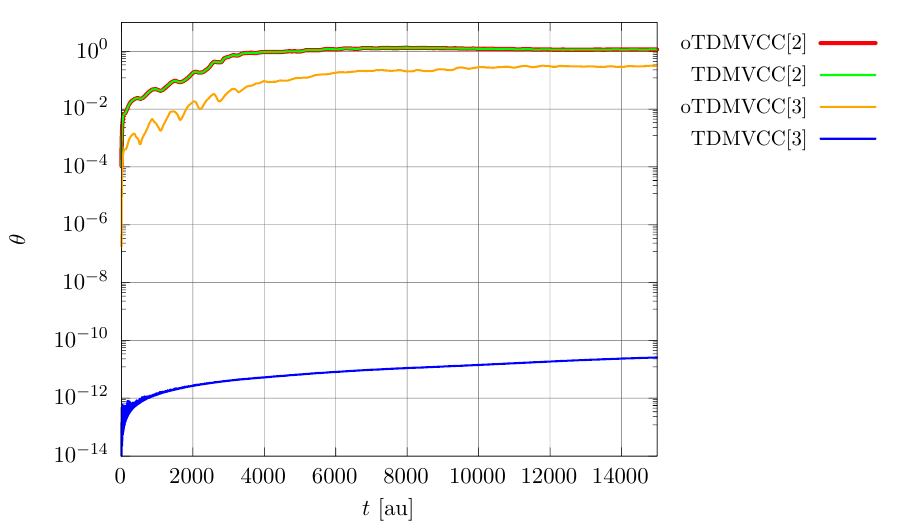}
        \label{fig:H2S_ketangle}
    \end{subfigure}


    \begin{subfigure}[c]{0.75\textwidth}
        \centering
        \caption{Bra angle}
        \includegraphics[width=\textwidth]{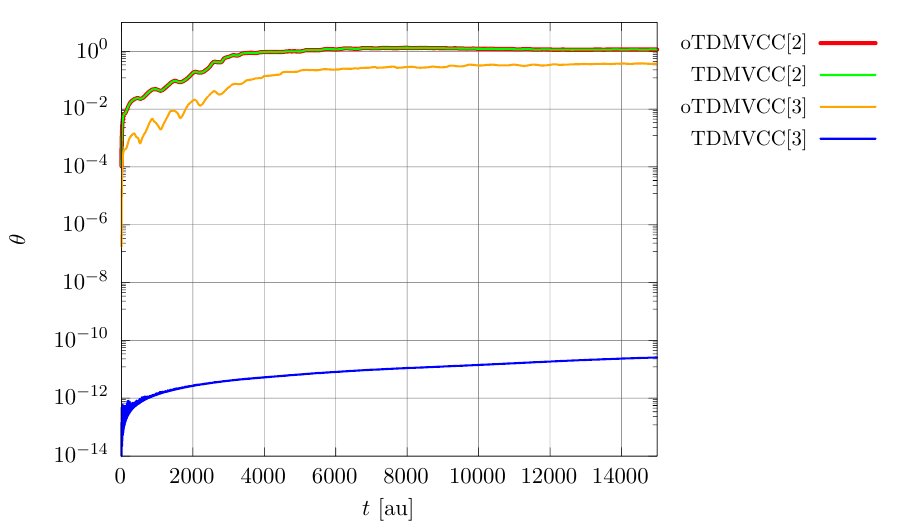}
        \label{fig:H2S_brangle}
    \end{subfigure}

    \caption{IVR of hydrogen sulfide at the oTDMVCC[2--3] and TDMVCC[2--3] levels
    with $N = 8$ and $N_{\2\subA} = 8$ for all modes. (a) Hilbert space ket angle. (b) Hilbert space bra angle.}
    \label{fig:H2S_ketbraangle}
\end{figure}

\newpage

\subsubsection{\Aclp{acf}}

%
%
\begin{figure}[H]
    \centering

    \begin{subfigure}[c]{0.75\textwidth}
        \centering
        \caption{Autocorrelation function}
        \includegraphics[width=\textwidth]{autocorr_H2S_tdmvcc23_vs_tdfvci_t15000_points15001.pdf}
        \label{fig:H2S_autocorr}
    \end{subfigure}


    \begin{subfigure}[c]{0.75\textwidth}
        \centering
        \caption{Absolute error in autocorrelation function}
        \includegraphics[width=\textwidth]{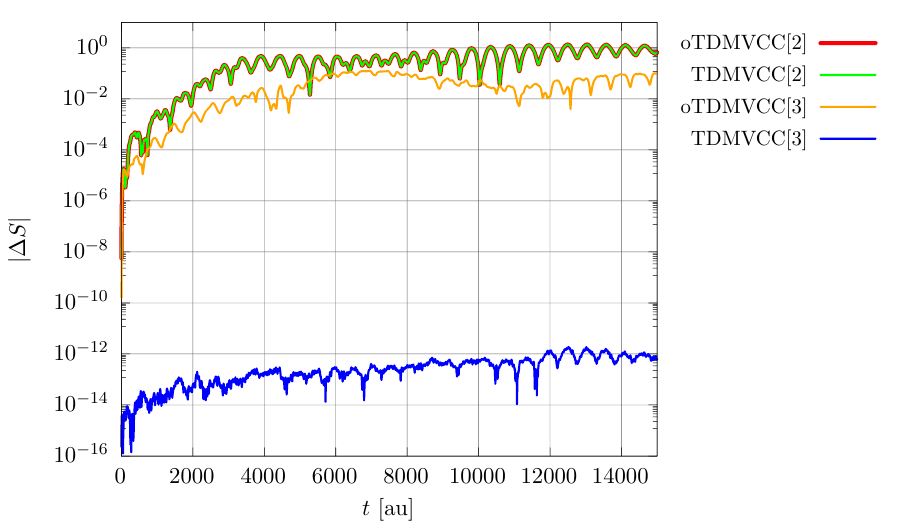}
        \label{fig:H2S_autocorr_diff}
    \end{subfigure}

    \caption{IVR of hydrogen sulfide at the oTDMVCC[2--3] and TDMVCC[2--3] levels
    with $N = 8$ and $N_{\2\subA} = 8$ for all modes. (a) \Acl{acf}. (b) Absolute error in the \acl{acf} (relative to TDFVCI).}
    \label{fig:H2S_autocorr_and_diff}
\end{figure}

\newpage

%
%
\begin{figure}[H]
    \centering
    \includegraphics[width=\textwidth]{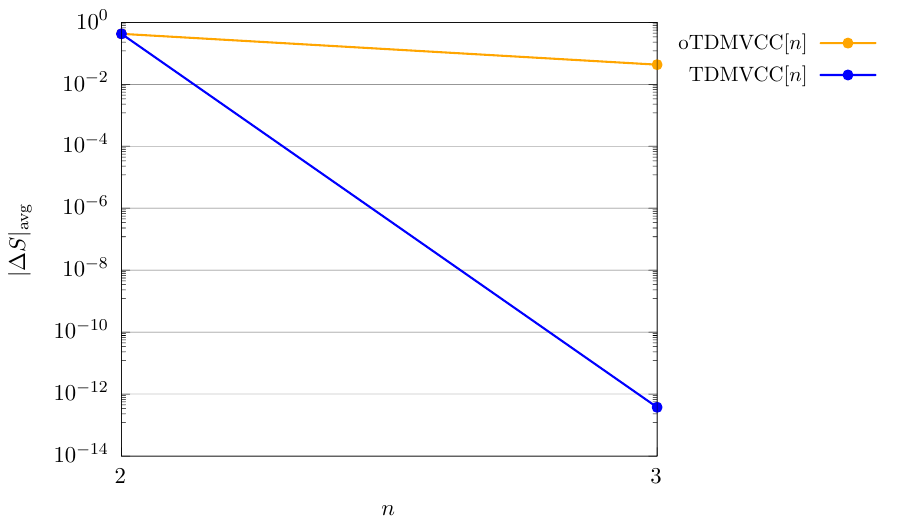}
    \caption{Average absolute error in the \acl{acf} for hydrogen sulfide at the oTDMVCC[2--3] and TDMVCC[2--3] levels
    with $N = 8$ and $N_{\2\subA} = 8$ for all modes. The errors are computed relative to TDFVCI.}
    \label{fig:H2S_autocorr_diff_avg}
\end{figure}

\newpage

\subsubsection{Expectation values}

%
%
\begin{figure}[H]
    \centering

    \begin{subfigure}[c]{0.75\textwidth}
        \centering
        \caption{}
        \includegraphics[width=\textwidth]{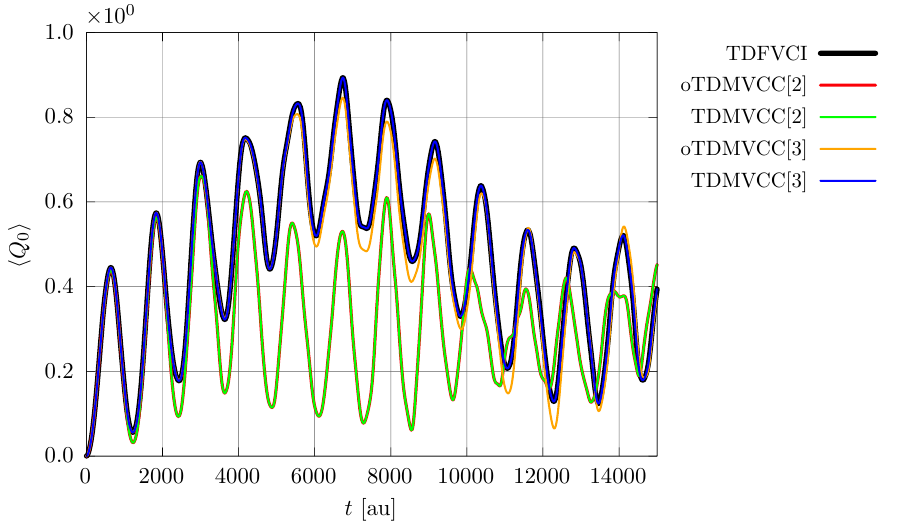}
        \label{fig:H2S_Q0}
    \end{subfigure}


    \begin{subfigure}[c]{0.75\textwidth}
        \centering
        \caption{}
        \includegraphics[width=\textwidth]{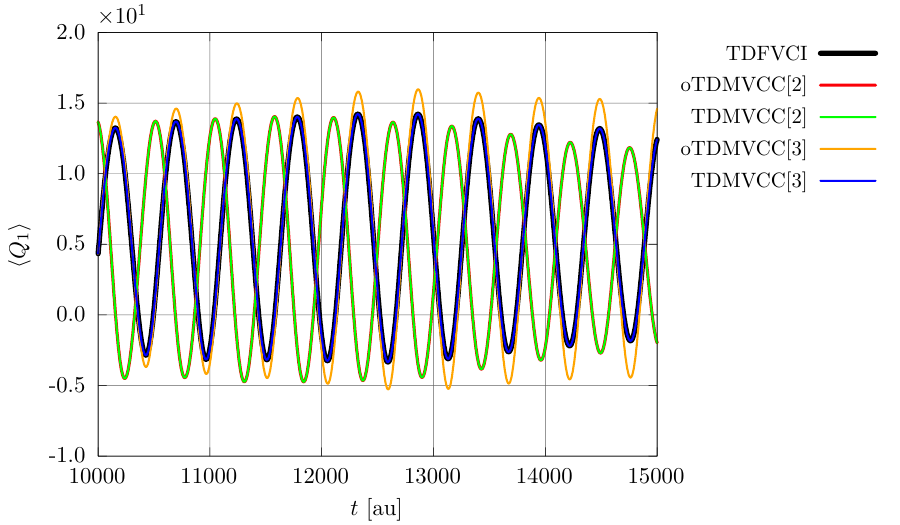}
        \label{fig:H2S_Q1}
    \end{subfigure}

    \caption{IVR of hydrogen sulfide at the oTDMVCC[2--3] and TDMVCC[2--3] levels
    with $N = 8$ and $N_{\2\subA} = 8$ for all modes. 
    Expectation value of (a) $Q_0$ (bend) and (b) $Q_1$ (symmetric stretch).
    Errors are calculated relative to TDFVCI.}

    \label{fig:H2S_Q0_Q1}
\end{figure}

\newpage

%
%
\begin{figure}[H]
    \centering

    \begin{subfigure}[c]{0.75\textwidth}
        \centering
        \caption{}
        \includegraphics[width=\textwidth]{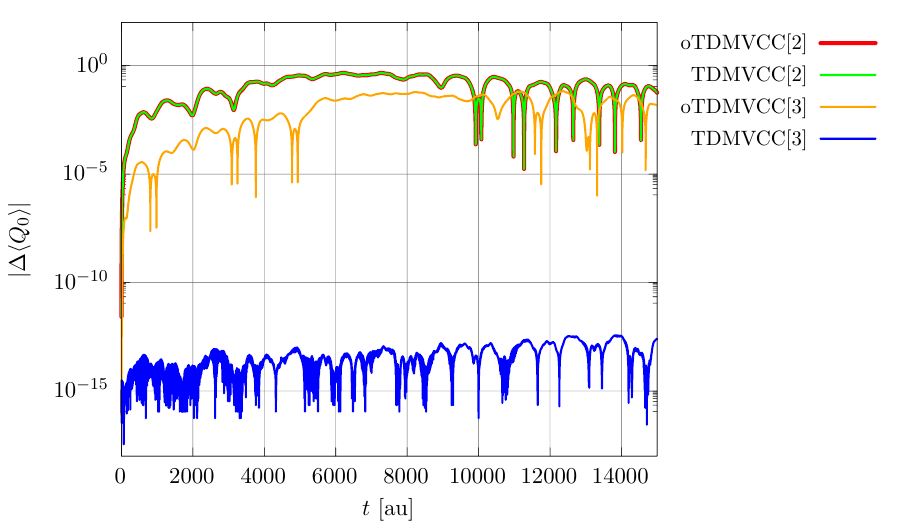}
        \label{fig:H2S_Q0_diff}
    \end{subfigure}


    \begin{subfigure}[c]{0.75\textwidth}
        \centering
        \caption{}
        \includegraphics[width=\textwidth]{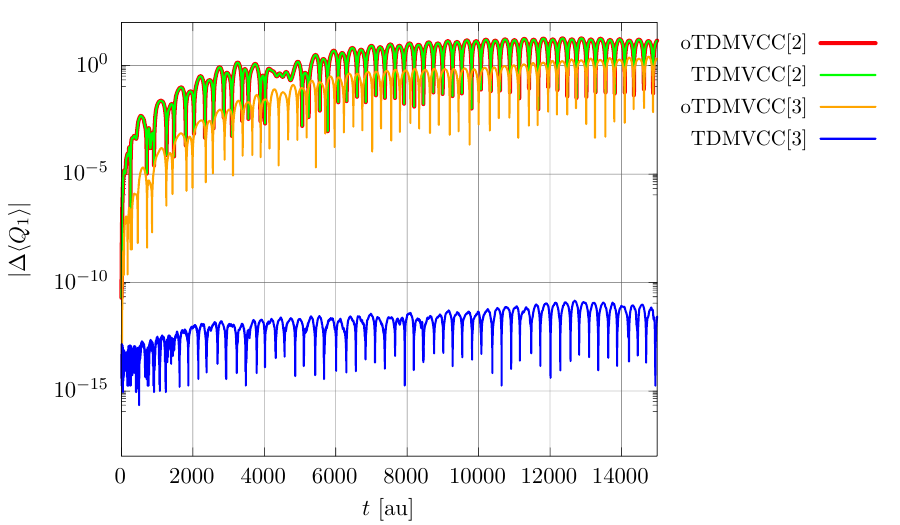}
        \label{fig:H2S_Q1_diff}
    \end{subfigure}

    \caption{IVR of hydrogen sulfide at the oTDMVCC[2--3] and TDMVCC[2--3] levels
    with $N = 8$ and $N_{\2\subA} = 8$ for all modes. 
    Absolute error in the expectation value of (a) $Q_0$ (bend) and (b) $Q_1$ (symmetric stretch).
    Errors are calculated relative to TDFVCI.}

    \label{fig:H2S_Q0_Q1_diff}
\end{figure}

\newpage

%
%
\begin{figure}[H]
    \centering

    \begin{subfigure}[c]{0.75\textwidth}
        \centering
        \caption{}
        \includegraphics[width=\textwidth]{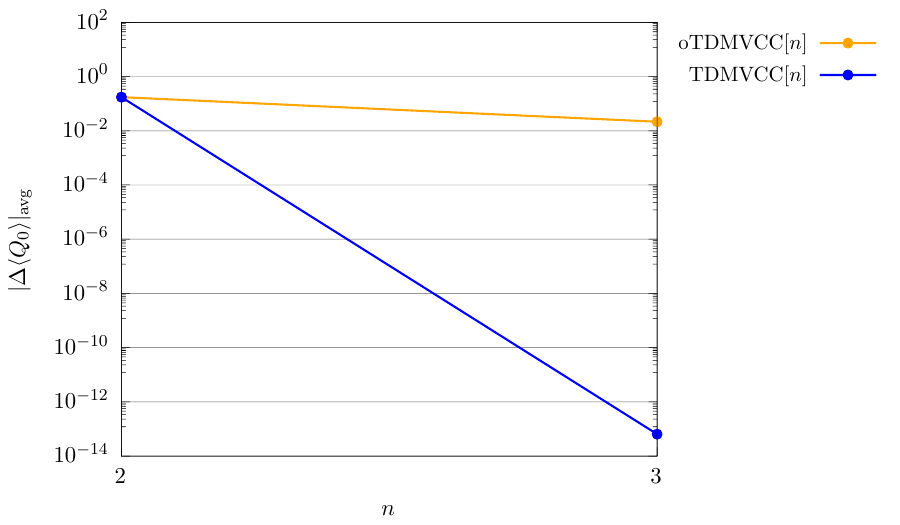}
        \label{fig:H2S_Q0_diff_avg}
    \end{subfigure}


    \begin{subfigure}[c]{0.75\textwidth}
        \centering
        \caption{}
        \includegraphics[width=\textwidth]{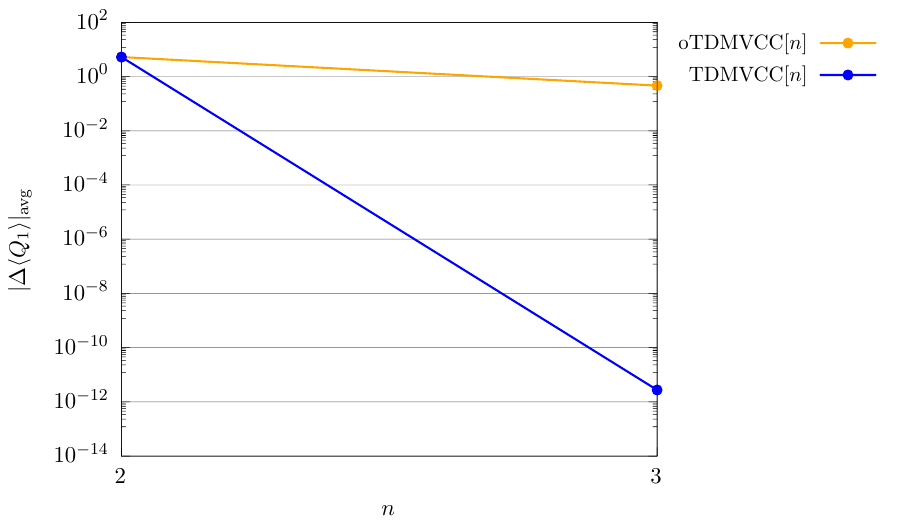}
        \label{fig:H2S_Q1_diff_avg}
    \end{subfigure}

    \caption{IVR of hydrogen sulfide at the oTDMVCC[2--3] and TDMVCC[2--3] levels
    with $N = 8$ and $N_{\2\subA} = 8$ for all modes. 
    Average absolute error in the expectation value of (a) $Q_0$ (bend) and (b) $Q_1$ (symmetric stretch).
    Errors are calculated relative to TDFVCI.}
    \label{fig:H2S_Q0_Q1_diff_avg}
\end{figure}

\newpage

\subsubsection{Imaginary part of expectation values}

%
%
\begin{figure}[H]
    \centering

    \begin{subfigure}[c]{0.75\textwidth}
        \centering
        \caption{}
        \includegraphics[width=\textwidth]{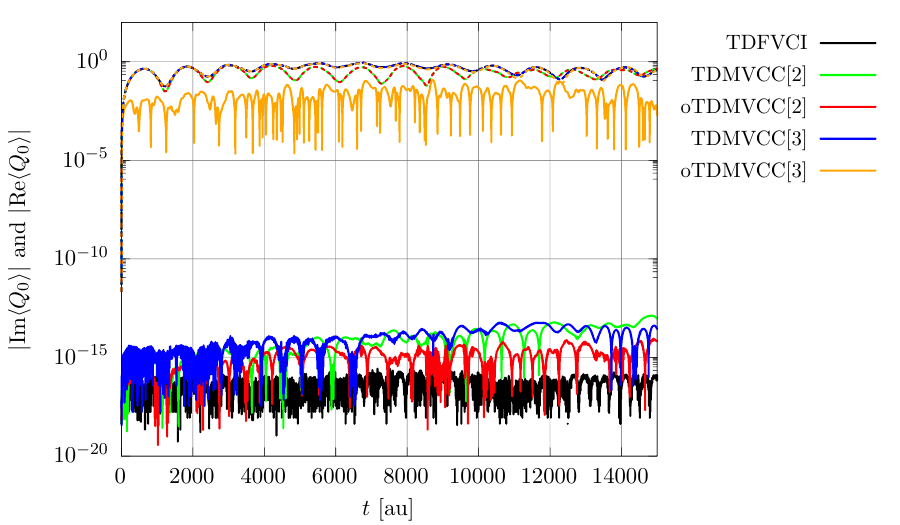}
        \label{fig:H2S_re_im_Q0}
    \end{subfigure}


    \begin{subfigure}[c]{0.75\textwidth}
        \centering
        \caption{}
        \includegraphics[width=\textwidth]{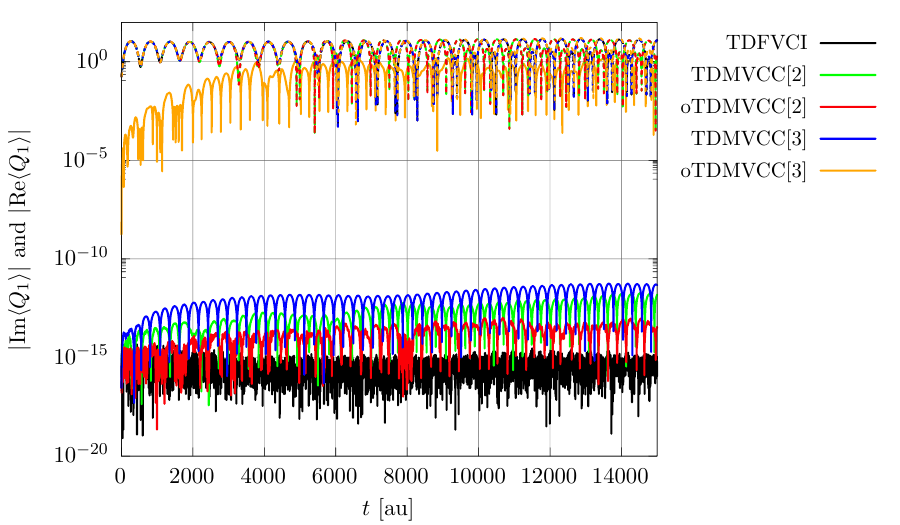}
        \label{fig:H2S_re_im_Q1}
    \end{subfigure}

    \caption{IVR of hydrogen sulfide at the oTDMVCC[2--3] and TDMVCC[2--3] levels
    with $N = 8$ and $N_{\2\subA} = 8$ for all modes. 
    Absolute value of $\mathrm{Re}\elm{\Psi'}{Q}{\Psi}$ (dashed) and $\mathrm{Im}\elm{\Psi'}{Q}{\Psi}$ (full) for (a) $Q_0$ (bend) and (b) $Q_1$ (symmetric stretch).
    Note that the dashed lines (real part) are almost on top of each other.}

    \label{fig:H2S_re_im_Q0_Q1}
\end{figure}

\newpage

%
%
\begin{figure}[H]
    \centering

    \begin{subfigure}[c]{0.75\textwidth}
        \centering
        \caption{}
        \includegraphics[width=\textwidth]{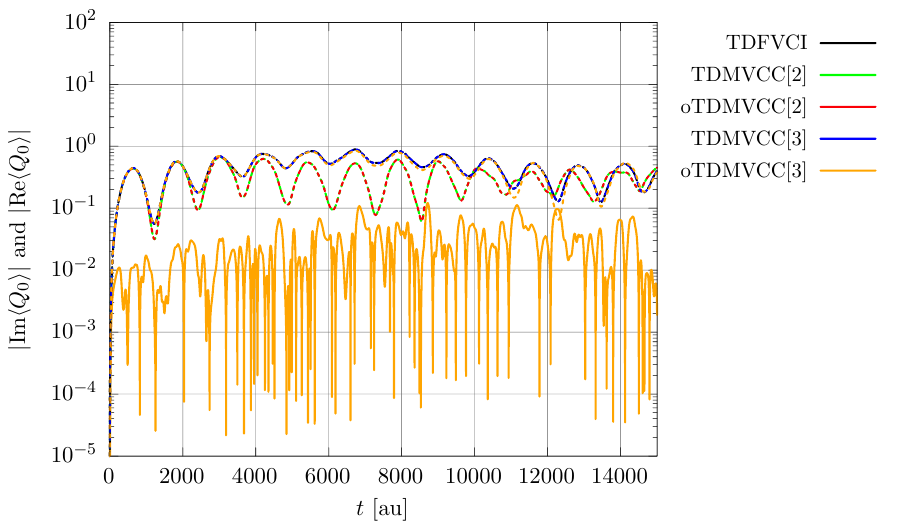}
        \label{fig:H2S_re_im_zoom_Q0}
    \end{subfigure}


    \begin{subfigure}[c]{0.75\textwidth}
        \centering
        \caption{}
        \includegraphics[width=\textwidth]{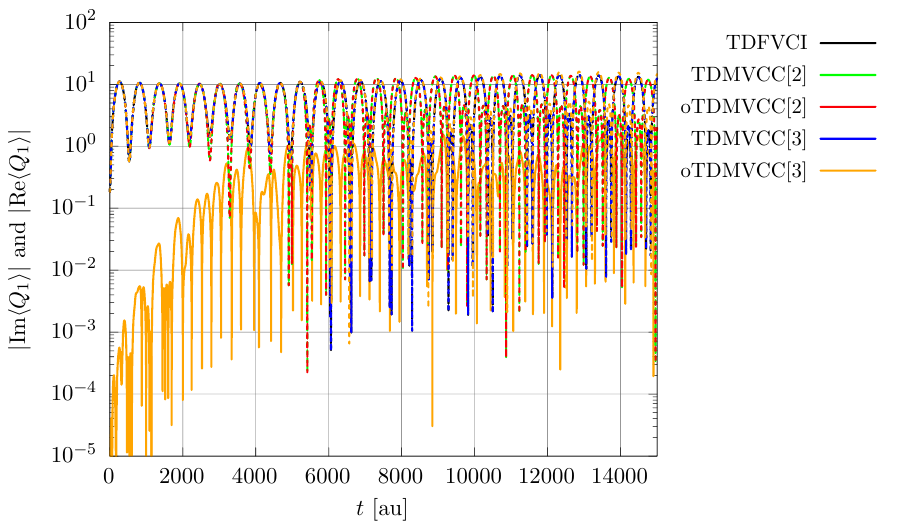}
        \label{fig:H2S_re_im_zoom_Q1}
    \end{subfigure}

    \caption{IVR of hydrogen sulfide at the oTDMVCC[2--3] and TDMVCC[2--3] levels
    with $N = 8$ and $N_{\2\subA} = 8$ for all modes. 
    Absolute value of $\mathrm{Re}\elm{\Psi'}{Q}{\Psi}$ (dashed) and $\mathrm{Im}\elm{\Psi'}{Q}{\Psi}$ (full) for (a) $Q_0$ (bend) and (b) $Q_1$ (symmetric stretch).
    Note that the dashed lines (real part) are almost on top of each other.}

    \label{fig:H2S_re_im_zoom_Q0_Q1}
\end{figure}

\newpage

%
%
\begin{figure}[H]
    \centering

    \begin{subfigure}[c]{0.75\textwidth}
        \centering
        \caption{}
        \includegraphics[width=\textwidth]{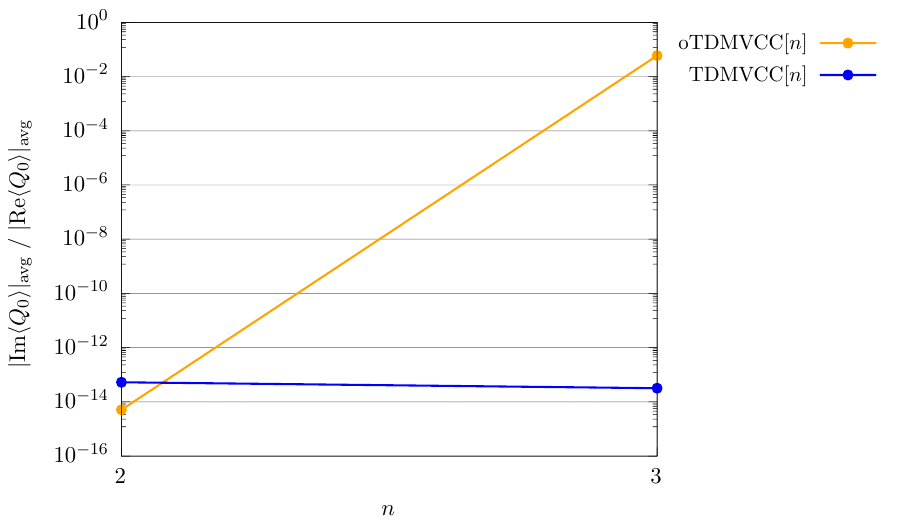}
        \label{fig:H2S_re_im_ratio_Q0}
    \end{subfigure}


    \begin{subfigure}[c]{0.75\textwidth}
        \centering
        \caption{}
        \includegraphics[width=\textwidth]{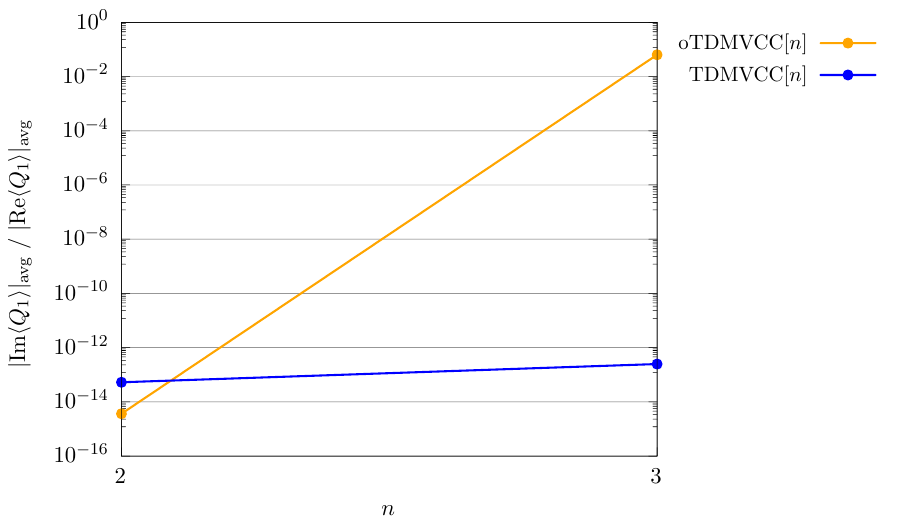}
        \label{fig:H2S_re_im_ratio_Q1}
    \end{subfigure}

    \caption{IVR of hydrogen sulfide at the oTDMVCC[2--3] and TDMVCC[2--3] levels
    with $N = 8$ and $N_{\2\subA} = 8$ for all modes. 
    Ratio of mean absolute values of $\mathrm{Re}\elm{\Psi'}{Q}{\Psi}$ (dashed) and $\mathrm{Im}\elm{\Psi'}{Q}{\Psi}$ (full) for (a) $Q_0$ (bend) and (b) $Q_1$ (symmetric stretch).}

    \label{fig:H2S_re_im_ratio_Q0_Q1}
\end{figure}

\newpage

\subsubsection{Energy conservation}

\begin{figure}[H]
    \centering

    \begin{subfigure}[c]{0.8\textwidth}
        \centering
        \caption{TDMVCC[2] and oTDMVCC[2]}
        \includegraphics[width=\textwidth]{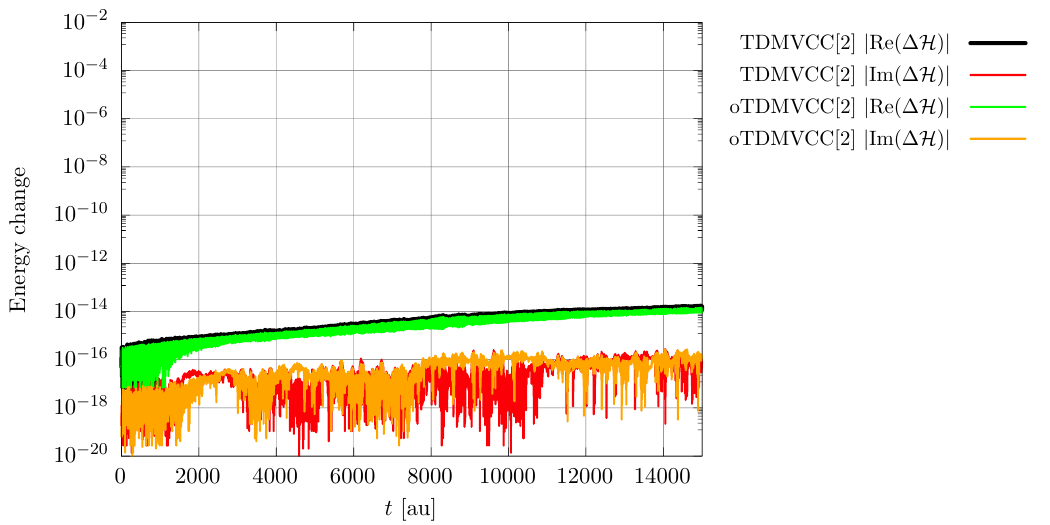}
        \label{fig:H2S_energy_tdmvcc2}
    \end{subfigure}


    \begin{subfigure}[c]{0.8\textwidth}
        \centering
        \caption{TDMVCC[3] and oTDMVCC[3]}
        \includegraphics[width=\textwidth]{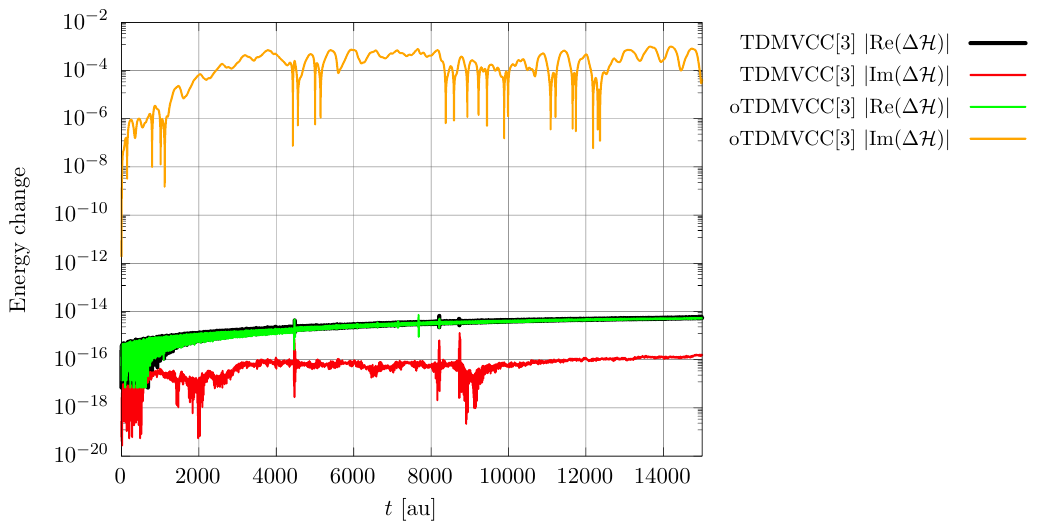}
        \label{fig:H2S_energy_tdmvcc3}
    \end{subfigure}

    \caption{IVR of hydrogen sulfide at the oTDMVCC[2--3] and TDMVCC[2--3] levels
    with $N = 8$ and $N_{\2\subA} = 8$ for all modes. 
    Absolute change in the energy function $\mathcal{H} = \elm{\Psi'}{H}{\Psi}$ for (a) TDMVCC[2]/oTDMVCC[2] and (b) TDMVCC[3]/oTDMVCC[3].}
    \label{fig:H2S_energy}
\end{figure}

\newpage

\subsubsection{Non-hermiticity in density matrices and mean fields}

\begin{figure}[H]
    \centering

    \begin{subfigure}[c]{0.8\textwidth}
        \centering
        \caption{Density}
        \includegraphics[width=\textwidth]{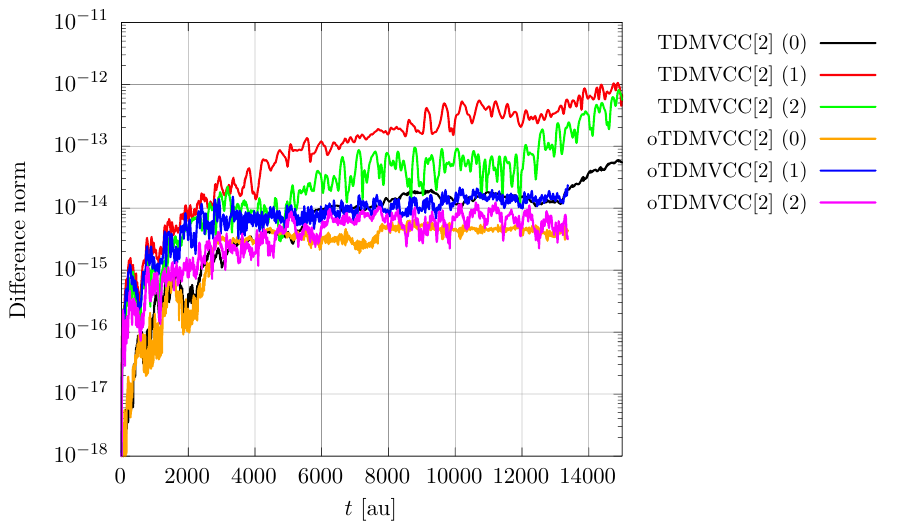}
        \label{fig:H2S_nonherm_tdmvcc2_density}
    \end{subfigure}


    \begin{subfigure}[c]{0.8\textwidth}
        \centering
        \caption{Mean field}
        \includegraphics[width=\textwidth]{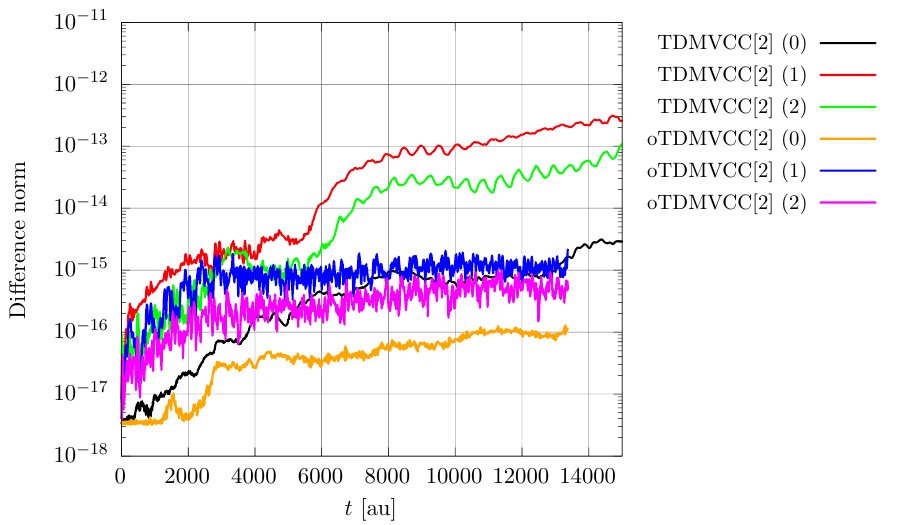}
        \label{fig:H2S_nonherm_tdmvcc2_meanfield}
    \end{subfigure}

    \caption{IVR of hydrogen sulfide at the oTDMVCC[2] and TDMVCC[2] levels
    with $N = 8$ and $N_{\2\subA} = 8$ for all modes. 
    Deviation from Hermiticity in 
    (a) one-mode density matrices ($|| \bm{\rho} -  \bm{\rho}^{\dagger} ||$) and 
    (b) half-transformed mean-field matrices ($|| \check{\mbf{F}} - \check{\mbf{F}}^{\prime \dagger} ||$).
    Mode indices are given in parentheses.}
    \label{fig:H2S_nonherm_tdmvcc2}
\end{figure}

\newpage 
\subsection{Basis set non-orthogonality}

As a measure of basis set non-orthogonality, we use
\begin{align}
    \eta_\mathrm{ket} &= || \mathbf{U}_{\!\subA}^{\dagger} \mathbf{U}_{\!\subA}^{\phantom{\dagger}}  - \mathbf{1} ||, \\
    \eta_\mathrm{bra} &= || \mathbf{W}_{\!\!\subA}^{\phantom{\dagger}}  \mathbf{W}_{\!\!\subA}^{\dagger} - \mathbf{1} ||,
\end{align}
for TDMVCC and 
\begin{align}
    \eta_\mathrm{ket} = \eta_\mathrm{bra} = || \mathbf{V}_{\!\!\subA}^{\dagger} \mathbf{V}_{\!\!\subA}^{\phantom{\dagger}} - \mathbf{1} ||
\end{align}
for oTDMVCC, where there is no need to distinguish between the bra and ket bases. These measures are calculated for each mode separately.

\newpage

%
%
\begin{figure}[H]
    \centering

    \begin{subfigure}[c]{0.85\textwidth}
        \centering
        \caption{Ket modals}
        \includegraphics[width=\textwidth]{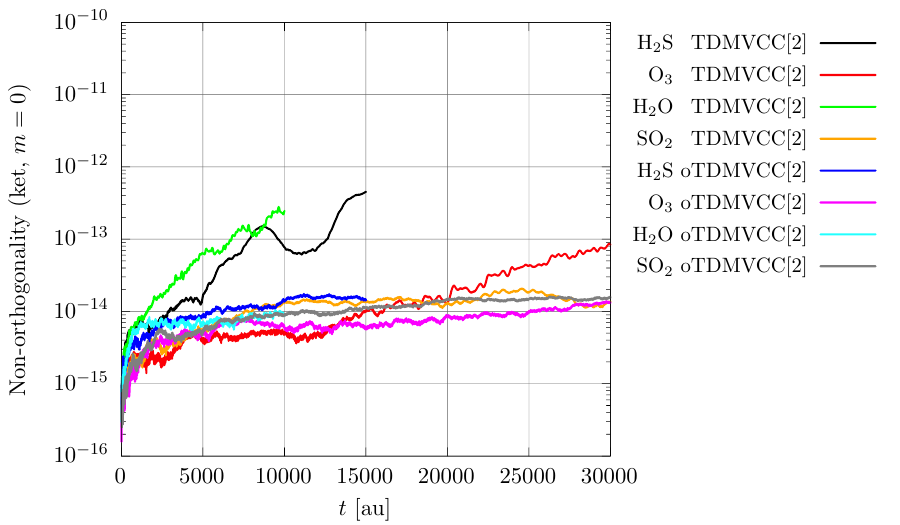}
        \label{fig:nonortho_tdmvcc2_0_ket}
    \end{subfigure}


    \begin{subfigure}[c]{0.85\textwidth}
        \centering
        \caption{Bra modals}
        \includegraphics[width=\textwidth]{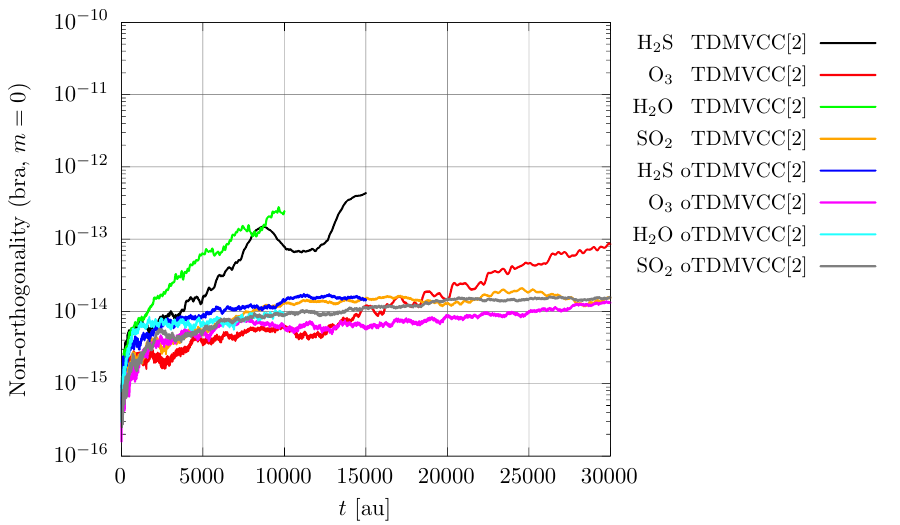}
        \label{fig:nonortho_tdmvcc2_0_bra}
    \end{subfigure}

    \caption{Non-orthogonality at the TDMVCC[2] and oTDMVCC[2] levels for $m = 0$ (bend).}
    \label{fig:nonortho_tdmvcc2_0_ket_bra}
\end{figure}

\newpage

%
%
\begin{figure}[H]
    \centering

    \begin{subfigure}[c]{0.85\textwidth}
        \centering
        \caption{Ket modals}
        \includegraphics[width=\textwidth]{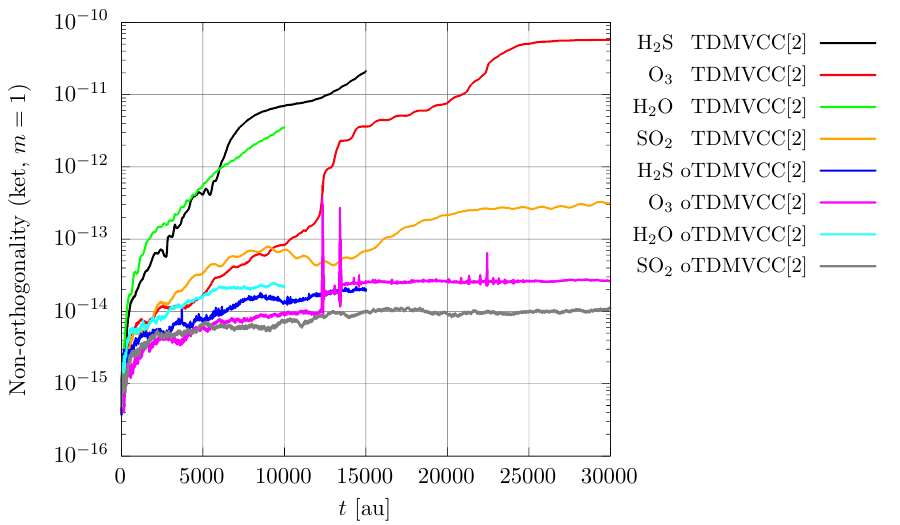}
        \label{fig:nonortho_tdmvcc2_1_ket}
    \end{subfigure}


    \begin{subfigure}[c]{0.85\textwidth}
        \centering
        \caption{Bra modals}
        \includegraphics[width=\textwidth]{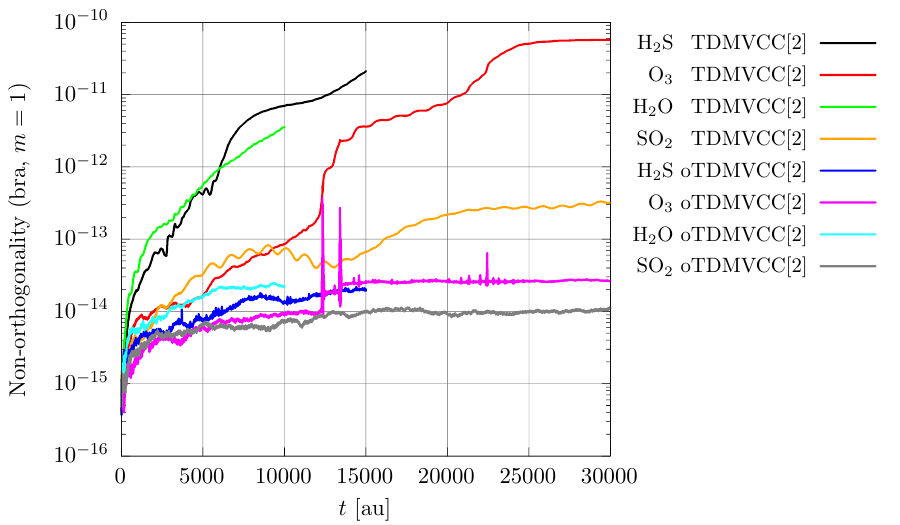}
        \label{fig:nonortho_tdmvcc2_1_bra}
    \end{subfigure}

    \caption{Non-orthogonality at the TDMVCC[2] and oTDMVCC[2] levels for $m = 1$ (symmetric stretch).}
    \label{fig:nonortho_tdmvcc2_1_ket_bra}
\end{figure}

\newpage

%
%
\begin{figure}[H]
    \centering

    \begin{subfigure}[c]{0.85\textwidth}
        \centering
        \caption{Ket modals}
        \includegraphics[width=\textwidth]{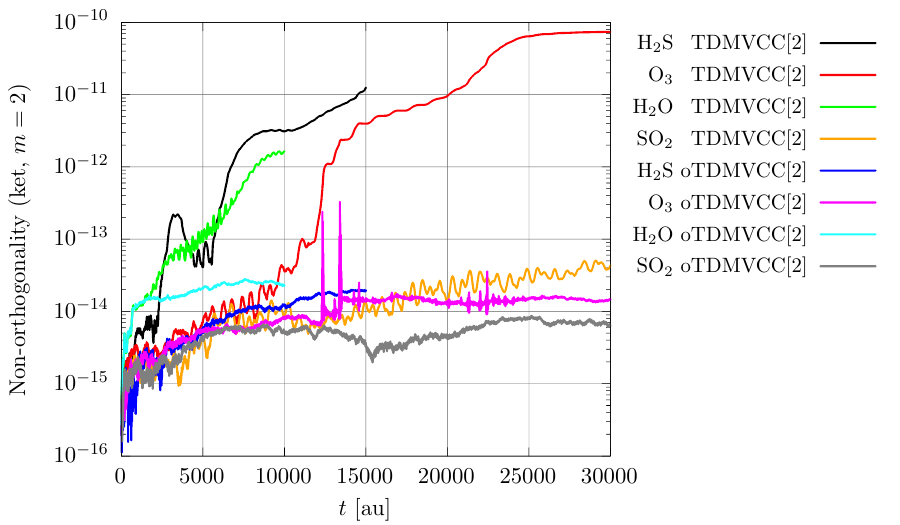}
        \label{fig:nonortho_tdmvcc2_2_ket}
    \end{subfigure}


    \begin{subfigure}[c]{0.85\textwidth}
        \centering
        \caption{Bra modals}
        \includegraphics[width=\textwidth]{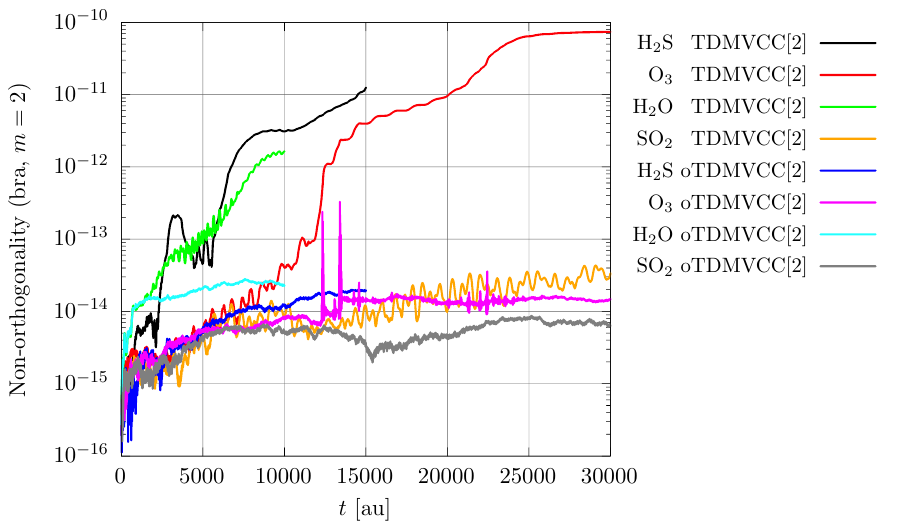}
        \label{fig:nonortho_tdmvcc2_2_bra}
    \end{subfigure}

    \caption{Non-orthogonality at the TDMVCC[2] and oTDMVCC[2] levels for $m = 2$ (asymmetric stretch).}
    \label{fig:nonortho_tdmvcc2_2_ket_bra}
\end{figure}

\newpage


%
%
\begin{figure}[H]
    \centering

    \begin{subfigure}[c]{0.85\textwidth}
        \centering
        \caption{Ket modals}
        \includegraphics[width=\textwidth]{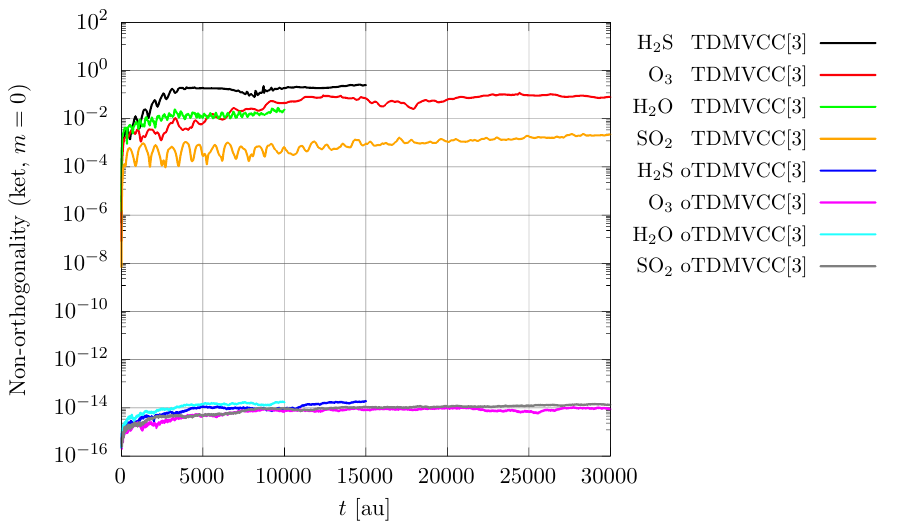}
        \label{fig:nonortho_tdmvcc3_0_ket}
    \end{subfigure}


    \begin{subfigure}[c]{0.85\textwidth}
        \centering
        \caption{Bra modals}
        \includegraphics[width=\textwidth]{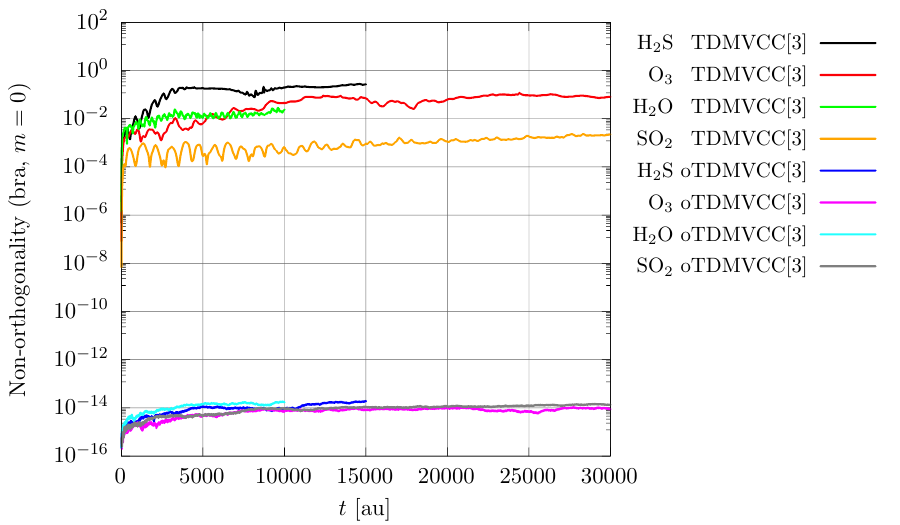}
        \label{fig:nonortho_tdmvcc3_0_bra}
    \end{subfigure}

    \caption{Non-orthogonality at the TDMVCC[3] and oTDMVCC[3] levels for $m = 0$ (bend).}
    \label{fig:nonortho_tdmvcc3_0_ket_bra}
\end{figure}

\newpage

%
%
\begin{figure}[H]
    \centering

    \begin{subfigure}[c]{0.85\textwidth}
        \centering
        \caption{Ket modals}
        \includegraphics[width=\textwidth]{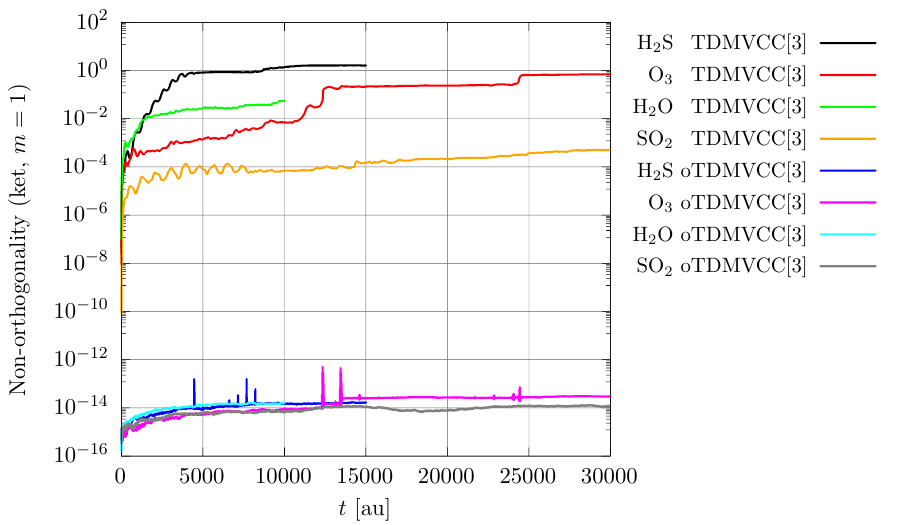}
        \label{fig:nonortho_tdmvcc3_1_ket}
    \end{subfigure}


    \begin{subfigure}[c]{0.85\textwidth}
        \centering
        \caption{Bra modals}
        \includegraphics[width=\textwidth]{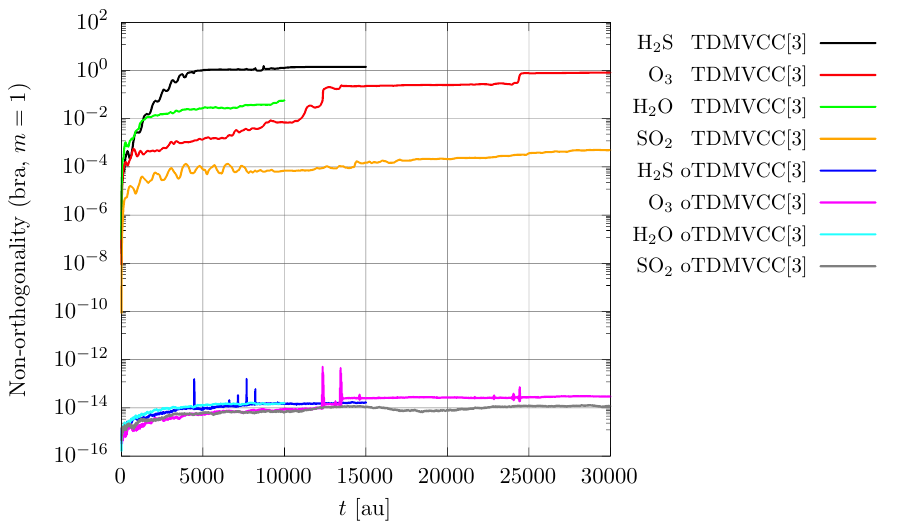}
        \label{fig:nonortho_tdmvcc3_1_bra}
    \end{subfigure}

    \caption{Non-orthogonality at the TDMVCC[3] and oTDMVCC[3] levels for $m = 1$ (symmetric stretch).}
    \label{fig:nonortho_tdmvcc3_1_ket_bra}
\end{figure}

\newpage

%
%
\begin{figure}[H]
    \centering

    \begin{subfigure}[c]{0.85\textwidth}
        \centering
        \caption{Ket modals}
        \includegraphics[width=\textwidth]{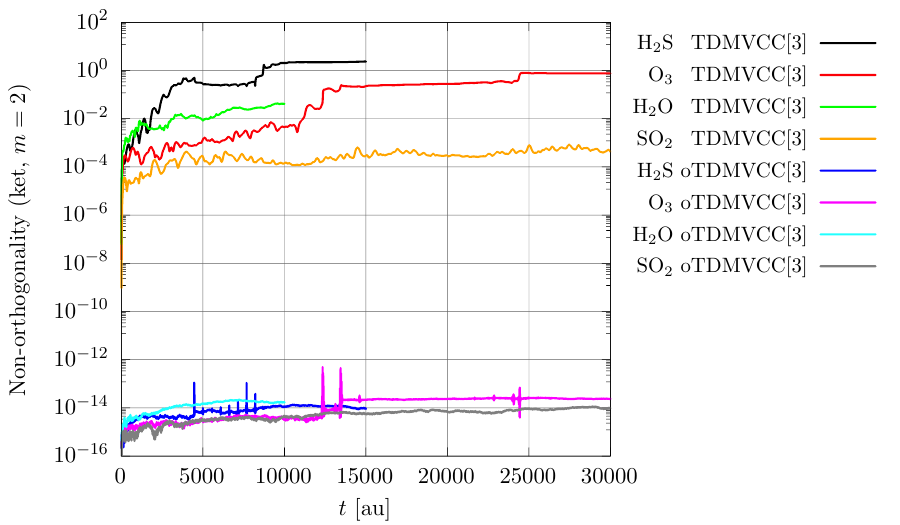}
        \label{fig:nonortho_tdmvcc3_2_ket}
    \end{subfigure}


    \begin{subfigure}[c]{0.85\textwidth}
        \centering
        \caption{Bra modals}
        \includegraphics[width=\textwidth]{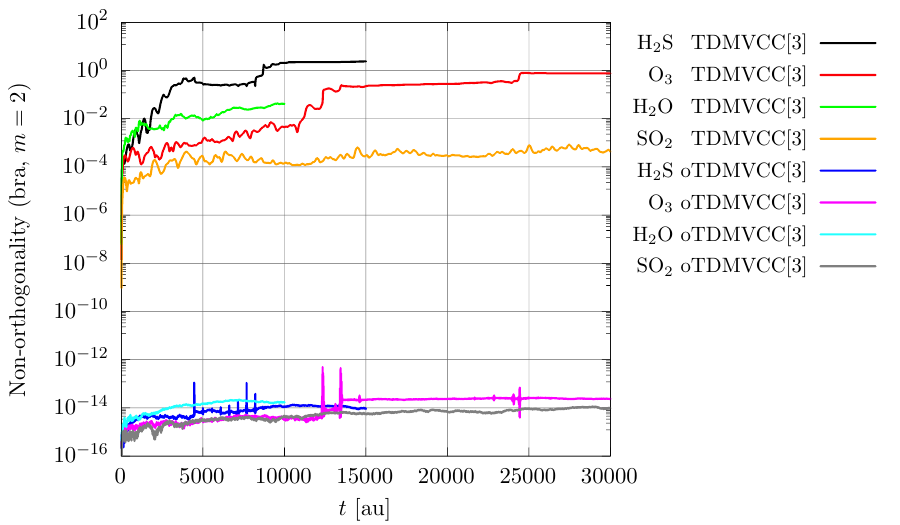}
        \label{fig:nonortho_tdmvcc3_2_bra}
    \end{subfigure}

    \caption{Non-orthogonality at the TDMVCC[3] and oTDMVCC[3] levels for $m = 2$ (asymmetric stretch).}
    \label{fig:nonortho_tdmvcc3_2_ket_bra}
\end{figure}

\newpage
\subsection{Are oTDMVCC[2] and TDMVCC[2] equivalent?}

%
%
\begin{figure}[H]
    \centering

    \begin{subfigure}[c]{1.0\textwidth}
        \centering
        \caption{Ket compared to ket}
        \includegraphics[width=\textwidth]{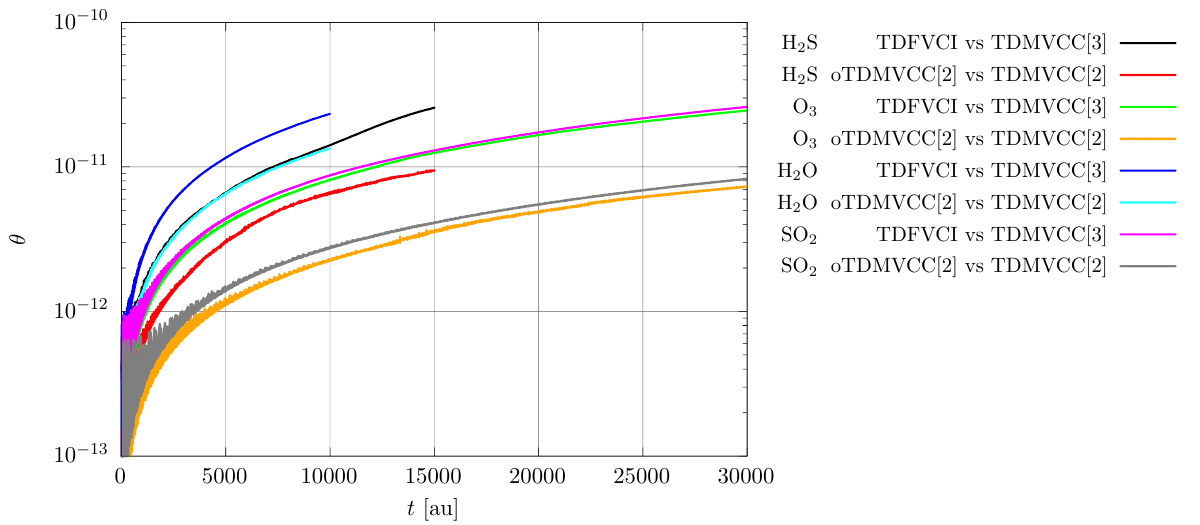}
        \label{fig:3mode_tdmvcc_vs_otdmvcc_ket_angle}
    \end{subfigure}


    \begin{subfigure}[c]{1.0\textwidth}
        \centering
        \caption{Bra compared to bra}
        \includegraphics[width=\textwidth]{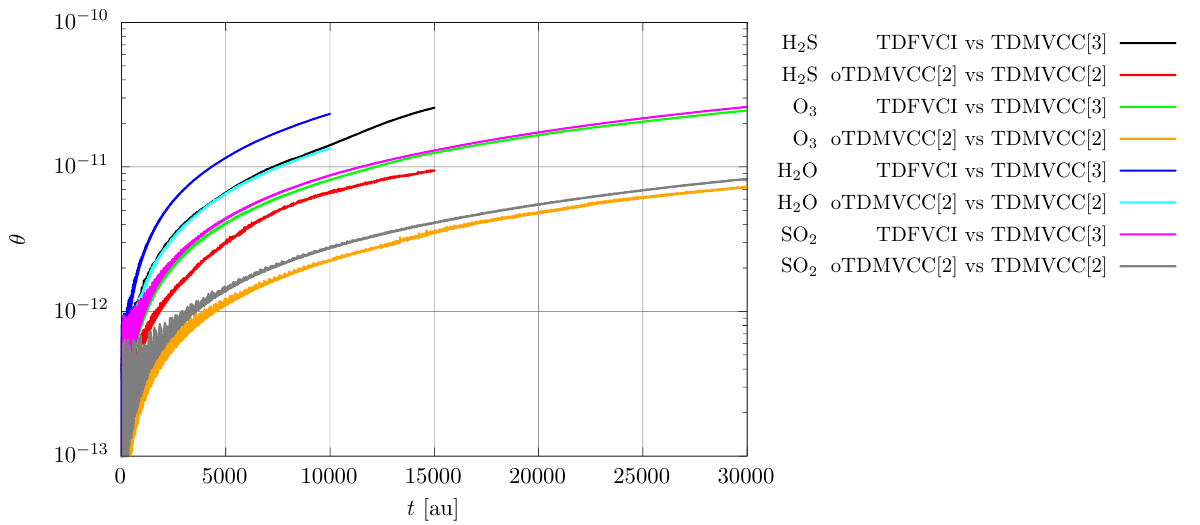}
        \label{fig:3mode_tdmvcc_vs_otdmvcc_bra_angle}
    \end{subfigure}

    \caption{Hilbert space angles comparing TDFVCI/TDMVCC[3] and oTDMVCC[2]/TDMVCC[2] for the triatomic molecules.
    If the angle between TDFVCI and TDMVCC[3] is taken as a benchmark for perfect agreement, then oTDMVCC[2] and TDMVCC[2] are identical
    for the calculations in question.}
    \label{fig:3mode_tdmvcc_vs_otdmvcc_ket_and_bra_angle}
\end{figure}

\newpage

\subsection{\textit{trans}-bithiophene}
\subsubsection{\Aclp{acf}}

\begin{figure}[H]
    \centering
    \includegraphics[width=\textwidth]{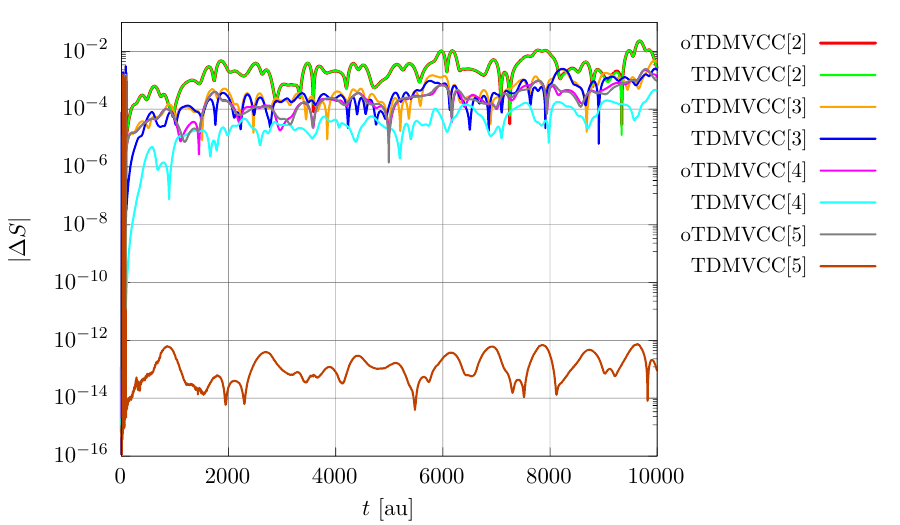}
    \caption{Absolute error in the \acl{acf} for a 5D \textit{trans}-bithiophene model at the oTDMVCC[2--5] and TDMVCC[2--5] levels
    with $N = 30$ and $N_{\2\subA} = 4$ for all modes. The errors are computed relative to MCTDH.}
    \label{fig:tbithio_autocorr_diff}
\end{figure}

\newpage

\subsubsection{Energy conservation}

\begin{figure}[H]
    \centering
    \includegraphics[width=\textwidth]{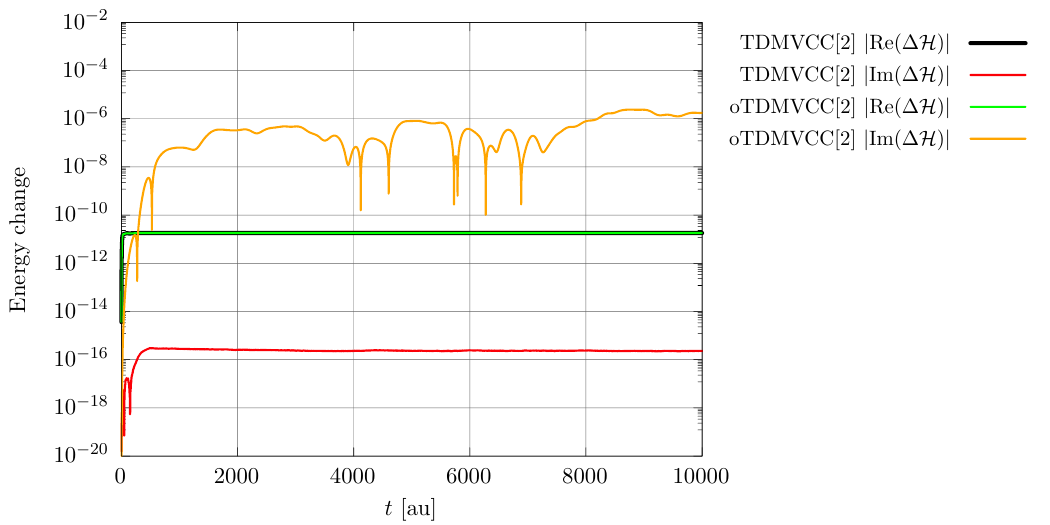}
    \caption{Absolute change in the energy function $\mathcal{H} = \elm{\Psi'}{H}{\Psi}$ for
    a 5D \textit{trans}-bithiophene model at the TDMVCC[2] and oTDMVCC[2] levels
    with $N = 30$ and $N_{\2\subA} = 4$ for all modes.}
    \label{fig:tbithio_energy_tdmvcc2}
\end{figure}

\begin{figure}[H]
    \centering
    \includegraphics[width=\textwidth]{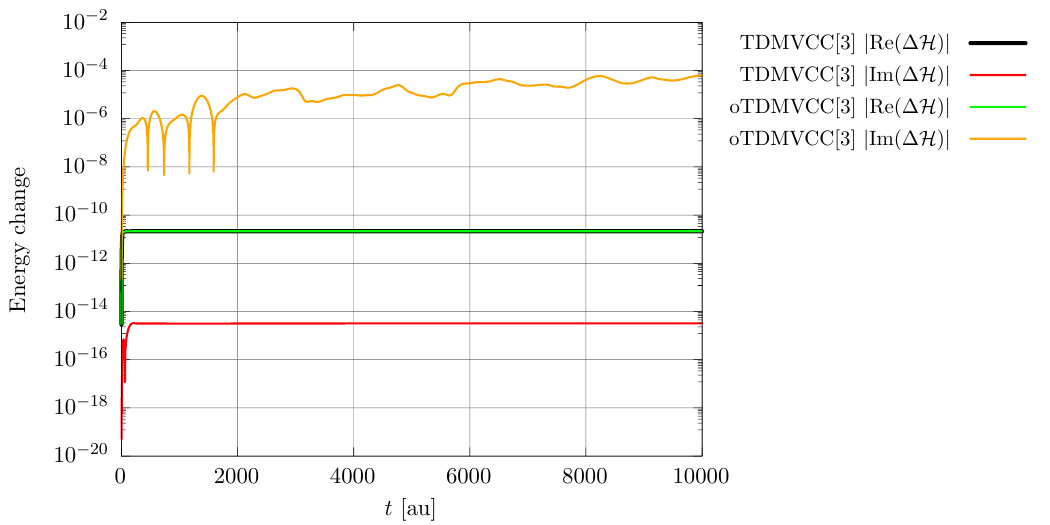}
    \caption{Absolute change in the energy function $\mathcal{H} = \elm{\Psi'}{H}{\Psi}$ for
    a 5D \textit{trans}-bithiophene model at the TDMVCC[3] and oTDMVCC[3] levels
    with $N = 30$ and $N_{\2\subA} = 4$ for all modes.}
    \label{fig:tbithio_energy_tdmvcc3}
\end{figure}
\newpage

\begin{figure}[H]
    \centering
    \includegraphics[width=\textwidth]{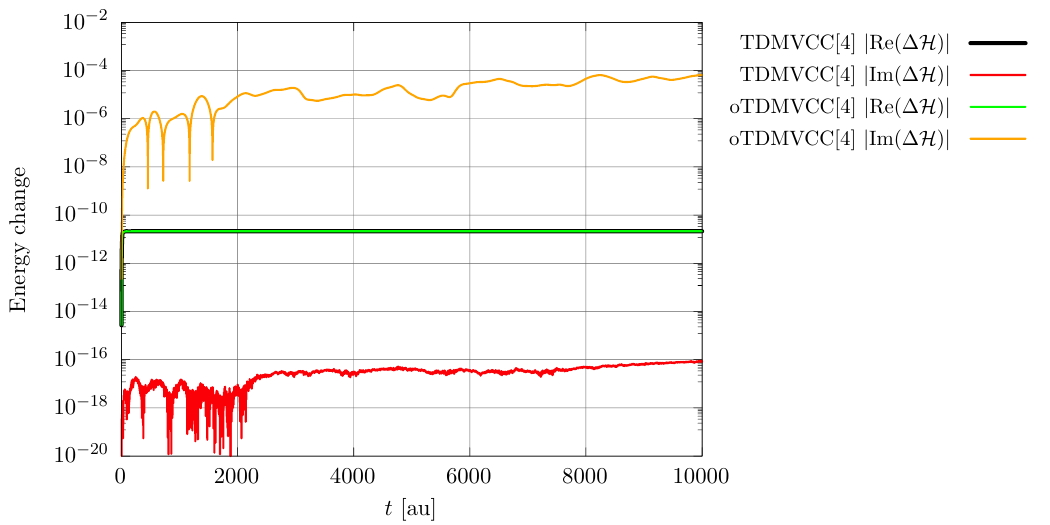}
    \caption{Absolute change in the energy function $\mathcal{H} = \elm{\Psi'}{H}{\Psi}$ for
    a 5D \textit{trans}-bithiophene model at the TDMVCC[4] and oTDMVCC[4] levels
    with $N = 30$ and $N_{\2\subA} = 4$ for all modes.}
    \label{fig:tbithio_energy_tdmvcc4}
\end{figure}

\begin{figure}[H]
    \centering
    \includegraphics[width=\textwidth]{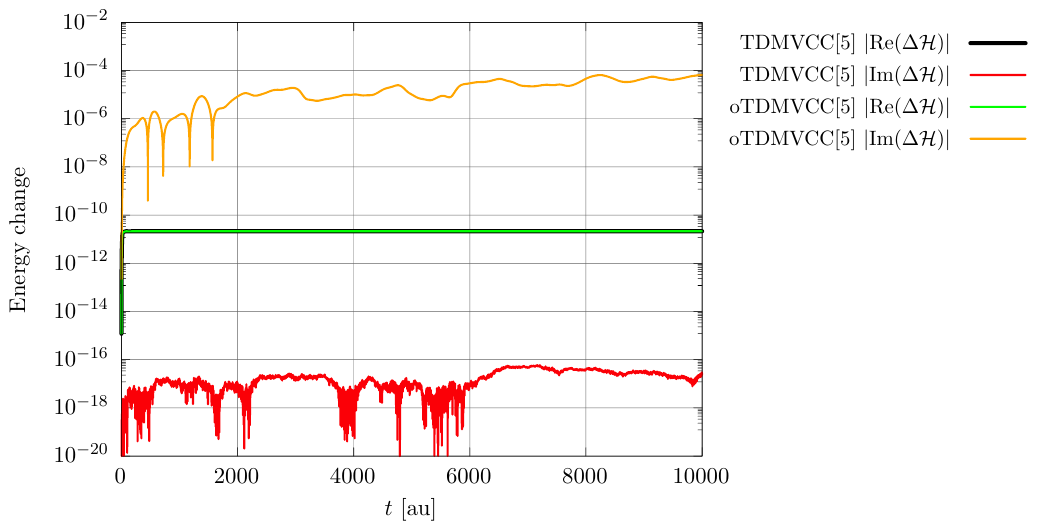}
    \caption{Absolute change in the energy function $\mathcal{H} = \elm{\Psi'}{H}{\Psi}$ for
    a 5D \textit{trans}-bithiophene model at the TDMVCC[5] and oTDMVCC[5] levels
    with $N = 30$ and $N_{\2\subA} = 4$ for all modes.}
    \label{fig:tbithio_energy_tdmvcc5}
\end{figure}
\newpage

\subsubsection{Non-hermiticity in density matrices and mean fields}

\begin{figure}[H]
    \centering

    \begin{subfigure}[c]{0.8\textwidth}
        \centering
        \caption{Density}
        \includegraphics[width=\textwidth]{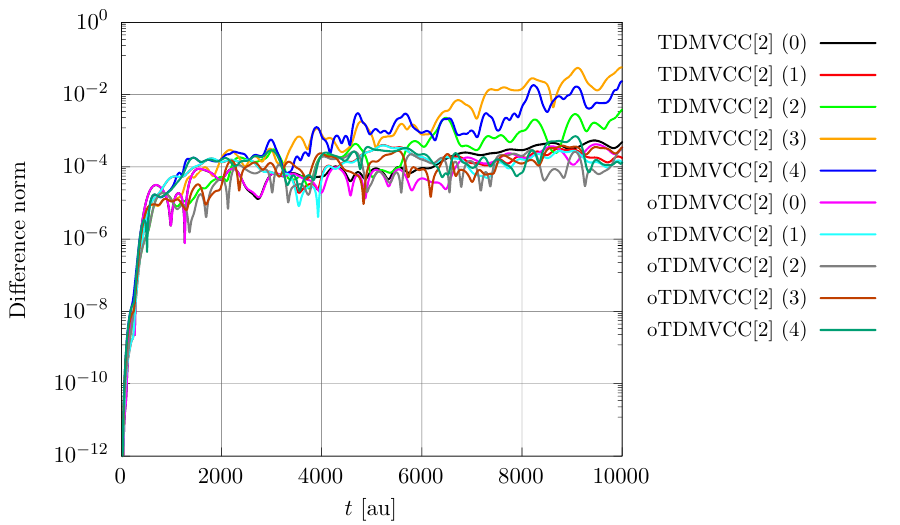}
        \label{fig:tbithio_nonherm_tdmvcc2_density}
    \end{subfigure}


    \begin{subfigure}[c]{0.8\textwidth}
        \centering
        \caption{Mean field}
        \includegraphics[width=\textwidth]{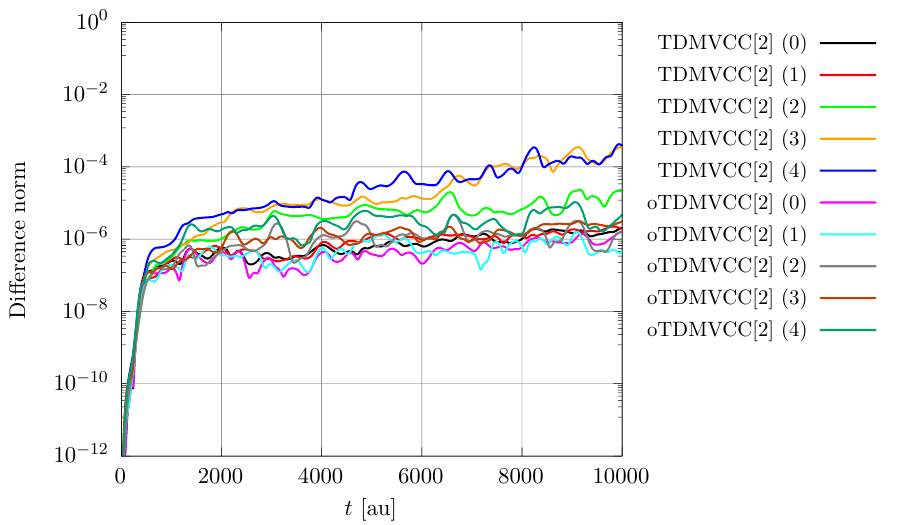}
        \label{fig:tbithio_nonherm_tdmvcc2_meanfield}
    \end{subfigure}

    \caption{5D \textit{trans}-bithiophene model at the TDMVCC[2] and oTDMVCC[2] levels
    with $N = 30$ and $N_{\2\subA} = 4$ for all modes
    Deviation from Hermiticity in 
    (a) one-mode density matrices ($|| \bm{\rho} -  \bm{\rho}^{\dagger} ||$) and 
    (b) half-transformed mean-field matrices ($|| \check{\mbf{F}} - \check{\mbf{F}}^{\prime \dagger} ||$).
    Mode indices are given in parentheses.}
    \label{fig:tbithio_nonherm_tdmvcc2}
\end{figure}

\newpage
\subsubsection{Expectation values: \textit{Q}\textsubscript{12}}

%
%
\begin{figure}[H]
    \centering

    \begin{subfigure}[c]{0.75\textwidth}
        \centering
        \caption{Expectation value}
        \includegraphics[width=\textwidth]{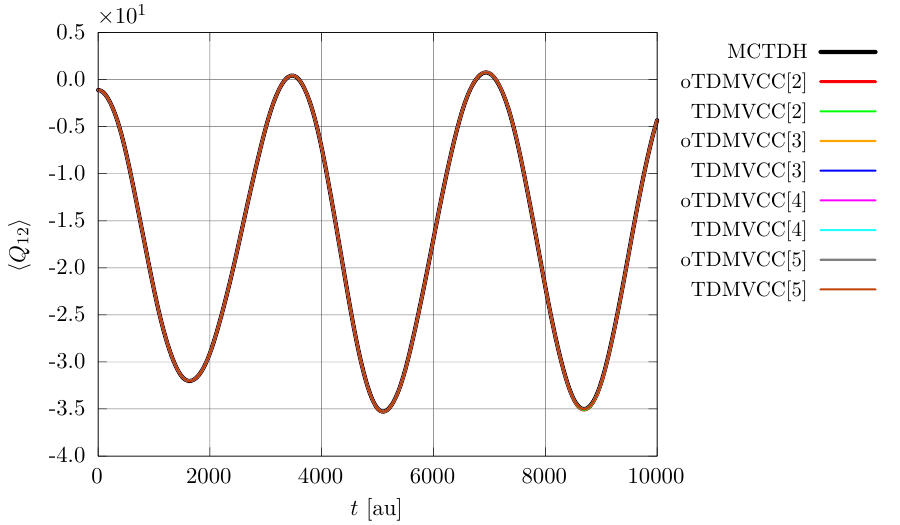}
        \label{fig:tbithio2345_Q12}
    \end{subfigure}


    \begin{subfigure}[c]{0.75\textwidth}
        \centering
        \caption{Absolute error in expectation value}
        \includegraphics[width=\textwidth]{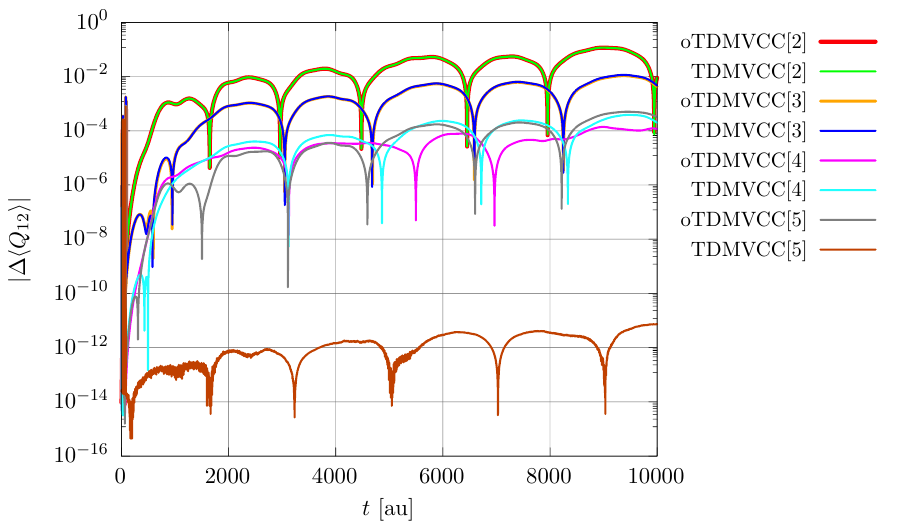}
        \label{fig:tbithio2345_Q12_diff}
    \end{subfigure}

    \caption{5D \textit{trans}-bithiophene model at the oTDMVCC[2--5] and TDMVCC[2--5] levels
    with $N = 30$ and $N_{\2\subA} = 4$ for all modes. (a) Expectation value of $Q_{12}$. 
    (b) Absolute error in the expectation value (relative to MCTDH).}
    \label{fig:tbithio2345_Q12_and_diff}
\end{figure}

\newpage

%
%
\begin{figure}[H]
    \centering
    \includegraphics[width=\textwidth]{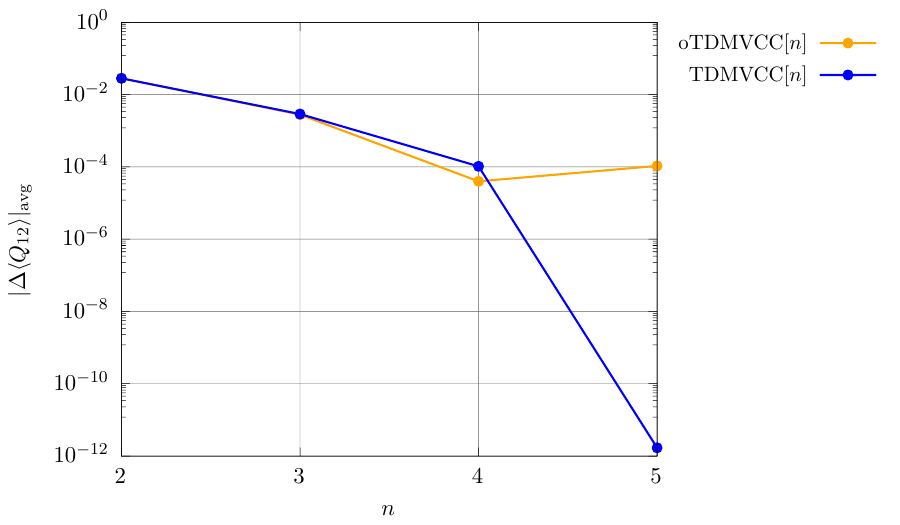}
    \caption{Average absolute error in the expectation value of $Q_{12}$ for a 5D \textit{trans}-bithiophene model at the oTDMVCC[2--5] and TDMVCC[2--5] levels
    with $N = 4$ and $N_{\2\subA} = 4$ for all modes. The errors are computed relative to MCTDH.}
    \label{fig:tbithio2345_Q12_diff_avg}
\end{figure}

\newpage
\subsubsection{Expectation values: \textit{Q}\textsubscript{19}}

%
%
\begin{figure}[H]
    \centering

    \begin{subfigure}[c]{0.75\textwidth}
        \centering
        \caption{Expectation value}
        \includegraphics[width=\textwidth]{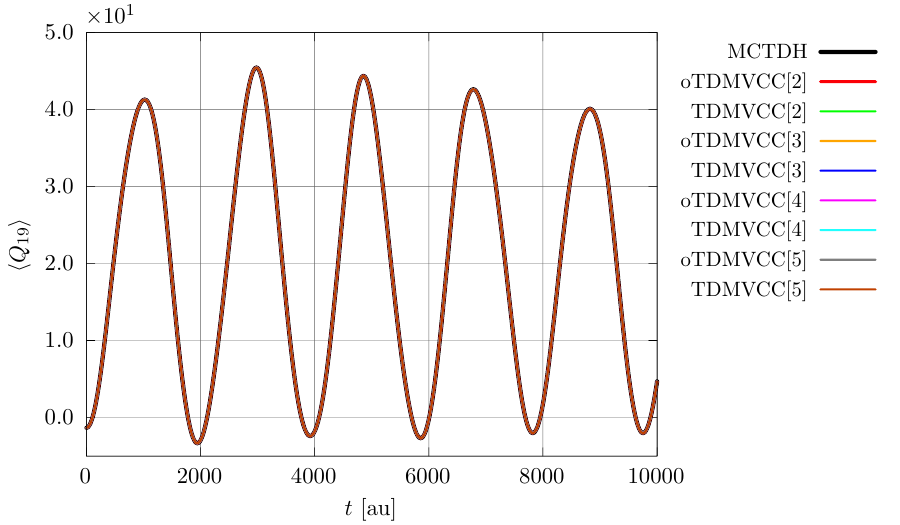}
        \label{fig:tbithio2345_Q19}
    \end{subfigure}


    \begin{subfigure}[c]{0.75\textwidth}
        \centering
        \caption{Absolute error in expectation value}
        \includegraphics[width=\textwidth]{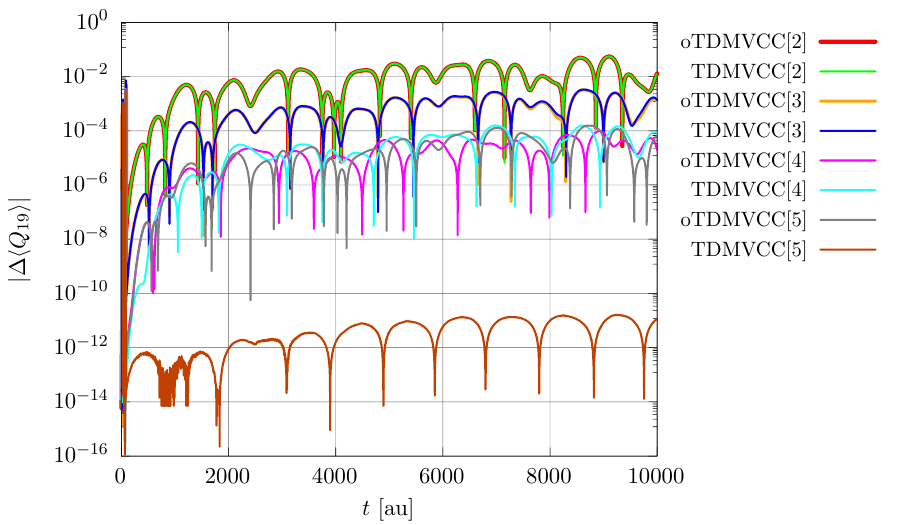}
        \label{fig:tbithio2345_Q19_diff}
    \end{subfigure}

    \caption{5D \textit{trans}-bithiophene model at the oTDMVCC[2--5] and TDMVCC[2--5] levels
    with $N = 30$ and $N_{\2\subA} = 4$ for all modes. (a) Expectation value of $Q_{19}$. 
    (b) Absolute error in the expectation value (relative to MCTDH).}
    \label{fig:tbithio2345_Q19_and_diff}
\end{figure}

\newpage

%
%
\begin{figure}[H]
    \centering
    \includegraphics[width=\textwidth]{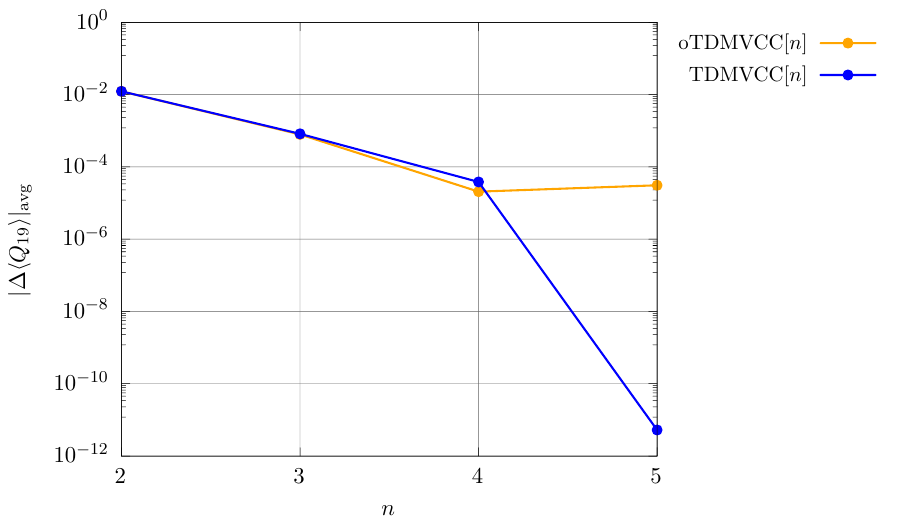}
    \caption{Average absolute error in the expectation value of $Q_{19}$ for a 5D \textit{trans}-bithiophene model at the oTDMVCC[2--5] and TDMVCC[2--5] levels
    with $N = 4$ and $N_{\2\subA} = 4$ for all modes. The errors are computed relative to MCTDH.}
    \label{fig:tbithio2345_Q19_diff_avg}
\end{figure}

\newpage
\subsubsection{Expectation values: \textit{Q}\textsubscript{34}}

%
%
\begin{figure}[H]
    \centering

    \begin{subfigure}[c]{0.75\textwidth}
        \centering
        \caption{Expectation value}
        \includegraphics[width=\textwidth]{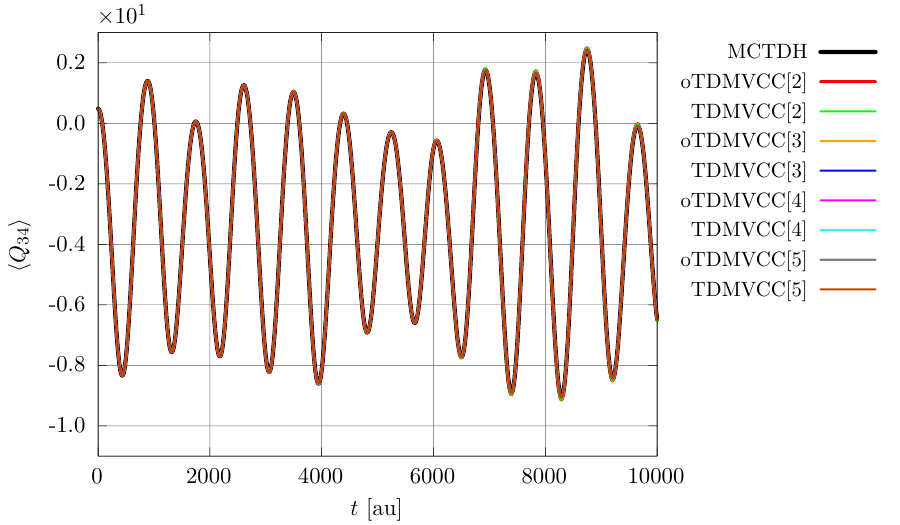}
        \label{fig:tbithio2345_Q34}
    \end{subfigure}


    \begin{subfigure}[c]{0.75\textwidth}
        \centering
        \caption{Absolute error in expectation value}
        \includegraphics[width=\textwidth]{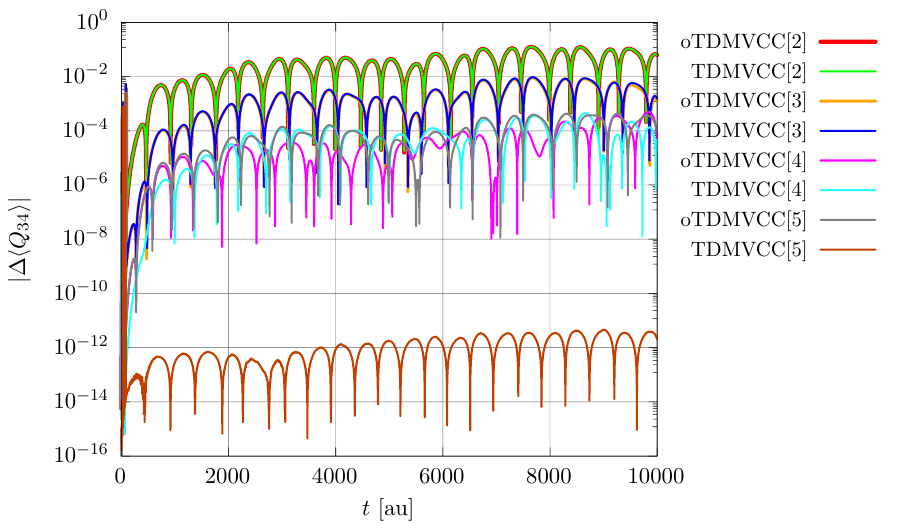}
        \label{fig:tbithio2345_Q34_diff}
    \end{subfigure}

    \caption{5D \textit{trans}-bithiophene model at the oTDMVCC[2--5] and TDMVCC[2--5] levels
    with $N = 30$ and $N_{\2\subA} = 4$ for all modes. (a) Expectation value of $Q_{34}$. 
    (b) Absolute error in the expectation value (relative to MCTDH).}
    \label{fig:tbithio2345_Q34_and_diff}
\end{figure}

\newpage

%
%
\begin{figure}[H]
    \centering
    \includegraphics[width=\textwidth]{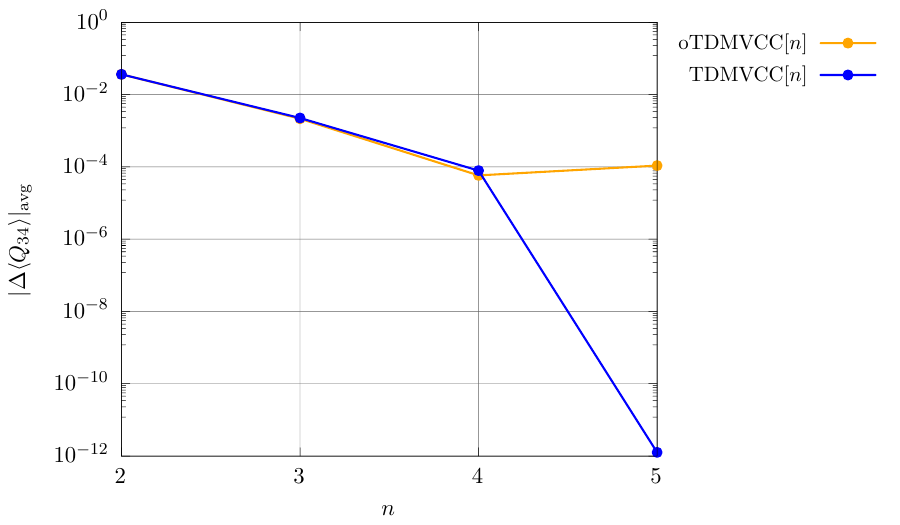}
    \caption{Average absolute error in the expectation value of $Q_{34}$ for a 5D \textit{trans}-bithiophene model at the oTDMVCC[2--5] and TDMVCC[2--5] levels
    with $N = 4$ and $N_{\2\subA} = 4$ for all modes. The errors are computed relative to MCTDH.}
    \label{fig:tbithio2345_Q34_diff_avg}
\end{figure}

\newpage
\subsubsection{Expectation values: \textit{Q}\textsubscript{41}}

%
%
\begin{figure}[H]
    \centering

    \begin{subfigure}[c]{0.75\textwidth}
        \centering
        \caption{Expectation value}
        \includegraphics[width=\textwidth]{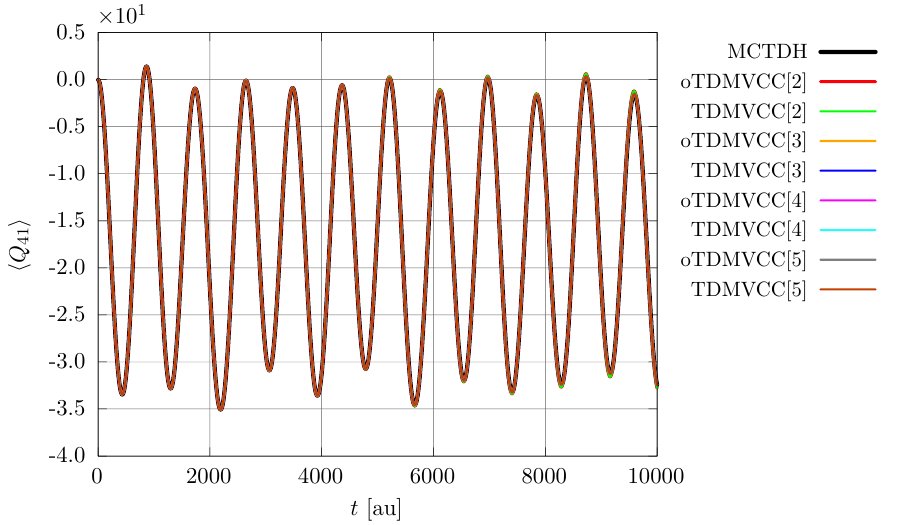}
        \label{fig:tbithio2345_Q41}
    \end{subfigure}


    \begin{subfigure}[c]{0.75\textwidth}
        \centering
        \caption{Absolute error in expectation value}
        \includegraphics[width=\textwidth]{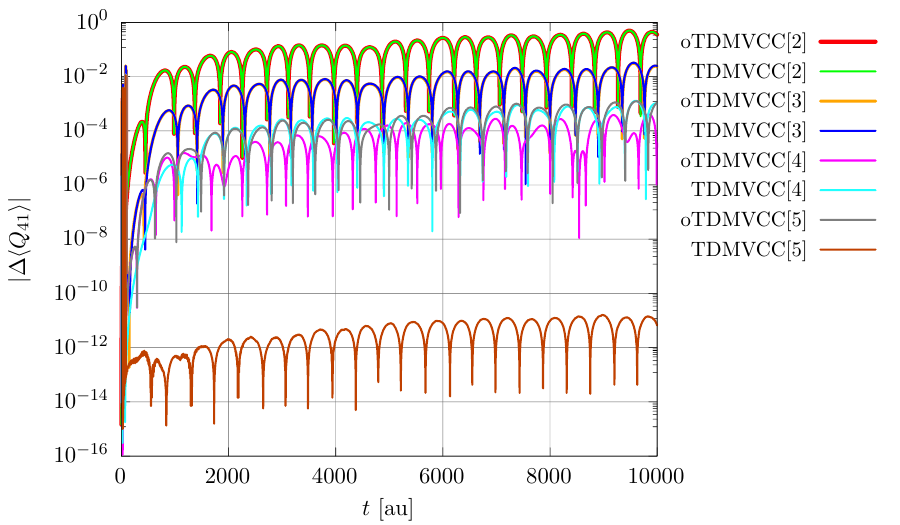}
        \label{fig:tbithio2345_Q41_diff}
    \end{subfigure}

    \caption{5D \textit{trans}-bithiophene model at the oTDMVCC[2--5] and TDMVCC[2--5] levels
    with $N = 30$ and $N_{\2\subA} = 4$ for all modes. (a) Expectation value of $Q_{41}$. 
    (b) Absolute error in the expectation value (relative to MCTDH).}
    \label{fig:tbithio2345_Q41_and_diff}
\end{figure}

\newpage

%
%
\begin{figure}[H]
    \centering
    \includegraphics[width=\textwidth]{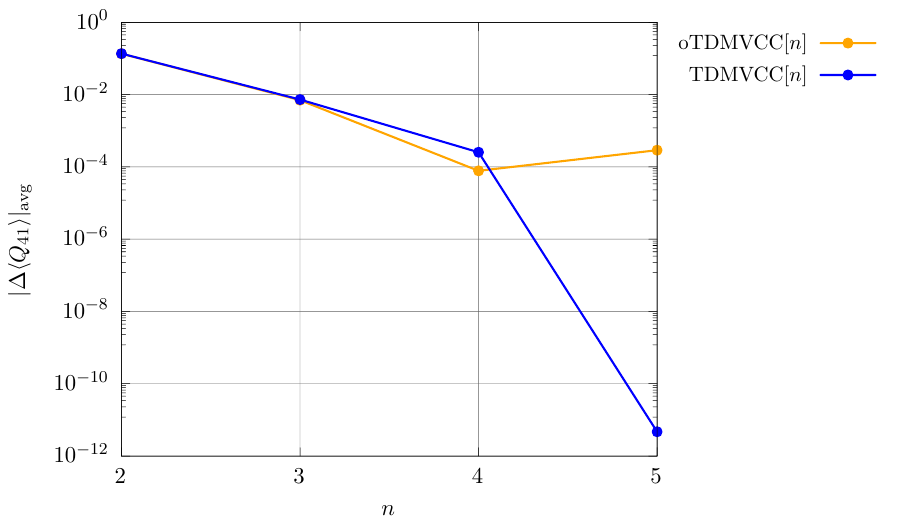}
    \caption{Average absolute error in the expectation value of $Q_{41}$ for a 5D \textit{trans}-bithiophene model at the oTDMVCC[2--5] and TDMVCC[2--5] levels
    with $N = 4$ and $N_{\2\subA} = 4$ for all modes. The errors are computed relative to MCTDH.}
    \label{fig:tbithio2345_Q41_diff_avg}
\end{figure}

\newpage
 
\subsubsection{Imaginary part of expectation values: \textit{Q}\textsubscript{10}}

%
%
\begin{figure}[H]
    \centering

    \begin{subfigure}[c]{0.75\textwidth}
        \centering
        \caption{}
        \includegraphics[width=\textwidth]{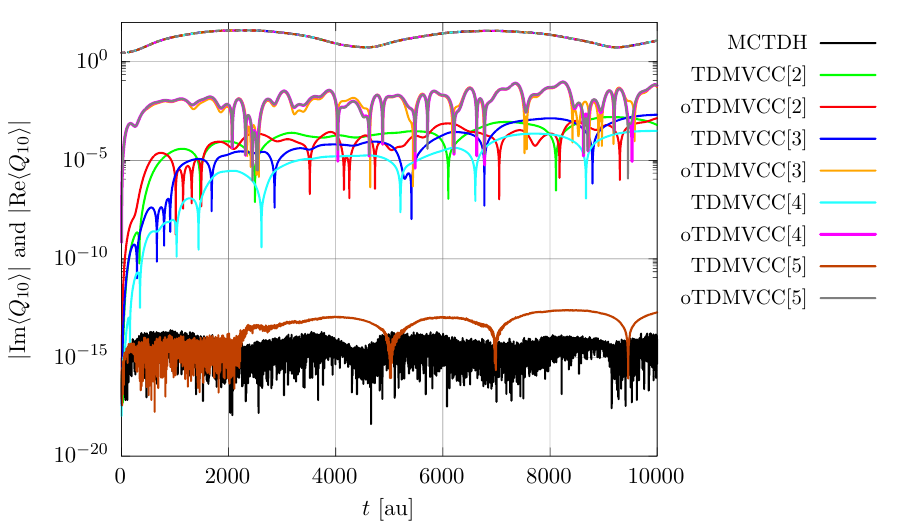}
        \label{fig:tbithio2345_Q10_re_im}
    \end{subfigure}


    \begin{subfigure}[c]{0.75\textwidth}
        \centering
        \caption{}
        \includegraphics[width=\textwidth]{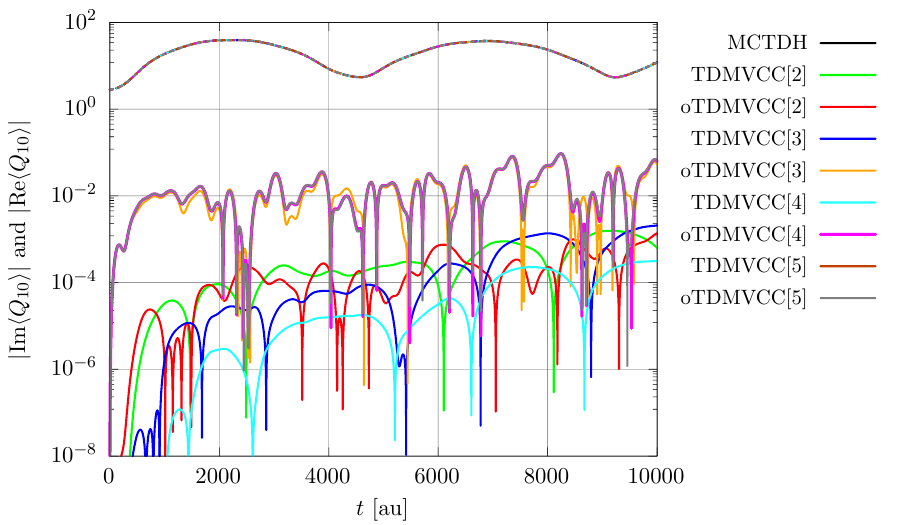}
        \label{fig:tbithio2345_Q10_re_im_zoom}
    \end{subfigure}

    \caption{5D \textit{trans}-bithiophene model at the oTDMVCC[2--5] and TDMVCC[2--5] levels
    with $N = 30$ and $N_{\2\subA} = 4$ for all modes.
    Absolute value of $\mathrm{Re}\elm{\Psi'}{Q}{\Psi}$ (dashed) and $\mathrm{Im}\elm{\Psi'}{Q}{\Psi}$ (full) for $Q_{10}$.
    (a) Full data set. 
    (b) Excerpt.
    Note that the dashed lines (real part) are almost on top of each other.}
    \label{fig:tbithio2345_Q10_re_im_full_and_zoom}
\end{figure}

\newpage

%
%
\begin{figure}[H]
    \centering
    \includegraphics[width=\textwidth]{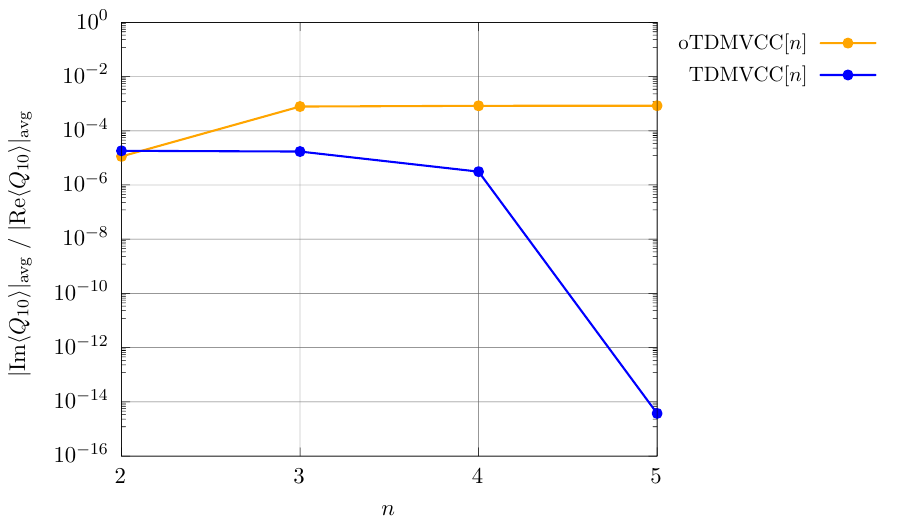}
    \caption{Ratio of mean absolute values of $\mathrm{Re}\elm{\Psi'}{Q}{\Psi}$ (dashed) and
    $\mathrm{Im}\elm{\Psi'}{Q}{\Psi}$ (full) for $Q_{10}$.}
    \label{fig:tbithio2345_Q10_re_im_ratio}
\end{figure}

\newpage
\subsubsection{Imaginary part of expectation values: \textit{Q}\textsubscript{12}}

%
%
\begin{figure}[H]
    \centering

    \begin{subfigure}[c]{0.75\textwidth}
        \centering
        \caption{}
        \includegraphics[width=\textwidth]{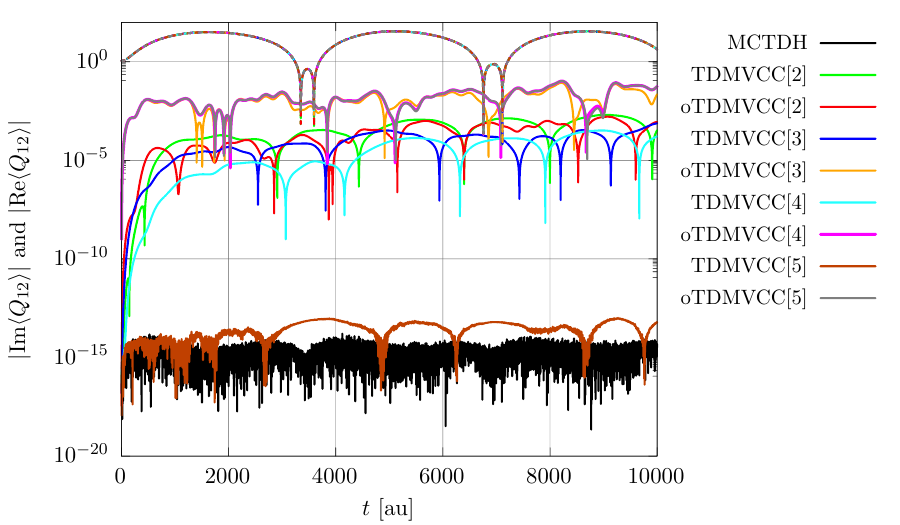}
        \label{fig:tbithio2345_Q12_re_im}
    \end{subfigure}


    \begin{subfigure}[c]{0.75\textwidth}
        \centering
        \caption{}
        \includegraphics[width=\textwidth]{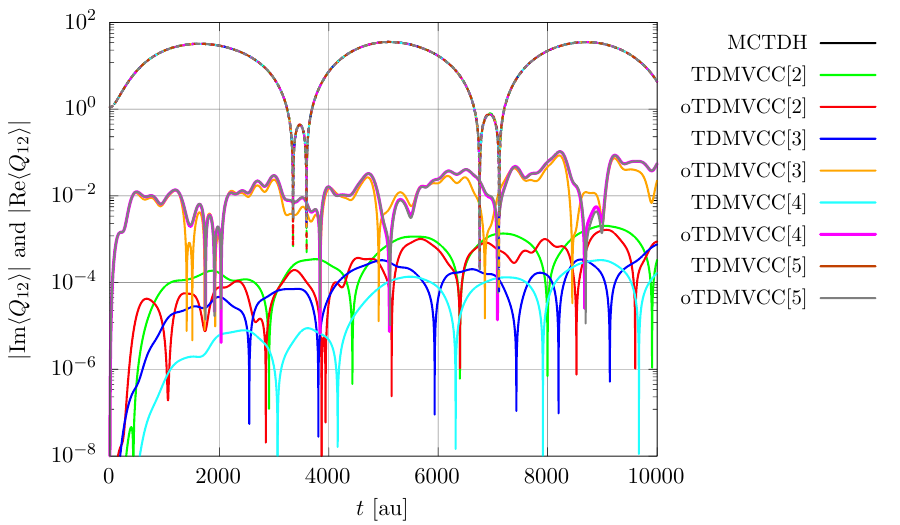}
        \label{fig:tbithio2345_Q12_re_im_zoom}
    \end{subfigure}

    \caption{5D \textit{trans}-bithiophene model at the oTDMVCC[2--5] and TDMVCC[2--5] levels
    with $N = 30$ and $N_{\2\subA} = 4$ for all modes.
    Absolute value of $\mathrm{Re}\elm{\Psi'}{Q}{\Psi}$ (dashed) and $\mathrm{Im}\elm{\Psi'}{Q}{\Psi}$ (full) for $Q_{12}$.
    (a) Full data set. 
    (b) Excerpt.
    Note that the dashed lines (real part) are almost on top of each other.}
    \label{fig:tbithio2345_Q12_re_im_full_and_zoom}
\end{figure}

\newpage

%
%
\begin{figure}[H]
    \centering
    \includegraphics[width=\textwidth]{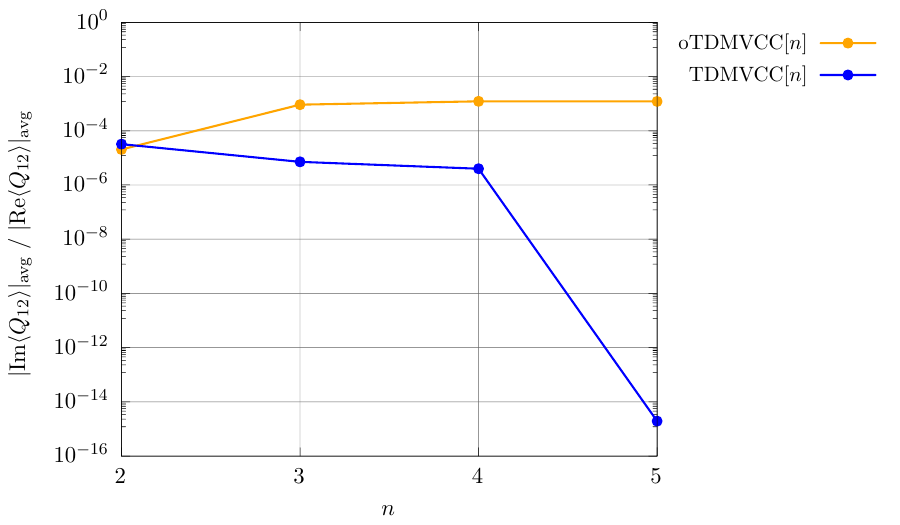}
    \caption{Ratio of mean absolute values of $\mathrm{Re}\elm{\Psi'}{Q}{\Psi}$ (dashed) and
    $\mathrm{Im}\elm{\Psi'}{Q}{\Psi}$ (full) for $Q_{12}$.}
    \label{fig:tbithio2345_Q12_re_im_ratio}
\end{figure}

\newpage
\subsubsection{Imaginary part of expectation values: \textit{Q}\textsubscript{19}}

%
%
\begin{figure}[H]
    \centering

    \begin{subfigure}[c]{0.75\textwidth}
        \centering
        \caption{}
        \includegraphics[width=\textwidth]{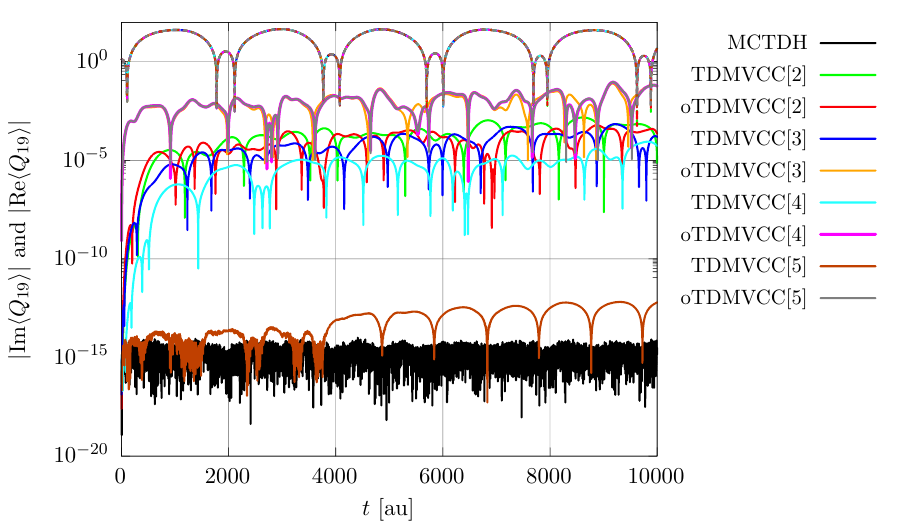}
        \label{fig:tbithio2345_Q19_re_im}
    \end{subfigure}


    \begin{subfigure}[c]{0.75\textwidth}
        \centering
        \caption{}
        \includegraphics[width=\textwidth]{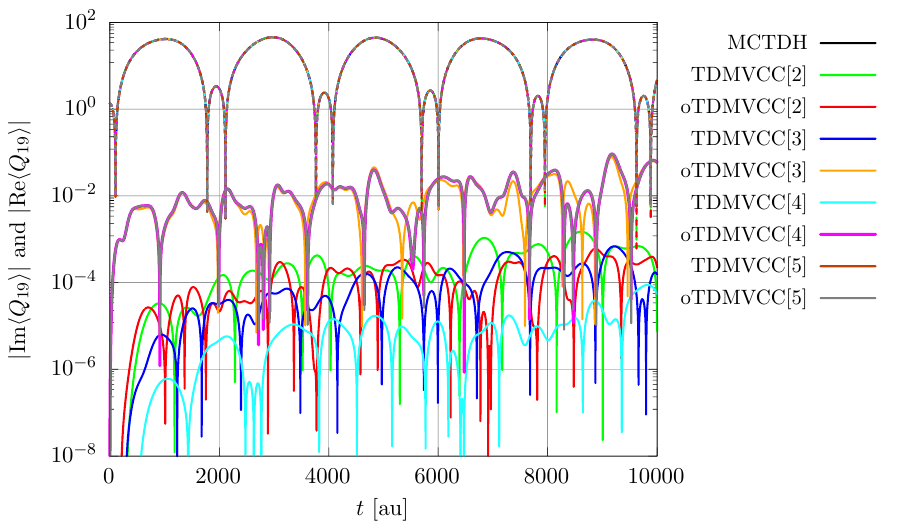}
        \label{fig:tbithio2345_Q19_re_im_zoom}
    \end{subfigure}

    \caption{5D \textit{trans}-bithiophene model at the oTDMVCC[2--5] and TDMVCC[2--5] levels
    with $N = 30$ and $N_{\2\subA} = 4$ for all modes.
    Absolute value of $\mathrm{Re}\elm{\Psi'}{Q}{\Psi}$ (dashed) and $\mathrm{Im}\elm{\Psi'}{Q}{\Psi}$ (full) for $Q_{19}$.
    (a) Full data set. 
    (b) Excerpt.
    Note that the dashed lines (real part) are almost on top of each other.}
    \label{fig:tbithio2345_Q19_re_im_full_and_zoom}
\end{figure}

\newpage

%
%
\begin{figure}[H]
    \centering
    \includegraphics[width=\textwidth]{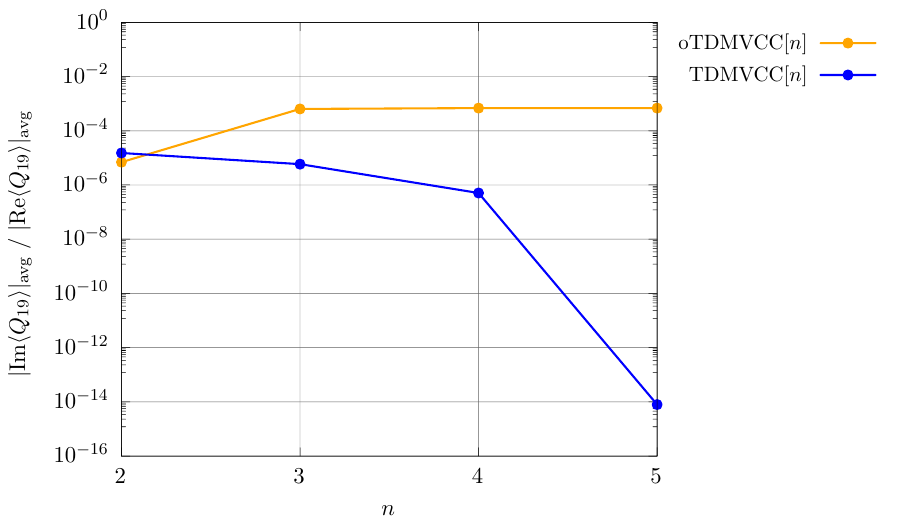}
    \caption{Ratio of mean absolute values of $\mathrm{Re}\elm{\Psi'}{Q}{\Psi}$ (dashed) and
    $\mathrm{Im}\elm{\Psi'}{Q}{\Psi}$ (full) for $Q_{19}$.}
    \label{fig:tbithio2345_Q19_re_im_ratio}
\end{figure}

\newpage 
\subsubsection{Imaginary part of expectation values: \textit{Q}\textsubscript{34}}

%
%
\begin{figure}[H]
    \centering

    \begin{subfigure}[c]{0.75\textwidth}
        \centering
        \caption{}
        \includegraphics[width=\textwidth]{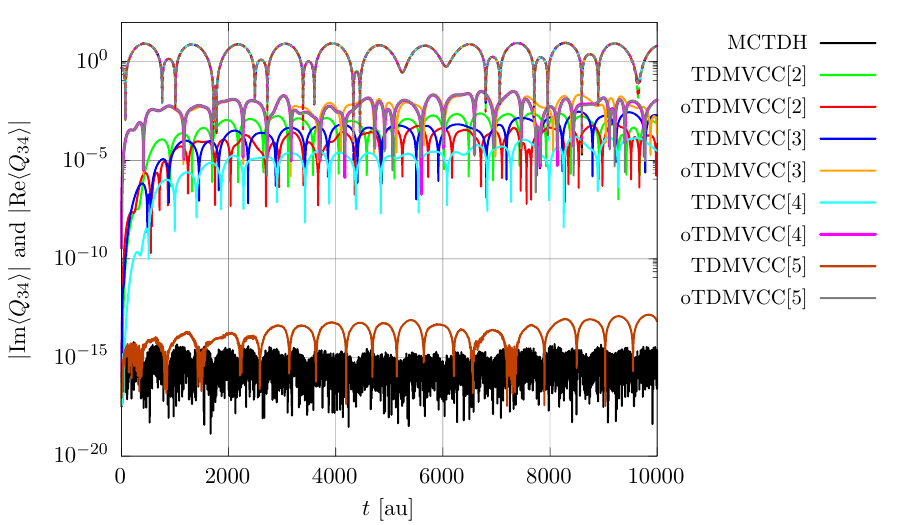}
        \label{fig:tbithio2345_Q34_re_im}
    \end{subfigure}


    \begin{subfigure}[c]{0.75\textwidth}
        \centering
        \caption{}
        \includegraphics[width=\textwidth]{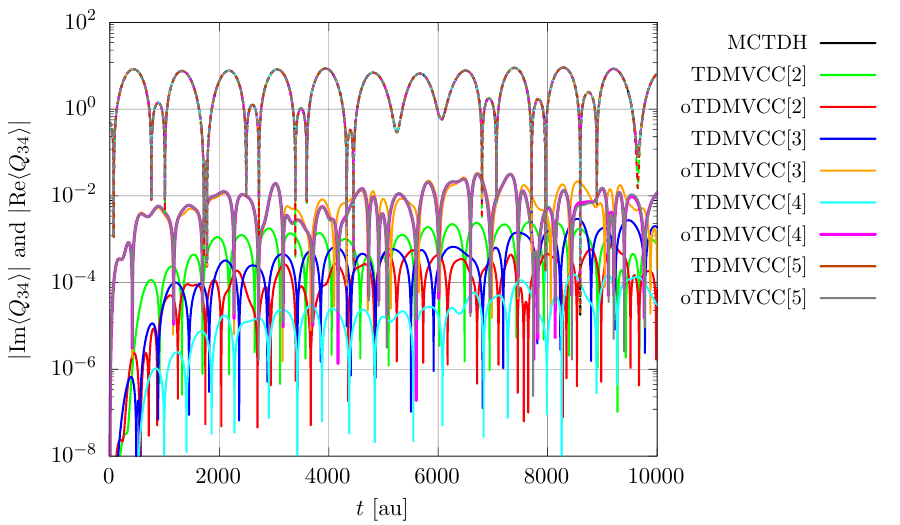}
        \label{fig:tbithio2345_Q34_re_im_zoom}
    \end{subfigure}

    \caption{5D \textit{trans}-bithiophene model at the oTDMVCC[2--5] and TDMVCC[2--5] levels
    with $N = 30$ and $N_{\2\subA} = 4$ for all modes.
    Absolute value of $\mathrm{Re}\elm{\Psi'}{Q}{\Psi}$ (dashed) and $\mathrm{Im}\elm{\Psi'}{Q}{\Psi}$ (full) for $Q_{34}$.
    (a) Full data set. 
    (b) Excerpt.
    Note that the dashed lines (real part) are almost on top of each other.}
    \label{fig:tbithio2345_Q34_re_im_full_and_zoom}
\end{figure}

\newpage

%
%
\begin{figure}[H]
    \centering
    \includegraphics[width=\textwidth]{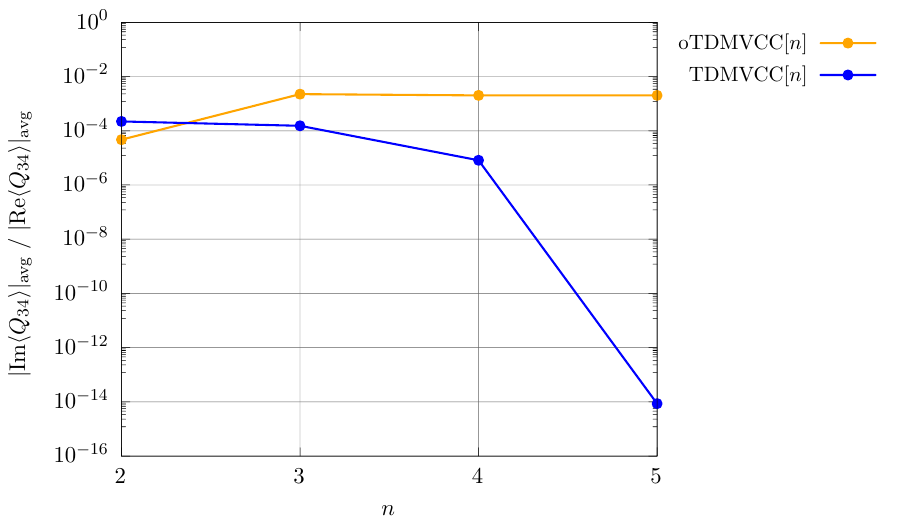}
    \caption{Ratio of mean absolute values of $\mathrm{Re}\elm{\Psi'}{Q}{\Psi}$ (dashed) and
    $\mathrm{Im}\elm{\Psi'}{Q}{\Psi}$ (full) for $Q_{34}$.}
    \label{fig:tbithio2345_Q34_re_im_ratio}
\end{figure}

\newpage 
\subsubsection{Imaginary part of expectation values: \textit{Q}\textsubscript{41}}

%
%
\begin{figure}[H]
    \centering

    \begin{subfigure}[c]{0.75\textwidth}
        \centering
        \caption{}
        \includegraphics[width=\textwidth]{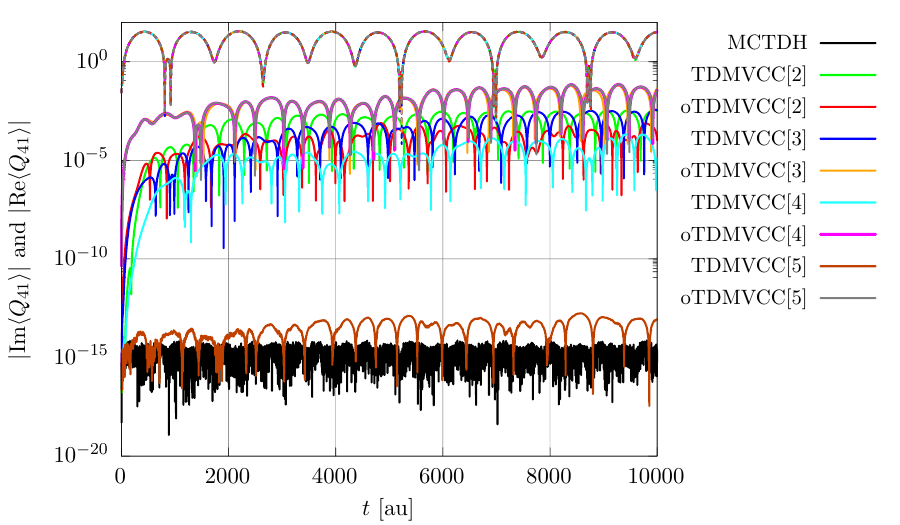}
        \label{fig:tbithio2345_Q41_re_im}
    \end{subfigure}


    \begin{subfigure}[c]{0.75\textwidth}
        \centering
        \caption{}
        \includegraphics[width=\textwidth]{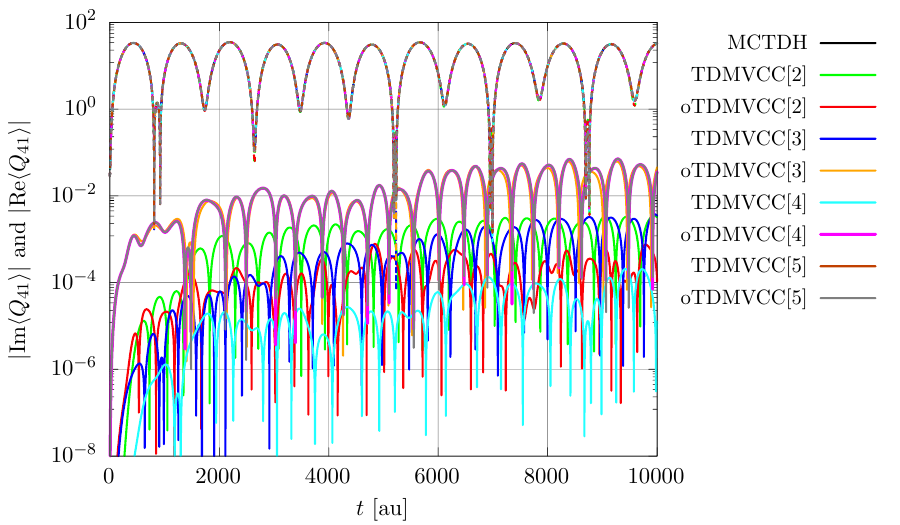}
        \label{fig:tbithio2345_Q41_re_im_zoom}
    \end{subfigure}

    \caption{5D \textit{trans}-bithiophene model at the oTDMVCC[2--5] and TDMVCC[2--5] levels
    with $N = 30$ and $N_{\2\subA} = 4$ for all modes.
    Absolute value of $\mathrm{Re}\elm{\Psi'}{Q}{\Psi}$ (dashed) and $\mathrm{Im}\elm{\Psi'}{Q}{\Psi}$ (full) for $Q_{41}$.
    (a) Full data set. 
    (b) Excerpt.
    Note that the dashed lines (real part) are almost on top of each other.}
    \label{fig:tbithio2345_Q41_re_im_full_and_zoom}
\end{figure}

\newpage

%
%
\begin{figure}[H]
    \centering
    \includegraphics[width=\textwidth]{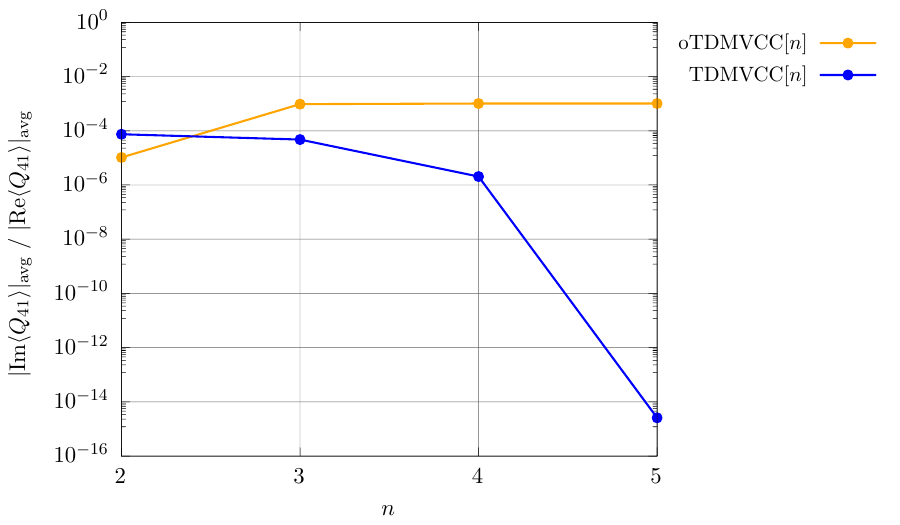}
    \caption{Ratio of mean absolute values of $\mathrm{Re}\elm{\Psi'}{Q}{\Psi}$ (dashed) and
    $\mathrm{Im}\elm{\Psi'}{Q}{\Psi}$ (full) for $Q_{41}$.}
    \label{fig:tbithio2345_Q41_re_im_ratio}
\end{figure}

\newpage